\documentclass[a4paper,11pt]{article}

\usepackage{jheppub}
\usepackage{amsmath,amssymb,amsthm}
\usepackage{mathrsfs}
\usepackage{hyperref}
\usepackage{faktor}
\usepackage{float}

\usepackage{wrapfig}
\usepackage{tikz-cd} 
\usepackage{tkz-graph}
\usepackage{arydshln}
\usepackage[skins,theorems]{tcolorbox}
\usepackage{fdsymbol}
\usepackage{lscape}
\usepackage{float}
\graphicspath{{Figures/}}

\usepackage{comment}
\usepackage{caption}
\usepackage{subcaption}
\usepackage{multicol}
\usepackage{multirow}
\usepackage{booktabs}
\usepackage{cancel}

\definecolor{amber}{rgb}{1.0, 0.49, 0.0}
\definecolor{Green}{rgb}{0.0, 0.5, 0.0}
\definecolor{purple}{rgb}{0.7,0,0.7}

\newcommand{\myboxy}[1]{\tikz[overlay]\node[fill=yellow!50,draw=black,inner sep=2pt, anchor=text, rectangle, rounded corners=1mm] {#1};\phantom{#1}}
\newcommand{\mybox}[1]{\textcolor{amber}{#1}}

\newtheorem{thm}{Theorem}

\newtheorem{dfn}[thm]{Definition}
\newtheorem{coj}[thm]{Conjecture}

\title{Permutohedra for knots and quivers}
\date{January 2021}
\author[1,3]{Jakub Jankowski} 
\author[2,3]{, Piotr Kucharski}
\author[3]{, H\'{e}lder Larragu\'{i}vel} 
\author[3]{, Dmitry Noshchenko}
\author[2,3]{, and Piotr Su{\l}kowski}

\affiliation[1]{Institute of Theoretical Physics, University of Wroc{\l}aw, PL-50204 Wroc{\l}aw,
Poland}
\affiliation[2]{Walter Burke Institute for Theoretical Physics, California Institute of Technology, \\ Pasadena, CA 91125, USA}
\affiliation[3]{Faculty of Physics, University of Warsaw, ul. Pasteura 5, 02-093 Warsaw, Poland}

\emailAdd{jakub.jankowski@uwr.edu.pl, piotrek@caltech.edu, helder.larraguivel@fuw.edu.pl, dmitry.noshchenko@fuw.edu.pl, psulkows@fuw.edu.pl}

\abstract{
The knots-quivers correspondence states that various characteristics of a~knot are encoded in the~corresponding quiver and the~moduli space of its representations. However, this correspondence is not a~bijection: more than one quiver may be assigned to a~given knot and encode the~same information. In this work we study this phenomenon systematically and show that it is generic rather than exceptional. First, we find conditions that characterize equivalent quivers. Then we show that equivalent quivers arise in families that have the~structure of permutohedra, and the~set of all equivalent quivers for a~given knot is parameterized by vertices of a~graph made of several permutohedra glued together. These graphs can be also interpreted as webs of dual 3d $\mathcal{N}=2$ theories. All these results are intimately related to properties of homological diagrams for knots, as well as to multi-cover skein relations that arise in counting of holomorphic curves with boundaries on Lagrangian branes in Calabi-Yau three-folds. 
\\
\\
\\
\\
\\
\\
\\
\\
\\
\\
\\
\\
\\
\\
\\
\\
CALT-2021-019
}

\begin{document}
\maketitle

\section{Introduction}

Knots and quivers play an~important role in high energy theoretical physics. Knots often arise in the~context of topological invariance and can be related to physical objects -- such as Wilson loops, defects, and Lagrangian branes -- in gauge theories and topological string theory. Quivers may encode interactions of BPS states assigned to their nodes, or the~structure of gauge theories. These two seemingly different entities have been recently related by the~so-called knots-quivers correspondence~\cite{KRSS1707short,KRSS1707long}, which identifies various characteristics of knots with those of quivers and moduli spaces of their representations. The~knots-quivers correspondence follows from properties of appropriately engineered brane systems in the~resolved conifold that represent knots, thus it is intimately related to topological string theory and Gromov-Witten theory~\cite{EKL1811,EKL1910}, and has been further generalized to branes in other Calabi-Yau manifolds~\cite{PS1811,Kimura:2020qns}, see also \cite{Bousseau:2020fus}. Other aspects and proofs (for two-bridge and arborescent knots and links) of the~knots-quivers correspondence are discussed in~\cite{PSS1802,SW1711,SW2004,Kuch2005}.

If there is a~correspondence between two types of objects, such as knots and quivers, an~important immediate question is how unique both sides of this correspondence are. Examples of two different quivers of the~same size that correspond to the~same knot were already identified in~\cite{KRSS1707long}, which means that the~knots-quivers correspondence is not a~bijection. In this paper we study this phenomenon systematically and find conditions that characterize equivalent quivers (i.e. different quivers that correspond to the~same knot). It turns out that these conditions lead to interesting local and global structure of the~set of equivalent quivers. We stress that equivalent quivers that we consider in this paper are of the~same size $m$, such that their nodes are in one-to-one correspondence with generators of HOMFLY-PT homology of a~given knot. One can always use certain $q$-identities to construct quivers of larger size that encode the~same generating functions of knot polynomials, however this phenomenon has already been studied (see~\cite{KRSS1707long,EKL1910}) and it is not of our primary interest.

Let us thus consider a~matrix $C$ of size $m$ (equal to the~number of HOMFLY-PT homology generators of a~given knot), such that entries $C_{ij}$ are numbers of arrows between nodes $i$ and~$j$ of a~symmetric quiver corresponding to this knot. We characterize the~local equivalence of quivers by showing that some of the~quivers equivalent to $C$ are encoded in matrices $C'$, such that $C$ and $C'$ differ only by a~transposition of two elements $C_{ab}$ and $C_{cd}$, whose values differ by one and which satisfy a~few additional conditions. From each such equivalent matrix $C'$ one can determine another set of equivalent matrices $C''$, etc. This procedure produces a~closed and connected network of equivalent quivers in a~finite number of steps. It follows that any two equivalent quivers from this network differ simply by a~sequence of transpositions of elements of their matrices.

\begin{figure}[h!]
    \centering
 \includegraphics[scale=0.4]{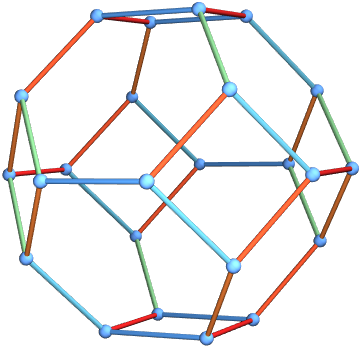}
    \caption{Permutohedron $\Pi_4$. Its vertices are labeled by permutations of elements $\{1,2,3,4\}$, and different colors of edges correspond to different types of transpositions $(i\ j)$ (for $1\leq i < j\leq 4$). Vertices connected by an~edge differ by one transposition of neighboring elements.}
    \label{fig:permutohedrapi4}
\end{figure}

\begin{figure}[t!]
    \centering
\includegraphics[scale=0.75]{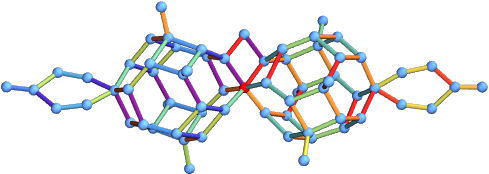}
    \caption{Permutohedra graph for $9_1$ torus knot. It consists of two series of permutohedra $\Pi_2$, $\Pi_3$ and $\Pi_4$ connected in the~middle, and several other permutohedra $\Pi_2$.}
    \label{fig:3DGraph91}
\end{figure}

\begin{figure}[t]
    \centering
\includegraphics[scale=0.75]{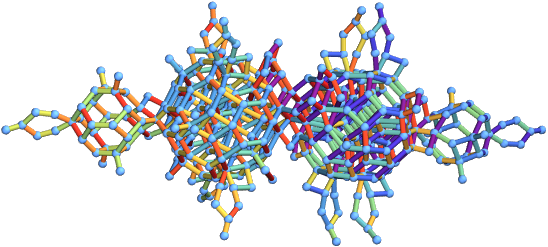}
    \caption{Permutohedra graph for $11_1$ torus knot. It consists of two series of permutohedra $\Pi_2$, $\Pi_3$, $\Pi_4$ and $\Pi_5$ connected in the~middle, and several other permutohedra $\Pi_2$ and $\Pi_3$.}
    \label{fig:3DGraph111}
\end{figure}

Furthermore, we find that the~network of such equivalent quivers has an~interesting global structure. We show that equivalent quivers arise in families that form permutohedra. Recall that a~permutohedron $\Pi_n$ is the~$(n-1)$-dimensional polytope, whose vertices are labeled by permutations $\sigma\in S_n$ and edges correspond to transpositions of adjacent elements. Permutohedron $\Pi_2$ consists of two vertices connected by an~edge, $\Pi_3$ is a~hexagon, and $\Pi_4$ is a~truncated octahedron shown in figure~\ref{fig:permutohedrapi4}. In our context, each vertex of a~permutohedron represents a~quiver matrix and each edge connects equivalent quivers (which are related by a~transposition of two appropriate elements). Every permutohedron arises from a~particular pattern of transpositions of elements of quiver matrices, or equivalently from some particular way of writing a~generating function of colored superpolynomials for a~given knot. For a~given knot, there are typically several ways of writing a~generating function of colored superpolynomials, which lead to different permutohedra connected by quivers they share. Examples of such graphs for torus knots $9_1$ and $11_1$ are shown in figures~\ref{fig:3DGraph91} and~\ref{fig:3DGraph111}, and we call them permutohedra graphs. 

We find that the~above mentioned conditions that characterize equivalent quivers have interesting interpretation in both knot theory and topological string theory. In the~knot theory these conditions are related to the~structure of the~(uncolored and $S^2$-colored) HOMFLY-PT homology of a~knot in question, and they have a~nice graphical manifestation at the~level of homological diagrams: they are the~center of mass conditions for homology generators. On the~other hand, these conditions can be also expressed in terms of multi-cover skein relations that arise in counting of holomorphic curves with boundaries on a~Lagrangian brane in Calabi-Yau three-folds. These connections provide a~new link between homological invariants of knots, Gromov-Witten theory, and moduli spaces of quiver representations. Moreover, equivalent quivers corresponding to a~given knot represent dual 3-dimensional theories with $\mathcal{N}=2$ supersymmetry, analogously as discussed in~\cite{AFGS1203,FGSS1209,Chung:2014qpa,EKL1811}. One can therefore interpret permutohedra graphs as webs of dual 3d $\mathcal{N}=2$ theories.

As mentioned above, the~appearance of permutohedra can be interpreted at the~level of generating functions of colored superpolynomials. More precisely, we show that each of them can be decomposed into a~piece that encodes a~given permutohedron, coupled to another piece that itself has a~structure of a~motivic generating function for a~smaller quiver that we refer to as a~prequiver. All equivalent quivers corresponding to a~given permutohedron are obtained from the~same prequiver in the~procedure of splitting that involves specifying some particular permutation -- so this is the~reason why permutohedra arise.

From the~above introductory remarks, or simply from figures~\ref{fig:3DGraph91} and~\ref{fig:3DGraph111}, it follows that the~appearance of equivalent quivers is not an~exception, but rather a~common and abundant phenomenon. This also means that one should regard as a~knot invariant the~whole set of equivalent quivers, rather than one particular quiver from this class; moduli spaces of all such equivalent quivers encode the~same information about the~corresponding knot. The~number of equivalent quivers that satisfy the~above mentioned conditions grows fast with the~size of the~homological diagram: it appears that the~unknot and trefoil are the~only knots such that corresponding quivers are unique, while some knots with 6 or 7 crossings already have over $100\, 000$ such equivalent quivers (see the~last column of table~\ref{tab:symmetries_table}). For a~given knot, the~number of equivalent quivers that we consider is of the~order of the~size of the~largest permutohedron in the~permutohedra graph. For example, we find that the~largest permutohedra for $(2,2p+1)$ torus knots are two $\Pi_p$, which means that the~number of equivalent quivers for this family grows factorially as $2 p!$. 

Apart from the~number of equivalent quivers, in table~\ref{tab:symmetries_table} we also present the~number of pairings and symmetries for various knots that we analyze in the~paper. By pairings we mean quadruples of generators in the~homological diagram that satisfy the~center of mass condition mentioned above; this is a~necessary, but not sufficient, condition of local equivalence (i.e. equivalence of quiver matrices that differ by one transposition of their elements). On the~other hand, by symmetries we mean quadruples of homology generators that satisfy sufficient conditions of local equivalence -- the~presence of symmetry means that an~appropriate transposition of matrix elements indeed produces an~equivalent quiver. In particular, we conjecture (and verify to high $p$) that the~numbers of pairings and symmetries for $(2,2p+1)$ torus knots are respectively $p^2 (p - 1)/2$ and $p (p^2 - 1)/3$. 

Finally, we also extend our analysis to quivers for knot complements~\cite{Kuch2005}, which encode $\hat{Z}$ invariants for knot complements (also referred to as $F_K$ invariants)~\cite{GM1904,Park1909,GGKPS20xx}. We show that for $(2,2p+1)$ torus knots  equivalence conditions that we find in this paper yield an~interesting relation between quivers discussed above (that arise in the~original knots-quivers correspondence) and quivers for knot complements.

\begin{table}[t!]
    \centering
    \begin{tabular}{||c|c|c|c|c||}
    \hline
       \multicolumn{2}{||c|}{Knot} & Pairings & Symmetries & Equivalent quivers  \\
    \hline \hline
        Unknot & $0_1$ & 0 & 0 & 1 \\
        \hline
        \multirow{9}{10em}{\centering{Torus knots $T_{2, 2 p + 1}$}} & $3_1$ & 0 & 0 & 1 \\
         & $5_1$ & 2 & 2 & 3 \\
         & $7_1$ & 9 & 8 & 13 \\
         & $9_1$ & 24 & 20 & 68 \\
         & $11_1$ & 50 & 40 & 405 \\
         & $13_1$ & 90 & 70 & 2\,684 \\
         & $15_1$ & 147 & 112 & 19\,557 \\
         & \vdots & \vdots & \vdots & \vdots \\
         & $(2p+1)_1$ & $ p^2 (p - 1)/2$ & $ p (p^2 - 1)/3$ & $\sim  2 p!$ \\
        \hline
        \multirow{3}{10em}{\centering{Twists knots $TK_{2 |p| + 2}$}} & $4_1$ & 1 & 1 & 2 \\
         & $6_1$ & 24 & 16 & 141 \\
         & $8_1$ & 105 & 61 & 36\,555 \\
        \hline
        \multirow{2}{10em}{\centering{Twists knots $TK_{2 p + 1}$}} & $5_2$ & 8 & 6 & 12 \\
         & $7_2$ & 52 & 34 & 1\,983 \\
        \hline
        \multirow{3}{10em}{\centering{Stand-alone examples}} & $6_2$ & 46 & 36 & 3\,534 \\
         & $6_3$ & 101 & 72 & 142\,368 \\
         & $7_3$ & 86 & 67 & 109\,636 \\
        
    \hline
    \end{tabular}
    \caption{The~number of pairings, symmetries, and equivalent quivers that we have found for $(2, 2 p + 1)$  torus knots, twist knots, and $6_2, 6_3, 7_3$ knots.}
    \label{tab:symmetries_table}
\end{table}

Note that in principle there might exist other equivalent quivers, which are not related by a~series of transpositions that we mentioned above (e.g. they might be related by a~cyclic permutation of length larger than 2, such that some transpositions of elements of quiver matrix, which arise from a~decomposition of such a~permutation, do not preserve the~partition function). However, based on the evidence discussed in what follows, we conjecture that such equivalent quivers do not arise.

This paper is structured as follows. Section~\ref{sec-pre} provides necessary background on knot homologies, knots-quivers correspondence, and multi-cover skein relations. In section~\ref{sec-thm} we focus on local equivalences and formulate the~local equivalence theorem, which states that appropriate transpositions of elements of a~given quiver matrix lead to equivalent quivers. In section~\ref{sec:global structure} we discuss how these local equivalences lead to the~global structure: we show that equivalent quivers arise in families that form permutohedra which are glued into larger graphs that parametrize all equivalent quivers for a~given knot. In section~\ref{SectionCaseStudies} we present examples of such a~global structure and illustrate how permutohedra of equivalent quivers arise and are glued together for various knots. In turn, in section~\ref{sec:local symmetries} we consider examples of local equivalences and determine them for some particular quivers for infinite families of $(2,2p+1)$ torus knots and twist knots, as well as $6_2, 6_3$ and $7_3$ knots. Section~\ref{sec-FK} reveals relations of our results to knot complement quivers and $F_K$ invariants. In the~appendix we present the~lists of all equivalent quiver matrices for knots $5_2$ and $7_1$, as well as particular choices of quiver matrices for infinite classes of twist knots.


\section{Prerequisites}  \label{sec-pre}

In this section~we summarize the~background material on knot homologies, knots-quivers correspondence, and multi-cover skein relations, as well as introduce the~notation that will be used throughout the~paper.


\subsection{Knot homologies}

The knots-quivers correspondence, which is of our main interest in this work, is inherently related to knot homologies. Let us therefore present first a~few basic facts about them. We are especially interested in colored HOMFLY-PT homologies, denoted $\mathcal{H}^R_{ijk}(K)$ for a~knot $K$, where $R$ is a~representation (labeled by a~Young diagram) referred to as the~color~\cite{DGR0505,GS}. In this paper we only consider symmetric representations $R=S^r$, and in various formulae we simply use  the~label $r$ instead of $S^r$. In particular, by $\mathscr{G}_r(K)$ we denote the~set of generators of the~$S^r$-colored homology. While explicit construction of colored HOMFLY-PT homologies has not been provided to date, strong constraints on their structure follow from conjectural properties of associated differentials that relate various generators. In particular, these constraints enable computation of colored superpolynomials and HOMFLY-PT polynomials for various knots. Colored superpolynomials are defined as follows:
\begin{equation}
    P_r(a,q,t) = \sum_{i,j,k} a^i q^j t^k \textrm{dim} \mathcal{H}^{S^r}_{ijk}(K) \equiv \sum_{i\in\mathscr{G}_r(K)} a^{a^{(r)}_i} q^{q^{(r)}_i} t^{t^{(r)}_i},
\end{equation}
where variables $a$ and $q$ are those that appear in HOMFLY-PT polynomials, $t$ is the~refinement (Poincar{\'e}) parameter, and we refer to triples $(a^{(r)}_i,q^{(r)}_i,t^{(r)}_i)$ as homological degrees of the~generator $i\in\mathscr{G}_r(K)$. In the~uncolored case $r=1$ we simply write $(a_i,q_i,t_i)\equiv (a^{(1)}_i,q^{(1)}_i,t^{(1)}_i)$. For a~large class of knots the~linear combination $t_i-a_i-q_i/2$ is constant for each $i\in \mathscr{G}_1(K)$; such knots are called \emph{thin} \cite{DGR0505}.

For a~given color $r$, it is useful to plot colored HOMFLY-PT generators on a~planar diagram, such that the~generator $i\in\mathscr{G}_r(K)$ is represented by a~dot in position $(q^{(r)}_i,a^{(r)}_i)$ (and possibly decorated by the~value $t^{(r)}_i$). The~structure of differentials mentioned above also imposes constraints on the~form of such diagrams. In particular, in the~uncolored case all generators are assembled into two types of structures, referred to as a~zig-zag and a~diamond~\cite{GS}. The~zig-zag consists of an~odd number of generators, while each diamond consists of four generators. The~homological diagram for each knot consists of one zig-zag and some number of diamonds. For example, homological diagrams for $(2,2p+1)$ torus knots consist of only one zig-zag made of $2p+1$ generators, while a~diagram for $4_1$ knot consists of one diamond and a~zig-zag made of only one dot. We will present examples of homological diagrams for these and other knots in what follows.

For $t=-1$ colored superpolynomials reduce to colored HOMFLY-PT polynomials that take form of the~Euler characteristic
\begin{equation}
    P_r(a,q) = P_r(a,q,-1) = \sum_{i,j,k} a^i q^j (-1)^k \textrm{dim} \mathcal{H}^{S^r}_{ijk}(K).
\end{equation}
We stress that by $P_r(a,q,t)$ and $P_r(a,q)$ we denote reduced polynomials (equal to 1 for the~unknot). We use this normalization throughout the~paper except section~\ref{sec-FK}, where using the~unreduced normalization is more appropriate. We also consider generating functions of colored superpolynomials and colored HOMFLY-PT polynomials for defined by
\begin{equation}
    P_K(x,a,q,t) = \sum_{r=0}^{\infty} \frac{x^r}{(q^2;q^2)_r} P_r(a,q,t), \qquad\quad P_K(x,a,q) = \sum_{r=0}^{\infty} \frac{x^r}{(q^2;q^2)_r} P_r(a,q).  \label{PK-generating}
\end{equation}
Including $q$-Pochhammer symbols $(q^2;q^2)_r=\prod_{i=1}^{r}(1-q^{2i})$ in denominators provides a~proper normalization for the~knots-quivers correspondence as defined in~\cite{KRSS1707short,KRSS1707long}.


\subsection{Knots-quivers correspondence}

The knots-quivers correspondence is the~statement that to a~given knot one can assign a~quiver in such a~way that various characteristics of the~knot are expressed in terms of invariants of this quiver (or invariants of moduli spaces of its representations). As already noticed in~\cite{KRSS1707long}, this correspondence is not a~bijection, and several quivers may correspond to the~same knot. In this work we explain how to identify all such equivalent quivers and reveal the~intricate structure they form. However, let us first present relevant background on quiver representation theory, and explain how it relates to knots.

A quiver $Q=(Q_0,Q_1)$ consists of a~set of nodes $Q_0$ and a~set of arrows $Q_1$. Each arrow connects either two different nodes, or a~node to itself -- in the~latter case it is called a~loop. We denote by $C_{ij}$ the~number of arrows from the~node $i$ to the~node $j$, and treat it as an~element of a~matrix $C$. Quivers that arise in knots-quivers correspondence are symmetric, which means that for each arrow $i\to j$ for $i,j\in Q_0$ there exists an~arrow in the~opposite direction $j\to i$; in this case the~matrix $C$ is symmetric.

In quiver representation theory one is interested in the~structure of moduli spaces of quiver representations. Let us consider a~symmetric quiver $Q$ with $m$ nodes and arrows determined by a~matrix $C$. We assign to each node $i$ a~complex vector space of dimension $d_i$; the~$m$-tuple $\boldsymbol{d}=(d_1,\ldots,d_m)$ is referred to as the~dimension vector. Furthermore, for such a~quiver we construct the~following generating series
\begin{equation}
    P_Q(\boldsymbol{x},q)= \sum_{\boldsymbol{d}}(-q)^{\boldsymbol{d}\cdot C \cdot \boldsymbol{d}}\frac{\boldsymbol{x}^{\boldsymbol{d}}}{(q^2;q^2)_{\boldsymbol{d}}} \equiv \sum_{d_1,\ldots,d_m\geq 0}(-q)^{\sum_{i,j=1}^m C_{ij} d_i d_j} \frac{x_1^{d_1}\cdots x_m^{d_m}}{(q^2;q^2)_{d_1}\cdots (q^2;q^2)_{d_m}} ,   \label{P-Q}
\end{equation}
where $\boldsymbol{x}=(x_1,\ldots,x_m)$ are referred to as quiver generating parameters. It turns out that this generating function encodes motivic Donaldson-Thomas invariants $\Omega_{d_1,\ldots,d_m;j}$ of quiver~$Q$, i.e. appropriately defined intersection~Betti numbers of moduli spaces of representations of $Q$, for all dimension vectors $\boldsymbol{d}$. These invariants are encoded in the~following product decomposition of~(\ref{P-Q}):
\begin{equation}
    P_Q(\boldsymbol{x},q) = \prod_{(d_1,\ldots,d_m)\neq 0} \prod_{j\in\mathbb{Z}} \prod_{k\geq 0} \Big(1 - (x_1^{d_1}\cdots x_m^{d_m}) q^{2k+j+1} \Big)^{(-1)^{j+1} \Omega_{d_1,\ldots,d_m;j}}.
\end{equation}
It was postulated in~\cite{KS1006} and proven in~\cite{Efi12} that motivic Donaldson-Thomas invariants $\Omega_{d_1,\ldots,d_m;j}$ are non-negative integers.

The knots-quivers correspondence was motivated by the~observation that generating series of colored knot polynomials~(\ref{PK-generating}) can be written in the~form~(\ref{P-Q}) for appropriate specialization of generating parameters $x_i$. This statement was proven in various examples in~\cite{KRSS1707long}, for two-bridge knots in~\cite{SW1711}, and for arborescent knots in~\cite{SW2004}. The~relation between~(\ref{PK-generating}) and~(\ref{P-Q}) has various interesting consequences. For example, it follows that Ooguri-Vafa invariants of a~knot~\cite{OV9912} are expressed in terms of motivic Donaldson-Thomas invariants; as the~latter invariants are proven to be integer, it follows that that Ooguri-Vafa invariants are also integer, as has been suspected for a~long time. On the~other hand, if all colored superpolynomials can be expressed in the~form~(\ref{P-Q}), it follows that all of them are encoded in a~finite number of parameters, i.e. the~elements of the~matrix~$C$ and additional parameters that arise in the~specialization of $x_i$. Let us now formulate the~knots-quivers correspondence in all details, in a~way appropriate for the~perspective of this work.  

\begin{dfn}\label{def:correspondence}
We say that the~quiver $Q$ corresponds to the~knot $K$ if $Q$ is symmetric and there exists a~bijection 
\begin{equation}\label{eq:KQbijection}
    Q_{0} \ni i 
    \longleftrightarrow
    i \in \mathscr{G}_1(K) 
\end{equation}
such that
\begin{equation}\label{eq:KQcorrespondence}
    \left.P_{Q}(\boldsymbol{x},q)\right|_{(-q)^{C_{ii}}x_{i}=xa^{a_{i}}q^{q_{i}}t^{t_{i}}}=P_{K}(x,a,q,t) \quad and \quad C_{ii}=t_{i}.
\end{equation}
\end{dfn}

\noindent The~substitution $(-q)^{C_{ii}}x_{i}=xa^{a_{i}}q^{q_{i}}t^{t_{i}}$ following the~bijection~(\ref{eq:KQbijection}) is called the~knots-quivers change of variables. Denoting $a^{a_{i}}q^{q_{i}-C_{ii}}(-t)^{C_{ii}}$ as $\lambda_{i}$, we can write it shortly as
\begin{equation} \label{eq:KQcorrespondence-x}
     x_{i}=x\lambda_{i} \qquad \textrm{or}\qquad \boldsymbol{x}=x\boldsymbol{\lambda}.
\end{equation}
The~above correspondence can be also reduced to the~level of HOMFLY-PT polynomials, simply by  putting $t=-1$ in the~knots-quivers change of variables. Note that the~above formulation differs from the~original one~\cite{KRSS1707short,KRSS1707long} that does not require bijectivity, only the~existence of $\{a_i,q_i\}_{i\in Q_0}$ allowing~(\ref{eq:KQcorrespondence}). In consequence, transformations enlarging the~quiver and preserving the~generating function --  forbidden by definition~\ref{def:correspondence} -- are allowed in~\cite{KRSS1707short,KRSS1707long}. Therefore, $Q$ corresponding to $K$ in the~sense of the~definition~\ref{def:correspondence} is the~minimal quiver in the~original sense of~\cite{KRSS1707short,KRSS1707long}. One can also define a~generalized knots-quivers correspondence~\cite{EKL1811}, which allows for $x_{i}=x^{n_{i}}\lambda_{i}$ (possibly with $n_i>1$), but we do not consider it here.


\subsection{Multi-cover skein relations and quivers}\label{sub:Skeins}

Let us now change perspective to that of curve counting for topological strings. It is natural to view holomorphic curves in a~Calabi-Yau three-fold with boundary on a~Lagrangian
$L$ as deforming Chern-Simons theory on $L$ (see~\cite{Wit92}).
In~\cite{ES1901} this perspective was used to give a~new mathematical
approach to open curve counts. Then,~\cite{EKL1910} showed that the
invariance of generalized holomorphic curve counts under bifurcations
of basic disks -- called \emph{multi-cover skein relation} -- generates
quiver degeneracies, i.e. implies the~existence of different quivers corresponding to the~same knot.

\begin{figure}[h]
\begin{centering}
\includegraphics[scale=0.45]{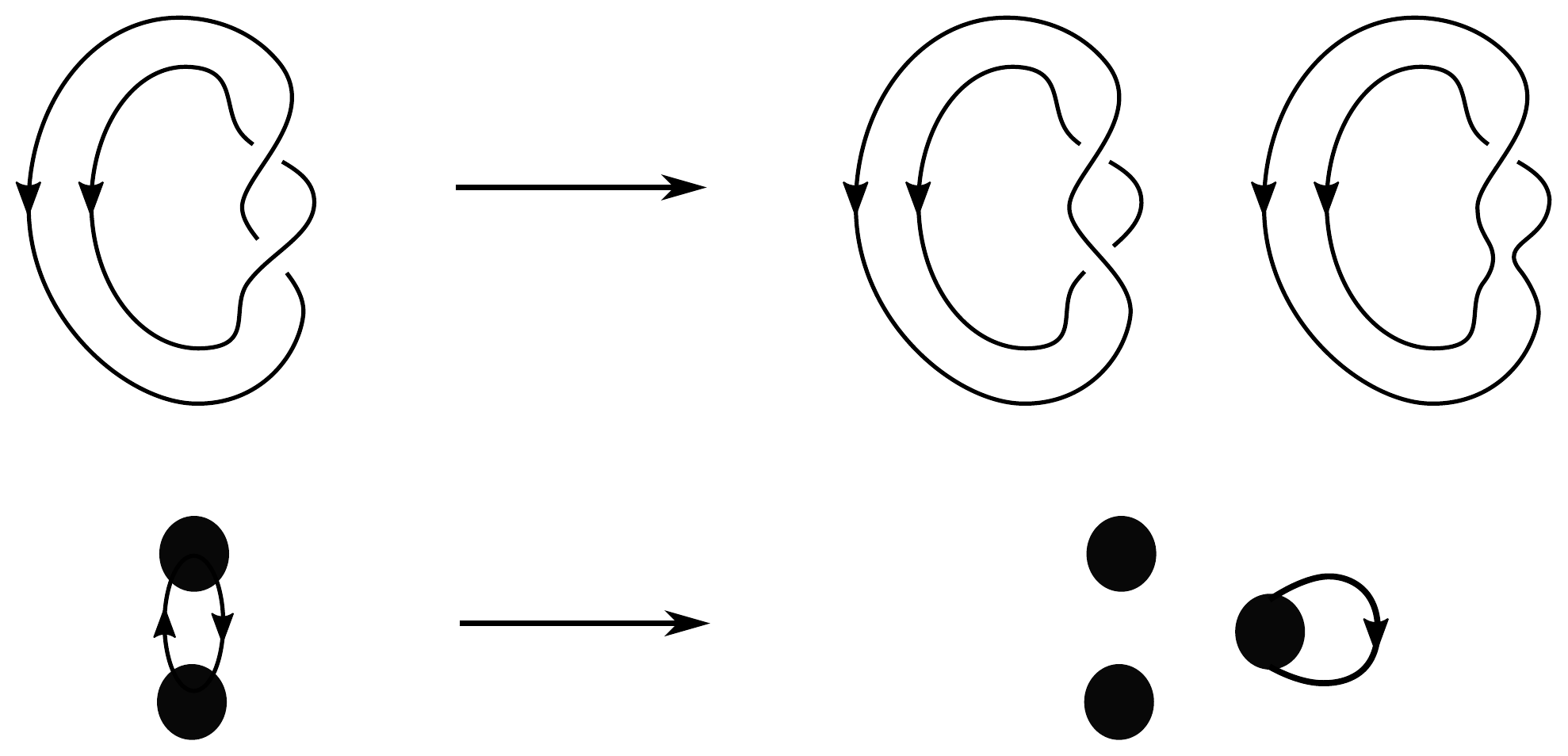}
\par\end{centering}
\caption{\label{fig:Multi-cover-skein-relation}Multi-cover skein relation
on linking disks (top) and dual quiver description (bottom)~\cite{EKL1910}.}
\end{figure}

One can visualize the~multi-cover skein relation as resolving the~intersection
between disk boundaries, see figure~\ref{fig:Multi-cover-skein-relation}.
Using the~language of~\cite{EKL1811}, it can be adapted to quivers
as the~equality of motivic generating series of two quivers shown
at the~bottom of figure~\ref{fig:Multi-cover-skein-relation},
where each basic disc corresponds to the~quiver node,
and the~linking number corresponds to the~number of arrows. Physically,
it corresponds to the~duality between two 3d $\mathcal{N}=2$ theories
and has an~interesting relations with the~wall-crossing from~\cite{KS0811,KS1006}.
More details can be found in~\cite{EKL1910}.

The phenomenon presented in figure~\ref{fig:Multi-cover-skein-relation}
is the~simplest example of \emph{unlinking}. From the~perspective of BPS states, it corresponds to reinterpreting the~bound state made
of two basic states as an~independent basic state. In terms of quivers,
it means removing one pair of arrows which encode the~interaction
leading to a~bound state and adding a~new node. Adapting~\cite{EKL1910}
to our notation, we define the~general case of unlinking in the~following way:

\begin{dfn}\label{def:unlinking}
Consider a~symmetric quiver $Q$ and fix $a,b\in Q_{0}$. The~unlinking of nodes $a,b$ is defined as a~transformation of $Q$ leading to a~new quiver $\widetilde{Q}$ such that:
\begin{itemize}
\item There is a~new node $n$: $\widetilde{Q}_{0}=Q_{0}\cup n$.
\item The~number of arrows of the~new quiver is given by
\begin{align}
\label{eq:unlinking arrows}
\widetilde{C}_{ab} & =C_{ab}-1, & \widetilde{C}_{nn} & =C_{aa}+2C_{ab}+C_{bb}-1,\\
\widetilde{C}_{in} & =C_{ai}+C_{bi}-\delta_{ai}-\delta_{bi}, & \widetilde{C}_{ij} & =C_{ij}\quad\textrm{for all other cases,} \nonumber
\end{align}
where $\delta_{ij}$ is a~Kronecker delta.
\end{itemize}
\end{dfn}
\noindent One can check that quivers on the~left- and right-hand side of figure~\ref{fig:Multi-cover-skein-relation}
correspond respectively to
\begin{equation}
C=\left[\begin{array}{cc}
C_{aa} & C_{ab}\\
C_{ba} & C_{bb}
\end{array}\right]=\left[\begin{array}{cc}
0 & 1\\
1 & 0
\end{array}\right]
\qquad \longrightarrow \qquad
\widetilde{C}=\left[\begin{array}{ccc}
\widetilde{C}_{aa} & \widetilde{C}_{ab} & \widetilde{C}_{an}\\
\widetilde{C}_{ba} & \widetilde{C}_{bb} & \widetilde{C}_{bn}\\
\widetilde{C}_{na} & \widetilde{C}_{nb} & \widetilde{C}_{nn}
\end{array}\right]=\left[\begin{array}{ccc}
0 & 0 & 0\\
0 & 0 & 0\\
0 & 0 & 1
\end{array}\right].
\end{equation}

For us, the~most important result of~\cite{EKL1910} is the~following statement:
\begin{thm}[Ekholm, Kucharski, Longhi]\label{thm:unlinking}
The~unlinking accompanied by the~substitution $x_{n}=q^{-1}x_{a}x_{b}$ preserves the~motivic generating function of the~quiver:
\begin{equation}
P_{Q}(\boldsymbol{x},q)=\left.P_{\widetilde{Q}}(\boldsymbol{x},q)\right|_{x_{n}=q^{-1}x_{a}x_{b}}.
\end{equation}
\end{thm}
\noindent In section~\ref{sec-proof} we use it to prove the~local equivalence theorem.

\section{Local equivalence of quivers}  \label{sec-thm}

In this section we show that for a~given quiver of size $m$ (equal to the~number of HOMFLY-PT generators of the~corresponding knot), encoded in a~symmetric matrix $C$, there exist equivalent quivers such that their matrices differ from $C$ only by a~transposition of two non-diagonal elements $C_{ab}$ and $C_{cd}$, as long as the~values of these two elements differ by~1 and certain extra conditions are met. This is the~phenomenon that we refer to as local equivalence of quivers. In the~next sections we show that these local equivalences give rise to an~intricate global structure whose building blocks are permutohedra, and provide various examples of this phenomenon.

We start from introducing an~equivalence relation that describes quiver degeneracies in a~natural way.  
\begin{dfn}\label{def:equivalence}
Assume that quiver $Q$ corresponds to the~knot $K$ and quiver $Q'$ corresponds to the~knot $K'$ in the~sense of the~definition~\ref{def:correspondence}. Then we define
\begin{equation}\label{eq:equivalence relation}
Q\sim Q' \Longleftrightarrow K \textrm{ and } K' \textrm{ have the~same colored HOMFLY-PT homology.}
\end{equation}
\end{dfn}
\noindent In the~rest of the~paper we refer to  the~simplest and most common version of~(\ref{eq:equivalence relation}), namely $K=K'$. However, each time we write that two (or more) quivers correspond to the~same knot, we keep in mind that another knot with the~same colored HOMFLY-PT homology would lead to the~same equivalence class of quivers.

\subsection{Analysis of possible equivalences}  \label{sec:possible equivalences}

Let us study when two quivers $Q$ and $Q'$ can correspond to the~same knot $K$. Using definition~\ref{def:correspondence}, we start from
\begin{equation}
P_{K}(x,a,q,t)=\left.P_{Q}(\boldsymbol{x},q)\right|_{\boldsymbol{x}=x\boldsymbol{\lambda}}=\left.P_{Q'}(\boldsymbol{x},q)\right|_{\boldsymbol{x}=x\boldsymbol{\lambda}'}\label{eq:KQ correspondece with lambdas}
\end{equation}
with 
\begin{equation}\label{eq:linear conditions}
    \lambda_{i}=\lambda'_{i}=a^{a_{i}}q^{q_{i}-C_{ii}}(-t)^{C_{ii}},\quad C_{ii}=t_i \quad \forall i\in Q_{0}=Q'_{0}.
\end{equation}
We will analyze equation ~(\ref{eq:KQ correspondece with lambdas}) order by order in $x$. The~linear one holds automatically, so let us focus on terms proportional to $x^2$:
\begin{equation}
\begin{split}\frac{P_{2}(a,q,t)x^{2}}{(1-q^{2})(1-q^{4})} & =\sum_{i\in Q_{0}}\frac{(-q)^{4C_{ii}}x^{2}\lambda_{i}^{2}}{(1-q^{2})(1-q^{4})} +\sum_{i,j\in Q_{0},i\neq j}\frac{(-q)^{C_{ii}+2C_{ij}+C_{jj}}x^{2}\lambda_{i}\lambda_{j}}{(1-q^{2})(1-q^{2})}\\
 & =\sum_{i\in Q'_{0}}\frac{(-q)^{4C_{ii}}x^{2}\lambda_{i}^{2}}{(1-q^{2})(1-q^{4})} +\sum_{i,j\in Q'_{0},i\neq j}\frac{(-q)^{C_{ii}+2C'_{ij}+C_{jj}}x^{2}\lambda_{i}\lambda_{j}}{(1-q^{2})(1-q^{2})},
\end{split}
\label{eq:second order}
\end{equation}
where we used~\eqref{eq:linear conditions} to write $\lambda_{i}=\lambda'_{i}$ and $C_{ii}=C'_{ii}$. In consequence, the~only difference between $Q$ and $Q'$ can appear in non-diagonal terms $C_{ij}$ and $C'_{ij}$. Since equation~\eqref{eq:second order} needs to hold for all $a$ and $t$ (which are independent from $C_{ij}$ and $C'_{ij}$), we require the~equality between coefficients of each monomial in these variables. The~only possibility of having $Q\neq Q'$ satisfying~\eqref{eq:KQ correspondece with lambdas} comes from $C_{ij}\neq C'_{ij}$ which however lead to the~same coefficient of each monomial in $a$ and $t$ on both sides. The~way $q$-monomials on both sides are matched can be described by permutations of terms in the~coefficient of each monomial in $a$ and $t$. 

Let us focus on the~simplest non-trivial case. We assume that each coefficient of monomials in $a$ and $t$ has only one term except from the~expression corresponding to $\lambda_{a}\lambda_{b}$ and $\lambda_{c}\lambda_{d}$. This means that we require $\lambda_{a}\lambda_{b}=q^{2s}\lambda_{c}\lambda_{d}$ for some $s\in \mathbb{Z}$ and $\lambda_{a}$, $\lambda_{b}$, $\lambda_{c}$, $\lambda_{d}$ being pairwise different. (Note that for thin knots we immediately know that $s=0$.) 
Therefore, we get $C_{ij}=C'_{ij}\; \forall i,j\in Q_0\backslash \{a,b,c,d\}$ and~\eqref{eq:second order} can be reduced to
\begin{equation}
    \lambda_{a}\lambda_{b}(-q)^{C_{aa}+C_{bb}}
    \bigg(q^{2C_{ab}}
    +q^{-2s+2C_{cd}}
    \bigg)
    = 
    \lambda_{a}\lambda_{b}(-q)^{C_{aa}+C_{bb}}
    \left(q^{2C'_{ab}}
    +q^{-2s+2C'_{cd}}
    \right),    
\end{equation}
where we used $C_{aa}+C_{bb}=C_{cc}+C_{dd}$ that comes from the~comparison of $t$~powers in $\lambda_{a}\lambda_{b}=q^{2s}\lambda_{c}\lambda_{d}$. In consequence, there is only one non-trivial way to satisfy (\ref{eq:second order}), namely
\begin{equation}
C'_{ab}= C_{cd}-s,\quad C'_{cd}= C_{ab}+s.
\label{eq:second_order_constraints_shifted}
\end{equation}
Using the~language of permutations of terms in the~generating function, this corresponds to the~transposition $\lambda_{a}\lambda_{b}(-q)^{C_{aa}+2C_{ab}+C_{bb}}$ $\leftrightarrow$ $\lambda_{c}\lambda_{d}(-q)^{C_{cc}+2C_{cd}+C_{dd}}$. For $s=0$ it translates to the~transposition of matrix entries $C_{ab}\leftrightarrow C_{cd}$.

Let us continue the~analysis of the~simplest non-trivial case and check what conditions come from the~cubic order of~(\ref{eq:KQ correspondece with lambdas}).
In order to save space, we start from examining where differences
between $\left.P_{Q}(\boldsymbol{x},q)\right|_{x_{i}=x\lambda_{i}}$
and $\left.P_{Q'}(\boldsymbol{x},q)\right|_{x_{i}=x\lambda_{i}}$
can arise. The~general formula reads
\begin{equation}
\begin{split}\frac{P_{3}(a,q,t)x^{3}}{(1-q^{2})(1-q^{4})(1-q^{6})} & =\sum_{i\in Q_{0}}\frac{(-q)^{9C_{ii}}x^{3}\lambda_{i}^{3}}{(1-q^{2})(1-q^{4})(1-q^{6})}\\
 & +\sum_{i,j\in Q{}_{0},i\neq j}\frac{(-q)^{4C_{ii}+4C_{ij}+C_{jj}}x^{3}\lambda_{i}^{2}\lambda_{j}}{(1-q^{2})(1-q^{4})(1-q^{2})}\\
 & +\sum_{i,j,k\in Q{}_{0},i\neq j\neq k}\frac{(-q)^{C_{ii}+2C_{ij}+C_{jj}+2C_{jk}+C_{kk}+2C_{ik}}x^{3}\lambda_{i}\lambda_{j}\lambda_{k}}{(1-q^{2})(1-q^{2})(1-q^{2})},
\end{split}
\label{eq:third order general formula}
\end{equation}
so we have to look for terms containing $\lambda_{a}\lambda_{b}$
or $\lambda_{c}\lambda_{d}$. They are given by
\begin{equation}
\begin{split}\frac{x^{3}\lambda_{a}\lambda_{b}}{(1-q^{2})(1-q^{4})(1-q^{6})} & \Big[(-q)^{4C_{aa}+4C'_{ab}+C_{bb}}\lambda_{a}+(-q)^{4C_{bb}+4C'_{ab}+C_{aa}}\lambda_{b}\\
 & +(1+q^{2})(-q)^{C_{aa}+2C'_{ab}+C_{bb}+2C_{bc}+C_{cc}+2C_{ac}}\lambda_{c}\\
 & +(1+q^{2})(-q)^{C_{aa}+2C'_{ab}+C_{bb}+2C_{bd}+C_{dd}+2C_{ad}}\lambda_{d}\\
 & +(1+q^{2})\sum_{i\in Q_{0}\backslash\{a,b,c,d\}}(-q)^{C_{aa}+2C'_{ab}+C_{bb}+2C_{bi}+C_{ii}+2C_{ai}}\lambda_{i}\Big]
\end{split}
\label{eq:third order ab}
\end{equation}
and
\begin{equation}
\begin{split}\frac{x^{3}\lambda_{c}\lambda_{d}}{(1-q^{2})(1-q^{4})(1-q^{6})} & \Big[(-q)^{4C_{cc}+4C'_{cd}+C_{dd}}\lambda_{c}+(-q)^{4C_{dd}+4C'_{cd}+C_{dd}}\lambda_{d}\\
 & +(1+q^{2})(-q)^{C_{cc}+2C'_{cd}+C_{dd}+2C_{ad}+C_{aa}+2C_{ac}}\lambda_{a}\\
 & +(1+q^{2})(-q)^{C_{cc}+2C'_{cd}+C_{dd}+2C_{bd}+C_{bb}+2C_{bc}}\lambda_{b}\\
 & +(1+q^{2})\sum_{i\in Q_{0}\backslash\{a,b,c,d\}}(-q)^{C_{cc}+2C'_{cd}+C_{dd}+2C_{di}+C_{ii}+2C_{ci}}\lambda_{i}\Big]
\end{split}
\label{eq:third order cd}
\end{equation}
for $\left.P_{Q'}(\boldsymbol{x},q)\right|_{\boldsymbol{x}=x\boldsymbol{\lambda}}$
and analogous terms without prime symbols for $\left.P_{Q}(\boldsymbol{x},q)\right|_{\boldsymbol{x}=x\boldsymbol{\lambda}}$.
Since $\lambda_{a}\lambda_{b}=q^{2s}\lambda_{c}\lambda_{d}$, imposing the~equality
between $\left.P_{Q'}(\boldsymbol{x},q)\right|_{\boldsymbol{x}=x\boldsymbol{\lambda}}$
and $\left.P_{Q}(\boldsymbol{x},q)\right|_{\boldsymbol{x}=x\boldsymbol{\lambda}}$ implies
conditions for sums of terms from both~(\ref{eq:third order ab})
and~(\ref{eq:third order cd}) for $\lambda_{a}$, $\lambda_{b}$,
$\lambda_{c}$, $\lambda_{d}$ and each $\lambda_{i}$, $i\in Q_{0}\backslash\{a,b,c,d\}$:
\begin{equation}
\begin{split}\lambda_{a} & \Big[(-q)^{4C_{aa}+4C'_{ab}+C_{bb}+2s}+(1+q^{2})(-q)^{C_{cc}+2C'_{cd}+C_{dd}+2C_{ad}+C_{aa}+2C_{ac}}\Big]\\
 & =\lambda_{a}\Big[(-q)^{4C_{aa}+4C_{ab}+C_{bb}+2s}+(1+q^{2})(-q)^{C_{cc}+2C_{cd}+C_{dd}+2C_{ad}+C_{aa}+2C_{ac}}\Big],
\end{split}
\label{eq:third order a}
\end{equation}
\begin{equation}
\begin{split}\lambda_{b} & \Big[(-q)^{4C_{bb}+4C'_{ab}+C_{aa}+2s}+(1+q^{2})(-q)^{C_{cc}+2C'_{cd}+C_{dd}+2C_{bd}+C_{bb}+2C_{bc}}\Big]\\
 & =\lambda_{b}\Big[(-q)^{4C_{bb}+4C_{ab}+C_{aa}+2s}+(1+q^{2})(-q)^{C_{cc}+2C_{cd}+C_{dd}+2C_{bd}+C_{bb}+2C_{bc}}\Big],
\end{split}
\label{eq:third order b}
\end{equation}
\begin{equation}
\begin{split}\lambda_{c} & \Big[(-q)^{4C_{cc}+4C'_{cd}+C_{dd}}+(1+q^{2})(-q)^{C_{aa}+2C'_{ab}+C_{bb}+2C_{bc}+C_{cc}+2C_{ac}+2s}\Big]\\
 & =\lambda_{c}\Big[(-q)^{4C_{cc}+4C_{cd}+C_{dd}}+(1+q^{2})(-q)^{C_{aa}+2C_{ab}+C_{bb}+2C_{bc}+C_{cc}+2C_{ac}+2s}\Big],
\end{split}
\label{eq:third order c}
\end{equation}
\begin{equation}
\begin{split}\lambda_{d} & \Big[(-q)^{4C_{dd}+4C'_{cd}+C_{cc}}+(1+q^{2})(-q)^{C_{aa}+2C'_{ab}+C_{bb}+2C_{bd}+C_{dd}+2C_{ad}+2s}\Big]\\
 & =\lambda_{d}\Big[(-q)^{4C_{dd}+4C_{cd}+C_{cc}}+(1+q^{2})(-q)^{C_{aa}+2C_{ab}+C_{bb}+2C_{bd}+C_{dd}+2C_{ad}+2s}\Big],
\end{split}
\label{eq:third order d}
\end{equation}
\begin{equation}
\begin{split}\lambda_{i} & \Big[(-q)^{C_{aa}+2C'_{ab}+C_{bb}+2C_{bi}+C_{ii}+2C_{ai}+2s}+(-q)^{C_{cc}+2C'_{cd}+C_{dd}+2C_{di}+C_{ii}+2C_{ci}}\Big]\\
 & =\lambda_{i}\Big[(-q)^{C_{aa}+2C_{ab}+C_{bb}+2C_{bi}+C_{ii}+2C_{ai}+2s}+(-q)^{C_{cc}+2C_{cd}+C_{dd}+2C_{di}+C_{ii}+2C_{ci}}\Big].
\end{split}
\label{eq:third order i}
\end{equation}
In each equation we have to match three $q$-monomials on both sides in a~non-trivial way. For example, in~\eqref{eq:third order a} we must take
\begin{align}
4C_{aa}+4C'_{ab}+C_{bb}+2s &=  C_{cc}+2C_{cd}+C_{dd}+2C_{ad}+C_{aa}+2C_{ac}+2, \nonumber\\
 C_{cc}+2C'_{cd}+C_{dd}+2C_{ad}+C_{aa}+2C_{ac} &= 4C_{aa}+4C_{ab}+C_{bb}+2s,\\
C_{cc}+2C'_{cd}+C_{dd}+2C_{ad}+C_{aa}+2C_{ac}+2 &= 
C_{cc}+2C_{cd}+C_{dd}+2C_{ad}+C_{aa}+2C_{ac}, \nonumber\\
&\textrm{or} \nonumber\\
4C_{aa}+4C'_{ab}+C_{bb}+2s &= C_{cc}+2C_{cd}+C_{dd}+2C_{ad}+C_{aa}+2C_{ac}, \nonumber\\
 C_{cc}+2C'_{cd}+C_{dd}+2C_{ad}+C_{aa}+2C_{ac} &=C_{cc}+2C_{cd}+C_{dd}+2C_{ad}+C_{aa}+2C_{ac}+2 ,\nonumber\\
C_{cc}+2C'_{cd}+C_{dd}+2C_{ad}+C_{aa}+2C_{ac}+2 &= 4C_{aa}+4C_{ab}+C_{bb}+2s.
\end{align}
Analogous matching for~equations for~(\ref{eq:third order b}-\ref{eq:third order d}), combined with $C_{aa}+C_{bb}=C_{cc}+C_{dd}$
and (\ref{eq:second_order_constraints_shifted}), leads to two possible ways for non-trivial pairwise cancellation:
\begin{equation}
\begin{aligned}
C_{ab}+s=&\ C_{cd}-1, \\
C_{aa}+C_{cd}=&\ C_{ad}+C_{ac}+s+1, \\
C_{bb}+C_{cd}=&\ C_{bd}+C_{bc}+s+1, \\
C_{ab}+C_{cc}+s=&\ C_{bc}+C_{ac}, \\
C_{ab}+C_{dd}+s=&\ C_{bd}+C_{ad}
\end{aligned}
\qquad
\text{or} \qquad
\begin{aligned}
C_{ab}+s=&\ C_{cd}+1, \\
C_{aa}+C_{cd}=&\ C_{ad}+C_{ac}+s, \\
C_{bb}+C_{cd}=&\ C_{bd}+C_{bc}+s, \\
C_{ab}+C_{cc}+s=&\ C_{bc}+C_{ac}+1, \\
C_{ab}+C_{dd}+s=&\ C_{bd}+C_{ad}+1.
\end{aligned}
\label{third_order_constraints_shifted_a}
\end{equation}
Combining~(\ref{third_order_constraints_shifted_a}) with $C_{aa}+C_{bb}=C_{cc}+C_{dd}$, we deduce that $s=0$. Putting it in equations (\ref{eq:third order a})-(\ref{eq:third order i}) and performing the~analogous matching of terms, we learn that:
\begin{align}
C_{cd} & =C_{ab}-1, &
C_{ci}+C_{di} & =C_{ai}+C_{bi}-\delta_{ai}-\delta_{bi}\quad \forall i\in Q_{0}
\label{eq:conditions for cd smaller than ab}\\
\textrm{ or }\qquad\quad
C_{ab}  &=C_{cd}-1,&
C_{ai}+C_{bi} & =C_{ci}+C_{di}-\delta_{ci}-\delta_{di}\quad \forall i\in Q_{0}.
\label{eq:conditions for ab smaller than cd}
\end{align}
These conditions are required for the~transposition $C_{ab}\leftrightarrow C_{cd}$ to lead to an equivalent quiver.

Now, let us slightly modify our assumptions to $\lambda_a=q^{2s_1}\lambda_c$, $\lambda_b=q^{2s_2} \lambda_d$, and requirement that 
$q^{2C_{ab}}\lambda_a\lambda_b+q^{2C_{cd}}\lambda_c\lambda_d+q^{2C_{ad}}\lambda_a\lambda_d+q^{2C_{bc}}\lambda_b\lambda_c$ 
corresponds to the~only monomial in $a$ and $t$ which coefficient has more than one $q$-monomial at the~level of $x^2$. Let us consider all types of permutations of these terms by focusing on which is equal to $q^{2C_{ab}}\lambda_a\lambda_b$ in $P_{Q'}$. If it is $q^{2C'_{ab}}\lambda_a\lambda_b$, then $C_{ab}=C'_{ab}$, if it is $q^{2C'_{cd}}\lambda_c\lambda_d$, then we have a~situation that was described earlier in this section. The~only truly different case comes from equating $q^{2C_{ab}}\lambda_a\lambda_b$ with $q^{2C'_{ad}}\lambda_a\lambda_d$ or $q^{2C'_{bc}}\lambda_b\lambda_c$. 
In the~first case the~analogs of equations~\eqref{eq:third order a} and~\eqref{eq:third order i} imply $s=0$ and $C_{bi}=C_{di}$ for every $i \in Q_0\backslash\{a, b, d \}$. This means that nodes $b$ and $d$ are indistinguishable and the~transposition $C_{ab}\leftrightarrow C_{ad}$ can be understood as a~relabeling~$b \leftrightarrow d$. The~second case is completely analogous and can be understood as  a~relabeling~$a \leftrightarrow c$.

Now we would like to analyze the~possibility of composing transpositions satisfying conditions (\ref{eq:conditions for cd smaller than ab}) or (\ref{eq:conditions for ab smaller than cd}) into a~bigger cycle. Let us therefore assume that $\lambda_a\lambda_b=\lambda_c\lambda_d=\lambda_e\lambda_f$, all lambdas -- as well as $C_{ab}$, $C_{cd}$, $C_{ef}$ -- are pairwise different, and equations (\ref{eq:conditions for cd smaller than ab}) or (\ref{eq:conditions for ab smaller than cd}) -- as well as their counterparts for $c,d,e,f$ -- are satisfied. Among them there is an~equation $C_{ac} + C_{bc} = C_{cc} + C_{cd}$ (if $C_{ab}<C_{cd}$) or $C_{ac} + C_{bc} = C_{cc} + C_{cd}-1$ (if $C_{ab}>C_{cd}$) which becomes violated after the~transposition $C_{cd}\leftrightarrow C_{ef}$. Similarly, performing the~transposition $C_{ab} \leftrightarrow C_{cd}$ causes the~violation of an~analogous equation required for $C_{cd}\leftrightarrow C_{ef}$. In consequence, we see that after composing transpositions which preserve the~generating function into a~bigger cycle, we always get an~inequivalent quiver. Moreover, an~analogous argument implies that the~composition of transpositions  $C_{ab} \leftrightarrow C_{cd}$ and 
$C_{de}\leftrightarrow C_{fg}$ (both of which involve the~same node $d$) leads to an~inequivalent quiver.

We have not yet excluded all non-trivial ways of matching terms in~\eqref{eq:second order} -- for example one may think about a~permutation that leads to an~equivalent quiver, but is composed of transpositions that change the~partition function. However, based on the evidence discussed below, it appears that such permutations are little likely to arise, and thus we make the~following conjecture:

\begin{coj}\label{coj:necessary conditions}
Consider a~quiver $Q$ corresponding to the~knot $K$. If there exists another
symmetric quiver $Q'$ such that $Q'\sim Q$ in the~sense of the~definition~\ref{def:equivalence}, then either $Q'=Q$ or they are related by a~sequence of disjoint transpositions, each exchanging non-diagonal elements 
    \begin{equation}\label{eq:transposition 1}
        C_{ab}\leftrightarrow C_{cd}, \qquad C_{ba}\leftrightarrow C_{dc},
    \end{equation}
for some pairwise different $a,b,c,d,\in  Q_{0}$, such that
    \begin{equation}\label{eq:center of mass contition 1}
         \lambda_{a}\lambda_{b} = \lambda_{c}\lambda_{d}
    \end{equation}
and 
    \begin{equation}\label{eq:theorem first case 1}
        C_{ab} = C_{cd}-1,\qquad\quad
        C_{ai}+C_{bi}=C_{ci}+C_{di}-\delta_{ci}-\delta_{di},\quad \forall i\in Q_{0},
    \end{equation}
    or
    \begin{equation}\label{eq:theorem second case 1}
        C_{cd} = C_{ab}-1,\qquad\quad
        C_{ci}+C_{di}=C_{ai}+C_{bi}-\delta_{ai}-\delta_{bi},\quad \forall i\in Q_{0}.
    \end{equation}
\end{coj}
For the~simplest thin knots we verify this conjecture in the~following way. Since $a_i$ and $t_i$ fix $q_i$ and~$C_{ii}$, permutations of terms in coefficients of monomials in $a$ and $t$ are in one-to-one correspondence with permutations of $C_{ij}$. Therefore, we just need to find all incident products $\lambda_a \lambda_b=\lambda_c \lambda_d=\lambda_e \lambda_f=\ldots$ and for each of them check all permutations of the~set~$\{C_{ab},C_{cd},C_{ef},\ldots\}$. Using this procedure, we verified conjecture \ref{coj:necessary conditions} for quivers corresponding to $3_1$, $4_1$, and $5_1$ knot. 

For thin knots we can also give another general argument supporting conjecture \ref{coj:necessary conditions} --  we can exclude those 3-cycles that are not necessarily composed of transpositions preserving the~generating function. To this end, let us assume that  $\lambda_a \lambda_b=\lambda_c \lambda_d=\lambda_e \lambda_f$, these terms are the~only instance of multiple $q$-monomials in the~coefficient of $a$ and $t$ monomials in (\ref{eq:second order}), and $Q'$ arises from $Q$ by performing the~3-cycle $(C_{ab}\; C_{cd}\; C_{ef})$ or $(C_{ab}\; C_{ef}\; C_{cd})$ with $C_{ab},\,C_{cd},\,C_{ef}$ being all distinct. Then, in the~qubic term (\ref{eq:third order general formula}), we have multiple ways to cancel the~terms in front of $\lambda_a,\lambda_b,\ldots,\lambda_f$. In total, it results in $44^3$ non-trivial systems of $30$ linear equations, which we treated with the~help of computer and confirmed that together with the~center of mass conditions they cannot be satisfied in a~non-trivial way.

In the~next section we formulate and prove the~theorem which is an~analog of conjecture~\ref{coj:necessary conditions} with a~reversed direction of implication. Together, they provide a~complete description of quiver equivalences.

\subsection{Local equivalence theorem} 

\begin{thm}\label{thm:main}
Consider a~quiver $Q$ corresponding to the~knot $K$ and another
symmetric quiver $Q'$ such that $Q'_0=Q_0$ and $\lambda'_i=\lambda_i\;\forall i\in Q_0$ ($\lambda_i$ comes from the~knots-quivers change of variables).
If $Q$ and $Q'$ are related by a~sequence of disjoint transpositions, each exchanging non-diagonal elements 
    \begin{equation}\label{eq:transposition}
         C_{ab}\leftrightarrow C_{cd}, \qquad C_{ba}\leftrightarrow C_{dc},
    \end{equation}
for some pairwise different $a,b,c,d,\in  Q_{0}$, such that
    \begin{equation}\label{eq:center of mass contition}
         \lambda_{a}\lambda_{b} = \lambda_{c}\lambda_{d}
    \end{equation}
and 
    \begin{equation}\label{eq:theorem first case}
        C_{ab} = C_{cd}-1,\qquad\quad
        C_{ai}+C_{bi}=C_{ci}+C_{di}-\delta_{ci}-\delta_{di},\quad \forall i\in Q_{0},
    \end{equation}
    or
    \begin{equation}\label{eq:theorem second case}
        C_{cd} = C_{ab}-1,\qquad\quad
        C_{ci}+C_{di}=C_{ai}+C_{bi}-\delta_{ai}-\delta_{bi},\quad \forall i\in Q_{0},
    \end{equation}
then $Q$ and $Q'$ are equivalent in the~sense of the~definition~\ref{def:equivalence}.
\end{thm}

In order to apply this theorem to various knots and quivers, we usually start from looking for $\lambda_a,\lambda_b,\lambda_c,\lambda_d$ that satisfy the~condition $\lambda_a\lambda_b=\lambda_c\lambda_d$. We call a~quadruple of pairwise different $a,b,c,d\in  Q_{0}$ such that this equation holds a~\emph{pairing}. Note that only some pairings generate transpositions~\eqref{eq:transposition} leading to equivalent quiver -- if this is the~case, we call them \emph{symmetries}. If a~symmetry is consistent with constraints~\eqref{eq:theorem first case} or~\eqref{eq:theorem second case} we call it \emph{non-trivial}; if it follows from $C'_{ij}=C_{ij}$ we  call it \emph{trivial}.

\begin{figure}[h!]
    \centering
    \tikzset{every picture/.style={line width=0.75pt}} 

\begin{tikzpicture}[x=0.75pt,y=0.75pt,yscale=-1,xscale=1]

\draw [line width=1.5]    (1992.97,8522.29) -- (2042.47,8472.79) ;
\draw  [draw opacity=0] (1918.73,8398.55) -- (2067.26,8398.55) -- (2067.26,8547.66) -- (1918.73,8547.66) -- cycle ; \draw  [color={rgb, 255:red, 155; green, 155; blue, 155 }  ,draw opacity=1 ] (1918.73,8398.55) -- (1918.73,8547.66)(1968.22,8398.55) -- (1968.22,8547.66)(2017.72,8398.55) -- (2017.72,8547.66)(2067.22,8398.55) -- (2067.22,8547.66) ; \draw  [color={rgb, 255:red, 155; green, 155; blue, 155 }  ,draw opacity=1 ] (1918.73,8398.55) -- (2067.26,8398.55)(1918.73,8448.04) -- (2067.26,8448.04)(1918.73,8497.54) -- (2067.26,8497.54)(1918.73,8547.04) -- (2067.26,8547.04) ; \draw  [color={rgb, 255:red, 155; green, 155; blue, 155 }  ,draw opacity=1 ]  ;
\draw [line width=1.5]    (1992.97,8522.29) -- (1943.48,8472.79) ;
\draw [line width=1.5]    (1943.48,8472.79) -- (1992.97,8423.3) ;
\draw [line width=1.5]    (2042.47,8472.79) -- (1992.97,8423.3) ;
\draw  [fill={rgb, 255:red, 0; green, 0; blue, 0 }  ,fill opacity=1 ] (1941.57,8469.61) .. controls (1943.32,8468.56) and (1945.6,8469.13) .. (1946.66,8470.88) .. controls (1947.71,8472.64) and (1947.14,8474.92) .. (1945.38,8475.97) .. controls (1943.63,8477.03) and (1941.35,8476.46) .. (1940.3,8474.7) .. controls (1939.24,8472.94) and (1939.81,8470.67) .. (1941.57,8469.61) -- cycle ;
\draw  [fill={rgb, 255:red, 0; green, 0; blue, 0 }  ,fill opacity=1 ] (2040.56,8469.61) .. controls (2042.32,8468.56) and (2044.6,8469.13) .. (2045.65,8470.88) .. controls (2046.7,8472.64) and (2046.13,8474.92) .. (2044.38,8475.97) .. controls (2042.62,8477.03) and (2040.34,8476.46) .. (2039.29,8474.7) .. controls (2038.24,8472.94) and (2038.81,8470.67) .. (2040.56,8469.61) -- cycle ;
\draw  [fill={rgb, 255:red, 0; green, 0; blue, 0 }  ,fill opacity=1 ] (1991.06,8420.12) .. controls (1992.82,8419.06) and (1995.1,8419.63) .. (1996.15,8421.39) .. controls (1997.21,8423.14) and (1996.64,8425.42) .. (1994.88,8426.48) .. controls (1993.13,8427.53) and (1990.85,8426.96) .. (1989.79,8425.2) .. controls (1988.74,8423.45) and (1989.31,8421.17) .. (1991.06,8420.12) -- cycle ;
\draw  [fill={rgb, 255:red, 0; green, 0; blue, 0 }  ,fill opacity=1 ] (1991.06,8519.11) .. controls (1992.82,8518.06) and (1995.1,8518.62) .. (1996.15,8520.38) .. controls (1997.21,8522.14) and (1996.64,8524.41) .. (1994.88,8525.47) .. controls (1993.13,8526.52) and (1990.85,8525.95) .. (1989.79,8524.2) .. controls (1988.74,8522.44) and (1989.31,8520.16) .. (1991.06,8519.11) -- cycle ;
\draw [color={rgb, 255:red, 74; green, 74; blue, 74 }  ,draw opacity=1 ] [dash pattern={on 4.5pt off 4.5pt}]  (1992.97,8472.79) -- (1946.48,8472.79) ;
\draw [shift={(1943.48,8472.79)}, rotate = 360] [fill={rgb, 255:red, 74; green, 74; blue, 74 }  ,fill opacity=1 ][line width=0.08]  [draw opacity=0] (10.72,-5.15) -- (0,0) -- (10.72,5.15) -- (7.12,0) -- cycle    ;
\draw [color={rgb, 255:red, 74; green, 74; blue, 74 }  ,draw opacity=1 ] [dash pattern={on 4.5pt off 4.5pt}]  (1992.97,8472.79) -- (1992.97,8519.29) ;
\draw [shift={(1992.97,8522.29)}, rotate = 270] [fill={rgb, 255:red, 74; green, 74; blue, 74 }  ,fill opacity=1 ][line width=0.08]  [draw opacity=0] (10.72,-5.15) -- (0,0) -- (10.72,5.15) -- (7.12,0) -- cycle    ;
\draw [color={rgb, 255:red, 74; green, 74; blue, 74 }  ,draw opacity=1 ] [dash pattern={on 4.5pt off 4.5pt}]  (1992.97,8472.79) -- (2039.47,8472.79) ;
\draw [shift={(2042.47,8472.79)}, rotate = 180] [fill={rgb, 255:red, 74; green, 74; blue, 74 }  ,fill opacity=1 ][line width=0.08]  [draw opacity=0] (10.72,-5.15) -- (0,0) -- (10.72,5.15) -- (7.12,0) -- cycle    ;
\draw [color={rgb, 255:red, 74; green, 74; blue, 74 }  ,draw opacity=1 ][fill={rgb, 255:red, 0; green, 0; blue, 0 }  ,fill opacity=1 ] [dash pattern={on 4.5pt off 4.5pt}]  (1992.97,8472.79) -- (1992.97,8426.3) ;
\draw [shift={(1992.97,8423.3)}, rotate = 450] [fill={rgb, 255:red, 74; green, 74; blue, 74 }  ,fill opacity=1 ][line width=0.08]  [draw opacity=0] (10.72,-5.15) -- (0,0) -- (10.72,5.15) -- (7.12,0) -- cycle    ;
\draw  (1918.73,8547.04) -- (2067.22,8547.04)(1918.73,8398.55) -- (1918.73,8547.04) -- cycle (2060.22,8542.04) -- (2067.22,8547.04) -- (2060.22,8552.04) (1913.73,8405.55) -- (1918.73,8398.55) -- (1923.73,8405.55)  ;
\draw  [fill={rgb, 255:red, 0; green, 0; blue, 0 }  ,fill opacity=1 ] (1991.06,8469.61) .. controls (1992.82,8468.56) and (1995.1,8469.13) .. (1996.15,8470.88) .. controls (1997.21,8472.64) and (1996.64,8474.92) .. (1994.88,8475.97) .. controls (1993.13,8477.03) and (1990.85,8476.46) .. (1989.79,8474.7) .. controls (1988.74,8472.94) and (1989.31,8470.67) .. (1991.06,8469.61) -- cycle ;

\draw (1939.79,8472.79) node [anchor=east] [inner sep=0.75pt]  [color={rgb, 255:red, 0; green, 0; blue, 255 }  ,opacity=1 ]  {$\lambda _{3}$};
\draw (1992.97,8522.68+2) node [anchor=north] [inner sep=0.75pt]  [color={rgb, 255:red, 0; green, 0; blue, 255 }  ,opacity=1 ]  {$\lambda _{5}$};
\draw (2044.91,8472.79) node [anchor=west] [inner sep=0.75pt]  [color={rgb, 255:red, 0; green, 0; blue, 255 }  ,opacity=1 ]  {${\lambda _{4}}$};
\draw (1992.97,8418.43) node [anchor=south] [inner sep=0.75pt]  [color={rgb, 255:red, 0; green, 0; blue, 255 }  ,opacity=1 ]  {$\lambda _{2}$};
\draw (2017.72,8475.91-5) node [anchor=south] [inner sep=0.75pt]  [font=\scriptsize,color={rgb, 255:red, 74; green, 74; blue, 74 }  ,opacity=1 ]  {$\vec{v}_{a}$};
\draw (1968.22,8475.91-5) node [anchor=south] [inner sep=0.75pt]  [font=\scriptsize,color={rgb, 255:red, 74; green, 74; blue, 74 }  ,opacity=1 ]  {$\vec{v}_{b}$};
\draw (1990.32+5,8448.04) node [anchor=west] [inner sep=0.75pt]  [font=\scriptsize,color={rgb, 255:red, 74; green, 74; blue, 74 }  ,opacity=1 ]  {$\vec{v}_{d}$};
\draw (1995.47-5,8497.54) node [anchor=east] [inner sep=0.75pt]  [font=\scriptsize,color={rgb, 255:red, 74; green, 74; blue, 74 }  ,opacity=1 ]  {$\vec{v}_{c}$};
\draw (2064.72,8543.61+10) node [anchor=north west][inner sep=0.75pt]    {$q$};
\draw (1921.22,8401.98-10) node [anchor=south east] [inner sep=0.75pt]    {$a$};
\draw (1944.16,8560.89) node [anchor=north] [inner sep=0.75pt]    {$-2$};
\draw (1993.66,8560.89) node [anchor=north] [inner sep=0.75pt]    {$0$};
\draw (2042.47,8560.89) node [anchor=north] [inner sep=0.75pt]    {$2$};
\draw (1905.86,8522.29) node [anchor=east] [inner sep=0.75pt]    {$-2$};
\draw (1904.61,8472.79) node [anchor=east] [inner sep=0.75pt]    {$0$};
\draw (1904.61,8423.3) node [anchor=east] [inner sep=0.75pt]    {$2$};
\draw (1995.74,8474.47-5) node [anchor=south east] [inner sep=0.75pt]  [color={rgb, 255:red, 0; green, 0; blue, 255 }  ,opacity=1 ]  {${\lambda _{1}}$};

\end{tikzpicture}
    \caption{The set of generators of the~uncolored HOMFLY-PT homology for $4_1$ knot and the~parallelogram corresponding to the~pairing $\lambda_2\lambda_5=\lambda_3\lambda_4$
    }
    \label{fig:parallelogram_explained}
\end{figure}
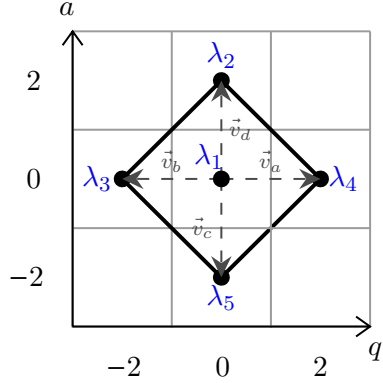

Furthermore, symmetries of quivers are tightly related to homological diagrams for knots, providing a~neat illustration of the~aforementioned conditions.
After the~change of variables (\ref{eq:KQcorrespondence-x}), each pairing $\lambda_a\lambda_b=\lambda_c\lambda_d$  gives the~vector identity $\vec{v}_a+\vec{v}_b=\vec{v}_c+\vec{v}_d$, where $\vec{v}_i=(q_i,a_i)$ is a~vector of homological degrees of the~generator $i$. This identity can be interpreted as a~requirement that the~centers of mass for pairs of nodes $\{a,b\}$ and $\{c,d\}$ coincide (assuming that masses of all nodes are equal).
We visualize it as a~parallelogram with the~diagonals $ab$ and $cd$, see figure~\ref{fig:parallelogram_explained}.

The remaining constraints~(\ref{eq:theorem first case}) or~(\ref{eq:theorem second case}) also have a~nice pictorial representation in terms of generators  of the~$S^{r}$-colored HOMFLY-PT homology. The~case $r=1$ corresponds to the~uncolored homology, encoded in the~linear term of the~quiver generating series and thus depending only on the~numbers of loops in $Q$. It suits well for visualizing the~pairing, but not the~rest of constraints. However, the~case $r=2$ involves the~quadratic term of the~quiver series and therefore depends on all entries of the~quiver matrix.
Moreover, there exists a~well-defined surjective map $Q_0\times Q_0 \rightarrow \mathscr{G}_{2}$ coming from the~knots-quivers change of variables. 

\begin{figure}[h!]
    \centering
    \input{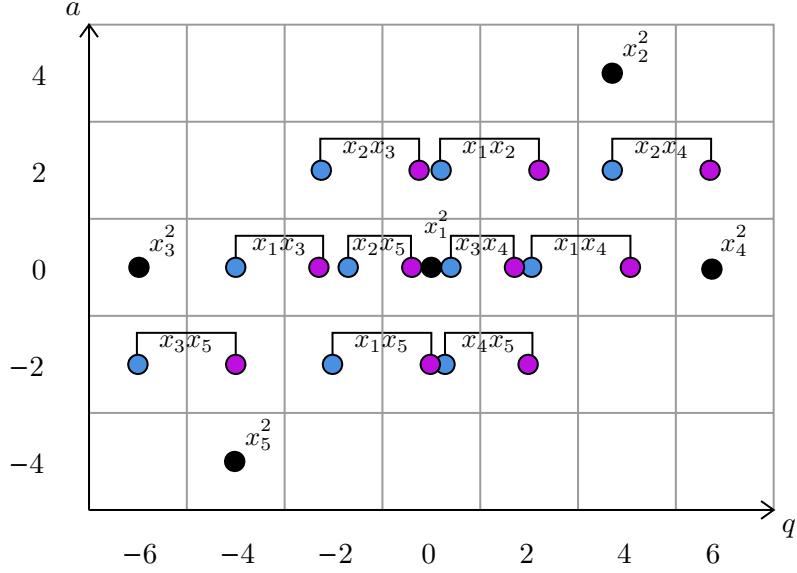}
    \caption{The set of generators of the~$S^2$-colored HOMFLY-PT homology for $4_1$ knot
    (the labels $x_ix_j$ are consistent with the~labels in figure~\ref{fig:parallelogram_explained}).}
    \label{fig:colored_homology_2}
\end{figure}

\begin{figure}[H]
    \centering
    \input{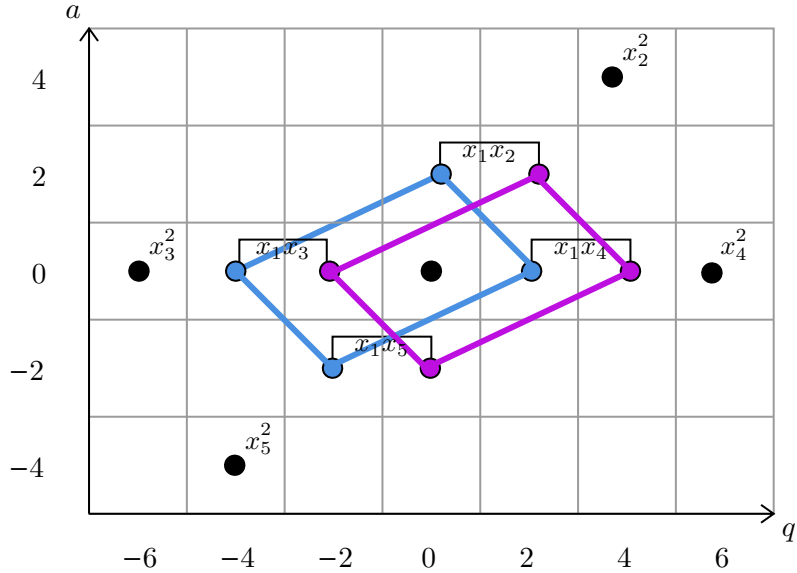}
    \caption{The constraint $C_{12}+C_{15}=C_{13}+C_{14}$ as a~parallelogram rule. 
    There are cancellations
    when equating the~sums of the~$q$- and $a$-degrees of $x_1x_2,x_1x_5$ and $x_1x_3,x_1x_4$, since $\lambda_2\lambda_5=\lambda_3\lambda_4$ implies $q_2+q_5=q_3+q_4$ and $a_2+a_5=a_3+a_4$.
    The~constraint holds only if the~corresponding sums of vectors agree
    (simultaneously for the~blue and purple quadruples of nodes).
    }
    \label{fig:colored_homology_3}
\end{figure}

For example, the~$S^2$-colored homology for $4_1$ knot is shown in figure~\ref{fig:colored_homology_2}. There are 3~kinds of generators: 5~black nodes are in one-to-one correspondence with $x_i^2,\ i=1\dots 5$. Blue and purple nodes correspond to $x_ix_j$ with $i\neq j$,
and for each pair $(i,j)$ there are exactly 2 generators, which we connect by an~arc. The~distinction between blue and purple nodes is justified by taking the~common denominator in the~quadratic term of the~quiver series. Each term $x_ix_j$ is multiplied by $(1+q^2)$, therefore contributing twice
to the~colored superpolynomial. The~blue node has the~$q$-degree $q_i+q_j+C_{ii}+2C_{ij}+C_{jj}$, while the~purple one is shifted by two: $q_i+q_j+C_{ii}+2C_{ij}+C_{jj}+2$.
Having in mind the~pairing condition inducing cancellations of all terms except those corresponding to arrows between different nodes ($2C_{ij}$), we can visualize any constraint of the~form $C_{is}+C_{js}=C_{ks}+C_{ls}$ as a~parallelogram connecting nodes with the~same color.
For example, the~constraint $C_{12}+C_{15}=C_{13}+C_{14}$ is visualized in figure~\ref{fig:colored_homology_3}. 

\subsection{Proof of the~local equivalence theorem}  \label{sec-proof}

Let us prove the~theorem~\ref{thm:main}. Since disjoint transpositions described there are independent, we 
can consider a~general form of one such transposition and show that it preserves the~generating function. This automatically implies that if $Q$ and $Q'$ are connected by a~sequence of such transformations, then they correspond to the~same knot. 

Therefore, without loss of generality, we assume that $Q$ corresponds to $K$, $Q'_0=Q_0$, $\lambda'_i=\lambda_i$  $\forall i\in Q_0$, and we have $C'_{ij}=C_{ij}$ except one transposition $C_{ab}\leftrightarrow C_{cd}$ for some pairwise different $a,b,c,d\in Q_{0}$.
We also require 
\begin{equation}
    \lambda_{a}\lambda_{b}=\lambda_{c}\lambda_{d},
    \quad 
    C_{cd}=C_{ab}-1, \quad C_{ci}+C_{di}=C_{ai}+C_{bi}-\delta_{ai}-\delta_{bi},\quad i\in Q_{0}
\end{equation}
and analogous constraints for $C'$ (the case $C_{ab}=C_{cd}-1$ can be covered by changing labels $ab\leftrightarrow cd$ in the~whole argument). 

We want to show that $Q'$ also corresponds to $K$. We will do it by connecting $Q'$ with $Q$ by transformations preserving the~motivic generating functions, namely unlinking nodes $a,b$ in $Q$ and nodes $c,d$ in $Q'$ (the invariance of generating function under these transformations is assured by theorem~\ref{thm:unlinking}).

From definition~\ref{def:unlinking} we have
\begin{align}
\widetilde{C}_{ij} & =C_{ij}\quad\forall i,j\in Q_{0}\backslash\{a,b\} & \widetilde{C}\,'_{ij} & =C'_{ij}\quad\forall i,j\in Q_{0}\backslash\{c,d\}\nonumber\\
\widetilde{C}_{ab} & =C_{ab}-1 & \widetilde{C}\,'_{cd} & =C'_{cd}-1 \\
\widetilde{C}_{in} & =C_{ai}+C_{bi}-\delta_{ai}-\delta_{bi}, & \widetilde{C}\,'_{in} & =C'_{ci}+C'_{di}-\delta_{ci}-\delta_{di},\nonumber\\
\widetilde{C}_{nn} & =C_{aa}+2C_{ab}+C_{bb}-1, & \widetilde{C}\,'_{nn} & =C'_{cc}+2C'_{cd}+C'_{dd}-1. \nonumber
\end{align}
In consequence 
\begin{align}
\widetilde{C}\,'_{ab} & =C'_{ab}=C_{cd}=C_{ab}-1=\widetilde{C}_{ab},\nonumber \\
\widetilde{C}\,'_{cd} & =C'_{cd}-1=C_{ab}-1=C_{cd}=\widetilde{C}{}_{cd},\nonumber \\
\widetilde{C}\,'_{an} & =C'_{ac}+C'_{ad}=C_{ac}+C_{ad}=C_{aa}+C_{ab}-1=\widetilde{C}_{an},\nonumber \\
\widetilde{C}\,'_{bn} & =C'_{bc}+C'_{bd}=C_{bc}+C_{bd}=C_{ab}+C_{bb}-1=\widetilde{C}_{bn}, \nonumber\\
\widetilde{C}'_{cn} & =C'_{cc}+C'_{cd}-1=C'_{ac}+C'_{bc}=C_{ac}+C_{bc}=\widetilde{C}_{cn}, \\
\widetilde{C}\,'_{dn} & =C'_{cd}+C'_{dd}-1=C'_{ad}+C'_{bd}=C_{ad}+C_{bd}=\widetilde{C}_{dn},\nonumber \\
\widetilde{C}\,'_{in} & =C'_{ci}+C'_{di}=C_{ci}+C_{di}=C_{ai}+C_{bi}=\widetilde{C}_{in},\quad\forall i\in Q_{0}\backslash\{a,b,c,d\},\nonumber \\
\widetilde{C}\,'_{nn} & =C'_{cc}+2C'_{cd}+C'_{dd}-1=C_{cc}+2C_{ab}+C_{dd}-1=C_{cc}+2C_{ab}+C_{dd}-1=\widetilde{C}_{nn},\nonumber \\
\widetilde{C}\,'_{ij} & =C'_{ij}=C_{ij}=\widetilde{C}_{ij}\quad\textrm{for all other cases,}\nonumber 
\end{align}
which can be summarized simply as $\widetilde{Q}\,'=\widetilde{Q}$. 

In our unlinking of $Q'$ and $Q$ we have a~freedom to choose the~knots-quivers change of variables for the~new nodes (for the~old ones we have $\lambda'_{i}=\lambda_{i}$). We take
\begin{equation}
\widetilde{\lambda}'_{n}=q^{-1}\lambda_{c}\lambda_{d}=q^{-1}\lambda_{a}\lambda_{b}=\widetilde{\lambda}_{n},
\end{equation}
and use theorem~\ref{thm:unlinking} to get
\[
\left.P_{Q'}(\boldsymbol{x},q)\right|_{x_{i}=x\lambda'_{i}}=\left.P_{\widetilde{Q}'}(\boldsymbol{x},q)\right|_{x_{i}=x\lambda'_{i},\;x_{n}=x\widetilde{\lambda}'_{n}}=\left.P_{\widetilde{Q}}(\boldsymbol{x},q)\right|_{x_{i}=x\lambda_{i},\;x_{n}=x\widetilde{\lambda}_{n}}=\left.P_{Q}(\boldsymbol{x},q)\right|_{x_{i}=x\lambda_{i}}.
\]
Therefore
\begin{equation}
    \left.P_{Q'}(\boldsymbol{x},q)\right|_{\boldsymbol{x}=x\boldsymbol{\lambda}'}=\left.P_{Q}(\boldsymbol{x},q)\right|_{\boldsymbol{x}=x\boldsymbol{\lambda}}=P_{K}(x,a,q,t),
\end{equation}
so $Q'$ also corresponds to $K$, as we wanted to show.


\section{Global structure and permutohedra graphs}
\label{sec:global structure}

In the~previous section we found transformations that produce equivalent quivers and conditions they satisfy. This fact enables us to systematically determine equivalent quivers for a~given knot: starting from some particular quiver we can consider all possible transpositions of its matrix elements, and identify those that satisfy conditions of theorem~\ref{thm:main} and thus yield equivalent quivers. Repeating this procedure for each newly found equivalent quiver, after a~finite number of steps we obtain a~closed and connected network with an~intricate structure. (Recall that in principle there might exist other equivalent quivers, which are not related by a~series of transpositions from theorem~\ref{thm:main} -- e.g. they might be related by a~cyclic permutation of length larger than 2, such that some transpositions of elements of quiver matrix, which arise from a~decomposition of such a~permutation, do not preserve the~partition function. However, we conjectured that such equivalent quivers do not arise, and we do not focus on them in the~rest of this work.)

In order to reveal the~structure of the~network of equivalent quivers mentioned above, it is of advantage to assemble these quivers in one graph, such that each vertex of this graph corresponds to one quiver, and two vertices are connected by an~edge if two corresponding quivers differ by one transposition of non-diagonal elements. Examples of such graphs are shown in figures~\ref{fig:3DGraph91} and~\ref{fig:3DGraph111} (for knots $9_1$ and $11_1$), and in section~\ref{SectionCaseStudies} for several other knots. One immediately observes that these graphs are built from smaller building blocks, which are combinatorial structures known as permutohedra. Various permutohedra are glued to each other and form a~connected graph representing all equivalent quivers, which we refer to as a~permutohedra graph in what follows. In this section we explain why equivalent quivers arise in families that form permutohedra, and how their structure follows from local properties revealed in theorem~\ref{thm:main}. In the~next section we illustrate these structures in detail in several explicit examples. 


\subsection{Permutohedra -- what they are and why they arise}

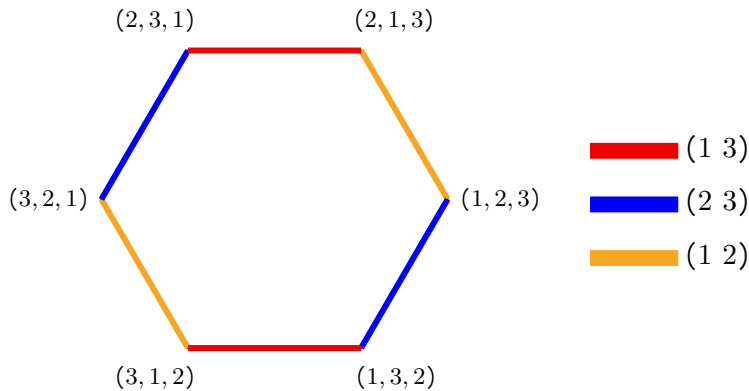
\begin{figure}[t]
    \centering
    \tikzset{every picture/.style={line width=0.75pt}} 

\begin{tikzpicture}[x=0.75pt,y=0.75pt,yscale=-1,xscale=1]

\draw   (1602.26,147.45) -- (1559.06,222.28) -- (1472.65,222.28) -- (1429.45,147.45) -- (1472.65,72.62) -- (1559.06,72.62) -- cycle ;
\draw [color={rgb, 255:red, 245; green, 166; blue, 35 }  ,draw opacity=1 ][line width=2.25]    (1602.26,147.45) -- (1559.06,72.62) ;
\draw [color={rgb, 255:red, 245; green, 166; blue, 35 }  ,draw opacity=1 ][line width=2.25]    (1472.65,222.28) -- (1429.45,147.45) ;
\draw [color={rgb, 255:red, 0; green, 0; blue, 255 }  ,draw opacity=1 ][line width=2.25]    (1559.06,222.28) -- (1602.26,147.45) ;
\draw [color={rgb, 255:red, 0; green, 0; blue, 255 }  ,draw opacity=1 ][line width=2.25]    (1429.45,147.45) -- (1472.65,72.62) ;
\draw [color={rgb, 255:red, 240; green, 0; blue, 0 }  ,draw opacity=1 ][line width=2.25]    (1472.65,72.62) -- (1559.06,72.62) ;
\draw [color={rgb, 255:red, 240; green, 0; blue, 0 }  ,draw opacity=1 ][line width=2.25]    (1472.65,222.28) -- (1518.92,222.28) -- (1559.06,222.28) ;
\draw [color={rgb, 255:red, 240; green, 0; blue, 0 }  ,draw opacity=1 ][line width=6]    (1673.35,123.07) -- (1717.42,123.07) ;
\draw [color={rgb, 255:red, 0; green, 0; blue, 255 }  ,draw opacity=1 ][line width=6]    (1673.35,150.04) -- (1717.42,150.04) ;
\draw [color={rgb, 255:red, 245; green, 166; blue, 35 }  ,draw opacity=1 ][line width=6]    (1673.35,177.02) -- (1717.42,177.02) ;

\draw (1607.26,147.45) node [anchor=west] [inner sep=0.75pt]  [font=\footnotesize]  {$( 1,2,3)$};
\draw (1554.06,57.62) node [anchor=west] [inner sep=0.75pt]  [font=\footnotesize]  {$( 2,1,3)$};
\draw (1555.06,237.28) node [anchor=west] [inner sep=0.75pt]  [font=\footnotesize]  {$( 1,3,2)$};
\draw (1477.65,57.62) node [anchor=east] [inner sep=0.75pt]  [font=\footnotesize]  {$( 2,3,1)$};
\draw (1424.45,147.45) node [anchor=east] [inner sep=0.75pt]  [font=\footnotesize]  {$( 3,2,1)$};
\draw (1477.65,237.28) node [anchor=east] [inner sep=0.75pt]  [font=\footnotesize]  {$( 3,1,2)$};
\draw (1719.35,122.07) node [anchor=west] [inner sep=0.75pt]    {$( 1\ 3)$};
\draw (1719.35,149.04) node [anchor=west] [inner sep=0.75pt]    {$( 2\ 3)$};
\draw (1719.35,176.02) node [anchor=west] [inner sep=0.75pt]    {$( 1\ 2)$};

\end{tikzpicture}
    \caption{Permutohedron $\Pi_3$. Each vertex represents a~particular permutation of 3~elements. Two vertices are connected by an~edge if corresponding permutations differ by a~flip of immediate neighbors. There are 3 types of flips, $(1\ 2), (2\ 3)$ and $(1\ 3)$, which are represented by different colors in the~figure.}
    \label{fig:hexagon}
\end{figure}

\begin{figure}[h!]
    \centering
 \input{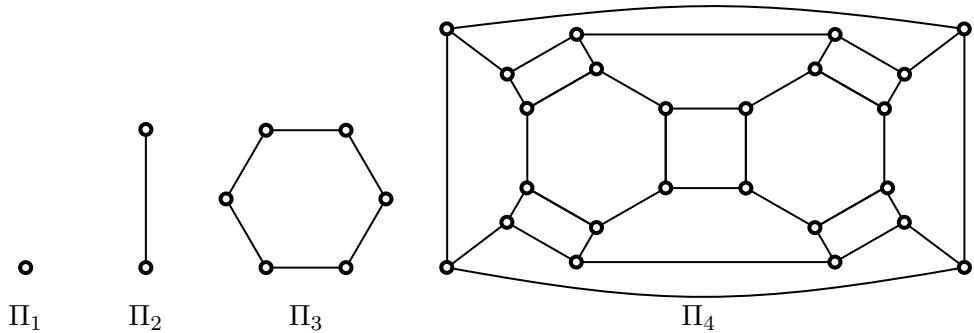}
    \caption{Planar realizations of permutohedra $\Pi_n$ of orders 1,2,3,4. One quadrangular face of $\Pi_4$ is represented by an~external region. Three-dimensional representation of permutohedron $\Pi_4$ is shown in figure~\ref{fig:permutohedrapi4}.}
    \label{fig:permutohedra}
\end{figure}

To start with, recall that a~permutohedron  of order $n$, denoted $\Pi_n$, is an~$(n-1)$-dimensional polytope whose vertices represent permutations of $n$ objects $\{1,\ldots,n\}$ and edges correspond to flips (transpositions) of adjacent neighbors~\cite{ziegler_lectures_1995,aguiar_hopf_2017}. The~permutohedron $\Pi_n$ has thus $n!$ vertices and each vertex has $n-1$ immediate neighbors. $\Pi_n$ has also $(n-1)n!/2$ edges; each edge corresponds to one of $n(n-1)/2$ types of flips $(i\ j)$ (for $1\leq i<j\leq n$). We call these operations flips in order to distinguish them from transpositions of elements of quiver matrices; as we will see, transpositions in quiver matrices are simply manifestations of certain underlying flips. The~permutohedron $\Pi_3$ is a~hexagon, see figure~\ref{fig:hexagon}. $\Pi_4$ is a~(3-dimensional) truncated octahedron that consists of $4!=24$ vertices. It has 36 edges of 6~different types, such that 3 edges meet at each vertex, and its faces form 6~quadrangles and 8~hexagons, see figure~\ref{fig:permutohedrapi4}. Planar realizations of $\Pi_n$ for $n=1,2,3,4$ are shown in figure~\ref{fig:permutohedra}. 

Let us explain now why certain families of equivalent quivers form permutohedra. To get some intuition, it is of advantage to understand it first as a~consequence of a~particular structure of generating functions of colored superpolynomials; in section~\ref{sec:local-global} we show how this structure arises from the~local properties revealed in theorem~\ref{thm:main}. We find that instead of writing a~generating function of colored superpolynomials in a~form of the~generating series~(\ref{eq:KQcorrespondence}) for a~quiver of size $m$, it can be written in an~intermediate form 
\begin{equation}  \label{PK-Pi}
    P_K(x,a,q,t) = \sum_{\check{d}_1,\ldots,\check{d}_{m-n}\geq 0}(-q)^{\sum_{i,j}\check{C}_{ij}\check{d}_i\check{d}_j}\frac{\check{x}_1^{\check{d}_1}\cdots \check{x}_{m-n}^{\check{d}_{m-n}}}{(q^{2};q^{2})_{\check{d}_1}\cdots (q^{2};q^{2})_{\check{d}_{m-n}}} 
    \Pi_{\check{d}_1,\ldots,\check{d}_n} \Big|_{\check{x}_i=x\check{\lambda}_i},
\end{equation}
for $2n\leq m$ and with the~following properties. The~first terms under the~sum take the~same form as the~summand in the~usual quiver generating series~(\ref{P-Q}), however they are associated to a~novel quiver of size $m-n$ that we call a~prequiver and denote its matrix by~$\check{C}$. Then, it is the~factor $\Pi_{\check{d}_1,\ldots,\check{d}_n}$ which is responsible for the~appearance of all equivalent quivers associated to a~particular permutohedron; note that it has only $n$ labels $\check{d}_1,\ldots,\check{d}_n$,  and we require that (combined with the~first $n$ $q$-Pochhammers from the~denominator) it has the~structure 
\begin{equation}    \label{prequiver-permute}
    \frac{\Pi_{\check{d}_1,\ldots,\check{d}_n}}{(q^{2};q^{2})_{\check{d}_1}\cdots (q^{2};q^{2})_{\check{d}_{n}}} = \sum_{\check{d}_1=\alpha_1+\beta_1}\cdots \sum_{\check{d}_n=\alpha_n+\beta_n} \frac{ (-q)^{2\sum_{i<j}\beta_{i}\alpha_{j}+ \pi_2(\alpha_1,\ldots,\alpha_n;\beta_1,\ldots,\beta_n)}\kappa^{\beta_1+\ldots+\beta_n}}{(q^{2};q^{2})_{\alpha_{1}}(q^{2};q^{2})_{\beta_{1}}\cdots(q^{2};q^{2})_{\alpha_{n}}(q^{2};q^{2})_{\beta_{n}}},
\end{equation}
where $\pi_2(\alpha_1,\ldots,\alpha_n;\beta_1,\ldots,\beta_n)$ is a~purely quadratic polynomial in $\alpha_i$'s, $\beta_j$'s, and other $\check{d}_{k}$'s (for $k>n)$, that is symmetric in $(\alpha_1,\ldots,\alpha_n)$ and (independently) in $(\beta_1,\ldots,\beta_n)$; $\kappa$ is an~extra parameter. Furthermore, we impose the~invariance of the~above expression under any permutation $\sigma\in S_n$ of indices $\{1,\ldots,n\}$, so that the~whole $\Pi_{\check{d}_1,\ldots,\check{d}_n}$ is symmetric in all~$\check{d}_1,\ldots,\check{d}_n$. Note that most of the~above expression on the~right-hand side, i.e. the~terms symmetric in $\alpha_i$'s and $\beta_j$'s, as well as the~defining relations $\check{d}_i=\alpha_i+\beta_i$, are already invariant under permutation of the~indices. The~only non-invariant term is $\sum_{i<j} \beta_i \alpha_j$, so in other words we impose that the~above expression is invariant if we replace this term by $\sum_{i<j} \beta_{\sigma(i)} \alpha_{\sigma(j)}$, for any permutation $\sigma$.

Below we provide specific forms of $\Pi_{\check{d}_1,\ldots,\check{d}_n}$, including symmetric polynomials $\pi_2$, that have the~above properties. At this stage let us stress that it is the~form of the~term $\sum_{i<j} \beta_{\sigma(i)} \alpha_{\sigma(j)}$ that uniquely determines a~permutation $\sigma$ and is responsible for the~appearance of a~permutohedron. First, a~permutation $\sigma$ is determined by a~set of its inversions, i.e. a~set of all pairs $(\sigma(i),\sigma(j))$, such that $i<j$ and $\sigma(i)>\sigma(j)$. We can therefore treat symbols $\beta$ and $\alpha$ as determining respectively the~first and the~second element of a~given pair $(\sigma(i),\sigma(j))$. For example, the~term $\sum_{i<j} \beta_{i} \alpha_{j}$ encodes the~trivial permutation. Any other permutation can be uniquely encoded by inverting labels in appropriate summands in $\sum_{i<j} \beta_{i} \alpha_{j}$. Therefore, if we insist that~(\ref{prequiver-permute}) is invariant under all permutations of indices $\{1,\ldots,n\}$, this means that in fact we can consider $n!$ expressions that are in one-to-one correspondence with permutations encoded in the~terms $\sum_{i<j} \beta_{\sigma(i)} \alpha_{\sigma(j)}$, and can be associated to vertices of a~permutohedron $\Pi_n$. Such a~permutohedron has $n(n-1)/2$ types of edges (denoted by different colors in various figures in this paper), which correspond to all transpositions $(k\ l)$, for $1\leq k < l\leq n$. However, at a~given vertex, corresponding to the~permutation $\sigma$ and the~term $\sum_{i<j} \beta_{\sigma(i)} \alpha_{\sigma(j)}$, only $n-1$ edges meet. They correspond to transpositions of adjacent elements that change only one summand in the~expression $\sum_{i<j} \beta_{\sigma(i)} \alpha_{\sigma(j)}$. Let us see it on the~example of a~vertex corresponding to the~trivial permutation, represented by $\sum_{i<j} \beta_{i} \alpha_{j}$, and  $n-1$ edges corresponding to transpositions of neighboring elements $\tau=(k\ (k+1)),\;k=1,\ldots,n-1$. In that case the~only difference between $\sum_{i<j} \beta_{i} \alpha_{j}$ and $\sum_{i<j} \beta_{\tau(i)} \alpha_{\tau(j)}$ amounts to replacing precisely one summand $\beta_k\alpha_{k+1}$ by $\beta_{k+1}\alpha_k$. This is why a~transformation of one term $\beta_k\alpha_{k+1}$ into $\beta_{k+1}\alpha_k$ (for $k=1,\ldots,n-1$) in (\ref{prequiver-permute}) is represented by one edge of a~permutohedron. Similarly, $n-1$ edges meeting at any other vertex that represents a~permutation $\sigma$, correspond to those transpositions $(k\ l)$ that affect precisely one term in $\sum_{i<j} \beta_{\sigma(i)} \alpha_{\sigma(j)}$. All this is also a~manifestation of the~well known fact that a~permutohedron is the~Hasse diagram of a~set of appropriately ordered inversions.

Furthermore, let us explain how the~prequiver $\check{C}$ introduced in~(\ref{PK-Pi}), combined with $\Pi_{\check{d}_1,\ldots,\check{d}_n}$, gives rise to the~original quiver $C$ of size $m$ and a~number of its equivalent companions. First, in the~expression~(\ref{PK-Pi}) there are $(m-n)$ $q$-Pochhammers $(q^2;q^2)_{\check{d}_i}$. In~(\ref{prequiver-permute}), $n$ of them are combined with $\Pi_{\check{d}_1,\ldots,\check{d}_n}$ and get split into pairs $(q^2;q^2)_{\alpha_i}(q^2;q^2)_{\beta_i}$. This produces $n$ new $q$-Pochhammers, and altogether we get $m$ independent $q$-Pochhammers that correspond to $m$ nodes of a~quiver $C$ that we are after. The~prequiver term $(-q)^{\sum_{i,j}\check{C}_{ij}\check{d}_i\check{d}_j}$ in~(\ref{PK-Pi}) together with $(-q)^{2\sum_{i<j} \beta_i\alpha_j + \pi_2(\alpha_1,\ldots,\alpha_n;\beta_1,\ldots,\beta_n)}$ give rise to an~overall quadratic expression that defines the~full quiver matrix $C$. The~terms $\kappa^{\beta_1+\ldots+\beta_n}$ get absorbed into the~first $n$ generating parameters: $\check{x}_1^{\check{d}_1}\cdots \check{x}_{n}^{\check{d}_{n}} \kappa^{\beta_1+\ldots \beta_n}= \check{x}_1^{\alpha_1}(\check{x}_1\kappa)^{\beta_1}\cdots \check{x}_{n}^{\alpha_{n}}(\check{x}_{n}\kappa)^{\beta_{n}}$.

In this way we obtain a~quiver generating function for the~quiver of size $m$ encoded in a~matrix $C$ that we are interested in. To see it more clearly and to make contact with the~notation in~(\ref{P-Q}), we can rename summation variables: for example identify all $\check{d}_k$ ($k=n+1,\ldots,m-n$) with $d_{n+k}$, and let $d_{2i-1}\equiv \alpha_i$ and $d_{2i}\equiv \beta_i$; in addition, identify $\check{x}_k$ with $x_{n+k}$ for $k=n+1,\ldots,m-n$, and let $x_{2i-1}\equiv \check{x}_i$ and $x_{2i} \equiv \check{x}_i \kappa$. This gives rise to generating parameters as in~(\ref{eq:KQcorrespondence-x}). We refer to the~process of replacing first $n$ nodes by $2n$ nodes, which is a~manifestation of~(\ref{prequiver-permute}), as splitting, while the~remaining $(m-2n)$ nodes of the~quiver $C$ we call spectators. Under this relabeling, for a~vertex representing the~permutation $\sigma$, a~flip of the~term $\beta_k\alpha_{l}$ (in the~sum $\sum_{i<j} \beta_{\sigma(i)}\alpha_{\sigma(j)})$ into $\beta_{l}\alpha_k$ translates into a~flip of $d_{2k}d_{2l-1}$ into $d_{2l}d_{2k-1}$, which encodes a~transposition of elements $C_{2k,2l-1}$ and $C_{2l,2k-1}$ (that we considered in theorem~\ref{thm:main}) at the~level of the~matrix $C$. For each vertex there are $n-1$ of such transpositions, which on one hand correspond to $n-1$ equivalent matrices related by one transposition to a~given matrix $C$, and on the~other hand correspond to $n-1$ edges meeting at each vertex of a~permutohedron $\Pi_n$. Note that we can make any other identification of indices that would amount to a~permutation of all variables $d_i$, and thus would yield a~permutation of rows and columns of the~matrix $C$; in particular, in section~\ref{SectionCaseStudies} we identify a~prequiver part as corresponding to the~last $n$ rather than first $n$ indices as above. 

Let us also note the~following interesting feature. Not only the~generating function of colored HOMFLY-PT polynomials, but also the~generating function of colored superpolynomials is expected to take form~(\ref{PK-Pi}). This means that the~full dependence on the~parameter~$a$, as well as~$t$, is captured by the~parameter $\kappa$ that appears in the~factor $\Pi_{\check{d}_1,\ldots,\check{d}_n}$ in~(\ref{prequiver-permute}), and in $\check{\lambda}_i$ that enter the~identification of generating parameters $\check{x}_i=x\check{\lambda}_i$. Note that $\check{\lambda}_i$ are just a~subset of all $\lambda_j$, so that $\lambda_j=\check{\lambda}_i$ for appropriate values of $i$, and the~remaining $\lambda_j$ arise from a~simple rescaling $\lambda_j=\kappa\check{\lambda}_k$ (for appropriate $k$ and $j$). As we will see in what follows, $\kappa$ is a~monomial of the~form $\kappa= a^{\kappa_a}q^{\kappa_q}(-t)^{\kappa_t}$. Also note that $\check{\lambda}_i$ are different for various realizations~(\ref{PK-Pi}) (corresponding to various permutohedra) for a~given knot, because they correspond to various subsets of all $\lambda_i$ that are associated to the~nodes that arise in a~given prequiver. In consequence, the~values of $\kappa$ are also different for various representations~(\ref{PK-Pi}) of the~same knot. It would be interesting to understand better why a~dependence on~$a$ and~$t$ is simply captured by $\kappa= a^{\kappa_a}q^{\kappa_q}(-t)^{\kappa_t}$ and $\check{\lambda}_i$, and possibly how it arises from properties of HOMFLY-PT homology. 

To sum up, after above identifications we obtain a~family of quiver generating functions for various quivers $C$ of size $m$ in the~standard form~(\ref{P-Q}), and with parameters $x_i$ appropriate for the~knots-quivers correspondence. The~family of quivers that we obtain is parametrized by all permutations $\sigma\in S_n$: the~combinations $\sum_{i<j} \beta_{\sigma(i)} \alpha_{\sigma(j)}$ for various $\sigma$ that appear in the~exponent of $(-q)$ affect the~form of the~matrix $C$ that we read off from quadratic terms, and thus give rise to $n!$ different but equivalent quivers, labeled by permutations of $n$ elements. This is why we can assign these quivers to vertices of permutohedron $\Pi_n$. An~edge of such a~permutohedron that represents a~flip (transposition) of two elements from the~set $\{1,\ldots,n\}$, at the~same time corresponds to a~transposition of certain two elements $C_{2k,2l-1}$ and $C_{2l,2k-1}$ of the~matrix $C$ that we analyzed in theorem~\ref{thm:main}.

The above analysis focuses on one permutohedron. However, typically we can write a~generating function of colored superpolynomials for a~given knot in the~form~(\ref{PK-Pi}) in several different ways, with different prequivers and terms $\Pi_{\check{d}_1,\ldots,\check{d}_n}$ for various choices of nodes. This gives rise to several permutohedra that encode all equivalent quivers for a~given knot. Some of these quivers are common between two (or more) permutohedra, therefore we obtain a~large connected graph made of several permutohedra glued together.


\subsection{Permutohedra from colored superpolynomials}

Let us now provide an~explicit form of~(\ref{prequiver-permute}). We stress that expressions given below naturally occur in formulae for colored superpolynomials, so it is useful to understand their role from the~perspective of equivalent quivers. First, we consider a~special case that arises from the~identification  $\Pi_{\check{d}_1,\ldots,\check{d}_n} = (\xi;q^2)_{\check{d}_1+\ldots+\check{d}_n}$, which is indeed familiar from various expressions for colored superpolynomials. We then have 
\begin{align}
    \frac{(\xi;q^2)_{\check{d}_1+\ldots+\check{d}_n}}{(q^2;q^2)_{\check{d}_{1}}\cdots(q^2;q^2)_{\check{d}_{n}}}&=  \sum\limits_{\alpha_{1}+\beta_{1}=\check{d}_{1}} \cdots \sum\limits_{\alpha_{n}+\beta_{n}=\check{d}_{n}} (-q)^{\beta_{1}^2+\ldots+\beta_{n}^2+2\sum_{i=1}^{n-1} \beta_{i+1}(\check{d}_{1}+\ldots + \check{d}_{i})} \times \nonumber\\
    & \quad \times \frac{\big(\xi q^{-1}\big)^{\beta_{1}+\cdots+\beta_n}}{(q^2;q^2)_{\alpha_{1}}(q^2;q^2)_{\beta_{1}}\cdots (q^2;q^2)_{\alpha_{n}}(q^2;q^2)_{\beta_{n}}},
    \label{eq:qpoch-sum-general-3}
\end{align}
which is proven in~\cite{KRSS1707long}. The~left-hand side is explicitly symmetric in $\check{d}_1,\ldots,\check{d}_n$, so the~above equality proves that the~right-hand side is also invariant under permutations of $\{1,\ldots,n\}$. In the~exponent of $(-q)$ we have $\sum_{i=1}^{n-1} \beta_{i+1}(\check{d}_{1}+\ldots + \check{d}_{i})=\sum_{i>j} \beta_i \alpha_j+\sum_{i>j} \beta_i \beta_j$, so the~first term $\sum_{i>j} \beta_i \alpha_j$ is responsible for the~permutohedron structure, while $\sum_{i>j} \beta_i \beta_j$ is the~second elementary symmetric polynomial, which is symmetric in all $\beta_i$ in agreement with~(\ref{prequiver-permute}). If $\xi$ is just a~constant (independent of $\check{d}_k$'s), we identify $\kappa=\xi q^{-1}$. 

An interesting version of~(\ref{eq:qpoch-sum-general-3}), that also appears in expressions for colored superpolynomials, arises for 
\begin{equation}
\xi=\kappa q^{2(h_{n+1} \check{d}_{n+1} + \ldots + h_{m-n} \check{d}_{m-n}) + 2k( \check{d}_1+\ldots+\check{d}_n)+1},
\end{equation}
where $h_s$ are fixed coefficients. Substituting such $\xi$ to~(\ref{eq:qpoch-sum-general-3}) also produces an~exponent of $q$ that is a~quadratic function, symmetric in $\alpha_i$'s and $\beta_j$'s. For brevity, let us type the~corresponding version of~(\ref{eq:qpoch-sum-general-3}) that involves just two summation variables $\check{d}_{i}$ and $\check{d}_{j}$ (which would correspond to a~single transposition) and one spectator node corresponding to the~variable $\check{d}_{s}$ and the~coefficient $h_s$
\begin{align}  \label{eq:qpoch-sum-general-4}
\frac{(\kappa q^{2h_{s}\check{d}_{s}+2k(\check{d}_{i}+\check{d}_{j})+1};q^{2})_{\check{d}_{i}+\check{d}_{j}}}{(q^{2};q^{2})_{\check{d}_{i}}(q^{2};q^{2})_{\check{d}_{j}}}=\sum\limits _{\alpha_{i}+\beta_{i}=\check{d}_{i}} & \sum\limits _{\alpha_{j}+\beta_{j}=\check{d}_{j}}(-q)^{\beta_{i}^{2}+\beta_{j}^{2}+2\beta_{i}(\alpha_{j}+\beta_{j})} \kappa^{\beta_i+\beta_j}\nonumber\\
 & \times\frac{q^{(2h_{s}\check{d}_{s}+2k(\check{d}_{i}+\check{d}_{j}))(\beta_{i}+\beta_j)}}{(q^{2};q^{2})_{\alpha_{i}}(q^{2};q^{2})_{\beta_{i}}(q^{2};q^{2})_{\alpha_{j}}(q^{2};q^{2})_{\beta_{j}}}
\\
= \sum\limits _{\alpha_{i}+\beta_{i}=\check{d}_{i}} & \sum\limits _{\alpha_{j}+\beta_{j}=\check{d}_{j}}(-q)^{(2k+1)\beta_{i}^{2}+(2k+1)\beta_{j}^{2}+2(k+1)\beta_{i}\alpha_{j}+2(2k+1)\beta_{i}\beta_{j}}\nonumber\\
 & \times\frac{(-q)^{2k(\beta_{i}\alpha_{i}+\beta_{j}\alpha_{i}+\beta_{j}\alpha_{j})+2h_{s}(\beta_{i}\check{d}_{s}+\beta_{j}\check{d}_{s})}\kappa^{\beta_{i}+\beta_{j}}}{(q^{2};q^{2})_{\alpha_{i}}(q^{2};q^{2})_{\beta_{i}}(q^{2};q^{2})_{\alpha_{j}}(q^{2};q^{2})_{\beta_{j}}}.\nonumber
\end{align}
From the~powers of $(-q)$ in the~last two lines above one can read off appropriate elements of the~resulting matrix $C$. Note that using indices $i$ and $j$ is helpful in understanding the~invariance of the~right-hand side of the~above expression under a~flip: if we identify $i=1$ and $j=2$ or $i=2$ and $j=1$, then the~left-hand side is clearly invariant, while the~only change on the~right amounts respectively to replacing $\beta_1\alpha_2$ by $\beta_2 \alpha_1$. 

Finally, the~most general form of~(\ref{prequiver-permute}) arises from   introducing an~arbitrary number of spectators and a~parameter $l$ in addition to $k$ in~(\ref{eq:qpoch-sum-general-4}) as follows:
\begin{equation}   \label{kl-formula}
\begin{split}
 \frac{\Pi_{\check{d}_1,\ldots,\check{d}_n}}{(q^{2};q^{2})_{\check{d}_1}\cdots (q^{2};q^{2})_{\check{d}_{n}}}& =\sum\limits _{\alpha_{1}+\beta_{1}=\check{d}_{1}}\cdots\sum\limits _{\alpha_{n}+\beta_{n}=\check{d}_{n}} 
\frac{\kappa^{\beta_1+\ldots+\beta_n}}{(q^{2};q^{2})_{\alpha_{1}}(q^{2};q^{2})_{\beta_{1}}\cdots(q^{2};q^{2})_{\alpha_{n}}(q^{2};q^{2})_{\beta_{n}}}\\
&\phantom{=}\times(-q)^{2\sum_{i<j}\beta_{i}\alpha_{j}+(2\sum_{s=n+1}^{m-n} h_{s}\check{d}_{s}+2k(\alpha_{1}+\ldots+\alpha_{n})+l(\beta_{1}+\ldots+\beta_{n}))(\beta_{1}+\ldots\beta_{n})},
\end{split}
\end{equation}
which is also invariant under permutations of indices $1,\ldots,n$, affecting the~form of the~term $\sum_{i<j}\beta_i \alpha_j$. If $l=2k+1$, the~above expression reduces to~(\ref{eq:qpoch-sum-general-4}) (generalized to $n$ summations), and then it can be written concisely using the~$q$-Pochhammer symbol. For $l\neq 2k+1$ we do not know if there is such a~concise manifestly symmetric representation, however we do not necessarily need it --
the crucial property is invariance of the~above expression under permutations of indices $\{1,\ldots,n\}$. In what follows we prove that~(\ref{kl-formula}) is indeed invariant under such permutations.


\subsection{Permutohedra from local equivalence}\label{sec:local-global}

In turn, we now show how permutohedra arise from the~local equivalence of quivers revealed in theorem~\ref{thm:main}, and in particular explain how~(\ref{kl-formula}) arises from this theorem (and thus has the~required symmetry properties).

Suppose that conditions~(\ref{eq:theorem first case}) of the~theorem~\ref{thm:main} are satisfied, so that two quivers related by a~transposition of elements $C_{ab}$ and $C_{cd}$ are equivalent. We now write the~quiver matrix $C$ in a~form that automatically implements these conditions. To this end, we focus first on the~$4\times 4$ submatrix of $C$ with elements $C_{ij}$ for $i,j=a,b,c,d$, and rewrite is as follows:
\begin{equation}\label{eq: quiver4x4}
   \left(
    \begin{array}{cc:cc}
        C_{aa} & C_{ad} & C_{ac} & \mybox{C_{ab}} \\
        C_{ad} & C_{dd} & \mybox{C_{cd}} & C_{bd} \\
        \hdashline
        C_{ac} & \mybox{C_{cd}} & C_{cc} & C_{bc} \\
        \mybox{C_{ab}} & C_{bd} & C_{bc} & C_{bb}
    \end{array}
    \right) =     \left(
    \begin{array}{cc:cc}
        C_{aa} & C_{aa} + k & C_{ac} & \mybox{C_{ac} + k} \\
        C_{aa} + k & C_{aa} + l & \mybox{C_{ac} + k + 1} & C_{ac}+l \\
        \hdashline
        C_{ac} & \mybox{C_{ac} + k + 1} & C_{cc} & C_{cc} + k \\
        \mybox{C_{ac} + k} & C_{ac} + l & C_{cc} + k & C_{cc} + l
    \end{array}
    \right).  
\end{equation}
In order to get the~right-hand side we introduced two parameters $k,l\in \mathbb{Z}$, defined such that $C_{ad} = C_{aa} + k$ and $C_{dd} = C_{aa} + l$. From the~second equation in~(\ref{eq:theorem first case}) with $i=a$ we then get $C_{ab} = C_{ac}+ C_{ad}-C_{aa} = C_{ac}+k$. Similarly, the~second equation in~(\ref{eq:theorem first case}) with $i=b$ takes form $C_{ad} + C_{bd} =  C_{dd} + C_{cd} - 1$, and combined with the~first equation in~(\ref{eq:theorem first case}) and the~above relations it yields $C_{bd} = C_{ac} + l$. Analogously,~(\ref{eq:theorem first case}) with $i=c$ and $i=d$ implies respectively $C_{cb} =  C_{cc} + k$ and $C_{bb} =  C_{cc} + l$. The~right-hand side of~(\ref{eq: quiver4x4}) follows from these relations and we rewrite it further as
\begin{equation}\label{eq:matrix solution to constraints}
  \left(
        \begin{array}{cc}
        C_{aa} & C_{ac} \\
        C_{ac} & C_{cc}
        \end{array}
    \right)
    \otimes
    \left(
        \begin{array}{cc}
        1 & 1 \\
        1 & 1
        \end{array}
    \right) + \left(
        \begin{array}{cc}
        1 & 1 \\
        1 & 1
        \end{array}
    \right)
    \otimes
    \left(
        \begin{array}{cc}
        0 & k \\
        k & l
        \end{array}
        \right) + 
     \left[   \left(
        \begin{array}{cc}
        0 & 1 \\
        0 & 0
        \end{array}
    \right)
    \otimes
    \left(
        \begin{array}{cc}
        0 & 0 \\
        1 & 0
        \end{array}
        \right) +
    \left(
        \begin{array}{cc}
        0 & 0 \\
        1 & 0
        \end{array}
        \right)
    \otimes
    \left(
        \begin{array}{cc}
        0 & 1 \\
        0 & 0
        \end{array}
    \right) \right]
\end{equation}
The terms in this expression turn out to have familiar interpretation. The~fist matrix is (an appropriate part of) the~prequiver $\check{C}$. In particular, if we rename summation variables as $(d_a,d_d,d_c,d_b) = (\alpha_a,\beta_a,\alpha_c,\beta_c)$ and $\check{d}_a=\alpha_a+\beta_a$ and $\check{d}_c=\alpha_c+\beta_c$, consistently with earlier conventions, the~composition of these vectors with the~first term in (\ref{eq:matrix solution to constraints}) can be written as
\begin{equation*}
    \left(
         \begin{array}{c}
              d_a  \\
              d_d  \\
              d_c  \\
              d_b  \\
         \end{array}
    \right)^T  
    \left(
    \begin{array}{cc:cc}
        C_{aa} & C_{aa} & C_{ac} & C_{ac} \\
        C_{aa} & C_{aa} & C_{ac} & C_{ac} \\
        \hdashline
        C_{ac} & C_{ac} & C_{cc} & C_{cc} \\
        C_{ac} & C_{ac} & C_{cc} & C_{cc}
    \end{array}
    \right) 
    \left(
         \begin{array}{c}
              d_a  \\
              d_d  \\
              d_c  \\
              d_b  \\
         \end{array}
    \right)
    = 
     \left(
         \begin{array}{c}
              \check{d}_a  \\
              \check{d}_c  \\
         \end{array}
    \right)^T  
    \left(
    \begin{array}{cc}
        C_{aa} & C_{ac} \\
        C_{ac} & C_{cc} \\
    \end{array}
    \right)  
    \left(
         \begin{array}{c}
              \check{d}_a  \\
              \check{d}_c  \\
         \end{array}
    \right)
    = C_{aa} \check{d}_a^2 + 2 C_{ac} \check{d}_a \check{d}_c + C_{cc} \check{d}_c^2,
\end{equation*}
so that $(-q)$ raised to the~above power indeed provides the~contribution from the~prequiver (i.e. the~first factor in the~summand) in (\ref{PK-Pi}). Analogous contribution from the~second term in (\ref{eq:matrix solution to constraints}) takes form
\begin{equation*}
    \left(
         \begin{array}{c}
              d_a  \\
              d_d  \\
              d_c  \\
              d_b  \\
         \end{array}
    \right)^T  
    \left(
    \begin{array}{cc:cc}
        0 & k & 0 & k \\
        k & l & k & l \\
        \hdashline
        0 & k & 0 & k \\
        k & l & k & l
    \end{array}
    \right) 
    \left(
         \begin{array}{c}
              d_a  \\
              d_d  \\
              d_c  \\
              d_b  \\
         \end{array}
    \right)
    = \big(2k (\alpha_a + \alpha_c) + l (\beta_a + \beta_c)\big) (\beta_a + \beta_c),
\end{equation*}
which we recognize as $k$- and $l$-dependent contribution in (\ref{kl-formula}). Finally, analogous contribution from the~last term (in round brackets) in (\ref{eq:matrix solution to constraints}) takes form $2\beta_a\alpha_c$,
which is nothing but the~term in (\ref{kl-formula}) that is responsible for the~permutohedron structure. In this case it is $\Pi_2$ and the~flip $\tau=(a\ c)$, realized by $2\beta_{\tau(a)}\alpha_{\tau(c)}=2\beta_c\alpha_a$, corresponds to the~transposition of non-diagonal terms $C_{ab}\leftrightarrow C_{cd}$, which gives the~quiver matrix equivalent to~(\ref{eq: quiver4x4}):
\begin{equation}\label{eq: quiver4x4 equivalent}
    \left(
    \begin{array}{cc:cc}
        C_{aa} & C_{ad} & C_{ac} & \mybox{C_{cd}} \\
        C_{ad} & C_{dd} & \mybox{C_{ab}} & C_{bd} \\
        \hdashline
        C_{ac} & \mybox{C_{ab}} & C_{cc} & C_{bc} \\
        \mybox{C_{cd}} & C_{bd} & C_{bc} & C_{bb}
    \end{array}
    \right) =     \left(
    \begin{array}{cc:cc}
        C_{aa} & C_{aa} + k & C_{ac} & \mybox{C_{ac} + k +1} \\
        C_{aa} + k & C_{aa} + l & \mybox{C_{ac} + k} & C_{ac}+l \\
        \hdashline
        C_{ac} & \mybox{C_{ac} + k} & C_{cc} & C_{cc} + k \\
        \mybox{C_{ac} + k + 1} & C_{ac} + l & C_{cc} + k & C_{cc} + l
    \end{array}
    \right).  
\end{equation}

We already can see how the~local constraints of theorem~\ref{thm:main} give rise to the~expression~(\ref{kl-formula}). There is just one more term in (\ref{kl-formula}) that we should reconstruct: the~one that involves spectator nodes. To this end we enlarge (\ref{eq: quiver4x4}) by two rows and columns, still assuming that $C_{ab}$ and $C_{cd}$ can be exchanged, and write such a~matrix in the~form:
\begin{equation*}
\begin{small}
    \left(
    \begin{array}{cc:cc:cc}
        C_{aa} & C_{ad} & C_{ac} & \mybox{C_{ab}} & C_{ae} & C_{af} \\
        C_{ad} & C_{dd} & \mybox{C_{cd}} & C_{bd} & C_{de} & C_{df} \\
        \hdashline
        C_{ac} & \mybox{C_{cd}} & C_{cc} & C_{bc} & C_{ce} & C_{cf} \\
        \mybox{C_{ab}} & C_{bd} & C_{bc} & C_{bb} & C_{be} & C_{bf} \\
        \hdashline
        C_{ae} & C_{de} & C_{ce} & C_{be} & C_{ee} & C_{ef} \\
        C_{af} & C_{df} & C_{cf} & C_{bf} & C_{ef} & C_{ff} \\
    \end{array}
    \right)
    = \left(
    \begin{array}{cc:cc:cc}
        C_{aa} & C_{aa} + k & C_{ac} & \mybox{C_{ac} + k} & C_{ae} & C_{af} \\
        C_{aa} + k & C_{aa} + l & \mybox{C_{ac} + k + 1} & C_{ac} + l & C_{ae} + h_e & C_{af} + h_f \\
        \hdashline
        C_{ac} & \mybox{C_{ac} + k + 1} & C_{cc} & C_{cc} + k & C_{ce} & C_{cf} \\
        \mybox{C_{ac} + k} & C_{ac} + l & C_{cc} + k & C_{cc} + l & C_{ce} + h_e & C_{cf} + h_f \\
        \hdashline
        C_{ae} & C_{ae} + h_e & C_{ce} & C_{ce} + h_e & C_{ee} & C_{ef} \\
        C_{af} & C_{af} + h_f & C_{cf} & C_{cf} + h_f & C_{ef} & C_{ff} \\
    \end{array}
    \right)
    \end{small}
\end{equation*}
The top-left $4\times 4$ submatrix is expressed in terms of $k$ and $l$ in the~same way as in (\ref{eq: quiver4x4}). In addition, if we denote $C_{de} - C_{ae} = h_e$ and substitute to the~second constraint in (\ref{eq:theorem first case}) with $i=e$, we get $C_{be} = C_{ce} + h_e$. Analogously, for $C_{df} - C_{af} = h_f$ we get $C_{bf} = C_{cf} + h_f$, and altogether we obtain the~matrix on the~right. It follows that the~contribution of these extra rows and columns to the~quiver generating function reads $(-q)^{\sum_s h_s \check{d}_s}$, which yields an~appropriate term in (\ref{kl-formula}) that we were after.

To sum up, we have shown how the~formula (\ref{kl-formula}) arises from local constraints of theorem~\ref{thm:main} in the~presence of one symmetry, which thus yields a~permutohedron $\Pi_2$. Let us now illustrate how permutohedron $\Pi_3$ arises if we assume that in addition to the~symmetry involving $C_{ab}$ and $C_{cd}$, there is also another symmetry that involves $C_{be}$ and $C_{cf}$. Such two symmetries may exist in a~matrix of size $6\times 6$, which we write in the~form
\begin{equation}\label{eq:general 6x6 quiver with two symmetries relabeled}
\begin{small}
    \left(
    \begin{array}{cc:cc:cc}
        C_{aa} & C_{ad} & C_{ac} & \mybox{C_{ab}} & C_{ae} & \textcolor{red}{C_{af}} \\
        C_{ad} & C_{dd} & \mybox{C_{cd}} & C_{bd} & \textcolor{red}{C_{de}} & C_{df} \\
        \hdashline
        C_{ac} & \mybox{C_{cd}} & C_{cc} & C_{bc} & C_{ce} & \textcolor{blue}{C_{cf}} \\
        \mybox{C_{ab}} & C_{bd} & C_{bc} & C_{bb} & \textcolor{blue}{C_{be}} & C_{bf} \\
        \hdashline
        C_{ae} & \textcolor{red}{C_{de}} & C_{ce} & \textcolor{blue}{C_{be}} & C_{ee} & C_{ef} \\
        \textcolor{red}{C_{af}} & C_{df} & \textcolor{blue}{C_{cf}} & C_{bf} & C_{ef} & C_{ff} \\
    \end{array}
    \right)
    = \left(
    \begin{array}{cc:cc:cc}
        C_{aa} & C_{aa} + k & C_{ac} & \mybox{C_{ac} + k} & C_{ae} & \textcolor{red}{C_{ae} + k} \\
        C_{aa} + k & C_{aa} + l & \mybox{C_{ac} + k + 1} & C_{ac} + l & \textcolor{red}{C_{ae} + k + 1} & C_{ae} + l \\
        \hdashline
        C_{ac} & \mybox{C_{ac} + k + 1} & C_{cc} & C_{cc} + k & C_{ce} & \textcolor{blue}{C_{ce} + k} \\
        \mybox{C_{ac} + k} & C_{ac} + l & C_{cc} + k & C_{cc} + l & \textcolor{blue}{C_{ce} + k + 1} & C_{ce} + l \\
        \hdashline
        C_{ae} & \textcolor{red}{C_{ae} + k + 1} & C_{ce} & \textcolor{blue}{C_{ce} + k + 1} & C_{ee} & C_{ee} + k \\
        \textcolor{red}{C_{ae} + k} & C_{ae} + l & \textcolor{blue}{C_{ce} + k} & C_{ce} + l & C_{ee} + k & C_{ee} + l \\
    \end{array}
    \right)
    \end{small}
\end{equation}
where the~right-hand side is expressed in terms of parameters $k$ and $l$ and arises from solving the~constraints of theorem~\ref{thm:main} analogously as above. Note that two symmetries of original quiver: $C_{ab}\leftrightarrow C_{cd}$ and $C_{be}\leftrightarrow C_{cf}$ correspond to transpositions $(1\ 2)$ and $(2\ 3)$ acting on the~element $(1,2,3)$; highlights in~\eqref{eq:general 6x6 quiver with two symmetries relabeled} match colors in figure~\ref{fig:hexagon}. After performing one of these transformations we obtain a~new quiver (with $+1$ in the~other highlighted entry, like in~\eqref{eq: quiver4x4 equivalent}), which also has two symmetries. One is an~inverse of the~transformation we just performed, the~other is a~transposition $C_{de}\leftrightarrow C_{af}$, denoted in red in (\ref{eq:general 6x6 quiver with two symmetries relabeled}). This behavior is perfectly consistent with the~structure of $\Pi_3$ -- the~new symmetry corresponds to transposition $(1\ 3)$, denoted in red in figure~\ref{fig:hexagon}. Using theorem~\ref{thm:main}, one can check that the~whole structure of $\Pi_3$ is preserved: there are six equivalent versions of the~matrix~\eqref{eq:general 6x6 quiver with two symmetries relabeled} connected by three symmetries, but only two of them can be applied to each representant of the~class.

Furthermore, the~right-hand side of (\ref{eq:general 6x6 quiver with two symmetries relabeled}) can be written in the~form
\[
\left(
        \begin{array}{ccc}
        C_{aa} & C_{ac} & C_{ae} \\
        C_{ac} & C_{cc} & C_{ce} \\
        C_{ae} & C_{ce} & C_{ee}
        \end{array}
    \right)
    \otimes
    \left(
        \begin{array}{cc}
        1 & 1 \\
        1 & 1
        \end{array}
    \right) +
    \left(
        \begin{array}{ccc}
        1 & 1 & 1 \\
        1 & 1 & 1 \\
        1 & 1 & 1
        \end{array}
    \right)
    \otimes
    \left(
        \begin{array}{cc}
        0 & k \\
        k & l
        \end{array}
    \right) 
    +
    \left(
        \begin{array}{ccc}
        0 & 1 & 1 \\
        0 & 0 & 1 \\
        0 & 0 & 0
        \end{array}
    \right)
    \otimes
    \left(
        \begin{array}{cc}
        0 & 0 \\
        1 & 0
        \end{array}
        \right)  +
    \left(
        \begin{array}{ccc}
        0 & 0 & 0 \\
        1 & 0 & 0 \\
        1 & 1 & 0
        \end{array}
    \right)
    \otimes
    \left(
    \begin{array}{cc}
        0 & 1 \\
        0 & 0
        \end{array}
    \right)
\]
where the~$3\times 3$ matrix in the~first term is a~prequiver. Straightforward generalization of the~above procedure to more symmetries leads to prequivers of arbitrary size and corresponding permutohedra, or -- equivalently -- the~general form of (\ref{kl-formula}):

\begin{dfn}\label{def:splitting}
    A~$(k,l)$-splitting of $n$ nodes with permutation $\sigma\in S_n$ in the~presence of $m-2n$ spectators (with corresponding integer shifts $h_s$) and with a~multiplicative factor $\kappa$ is defined as the~following transformation of a~quiver $\check{C}$ and a~change of variables $\boldsymbol{\check{\lambda}}$. For any two split nodes $i$ and $j$, $i<j$, and any spectator $s$, we transform the~matrix $\check{C}$ in the~following way (depending on the~presence of inversion in permutation $\sigma$):
\tikzset{every picture/.style={line width=0.75pt}} 

\begin{tikzpicture}[x=0.75pt,y=0.75pt,yscale=-1,xscale=1]

\draw    [-stealth](394,184.4+20) -- (472.06,164.89) ;
\draw    [-stealth](394,344.4-20) -- (472.06,363.91) ;

\draw (394,264.4) node    {$\left(\begin{array}{c:c:c:c:c}
\check{C}_{ss} & \cdots  & \check{C}_{si} & \cdots  & \check{C}_{sj}\\ \hdashline
\vdots  & \ddots  & \vdots  &  & \vdots \\ \hdashline
\check{C}_{is} & \cdots  & \check{C}_{ii} & \cdots  & \check{C}_{ij}\\ \hdashline
\vdots  &  & \vdots  & \ddots  & \vdots \\ \hdashline
\check{C}_{js} & \cdots  & \check{C}_{ji} & \cdots  & \check{C}_{jj}
\end{array}\right)$};
\draw (684,164.4) node    {$\left(\begin{array}{ c : c : c : c : c : c : c }
\check{C}_{ss} & \cdots  & \check{C}_{si} & \check{C}_{si} +h_{s} & \cdots  & \check{C}_{sj} & \check{C}_{sj} +h_{s}\\ \hdashline
\vdots  & \ddots  & \vdots  & \vdots  &  & \vdots  & \vdots \\ \hdashline
\check{C}_{is} & \cdots  & \check{C}_{ii} & \check{C}_{ii} +k & \cdots  & \check{C}_{ij} & \textcolor[rgb]{0.96,0.65,0.14}{\check{C}_{ij} +k}\\ \hdashline
\check{C}_{is} +h_{s} & \cdots  & \check{C}_{ii} +k & \check{C}_{ii} +l & \cdots  & \textcolor[rgb]{0.96,0.65,0.14}{\check{C}_{ij} +k+1} & \check{C}_{ij} +l\\ \hdashline
\vdots  &  & \vdots  & \vdots  & \ddots  & \vdots  & \vdots \\ \hdashline
\check{C}_{js} & \cdots  & \check{C}_{ji} & \textcolor[rgb]{0.96,0.65,0.14}{\check{C}_{ji} +k+1} & \cdots  & \check{C}_{jj} & \check{C}_{jj} +k\\ \hdashline
\check{C}_{js} +h_{s} & \cdots  & \textcolor[rgb]{0.96,0.65,0.14}{\check{C}_{ji} +k} & \check{C}_{ji} +l & \cdots  & \check{C}_{jj} +k & \check{C}_{jj} +l
\end{array}\right)$};
\draw (684,364.4) node    {$\left(\begin{array}{c:c:c:c:c:c:c}
\check{C}_{ss} & \cdots  & \check{C}_{si} & \check{C}_{si} +h_{s} & \cdots  & \check{C}_{sj} & \check{C}_{sj} +h_{s}\\ \hdashline
\vdots  & \ddots  & \vdots  & \vdots  &  & \vdots  & \vdots \\ \hdashline
\check{C}_{is} & \cdots  & \check{C}_{ii} & \check{C}_{ii} +k & \cdots  & \check{C}_{ij} & \textcolor[rgb]{0.96,0.65,0.14}{\check{C}_{ij} +k+1}\\ \hdashline
\check{C}_{is} +h_{s} & \cdots  & \check{C}_{ii} +k & \check{C}_{ii} +l & \cdots  & \textcolor[rgb]{0.96,0.65,0.14}{\check{C}_{ij} +k} & \check{C}_{ij} +l\\ \hdashline
\vdots  &  & \vdots  & \vdots  & \ddots  & \vdots  & \vdots \\ \hdashline
\check{C}_{js} & \cdots  & \check{C}_{ji} & \textcolor[rgb]{0.96,0.65,0.14}{\check{C}_{ji} +k} & \cdots  & \check{C}_{jj} & \check{C}_{jj} +k\\ \hdashline
\check{C}_{js} +h_{s} & \cdots  & \textcolor[rgb]{0.96,0.65,0.14}{\check{\textcolor[rgb]{0.96,0.65,0.14}{C}}_{ji} +k+1} & \check{C}_{ji} +l & \cdots  & \check{C}_{jj} +k & \check{C}_{jj} +l
\end{array}\right)$};
\draw (394+50,144.4) node    {$\sigma ( i) < \sigma ( j)$};
\draw (394+50,384.4) node    {$\sigma ( i)  >\sigma ( j)$};

\end{tikzpicture}
whereas for any permutation the~change of variables is transformed as follows:
\begin{equation*}
    \left(\begin{array}{c}
    \check{\lambda}_s\\
    \vdots\\
    \check{\lambda}_i\\
    \vdots\\
    \check{\lambda}_j
    \end{array}\right)
    \longrightarrow
    \left(\begin{array}{c}
    \check{\lambda}_s\\
    \vdots\\
    \check{\lambda}_i\\
    \check{\lambda}_i\kappa\\
    \vdots\\
    \check{\lambda}_j\\
    \check{\lambda}_j\kappa
    \end{array}\right).
\end{equation*}
\end{dfn}

Clearly, the~top right matrix above (corresponding to $\sigma(i)<\sigma(j)$ $\rightarrow$ no inversion) is encoded in the~quadratic terms in the~powers of $(-q)$ in~(\ref{kl-formula}). The~bottom right matrix (corresponding to $\sigma(i)>\sigma(j)$ $\rightarrow$ inversion) arises after exchanging labels $i$ and $j$ in~(\ref{kl-formula}). Moreover, in the~language of the~definition~\ref{def:splitting}, the~$(k,2k+1)$-splitting is a~manifestation of the~formula~(\ref{eq:qpoch-sum-general-4}). For $k=0$ it specializes to $(0,1)$-splitting that is a~manifestation of the~basic formula~(\ref{eq:qpoch-sum-general-3}) with $\xi=\kappa q$.

\begin{dfn}\label{def:prequiver}
    If the~inverse of splitting -- for any parameters from definition~\ref{def:splitting} --  can be applied to a~given quiver $C$ and associated~change of variables $\boldsymbol{\lambda}$, we call the~target of this operation a~prequiver $\check{C}$, and the~associated change of variables is denoted $\check{ \boldsymbol{\lambda}}$. Conversely, splitting the~nodes of a~prequiver produces the~quiver:
\begin{equation}
    \check{C}
    \;\longrightarrow\; C, \qquad \qquad 
    \boldsymbol{\check{\lambda}}\;\longrightarrow\; \boldsymbol{\lambda}.
    \end{equation}
    \end{dfn}
    
For clarity, let us see how $(k,l)$-splitting looks for a~full matrix in which we split the~first $n$ nodes in the~presence of
$m-2n$ spectators with shifts $h_{1},\ldots,h_{m-2n}$ and trivial permutation:
\begin{equation}
\begin{tiny}
\left[\begin{array}{c:c:c:c:ccc}
\check{C}_{11} & \check{C}_{12} & \ldots & \check{C}_{1n} & \check{C}_{1,n+1} & \ldots & \check{C}_{1,m-n}\\ \hdashline
\check{C}_{21} & \check{C}_{22} & \ldots & \check{C}_{2n} & \check{C}_{2,n+1} & \ldots & \check{C}_{2,m-n}\\ \hdashline
\vdots & \vdots & \ddots & \vdots & \vdots &  & \vdots\\ \hdashline
\check{C}_{n1} & \check{C}_{n2} & \ldots & \check{C}_{nn} & \check{C}_{n,n+1} & \ldots & \check{C}_{n,m-n}\\ \hdashline
\check{C}_{n+1,1} & \check{C}_{n+1,2} & \ldots & \check{C}_{n+,1n} & \check{C}_{n+1,n+1} & \ldots & \check{C}_{n+1,m-n}\\ 
\vdots & \vdots &  & \vdots & \vdots & \ddots & \vdots\\ 
\check{C}_{m-n,1} & \check{C}_{m-n,2} & \ldots & \check{C}_{m-n,n} & \check{C}_{m-n,n+1} & \ldots & \check{C}_{m-n,m-n}
\end{array}\right]\nonumber
\end{tiny}
\end{equation}
\[
\downarrow
\]
\begin{equation}
\begin{tiny}
\left[\begin{array}{cc:cc:c:cc:ccc}
\check{C}_{11} & \check{C}_{11}+k & \check{C}_{12} & \check{C}_{12}+k & \ldots & \check{C}_{1n} & \check{C}_{1n}+k & \check{C}_{1,n+1} & \ldots & \check{C}_{1,m-n}\\
\check{C}_{11}+k & \check{C}_{11}+l & \check{C}_{12}+k+1 & \check{C}_{12}+l & \ldots & \check{C}_{1n}+k+1 & \check{C}_{1n}+l & \check{C}_{1,n+1}+h_{1} & \ldots & \check{C}_{1,m-n}+h_{m-2n}\\ \hdashline
\check{C}_{21} & \check{C}_{21}+k+1 & \check{C}_{22} & \check{C}_{22}+k & \ldots & \check{C}_{2n} & \check{C}_{2n}+k & \check{C}_{2,n+1} & \ldots & \check{C}_{2,m-n}\\
\check{C}_{21}+k & \check{C}_{21}+l & \check{C}_{22}+k & \check{C}_{22}+l & \ldots & \check{C}_{2n}+k+1 & \check{C}_{2n}+l & \check{C}_{2,n+1}+h_{1} & \ldots & \check{C}_{2,m-n}+h_{m-2n}\\ \hdashline
\vdots & \vdots & \vdots & \vdots & \ddots & \vdots &  & \vdots &  & \vdots\\ \hdashline
\check{C}_{n1} & \check{C}_{n1}+k+1 & \check{C}_{n2} & \check{C}_{n2}+k+1 & \ldots & \check{C}_{nn} & \check{C}_{nn}+k & \check{C}_{n,n+1} & \ldots & \check{C}_{n,m-n}\\
\check{C}_{n1}+k & \check{C}_{n1}+l & \check{C}_{n2}+k & \check{C}_{n2}+l & \ldots & \check{C}_{nn}+k & \check{C}_{nn}+l & \check{C}_{n,n+1}+h_{1} & \ldots & \check{C}_{n,m-n}+h_{m-2n}\\ \hdashline
\check{C}_{n+1,1} & \check{C}_{n+1,1}+h_{1} & \check{C}_{n+1,2} & \check{C}_{n+1,2}+h_{1} & \ldots & \check{C}_{n+1,n} & \check{C}_{n+1,n}+h_{1} & \check{C}_{n+1,n+1} & \ldots & \check{C}_{n+1,m-n}\\
\vdots & \vdots & \vdots & \vdots &  & \vdots & \vdots & \vdots & \ddots & \vdots\\
\check{C}_{m-n,1} & \check{C}_{m-n,1}+h_{m-2n} & \check{C}_{m-n,2} & \check{C}_{m-n,2}+h_{m-2n} & \ldots & \check{C}_{m-n,n} & \check{C}_{m-n,n}+h_{m-2n} & \check{C}_{m-n,n+1} & \ldots & \check{C}_{m-n,m-n}
\end{array}\right].\nonumber
\end{tiny}
\end{equation}
It is straightforward to check that the~constraints from theorem~\ref{thm:main} are satisfied for the~above matrix, and that it is consistent with (\ref{PK-Pi}) and (\ref{kl-formula}).

\section{Examples -- global structure}\label{SectionCaseStudies}

In this section we analyze in detail equivalent quivers and the~structure of their permutohedra graphs for knots $3_1$, $4_1$, $5_1$, $5_2$, $6_1$, $7_1$, and the~whole series of $(2,2p+1)$ torus knots.



\subsection{Trefoil knot, \texorpdfstring{$3_1$}{31} \label{ssec-31}}

\begin{figure}[h!]
    \centering
    \tikzset{every picture/.style={line width=0.75pt}} 

\begin{tikzpicture}[x=0.75pt,y=0.75pt,yscale=-1,xscale=1]

\draw  [dash pattern={on 0.84pt off 2.51pt}]  (1707.46,1222) -- (1747.46,1182) ;
\draw  [dash pattern={on 0.84pt off 2.51pt}]  (1787.46,1222) -- (1747.46,1182) ;
\draw  [fill={rgb, 255:red, 208; green, 2; blue, 27 }  ,fill opacity=1 ][line width=1.5]  (1904.82,1159.14) .. controls (1907,1157.83) and (1909.82,1158.54) .. (1911.13,1160.72) .. controls (1912.44,1162.89) and (1911.73,1165.72) .. (1909.55,1167.03) .. controls (1907.38,1168.33) and (1904.55,1167.63) .. (1903.25,1165.45) .. controls (1901.94,1163.27) and (1902.64,1160.45) .. (1904.82,1159.14) -- cycle ;
\draw  [fill={rgb, 255:red, 0; green, 0; blue, 0 }  ,fill opacity=1 ][line width=1.5]  (1706.39,1220.21) .. controls (1707.37,1219.62) and (1708.65,1219.94) .. (1709.25,1220.93) .. controls (1709.84,1221.91) and (1709.52,1223.2) .. (1708.53,1223.79) .. controls (1707.54,1224.38) and (1706.26,1224.06) .. (1705.67,1223.07) .. controls (1705.08,1222.09) and (1705.4,1220.81) .. (1706.39,1220.21) -- cycle ;
\draw  [fill={rgb, 255:red, 0; green, 0; blue, 0 }  ,fill opacity=1 ][line width=1.5]  (1786.39,1220.21) .. controls (1787.37,1219.62) and (1788.65,1219.94) .. (1789.25,1220.93) .. controls (1789.84,1221.91) and (1789.52,1223.2) .. (1788.53,1223.79) .. controls (1787.54,1224.38) and (1786.26,1224.06) .. (1785.67,1223.07) .. controls (1785.08,1222.09) and (1785.4,1220.81) .. (1786.39,1220.21) -- cycle ;
\draw  [fill={rgb, 255:red, 0; green, 0; blue, 0 }  ,fill opacity=1 ][line width=1.5]  (1746.39,1180.21) .. controls (1747.37,1179.62) and (1748.65,1179.94) .. (1749.25,1180.93) .. controls (1749.84,1181.91) and (1749.52,1183.2) .. (1748.53,1183.79) .. controls (1747.54,1184.38) and (1746.26,1184.06) .. (1745.67,1183.07) .. controls (1745.08,1182.09) and (1745.4,1180.81) .. (1746.39,1180.21) -- cycle ;

\draw (1707.46,1212) node [anchor=south] [inner sep=0.75pt]  [color={rgb, 255:red, 0; green, 0; blue, 255 }  ,opacity=1 ]  {${\lambda _{1}}$};
\draw (1747.46,1172) node [anchor=south] [inner sep=0.75pt]  [color={rgb, 255:red, 0; green, 0; blue, 255 }  ,opacity=1 ]  {${\lambda _{3}}$};
\draw (1787.46,1212) node [anchor=south] [inner sep=0.75pt]  [color={rgb, 255:red, 0; green, 0; blue, 255 }  ,opacity=1 ]  {${\lambda _{2}}$};
\draw (1707.46,1232) node [anchor=north] [inner sep=0.75pt]  [color={rgb, 255:red, 0; green, 0; blue, 0 }  ,opacity=1 ]  {$0$};
\draw (1747.46,1192) node [anchor=north] [inner sep=0.75pt]  [color={rgb, 255:red, 0; green, 0; blue, 0 }  ,opacity=1 ]  {$3$};
\draw (1787.46,1232) node [anchor=north] [inner sep=0.75pt]  [color={rgb, 255:red, 0; green, 0; blue, 0 }  ,opacity=1 ]  {$2$};
\draw (1907.46,1172) node [anchor=north] [inner sep=0.75pt]    {$\begin{bmatrix}
0 & 1 & 1\\
1 & 2 & 2\\
1 & 2 & 3
\end{bmatrix}$};

\end{tikzpicture}
    \caption{Homology diagram and a~quiver matrix for $3_1$ knot.
    The~labels 0, 2 and 3 are $t$-degrees of generators, while $\textcolor{blue}{\lambda_i}$ arise in specialization of quiver generating parameters. For $3_1$ knot the~quiver is unique, so the~permutohedra graph consists of one vertex (shown in red).}
    \label{fig:symmetries_3_1}
\end{figure}
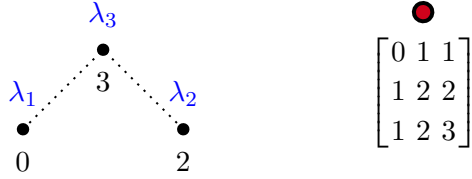

The~generating function of superpolynomials of the~knot $3_1$ is given by~\cite{FGS1205}
\begin{equation}\label{eq:trefoil P(x)}
    P_{3_1}(x,a,q,t)=\sum_{r=0}^{\infty} \frac{x^ra^{2r}q^{-2r}}{(q^2;q^2)_r}\sum_{k=0}^{r} \left[\begin{array}{c}
    r\\
    k
    \end{array}\right] q^{2k(r+1)}t^{2k} (-a^2q^{-2}t;q^2)_k,
\end{equation}
where we use the~$q$-binomial
\begin{equation}\label{eq:q-binomial}
    \left[\begin{array}{c}
    r\\
    k
    \end{array}\right]=\frac{(q^2;q^2)_r}{(q^2;q^2)_{r-k}(q^2;q^2)_k}.
\end{equation}
Linear order ($r=1$) of~\eqref{eq:trefoil P(x)} encodes the~uncolored superpolynomial $P_1(a,q,t)=a^2 q^{-2} + a^2 q^2 t^2 + a^4 t^3$. Its homological diagram consists of one zig-zag made of $3$ nodes, see figure~\ref{fig:symmetries_3_1}. 

Let us rederive the~trefoil quiver following section~\ref{sec:global structure}. We start from noticing that if we keep the~$q$-Pochhammer $(-a^2q^{-2}t;q^2)_k$ on the~side, the~remaining part of $P_{3_1}(x,a,q,t)$ can be easily rewritten in the~quiver form. First, we express the~$q$-binomial as in~\eqref{eq:q-binomial} and cancel $(q^2;q^2)_r$:
\begin{equation}
\label{eq:trefoil prequiver derivation}
    \sum_{r=0}^{\infty} \frac{x^ra^{2r}q^{-2r}}{(q^2;q^2)_r}\sum_{k=0}^{r} \left[\begin{array}{c}
    r\\
    k
    \end{array}\right] q^{2k(r+1)}t^{2k}
    =\sum_{r=0}^{\infty} x^ra^{2r}q^{-2r}\sum_{k=0}^{r} \frac{1}{(q^2;q^2)_{r-k}(q^2;q^2)_k} q^{2k(r+1)}t^{2k}.
\end{equation}
Then, we define new summation variables: $\check{d}_1=r-k$ and $\check{d}_2=k$, which allows us rewrite~\eqref{eq:trefoil prequiver derivation} as a~motivic generating function of a~prequiver:
\begin{equation}
\begin{split} \label{eq:trefoil prequiver}
    \sum_{\check{d}_1,\check{d}_2 \geq 0}(-q)^{2\check{d}_1\check{d}_2+2\check{d}_2^2} \frac{\left(xa^2q^{-2}\right)^{\check{d}_1}\left(xa^2(-t)^{2}\right)^{\check{d}_2}}{(q^2;q^2)_{\check{d}_1}(q^2;q^2)_{\check{d}_2}}
    &=\left. \sum_{\boldsymbol{\check{d}}}(-q)^{\boldsymbol{\check{d}}\cdot \check{C} \cdot \boldsymbol{\check{d}}}\frac{\boldsymbol{\check{x}}^{\boldsymbol{\check{d}}}}{(q^2;q^2)_{\boldsymbol{\check{d}}}}\right|_{\boldsymbol{\check{x}}=x\boldsymbol{\check{\lambda}}}, \\
    \check{C}=\left[\begin{array}{c:c}
    0 & 1\\
    \hdashline
    1 & 2
    \end{array}\right], \phantom{\qquad} & \phantom{\qquad} \boldsymbol{\check{\lambda}} = \left[\begin{array}{c}
    a^2q^{-2}\\
    a^2 (-t)^2
    \end{array}\right].
\end{split}    
\end{equation}
Now we put $(-a^2q^{-2}t;q^2)_k$ back with $k=\check{d}_2$ and apply a~variant of formula~\eqref{eq:qpoch-sum-general-3} for splitting one node (because only one $\check{d}_i$ enters $k$):
\begin{equation}
    \frac{(\xi;q^2)_{\check{d}_{i}}}{(q^2;q^2)_{\check{d}_{i}}}=  \sum\limits_{\alpha_{i}+\beta_{i}=\check{d}_{i}}  (-q)^{\beta_{i}^2}\frac{\left(\xi q^{-1}\right)^{\beta_{i}}}{(q^2;q^2)_{\alpha_{i}}(q^2;q^2)_{\beta_{i}}},
    \label{eq:qpoch-sum for 1 node}
\end{equation}
with $\xi=-a^2q^{-2}t$ and $i=2$. 
This leads to
\begin{equation}
\begin{split}
 \label{eq:trefoil quiver}
    P_{3_1}(x,a,q,t)=\sum_{d_1,\alpha_2,\beta_2 \geq 0} & \frac{\left(xa^2q^{-2}\right)^{d_1}\left(xa^2(-t)^{2}\right)^{\alpha_2}\left(xa^4q^{-3}(-t)^{2}\right)^{\beta_2}}{(q^2;q^2)_{d_1}(q^2;q^2)_{\alpha_2}(q^2;q^2)_{\beta_2}}\\ &\times (-q)^{2d_1\alpha_2+2d_1\beta_2+2\alpha_2^2+ 2\alpha_2\beta_2+3\beta_{2}^2},
\end{split}
\end{equation}
which is equal to $\left. P_Q(\boldsymbol{x},q)\right|_{\boldsymbol{x}=x\boldsymbol{\lambda}}$ for
\begin{equation}\label{eq:3_1 quiver}
    C=\left[\begin{array}{c:cc}
    0 & 1 & 1\\
    \hdashline
    1 & 2 & 2\\
    1 & 2 & 3
    \end{array}\right], \qquad  \qquad \boldsymbol{\lambda} = \left[\begin{array}{c}
    a^2q^{-2}\\
    a^2 (-t)^2\\
    a^4 q^{-3} (-t)^3
    \end{array}\right].
\end{equation}
This is the~quiver found in~\cite{KRSS1707short,KRSS1707long}; in the~language of definition~\ref{def:splitting} it arises from~\eqref{eq:trefoil prequiver} by $(0,1)$-splitting of the~second node, with trivial permutation $\sigma(2)=2$, $h_1=0$, and $\kappa=-a^2q^{-3}t$:
\begin{equation}\label{eq:3_1 splitting}
\begin{split}
    \check{C}=\left[\begin{array}{c:c}
    0 & 1\\
    \hdashline
    1 & 2
    \end{array}\right]
    &\;\longrightarrow\;
    C=\left[\begin{array}{c:cc}
    0 & 1 & 1+0\\
    \hdashline
    1 & 2 & 2+0\\
    1+0 & 2+0 & 2+1
    \end{array}\right], \\ 
    \boldsymbol{\check{\lambda}} = \left[\begin{array}{c}
    a^2q^{-2}\\
    a^2 (-t)^2
    \end{array}\right] 
    & \; \longrightarrow \;
    \boldsymbol{\lambda} = \left[\begin{array}{c}
    a^2q^{-2}\\
    a^2 (-t)^2\\
    a^2 (-t)^2 \times a^2q^{-3}(-t)
    \end{array}\right].
\end{split}
\end{equation}

In the~above process we did not have to make any choices, therefore we expect that the~above quiver is unique. This is indeed the~case: since the~trefoil knot is thin, all quiver equivalences come from permutations of non-diagonal matrix entries, but there are no possible pairings that could lead to non-trivial permutations of non-diagonal entries. In consequence, the~conjecture~\ref{coj:necessary conditions} holds for the~trefoil knot.


\subsection{Figure-eight knot, \texorpdfstring{$4_1$}{41}}\label{sec:figure-eight}

\begin{figure}[h!]
    \centering
\tikzset{every picture/.style={line width=0.75pt}} 

\begin{tikzpicture}[x=0.75pt,y=0.75pt,yscale=-1,xscale=1]

\draw [color={rgb, 255:red, 31; green, 160; blue, 144 }  ,draw opacity=1 ][line width=1.5]    (1707.46,1209) -- (1747.46,1169) ;
\draw [color={rgb, 255:red, 31; green, 160; blue, 144 }  ,draw opacity=1 ][line width=1.5]    (1707.46,1209) -- (1747.46,1249) ;
\draw [color={rgb, 255:red, 31; green, 160; blue, 144 }  ,draw opacity=1 ][line width=1.5]    (1747.46,1249) -- (1787.46,1209) ;
\draw [color={rgb, 255:red, 31; green, 160; blue, 144 }  ,draw opacity=1 ][line width=1.5]    (1747.46,1169) -- (1787.46,1209) ;
\draw [color={rgb, 255:red, 31; green, 160; blue, 144 }  ,draw opacity=1 ][line width=2.25]    (1627.46,1301) -- (1867.46,1301) ;
\draw  [fill={rgb, 255:red, 208; green, 2; blue, 27 }  ,fill opacity=1 ][line width=1.5]  (1865.09,1297.06) .. controls (1867.27,1295.75) and (1870.09,1296.46) .. (1871.4,1298.63) .. controls (1872.71,1300.81) and (1872,1303.64) .. (1869.83,1304.94) .. controls (1867.65,1306.25) and (1864.82,1305.54) .. (1863.52,1303.37) .. controls (1862.21,1301.19) and (1862.92,1298.36) .. (1865.09,1297.06) -- cycle ;
\draw  [fill={rgb, 255:red, 0; green, 0; blue, 0 }  ,fill opacity=1 ][line width=1.5]  (1706.39,1207.21) .. controls (1707.37,1206.62) and (1708.65,1206.94) .. (1709.25,1207.93) .. controls (1709.84,1208.91) and (1709.52,1210.2) .. (1708.53,1210.79) .. controls (1707.54,1211.38) and (1706.26,1211.06) .. (1705.67,1210.07) .. controls (1705.08,1209.09) and (1705.4,1207.81) .. (1706.39,1207.21) -- cycle ;
\draw  [fill={rgb, 255:red, 0; green, 0; blue, 0 }  ,fill opacity=1 ][line width=1.5]  (1786.39,1207.21) .. controls (1787.37,1206.62) and (1788.65,1206.94) .. (1789.25,1207.93) .. controls (1789.84,1208.91) and (1789.52,1210.2) .. (1788.53,1210.79) .. controls (1787.54,1211.38) and (1786.26,1211.06) .. (1785.67,1210.07) .. controls (1785.08,1209.09) and (1785.4,1207.81) .. (1786.39,1207.21) -- cycle ;
\draw  [fill={rgb, 255:red, 0; green, 0; blue, 0 }  ,fill opacity=1 ][line width=1.5]  (1746.39,1167.21) .. controls (1747.37,1166.62) and (1748.65,1166.94) .. (1749.25,1167.93) .. controls (1749.84,1168.91) and (1749.52,1170.2) .. (1748.53,1170.79) .. controls (1747.54,1171.38) and (1746.26,1171.06) .. (1745.67,1170.07) .. controls (1745.08,1169.09) and (1745.4,1167.81) .. (1746.39,1167.21) -- cycle ;
\draw  [fill={rgb, 255:red, 0; green, 0; blue, 0 }  ,fill opacity=1 ][line width=1.5]  (1746.39,1202.21) .. controls (1747.37,1201.62) and (1748.65,1201.94) .. (1749.25,1202.93) .. controls (1749.84,1203.91) and (1749.52,1205.2) .. (1748.53,1205.79) .. controls (1747.54,1206.38) and (1746.26,1206.06) .. (1745.67,1205.07) .. controls (1745.08,1204.09) and (1745.4,1202.81) .. (1746.39,1202.21) -- cycle ;
\draw  [fill={rgb, 255:red, 255; green, 255; blue, 255 }  ,fill opacity=1 ][line width=1.5]  (1625.09,1297.06) .. controls (1627.27,1295.75) and (1630.09,1296.46) .. (1631.4,1298.63) .. controls (1632.71,1300.81) and (1632,1303.64) .. (1629.83,1304.94) .. controls (1627.65,1306.25) and (1624.82,1305.54) .. (1623.52,1303.37) .. controls (1622.21,1301.19) and (1622.92,1298.36) .. (1625.09,1297.06) -- cycle ;
\draw  [fill={rgb, 255:red, 0; green, 0; blue, 0 }  ,fill opacity=1 ][line width=1.5]  (1746.39,1247.21) .. controls (1747.37,1246.62) and (1748.65,1246.94) .. (1749.25,1247.93) .. controls (1749.84,1248.91) and (1749.52,1250.2) .. (1748.53,1250.79) .. controls (1747.54,1251.38) and (1746.26,1251.06) .. (1745.67,1250.07) .. controls (1745.08,1249.09) and (1745.4,1247.81) .. (1746.39,1247.21) -- cycle ;

\draw (1707.46,1199) node [anchor=south] [inner sep=0.75pt]  [color={rgb, 255:red, 0; green, 0; blue, 255 }  ,opacity=1 ]  {${\lambda _{3}}$};
\draw (1747.46,1159) node [anchor=south] [inner sep=0.75pt]  [color={rgb, 255:red, 0; green, 0; blue, 255 }  ,opacity=1 ]  {${\lambda _{5}}$};
\draw (1787.46,1199) node [anchor=south] [inner sep=0.75pt]  [color={rgb, 255:red, 0; green, 0; blue, 255 }  ,opacity=1 ]  {${\lambda _{4}}$};
\draw (1707.46,1219) node [anchor=north] [inner sep=0.75pt]  [color={rgb, 255:red, 0; green, 0; blue, 0 }  ,opacity=1 ]  {$-1$};
\draw (1748.53,1175.79) node [anchor=north] [inner sep=0.75pt]  [color={rgb, 255:red, 0; green, 0; blue, 0 }  ,opacity=1 ]  {$2$};
\draw (1787.46,1219) node [anchor=north] [inner sep=0.75pt]  [color={rgb, 255:red, 0; green, 0; blue, 0 }  ,opacity=1 ]  {$1$};
\draw (1877.46,1301) node [anchor=west] [inner sep=0.75pt]    {$\begin{bmatrix}
    0  &- 1 & -1 &  0 &  0\\
    -1 & -2 & -2 & -1 & -1\\
    -1 & -2 & -1 &  0 &  0\\
    0  & -1 &  0 &  1 &  1\\
    0  & -1 &  0 &  1 &  2
\end{bmatrix}$};
\draw (1747.46,1239) node [anchor=south] [inner sep=0.75pt]  [color={rgb, 255:red, 0; green, 0; blue, 255 }  ,opacity=1 ]  {${\lambda _{2}}$};
\draw (1747.46,1254) node [anchor=north] [inner sep=0.75pt]  [color={rgb, 255:red, 0; green, 0; blue, 0 }  ,opacity=1 ]  {$-2$};
\draw (1617.46,1301) node [anchor=east] [inner sep=0.75pt]    {
$\begin{bmatrix}
    0  &- 1 & -1 &  0 &  0\\
    -1 & -2 & -2 & -1 & \myboxy{0}\\
    -1 & -2 & -1 & \myboxy{-1} & 0\\
    0  & -1 & \myboxy{-1} &  1 & 1\\
    0  &  \myboxy{0} &  0 &  1 & 2
\end{bmatrix}$};
\draw (1747.46,1306) node [anchor=north] [inner sep=0.75pt]  [color={rgb, 255:red, 0; green, 0; blue, 255 }  ,opacity=1 ]  {$\lambda _{2} \lambda _{5} =\lambda _{3} \lambda _{4}$};
\draw (1747.46,1202) node [anchor=south] [inner sep=0.75pt]  [color={rgb, 255:red, 0; green, 0; blue, 255 }  ,opacity=1 ]  {${\lambda _{1}}$};
\draw (1747.46,1209) node [anchor=north] [inner sep=0.75pt]  [color={rgb, 255:red, 0; green, 0; blue, 0 }  ,opacity=1 ]  {$0$};

\end{tikzpicture}
    \caption{Homological diagram for $4_1$ knot, with labels $\lambda_i$ assigned to various nodes (top). In the~bottom the~two equivalent quivers are shown, which differ by a~transposition of elements $C_{2,5}$ and $C_{3,4}$ of the~quiver matrix (shown in yellow, together with their symmetric companions). The~positions of these elements are encoded in combinations $\lambda_2\lambda_5$ and $\lambda_3\lambda_4$, which are equal to each other (satisfy the~center of mass condition). The~permutohedra graph is given by $\Pi_2$ that consists of two vertices connected by one edge. 
    }
    \label{fig:symmetries_4_1}
\end{figure}
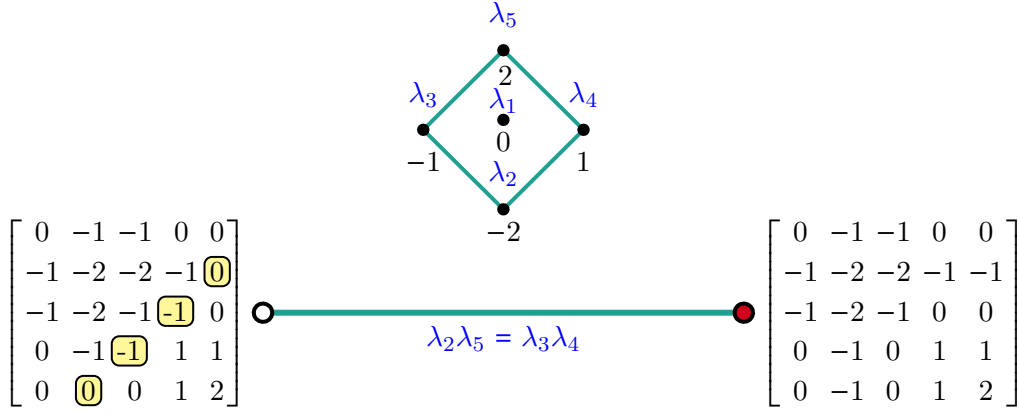

For the~figure-eight knot two corresponding quivers have been already found in~\cite{KRSS1707long,EKL1910}. Let us rederive this result and check that there are no other equivalent quivers. The~generating function of superpolynomials of the~figure-eight knot reads~\cite{FGS1205}:
\[
    P_{4_1}(x,a,q,t)= \sum_{r=0}^{\infty}\sum_{k=0}^{r} \frac{x^r (-1)^k a^{-2k} t^{-2k} q^{-k^2+3k}(q^{-2r};q^2)_k}{(q^2;q^2)_r (q^{2};q^2)_k} (-a^{2}q^{-2}t;q^{2})_k(-a^{2}q^{2r}t^{3};q^{2})_k.    
\]
For $r=1$ we obtain the~superpolynomial $P_1(a,q,t) = 1 + a^{-2} t^{-2} + q^{-2} t^{-1} + q^2 t + a^2 t^2$. The~corresponding homological diagram consists of a~degenerate zig-zag made of one node and a~diamond, see figure~\ref{fig:symmetries_4_1}. 

In order to find equivalent quivers we follow section~\ref{sec:global structure} again.
We use the~relation $(q^{-2r};q^2)_k = (-1)^k q^{-2rk+k(k-1)}\frac{(q^2;q^2)_r}{(q^2;q^2)_{r-k}}$, as well as~\eqref{eq:qpoch-sum for 1 node} for $(-a^{2}q^{2r}t^{3};q^{2})_k / (q^2;q^2)_k$, to rewrite
\begin{align} \label{eq:figure-eight prequiver}
    \sum_{0\leq k \leq r} \frac{x^r (-1)^k a^{-2k} t^{-2k} q^{-k^2+3k}(q^{-2r};q^2)_k}{(q^2;q^2)_r (q^{2};q^2)_k} (-a^{2}q^{2r}t^{3};q^{2})_k
    &=\left. \sum_{\boldsymbol{\check{d}}}\frac{(-q)^{\boldsymbol{\check{d}}\cdot \check{C} \cdot \boldsymbol{\check{d}}}\boldsymbol{\check{x}}^{\boldsymbol{\check{d}}}}{(q^2;q^2)_{\boldsymbol{\check{d}}}}\right|_{\boldsymbol{\check{x}}=x\boldsymbol{\check{\lambda}}},\nonumber\\
    \check{C}=\left[\begin{array}{c:c:c}
    0  & -1 & 0 \\
    \hdashline
    -1  & -2 & -1 \\
    \hdashline
    0  & -1 & 1
    \end{array}\right], \phantom{\qquad}  \phantom{\qquad}  \boldsymbol{\check{\lambda}} & = \left[\begin{array}{c}
    1 \\
    a^{-2} q^2 (-t)^{-2}\\
    q (-t)
    \end{array}\right],
\end{align}
where we substitute $r-k=\check{d}_1$ and $k=\check{d}_2+\check{d}_3$. In addition, we rewrite the~remaining term $(-a^2q^{-2}t;q^2)_k\equiv (-a^2q^{-2}t;q^2)_{\check{d}_2+\check{d}_3}$, using~\eqref{eq:qpoch-sum-general-3} for $n=2$:
\begin{equation}
\begin{split}
    \frac{(\xi;q^2)_{\check{d}_{i}+\check{d}_j}}{(q^2;q^2)_{\check{d}_{i}}(q^2;q^2)_{\check{d}_{j}}}=  \sum\limits_{\alpha_{i}+\beta_{i}=\check{d}_{i}}\sum\limits_{\alpha_{j}+\beta_{j}=\check{d}_{j}} & (-q)^{\beta_{i}^2+\beta_{j}^2+2\beta_{i}(\alpha_{j}+\beta_{j})} \\
    & \times \frac{\left(\xi q^{-1}\right)^{\beta_{i}}}{(q^2;q^2)_{\alpha_{i}}(q^2;q^2)_{\beta_{i}}} \frac{\left(\xi q^{-1}\right)^{\beta_{j}}}{(q^2;q^2)_{\alpha_{j}}(q^2;q^2)_{\beta_{j}}}.
    \label{eq:qpoch-sum}    
\end{split}
\end{equation}
Now the~two equivalent quivers arise from two possible specializations of $(i,j)$ in the~term $\beta_i\alpha_j$ in the~above expression. For $(i,j)=(2,3)$, from the~quadratic terms in the~exponent of $(-q)$ we read off the~following quiver matrix: 
\begin{equation}\label{eq:figure-eight quiver 1}
    C=\left[\begin{array}{c:cc:cc}
    0  &- 1 & -1 &  0 &  0\\
    \hdashline
    -1 & -2 & -2 & -1 & \myboxy{-1}\\
    -1 & -2 & -1 &  \myboxy{0} &  0\\
    \hdashline
    0  & -1 &  \myboxy{0} &  1 &  1\\
    0  & \myboxy{-1} &  0 &  1 &  2
    \end{array}\right], \qquad  \qquad \boldsymbol{\lambda} = \left[\begin{array}{c}
    1 \\
    a^{-2} q^2 (-t)^{-2}\\
    q^{-1} (-t)^{-1}\\
    q (-t)\\
    a^2 q^{-2} (-t)^2
    \end{array}\right]
\end{equation}
which is consistent with the~result in~\cite{KRSS1707long} (up to a~permutation of rows and columns) and corresponds to the~red dot in figure~\ref{fig:symmetries_4_1}. 
On the~other hand, setting $(i,j)=(3,2)$ yields
\begin{equation}\label{eq:figure-eight quiver 2}
    C=\left[\begin{array}{c:cc:cc}
    0  &- 1 & -1 &  0 &  0\\
    \hdashline
    -1 & -2 & -2 & -1 & \myboxy{0}\\
    -1 & -2 & -1 & \myboxy{-1} & 0\\
    \hdashline
    0  & -1 & \myboxy{-1} &  1 & 1\\
    0  &  \myboxy{0} &  0 &  1 & 2
    \end{array}\right], \qquad  \qquad \boldsymbol{\lambda} = \left[\begin{array}{c}
    1 \\
    a^{-2} q^2 (-t)^{-2}\\
    q^{-1} (-t)^{-1}\\
    q (-t)\\
    a^2 q^{-2} (-t)^2
    \end{array}\right]
\end{equation}
which is consistent with the~second, equivalent quiver found in~\cite{EKL1910}. The~two above quivers are also presented in figure~\ref{fig:symmetries_4_1} and they differ by a~transposition of elements shown in yellow. This transposition corresponds to a~single possible inversion encoded in the~term $\beta_i\alpha_j$ in~(\ref{eq:qpoch-sum}).  

In the~language of definition~\ref{def:splitting}, quivers~(\ref{eq:figure-eight quiver 1}) and~(\ref{eq:figure-eight quiver 2}) arise from the~prequiver~\eqref{eq:figure-eight prequiver} by $(0,1)$-splitting of nodes number 2 and 3. Since we split two nodes, there are 2 possible permutations. For the~identity permutation ($\sigma(2)=2$, $\sigma(3)=3$) we obtain~\eqref{eq:figure-eight quiver 1}
\begin{equation}\label{eq:4_1 splitting v1}
    \check{C}=\left[\begin{array}{c:c:c}
    0  & -1 & 0 \\
    \hdashline
    -1  & -2 & -1 \\
    \hdashline
    0  & -1 & 1
    \end{array}\right]
    \overset{\sigma(2)<\sigma(3)}{\longrightarrow}
    C=\left[\begin{array}{c:cc:cc}
    0  &- 1 & -1+0 &  0 &  0+0\\
    \hdashline
    -1 & -2 & -2+0 & -1 & \mybox{-1+0}\\
    -1+0 & -2+0 & -2+1 & \mybox{-1+0+1} & -1+1\\
    \hdashline
    0  & -1 &  \mybox{-1+0+1} &  1 &  1+0\\
    0+0  & \mybox{-1+0} &  -1+1 &  1+0 &  1+1
    \end{array}\right].
\end{equation}
On the~other hand, for a~transposition $\sigma=(2\ 3)$ (i.e. $\sigma(2)=3$, $\sigma(3)=2$) we get
\begin{equation}\label{eq:4_1 splitting v2}
    \check{C}=\left[\begin{array}{c:c:c}
    0  & -1 & 0 \\
    \hdashline
    -1  & -2 & -1 \\
    \hdashline
    0  & -1 & 1
    \end{array}\right]
    \overset{\sigma(2)>\sigma(3)}{\longrightarrow}
    C=\left[\begin{array}{c:cc:cc}
    0  &- 1 & -1+0 &  0 &  0+0\\
    \hdashline
    -1 & -2 & -2+0 & -1 & \mybox{-1+0+1}\\
    -1+0 & -2+0 & -2+1 & \mybox{-1+0} & -1+1\\
    \hdashline
    0  & -1 &  \mybox{-1+0} &  1 &  1+0\\
    0+0  & \mybox{-1+0+1} &  -1+1 &  1+0 &  1+1
    \end{array}\right].
\end{equation}
In both cases we have $h_1=0$ and $\kappa=-a^2q^{-3}t$.

The quiver matrices~\eqref{eq:figure-eight quiver 1} and~\eqref{eq:figure-eight quiver 2} are related by a~transposition of non-diagonal entries. The~condition $\lambda_2 \lambda_5=\lambda_3 \lambda_4$ from theorem~\ref{thm:main} is satisfied, so it is a~symmetry. The~permutohedra graph is given by $\Pi_2$ that consists of two vertices connected by an~edge, as shown in figure~\ref{fig:symmetries_4_1}. Since the~$4_1$ knot is thin, all equivalent quivers come from permutations of non-diagonal elements of $C$. However, we checked that there are no more pairings apart from $\lambda_2 \lambda_5=\lambda_3 \lambda_4$, so we found the~whole equivalence class and the~conjecture~\ref{coj:necessary conditions} holds for the~figure-eight knot.

\subsection{Cinquefoil knot, \texorpdfstring{$5_1$}{51}}

\begin{figure}[h!]
    \centering
\input{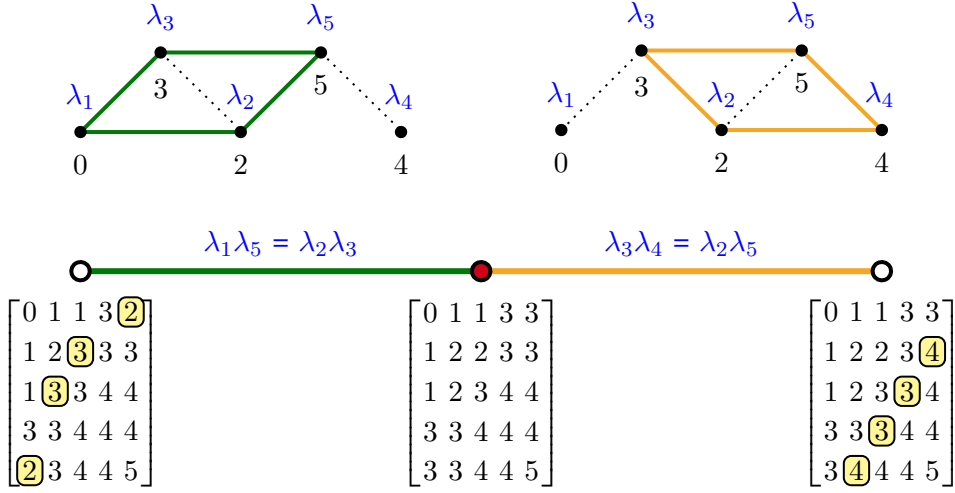}
  \caption{Two copies of the~homological diagram for $5_1$ knot are shown on top. On each copy we denoted a~parallellogram that encodes a~symmetry, i.e. a~transposition of two matrix elements that yields an~equivalent quiver. In total there are 3 equivalent quivers, shown in bottom, which correspond to 3 vertices of the~permutohedra graph. The~permutohedra graph is made of two $\Pi_2$ that share a~common vertex (in red).   
  }
    \label{fig:5_1matrixplot}
\end{figure}

In turn, we anlayze $5_1$ knot. The~generating function of its colored superpolynomials is given by~\cite{FGS1205}
\begin{align}\label{eq:cinquefoil P(x)}
    P_{5_1}(x,a,q,t)=\sum_{r=0}^{\infty} \frac{x^r a^{4r}q^{-4r}}{(q^2;q^2)_r}\sum_{0\leq k_2 \leq k_1 \leq r} \left[\begin{array}{c}
    r\\
    k_1
    \end{array}\right]
    \left[\begin{array}{c}
    k_1\\
    k_2
    \end{array}\right] &
    (-a^2q^{-2}t;q^2)_{k_1}\\
    \times &\, q^{2[(2r+1)(k_1+k_2)-r k_1-k_1 k_2]}t^{2(k_1+k_2)},    \nonumber
\end{align}
which for $r=1$ encodes the~superpolynoimal $P_1(a,q,t)=a^4 q^{-4} + a^4 t^2 + a^6 q^{-2} t^3 + a^4 q^{4} t^4  + a^6 q^{2} t^5 $. The~homological diagram is a~a~zig-zag made of 5 nodes, see figure~\ref{fig:5_1matrixplot}. 

In analogy to the~case of $4_1$, we rewrite the~summand in~(\ref{eq:cinquefoil P(x)}) as a~product of the~motivic generating series for the~prequiver and $(-a^2q^{-2}t;q^2)_{k_1}$ with $k_1=(k_1-k_2)+k_2=\check{d}_2+\check{d}_3$ 
\begin{equation}
\begin{split}
    P_{5_{1}}(x,a,q,t)&=\sum_{\boldsymbol{\check{d}}}(-q)^{\boldsymbol{\check{d}}\cdot\check{C}\cdot\boldsymbol{\check{d}}}\frac{\boldsymbol{x}^{\boldsymbol{\check{d}}}}{(q^{2};q^{2})_{\boldsymbol{\check{d}}}}(-a^{2}q^{-2}t;q^{2})_{\check{d}_{2}+\check{d}_{3}}\Big|_{\boldsymbol{\check{x}}=x\boldsymbol{\check{\lambda}}}\\
    \check{C}&=\left[\begin{array}{c:c:c}
    0 & 1 & 3\\
    \hdashline
    1 & 2 & 3\\
    \hdashline
    3 & 3 & 4
    \end{array}\right],  \qquad  \qquad \boldsymbol{\check{\lambda}} = \left[\begin{array}{c}
    a^4q^{-4}\\
    a^4q^{-2}(-t)^2\\
    a^4 (-t)^4
    \end{array}\right].    
\end{split}
\end{equation}
Then, the~application of~\eqref{eq:qpoch-sum-general-3} leads to $(0,1)$-splitting of nodes number 2 and 3 (the node number 1 is a~spectator with $h_1=0$; $\kappa=-a^2q^{-3}t$), which can be done in two ways. The~identity permutation $(\sigma(2)=2,\sigma(3)=3)$ yields
\begin{equation}\label{eq:5_1 quiver v1}
    C=\left[\begin{array}{c:cc:cc}
    0 & 1 & 1 & 3 & 3\\
    \hdashline
    1 & 2 & 2 & 3 & \myboxy{3}\\
    1 & 2 & 3 & \myboxy{4} & 4\\
    \hdashline
    3 & 3 & \myboxy{4} & 4 & 4\\
    3 & \myboxy{3} & 4 & 4 & 5
    \end{array}\right], \qquad \qquad   \boldsymbol{\lambda} = \left[\begin{array}{c}
    a^4q^{-4}\\
    a^4q^{-2}(-t)^2\\
    a^6q^{-5}(-t)^3\\
    a^4 (-t)^4\\
    a^6q^{-3}(-t)^5
    \end{array}\right]
\end{equation}
whereas the~transposition $\sigma=(2\ 3)$ gives
\begin{equation}\label{eq:5_1 quiver v2}
    C=\left[\begin{array}{c:cc:cc}
    0 & 1 & 1 & 3 & 3\\
    \hdashline
    1 & 2 & 2 & 3 & \myboxy{4}\\
    1 & 2 & 3 & \myboxy{3} & 4\\
    \hdashline
    3 & 3 & \myboxy{3} & 4 & 4\\
    3 & \myboxy{4} & 4 & 4 & 5
    \end{array}\right], \qquad \qquad   \boldsymbol{\lambda} = \left[\begin{array}{c}
    a^4q^{-4}\\
    a^4q^{-2}(-t)^2\\
    a^6q^{-5}(-t)^3\\
    a^4 (-t)^4\\
    a^6q^{-3}(-t)^5
    \end{array}\right].
\end{equation}
Comparing with theorem~\ref{thm:main}, it is clear that this symmetry comes from the~pairing $\lambda_3\lambda_4 = \lambda_2 \lambda_5$ (shown in orange in figure~\ref{fig:5_1matrixplot}). However, for the~cinquefoil knot we find another pairing $\lambda_1 \lambda_5 = \lambda_2 \lambda_3$ (shown in green in figure~\ref{fig:5_1matrixplot}), which also leads to a~non-trivial symmetry. Using definitions~\ref{def:splitting} and~\ref{def:prequiver} we can see that the~quiver from~\eqref{eq:5_1 quiver v1} admits not only the~inverse of $(0,1)$-splitting analyzed above, but also the~inverse of $(1,3)$-splitting.\footnote{In fact it admits also the~inverse of $(1,2)$-splitting with $h_1=0$ and $h_1=2$, but they capture the~same symmetries. This phenomenon is characteristic for all instances of splitting two nodes, when it is possible to interpret $\lambda_a \lambda_b=\lambda_c \lambda_d$ as $\lambda_a$, $\lambda_c$ coming from splitting node $a$ and $\lambda_d$, $\lambda_b$ coming from splitting node $d$ or $\lambda_a$, $\lambda_d$ coming from splitting node $a$ and $\lambda_c$, $\lambda_b$ coming from splitting node $c$.} 
More precisely, $P_{5_1}$~can be rewritten as 
\begin{equation}\label{eq:5_1 prequiver v2}
\begin{split}
    P_{5_{1}}(x,a,q,t)&=\left.\sum_{\boldsymbol{\check{d}}}(-q)^{\boldsymbol{\check{d}}\cdot\check{C}\cdot\boldsymbol{\check{d}}}\frac{\boldsymbol{\check{x}}^{\boldsymbol{\check{d}}}}{(q^{2};q^{2})_{\boldsymbol{\check{d}}}}(-a^{2}q^{2r}t^{3};q^{2})_{\check{d}_{2}+\check{d}_{3}}\right|_{\boldsymbol{\check{x}}=x\boldsymbol{\check{\lambda}}}\\
    \check{C}&=\left[\begin{array}{c:c:c}
    4 & 3 & 3\\
    \hdashline
    3 & 0 & 1\\
    \hdashline
    3 & 1 & 2
    \end{array}\right], \qquad  \qquad \boldsymbol{\check{\lambda}} = \left[\begin{array}{c}
    a^4 (-t)^4\\
    a^4q^{-4}\\
    a^4q^{-2}(-t)^2
    \end{array}\right],    
\end{split}
\end{equation}
which leads to~\eqref{eq:5_1 quiver v1} by $(1,3)$-splitting of nodes number 2 and 3 (the node number 1 is a~spectator with $h_1=1$) with permutation $\sigma=(2\ 3)$ and $\kappa=-a^2q^{-1}t^3$. This automatically implies that there exists another equivalent quiver, arising from  $(1,3)$-splitting of~\eqref{eq:5_1 prequiver v2} with the~trivial permutation
\begin{equation}\label{eq:5_1 quiver v3}
    C=\left[\begin{array}{c:cc:cc}
    4 & 3 & 4 & 3 & 4\\
    \hdashline
    3 & 0 & 1 & 1 & \myboxy{2}\\
    4 & 1 & 3 & \myboxy{3} & 4\\
    \hdashline
    3 & 1 & \myboxy{3} & 2 & 3\\
    4 & \myboxy{2} & 4 & 3 & 5
    \end{array}\right], \qquad \qquad   \boldsymbol{\lambda} = \left[\begin{array}{c}
    a^4 (-t)^4\\
    a^4q^{-4}\\
    a^6q^{-5}(-t)^3\\
    a^4q^{-2}(-t)^2\\
    a^6q^{-3}(-t)^5
    \end{array}\right],
\end{equation}
which is the~quiver on the~left-hand side in figure~\ref{fig:5_1matrixplot} (up to a~permutation of nodes).

To sum up, we have found 3 equivalent quivers for $5_1$, and from quiver~(\ref{eq:5_1 quiver v1}) we can obtain either of the~other two, by appropriate transpositions of elements of the~quiver matrix. However, since these transpositions are not disjoint, we cannot compose them. In consequence the~permutohedra graph, shown in figure~\ref{fig:5_1matrixplot}, consists of two permutohedra $\Pi_2$ that share a~common vertex (in red) that represents quiver~(\ref{eq:5_1 quiver v1}). Using an~argument analogous to the~one for the~figure-eight knot, we can check that since there are no pairings other than those depicted in figure~\ref{fig:5_1matrixplot}, we have found all equivalent quivers. In consequence, the~conjecture~\ref{coj:necessary conditions} holds for the~cinquefoil knot.


\subsection{\texorpdfstring{$5_2$}{52} knot}\label{52 knot}

The knot $5_2$ is a~more involved example. Having identified one quiver for this knot (e.g.~the~one found in~\cite{KRSS1707long}) and considering all possible local equivalences following theorem~\ref{thm:main}, we found 12 equivalent quivers for this knot (they are listed explicitly in appendix~\ref{appendix_matrices}). It turns out these quivers form an~interesting structure of three permutohedra $\Pi_3$ glued along their edges. Let us explain how this structure arises.

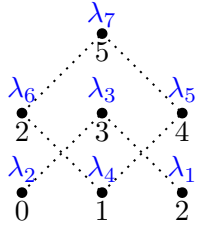
\begin{figure}[h]
    \centering
    \tikzset{every picture/.style={line width=0.75pt}} 

\begin{tikzpicture}[x=0.75pt,y=0.75pt,yscale=-1,xscale=1]

\draw  [dash pattern={on 0.84pt off 2.51pt}]  (147.04,507.18) -- (187.04,467.18) ;
\draw  [dash pattern={on 0.84pt off 2.51pt}]  (187.04,507.18) -- (227.04,547.18) ;
\draw  [dash pattern={on 0.84pt off 2.51pt}]  (147.04,547.18) -- (187.04,507.18) ;
\draw  [dash pattern={on 0.84pt off 2.51pt}]  (187.04,547.18) -- (227.04,507.18) ;
\draw  [dash pattern={on 0.84pt off 2.51pt}]  (147.04,507.18) -- (187.04,547.18) ;
\draw  [dash pattern={on 0.84pt off 2.51pt}]  (187.04,467.18) -- (227.04,507.18) ;
\draw  [fill={rgb, 255:red, 0; green, 0; blue, 0 }  ,fill opacity=1 ] (184.87,466.97) .. controls (184.99,465.77) and (186.06,464.89) .. (187.26,465.01) .. controls (188.46,465.13) and (189.33,466.2) .. (189.21,467.4) .. controls (189.09,468.6) and (188.02,469.47) .. (186.82,469.35) .. controls (185.62,469.24) and (184.75,468.17) .. (184.87,466.97) -- cycle ;
\draw  [fill={rgb, 255:red, 0; green, 0; blue, 0 }  ,fill opacity=1 ] (224.87,506.97) .. controls (224.99,505.77) and (226.06,504.89) .. (227.26,505.01) .. controls (228.46,505.13) and (229.33,506.2) .. (229.21,507.4) .. controls (229.09,508.6) and (228.02,509.47) .. (226.82,509.35) .. controls (225.62,509.24) and (224.75,508.17) .. (224.87,506.97) -- cycle ;
\draw  [fill={rgb, 255:red, 0; green, 0; blue, 0 }  ,fill opacity=1 ] (184.87,506.97) .. controls (184.99,505.77) and (186.06,504.89) .. (187.26,505.01) .. controls (188.46,505.13) and (189.33,506.2) .. (189.21,507.4) .. controls (189.09,508.6) and (188.02,509.47) .. (186.82,509.35) .. controls (185.62,509.24) and (184.75,508.17) .. (184.87,506.97) -- cycle ;
\draw  [fill={rgb, 255:red, 0; green, 0; blue, 0 }  ,fill opacity=1 ] (144.87,506.97) .. controls (144.99,505.77) and (146.06,504.89) .. (147.26,505.01) .. controls (148.46,505.13) and (149.33,506.2) .. (149.21,507.4) .. controls (149.09,508.6) and (148.02,509.47) .. (146.82,509.35) .. controls (145.62,509.24) and (144.75,508.17) .. (144.87,506.97) -- cycle ;
\draw  [fill={rgb, 255:red, 0; green, 0; blue, 0 }  ,fill opacity=1 ] (144.87,546.97) .. controls (144.99,545.77) and (146.06,544.89) .. (147.26,545.01) .. controls (148.46,545.13) and (149.33,546.2) .. (149.21,547.4) .. controls (149.09,548.6) and (148.02,549.47) .. (146.82,549.35) .. controls (145.62,549.24) and (144.75,548.17) .. (144.87,546.97) -- cycle ;
\draw  [fill={rgb, 255:red, 0; green, 0; blue, 0 }  ,fill opacity=1 ] (224.87,546.97) .. controls (224.99,545.77) and (226.06,544.89) .. (227.26,545.01) .. controls (228.46,545.13) and (229.33,546.2) .. (229.21,547.4) .. controls (229.09,548.6) and (228.02,549.47) .. (226.82,549.35) .. controls (225.62,549.24) and (224.75,548.17) .. (224.87,546.97) -- cycle ;
\draw  [fill={rgb, 255:red, 0; green, 0; blue, 0 }  ,fill opacity=1 ] (184.87,546.97) .. controls (184.99,545.77) and (186.06,544.89) .. (187.26,545.01) .. controls (188.46,545.13) and (189.33,546.2) .. (189.21,547.4) .. controls (189.09,548.6) and (188.02,549.47) .. (186.82,549.35) .. controls (185.62,549.24) and (184.75,548.17) .. (184.87,546.97) -- cycle ;

\draw (147.04,543.18) node [anchor=south] [inner sep=0.75pt]  [color={rgb, 255:red, 0; green, 0; blue, 255 }  ,opacity=1 ]  {$\lambda _{2}{}$};
\draw (227.04,543.18) node [anchor=south] [inner sep=0.75pt]  [color={rgb, 255:red, 0; green, 0; blue, 255 }  ,opacity=1 ]  {$\lambda _{1}$};
\draw (187.04,503.18) node [anchor=south] [inner sep=0.75pt]  [color={rgb, 255:red, 0; green, 0; blue, 255 }  ,opacity=1 ]  {$\lambda _{3}$};
\draw (227.04,503.18) node [anchor=south] [inner sep=0.75pt]  [color={rgb, 255:red, 0; green, 0; blue, 255 }  ,opacity=1 ]  {$\lambda _{5}$};
\draw (147.04,503.18) node [anchor=south] [inner sep=0.75pt]  [color={rgb, 255:red, 0; green, 0; blue, 255 }  ,opacity=1 ]  {$\lambda _{6}$};
\draw (186.82,465.35) node [anchor=south] [inner sep=0.75pt]  [color={rgb, 255:red, 0; green, 0; blue, 255 }  ,opacity=1 ]  {$\lambda _{7}$};
\draw (187.04,543.18) node [anchor=south] [inner sep=0.75pt]  [color={rgb, 255:red, 0; green, 0; blue, 255 }  ,opacity=1 ]  {$\lambda _{4}$};
\draw (147.04,550.18) node [anchor=north] [inner sep=0.75pt]  [color={rgb, 255:red, 0; green, 0; blue, 0 }  ,opacity=1 ]  {$0$};
\draw (187.04,510.18) node [anchor=north] [inner sep=0.75pt]  [color={rgb, 255:red, 0; green, 0; blue, 0 }  ,opacity=1 ]  {$3$};
\draw (227.04,550.18) node [anchor=north] [inner sep=0.75pt]  [color={rgb, 255:red, 0; green, 0; blue, 0 }  ,opacity=1 ]  {$2$};
\draw (187.04,550.18) node [anchor=north] [inner sep=0.75pt]  [color={rgb, 255:red, 0; green, 0; blue, 0 }  ,opacity=1 ]  {$1$};
\draw (187.04,470.18) node [anchor=north] [inner sep=0.75pt]  [color={rgb, 255:red, 0; green, 0; blue, 0 }  ,opacity=1 ]  {$5$};
\draw (227.04,510.18) node [anchor=north] [inner sep=0.75pt]  [color={rgb, 255:red, 0; green, 0; blue, 0 }  ,opacity=1 ]  {$4$};
\draw (147.04,510.18) node [anchor=north] [inner sep=0.75pt]  [color={rgb, 255:red, 0; green, 0; blue, 0 }  ,opacity=1 ]  {$2$};

\end{tikzpicture}
    \caption{Homology diagram for $5_2$ knot; labels $\lambda_i$ are consistent with~(\ref{eq:5_2 from KRSS reordered}).}
    \label{fig:52_homology}
\end{figure}

We start from the~following generating function of  superpolynomials~\cite{KRSS1707long}
\begin{align}
\nonumber
P_{5_2}(x,a,q,t)=\sum_{r=0}^\infty\frac{x^r }{(q^2;q^2)_r}\sum_{0\leq k_2\leq k_1\leq r} & \left[\begin{array}{c}
    r\\
    k_1
    \end{array}\right]
    \left[\begin{array}{c}
    k_1\\
    k_2
    \end{array}\right](-1)^{r+k_1} (-a^{2}q^{-2}t;q^{2})_{k_{1}}(-a^2q^{2r}t^{3};q^2)_{k_1}\\
&\times a^{2k_2}q^{k_1^2+k_1+2(k_2^2-k_2-rk_1)}t^{2k_2-r}~.
\label{eq:5_2 generating function}
\end{align}
At linear order we find the~superpolynomial $P_1(a,q,t) = a^2q^2t^2 + a^2q^{-2} + a^4 t^3 + a^2t + a^4q^2t^4 + a^4q^{-2} t^2 + a^6 t^5$. The~homological diagram consists of a~diamond and a~zig-zag of length three, see figure~\ref{fig:52_homology}.

The generating function~(\ref{eq:5_2 generating function}) can be rewritten in the~form 
\begin{equation}\label{eq:5_2 prequiver v1}
\begin{split}
    P_{5_{2}}(x,a,q,t)&=\sum_{\boldsymbol{\check{d}}}(-q)^{\boldsymbol{\check{d}}\cdot\check{C}\cdot\boldsymbol{\check{d}}}\frac{\boldsymbol{x}^{\boldsymbol{\check{d}}}}{(q^{2};q^{2})_{\boldsymbol{\check{d}}}}(-a^{2}q^{-2}t;q^{2})_{\check{d}_{2}+\check{d}_{3}+\check{d}_{4}}\Big|_{\boldsymbol{x}=x\boldsymbol{\check{\lambda}}},\\
    \check{C}&=\left[\begin{array}{c:c:c:c}
    0 & 0 & 1 & 1\\
    \hdashline
    0 & 1 & 1 & 2\\
    \hdashline
    1 & 1 & 2 & 2\\
    \hdashline
    1 & 2 & 2 & 4
    \end{array}\right], \qquad  \qquad \boldsymbol{\check{\lambda}} = \left[\begin{array}{c}
    a^2q^{-2}\\
    a^2q^{-1}(-t)\\
    a^2(-t)^2\\
    a^4q^{-2}(-t)^4
    \end{array}\right].    
\end{split}
\end{equation}
Then, (0,1)-splitting of nodes number 2, 3, 4 with trivial permutation, $h_1=0$, and $\kappa=-a^2q^{-3}t$ leads to
\begin{equation}
\label{eq:5_2 quiver}
    C=\left[\begin{array}{c:cc:cc:cc}
    0 & 0 & 0 & 1 & 1 & 1 & 1\\
    \hdashline
    0 & 1 & 1 & 1 & 1 & 2 & 2\\
    0 & 1 & 2 & 2 & 2 & 3 & 3\\
    \hdashline
    1 & 1 & 2 & 2 & 2 & 2 & 2\\
    1 & 1 & 2 & 2 & 3 & 3 & 3\\
    \hdashline
    1 & 2 & 3 & 2 & 3 & 4 & 4\\
    1 & 2 & 3 & 2 & 3 & 4 & 5\\
    \end{array}\right], \qquad  \qquad \boldsymbol{\lambda} = \left[\begin{array}{c}
    a^2q^{-2}\\
    a^2q^{-1}(-t)\\
    a^4q^{-4}(-t)^2\\
    a^2(-t)^2\\
    a^4q^{-3}(-t)^3\\
    a^4q^{-2}(-t)^4\\
    a^6q^{-5}(-t)^5
    \end{array}\right].
\end{equation}
Because the~splitting involves three nodes, it gives rise to a~permutohedron $\Pi_3$, which is a~hexagon.

Furthermore,~(\ref{eq:5_2 generating function}) can be rewritten in another form
\begin{align}
    P_{5_{2}}(x,a,q,t)&=\sum_{\boldsymbol{\check{d}}}(-q)^{\boldsymbol{\check{d}}\cdot\check{C}\cdot\boldsymbol{\check{d}}}\frac{\boldsymbol{x}^{\boldsymbol{\check{d}}}}{(q^{2};q^{2})_{\boldsymbol{\check{d}}}}\Pi_{\check{d}_2,\check{d}_3,\check{d}_4}\Big|_{\boldsymbol{x}=x\boldsymbol{\check{\lambda}}}\nonumber\\
    \check{C}&=\left[\begin{array}{c:c:c:c}
    1 & 0 & 1 & 1\\
\hdashline
0 & 0 & 1 & 1\\
\hdashline
1 & 1 & 2 & 2\\
\hdashline
1 & 1 & 2 & 3\\
\end{array}\right], \qquad  \qquad \boldsymbol{\check{\lambda}} = \left[\begin{array}{c}
    a^2q^{-1}(-t)\\
    a^2q^{-2}\\
    a^2(-t)^2\\
    a^4q^{-3}(-t)^3\\
    \end{array}\right],\label{eq:5_2 prequiver v2}
\end{align}
\begin{align}
    \Pi_{\check{d}_2,\check{d}_3,\check{d}_4}=\sum\limits _{\alpha_{2}+\beta_{2}=\check{d}_{2}}&\sum\limits _{\alpha_{3}+\beta_{3}=\check{d}_{3}}\sum\limits _{\alpha_{4}+\beta_{4}=\check{d}_{4}}(-q)^{2\check{d}_{1}(\beta_{2}+\beta_{3}+\beta_{4})+2(\beta_{2}+\beta_{3}+\beta_{4})^{2}+2(\beta_{2}\alpha_{3}+\beta_{2}\alpha_{4}+\beta_{3}\alpha_{4})}\nonumber\\
    & \times\frac{(a^{2}q^{-2}t^{2})^{(\beta_{2}+\beta_{3}+\beta_{4})}(q^{2};q^{2})_{\check{d}_{2}}(q^{2};q^{2})_{\check{d}_{3}}(q^{2};q^{2})_{\check{d}_{4}}}{(q^{2};q^{2})_{\alpha_{2}}(q^{2};q^{2})_{\beta_{2}}(q^{2};q^{2})_{\alpha_{3}}(q^{2};q^{2})_{\beta_{3}}(q^{2};q^{2})_{\alpha_{4}}(q^{2};q^{2})_{\beta_{4}}}.\nonumber
\end{align}
In this case the~factor $\Pi_{d_2,d_3,d_4}$ encodes $(0,2)$-splitting of the~last three nodes with trivial permutation, $h_1=1$, and $\kappa=a^2q^{-2}t^2$, which leads to a~rearrangment of  quiver~\eqref{eq:5_2 quiver}:
\begin{equation}
\label{eq:5_2 quiver reordered}
    C=\left[\begin{array}{c:cc:cc:cc}
    1 & 0 & 1 & 1 & 2 & 1 & 2\\
    \hdashline
    0 & 0 & 0 & 1 & \myboxy{1} & 1 & 1\\
    1 & 0 & 2 & \myboxy{2} & 3 & 2 & 3\\
    \hdashline
    1 & 1 & \myboxy{2} & 2 & 2 & 2 & 2\\
    2 & \myboxy{1} & 3 & 2 & 4 & 3 & 4\\
    \hdashline
    1 & 1 & 2 & 2 & 3 & 3 & 3\\
    2 & 1 & 3 & 2 & 4 & 3 & 5\\
    \end{array}\right], \qquad  \qquad \boldsymbol{\lambda} = \left[\begin{array}{c}
    a^2q^{-1}(-t)\\
    a^2q^{-2}\\
    a^2(-t)^2\\
    a^4q^{-4}(-t)^2\\
    a^4q^{-2}(-t)^4\\
    a^4q^{-3}(-t)^3\\
    a^6q^{-5}(-t)^5
    \end{array}\right].
\end{equation}
This means that the~corresponding permutohedron is also $\Pi_3$, and one of its vertices corresponding to the~above matrix is shared with the~previous permutohedron (there is also another quiver common to these two permutohedra). Note that $(0,2)$-splitting of prequiver~\eqref{eq:5_2 prequiver v2} with permutation $\sigma=(2\ 3)$ yields the~quiver for $5_2$ knot found in~\cite{KRSS1707long}:
\begin{equation}
\label{eq:5_2 from KRSS}
    C=\left[\begin{array}{c:cc:cc:cc}
    1 & 0 & 1 & 1 & 2 & 1 & 2\\
    \hdashline
    0 & 0 & 0 & 1 & \myboxy{2} & 1 & 1\\
    1 & 0 & 2 & \myboxy{1} & 3 & 2 & 3\\
    \hdashline
    1 & 1 & \myboxy{1} & 2 & 2 & 2 & 2\\
    2 & \myboxy{2} & 3 & 2 & 4 & 3 & 4\\
    \hdashline
    1 & 1 & 2 & 2 & 3 & 3 & 3\\
    2 & 1 & 3 & 2 & 4 & 3 & 5\\
    \end{array}\right], \qquad  \qquad \boldsymbol{\lambda} = \left[\begin{array}{c}
    a^2q^{-1}(-t)\\
    a^2q^{-2}\\
    a^2(-t)^2\\
    a^4q^{-4}(-t)^2\\
    a^4q^{-2}(-t)^4\\
    a^4q^{-3}(-t)^3\\
    a^6q^{-5}(-t)^5
    \end{array}\right].
\end{equation}
Elements that are transposed between~\eqref{eq:5_2 quiver reordered} and~(\ref{eq:5_2 from KRSS})
are highlighted in yellow.

Furthermore, the~quiver~\eqref{eq:5_2 from KRSS} also admits the~inverse of another splitting, which corresponds to the~following rewriting of~(\ref{eq:5_2 generating function}):
\begin{equation}\label{eq:5_2 prequiver v3}
\begin{split}
    P_{5_{2}}(x,a,q,t)&=\left.\sum_{\boldsymbol{\check{d}}}(-q)^{\boldsymbol{\check{d}}\cdot\check{C}\cdot\boldsymbol{\check{d}}}\frac{\boldsymbol{x}^{\boldsymbol{\check{d}}}}{(q^{2};q^{2})_{\boldsymbol{\check{d}}}}(-a^{2}q^{2r}t^{3};q^{2})_{\check{d}_{2}+\check{d}_{3}+\check{d}_{4}}\right|_{\boldsymbol{x}=x\boldsymbol{\check{\lambda}}}\\
    \check{C}&=\left[\begin{array}{c:c:c:c}
2 & 1 & 1 & 1\\
\hdashline
1 & 0 & 0 & 0\\
\hdashline
1 & 0 & 1 & 1\\
\hdashline
1 & 0 & 1 & 2
\end{array}\right] \qquad  \qquad \boldsymbol{\check{\lambda}} = \left[\begin{array}{c}
    a^2(-t)^2\\
    a^2q^{-2}\\
    a^2q^{-1}(-t)\\
    a^4q^{-4}(-t)^2
    \end{array}\right].
\end{split}
\end{equation}
In this case $(1,3)$-splitting of the~last three nodes with permutation $\sigma=(2\ 3)$, $h_1=1$, and $\kappa=-a^2q^{-1}t^3$ leads to
\begin{equation}
\label{eq:5_2 from KRSS reordered}
    C=\left[\begin{array}{c:cc:cc:cc}
    2 & 1 & 2 & 1 & 2 & 1 & 2\\
    \hdashline
    1 & 0 & 1 & 0 & 2 & 0 & 1\\
    2 & 1 & 3 & 1 & 3 & 2 & 3\\
    \hdashline
    1 & 0 & 1 & 1 & 2 & 1 & 2\\
    2 & 2 & 3 & 2 & 4 & 3 & 4\\
    \hdashline
    1 & 0 & 2 & 1 & 3 & 2 & 3\\
    2 & 1 & 3 & 2 & 4 & 3 & 5\\
    \end{array}\right], \qquad  \qquad \boldsymbol{\lambda} = \left[\begin{array}{c}
    a^2(-t)^2\\
    a^2q^{-2}\\
    a^4q^{-3}(-t)^3\\
    a^2q^{-1}(-t)\\    
    a^4q^{-2}(-t)^4\\
    a^4q^{-4}(-t)^2\\
    a^6q^{-5}(-t)^5
    \end{array}\right],
\end{equation}
which is also a~reordering of ~\eqref{eq:5_2 from KRSS}. This means that~(\ref{eq:5_2 prequiver v3}) captures the~third permutohedron $\Pi_3$, and the~quiver~(\ref{eq:5_2 from KRSS}) (or its reordered version~(\ref{eq:5_2 from KRSS reordered})) corresponds to the~vertex that is shared with the~previous $\Pi_3$.

\begin{figure}[h!]
    \centering
    \includegraphics[scale=0.26]{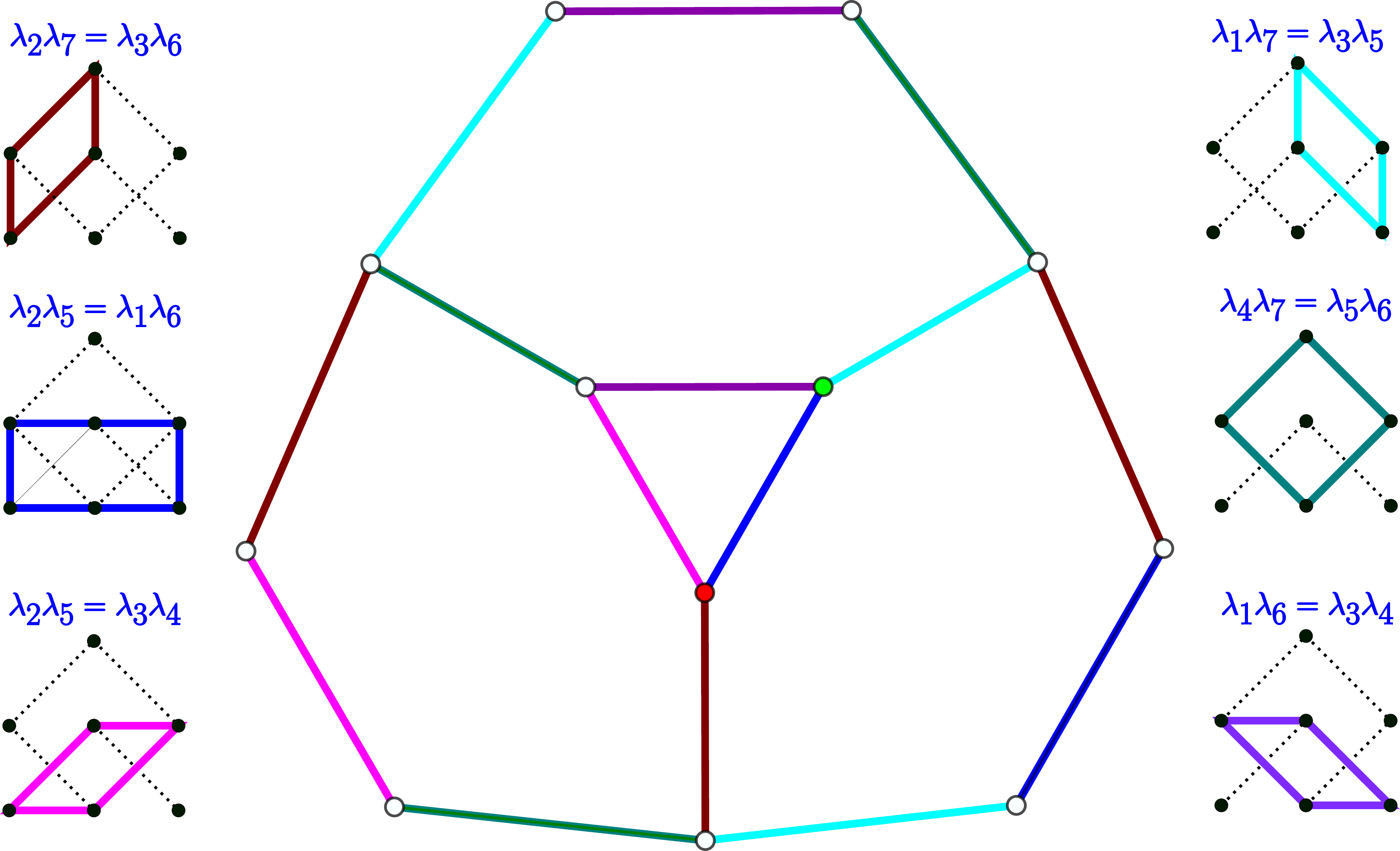}
    \vspace{0.5cm}
    \tikzset{every picture/.style={line width=0.75pt}} 

\begin{tikzpicture}[x=0.75pt,y=0.75pt,yscale=-1,xscale=1]

\draw  [fill={rgb, 255:red, 255; green, 255; blue, 255 }  ,fill opacity=1 ] (723.54,90.1) -- (716.76,101.83) -- (703.21,101.83) -- (696.43,90.1) -- (703.21,78.36) -- (716.76,78.36) -- cycle ;
\draw  [fill={rgb, 255:red, 155; green, 155; blue, 155 }  ,fill opacity=1 ] (764.2,90.1) -- (757.42,101.83) -- (743.87,101.83) -- (737.09,90.1) -- (743.87,78.36) -- (757.42,78.36) -- cycle ;
\draw   (743.87,54.88) -- (737.09,66.62) -- (723.54,66.62) -- (716.76,54.88) -- (723.54,43.15) -- (737.09,43.15) -- cycle ;
\draw  [dash pattern={on 0.84pt off 2.51pt}]  (723.54,66.62) -- (716.76,78.36) ;
\draw  [dash pattern={on 0.84pt off 2.51pt}]  (716.76,54.88) -- (703.21,78.36) ;
\draw  [dash pattern={on 0.84pt off 2.51pt}]  (737.09,66.62) -- (743.87,78.36) ;
\draw  [dash pattern={on 0.84pt off 2.51pt}]  (743.87,54.88) -- (757.42,78.36) ;
\draw  [dash pattern={on 0.84pt off 2.51pt}]  (743.87,101.83) -- (716.76,101.83) ;
\draw  [dash pattern={on 0.84pt off 2.51pt}]  (737.09,90.1) -- (723.54,90.1) ;
\draw   (603.54,90.1) -- (596.76,101.83) -- (583.21,101.83) -- (576.43,90.1) -- (583.21,78.36) -- (596.76,78.36) -- cycle ;
\draw   (644.2,90.1) -- (637.42,101.83) -- (623.87,101.83) -- (617.09,90.1) -- (623.87,78.36) -- (637.42,78.36) -- cycle ;
\draw  [fill={rgb, 255:red, 155; green, 155; blue, 155 }  ,fill opacity=1 ] (623.87,54.88) -- (617.09,66.62) -- (603.54,66.62) -- (596.76,54.88) -- (603.54,43.15) -- (617.09,43.15) -- cycle ;
\draw  [dash pattern={on 0.84pt off 2.51pt}]  (603.54,66.62) -- (596.76,78.36) ;
\draw  [dash pattern={on 0.84pt off 2.51pt}]  (596.76,54.88) -- (583.21,78.36) ;
\draw  [dash pattern={on 0.84pt off 2.51pt}]  (617.09,66.62) -- (623.87,78.36) ;
\draw  [dash pattern={on 0.84pt off 2.51pt}]  (623.87,54.88) -- (637.42,78.36) ;
\draw  [dash pattern={on 0.84pt off 2.51pt}]  (623.87,101.83) -- (596.76,101.83) ;
\draw  [dash pattern={on 0.84pt off 2.51pt}]  (617.09,90.1) -- (603.54,90.1) ;
\draw  [fill={rgb, 255:red, 155; green, 155; blue, 155 }  ,fill opacity=1 ] (841.54,90.1) -- (834.76,101.83) -- (821.21,101.83) -- (814.43,90.1) -- (821.21,78.36) -- (834.76,78.36) -- cycle ;
\draw  [fill={rgb, 255:red, 255; green, 255; blue, 255 }  ,fill opacity=1 ] (882.2,90.1) -- (875.42,101.83) -- (861.87,101.83) -- (855.09,90.1) -- (861.87,78.36) -- (875.42,78.36) -- cycle ;
\draw   (861.87,54.88) -- (855.09,66.62) -- (841.54,66.62) -- (834.76,54.88) -- (841.54,43.15) -- (855.09,43.15) -- cycle ;
\draw  [dash pattern={on 0.84pt off 2.51pt}]  (841.54,66.62) -- (834.76,78.36) ;
\draw  [dash pattern={on 0.84pt off 2.51pt}]  (834.76,54.88) -- (821.21,78.36) ;
\draw  [dash pattern={on 0.84pt off 2.51pt}]  (855.09,66.62) -- (861.87,78.36) ;
\draw  [dash pattern={on 0.84pt off 2.51pt}]  (861.87,54.88) -- (875.42,78.36) ;
\draw  [dash pattern={on 0.84pt off 2.51pt}]  (861.87,101.83) -- (834.76,101.83) ;
\draw  [dash pattern={on 0.84pt off 2.51pt}]  (855.09,90.1) -- (841.54,90.1) ;

\draw (610,108) node [anchor=north] [inner sep=0.75pt]    {(\ref{eq:5_2 prequiver v1})};
\draw (730,108) node [anchor=north] [inner sep=0.75pt]    {(\ref{eq:5_2 prequiver v2})};
\draw (850,108) node [anchor=north] [inner sep=0.75pt]    {(\ref{eq:5_2 prequiver v3})};

\end{tikzpicture}
    \caption{The permutohedra graph for $5_2$ knot consists of three $\Pi_3$ (shown schematically in bottom together with the~formulas they correspond to) glued along common edges. The~edges in this graph correspond to 6 types of transpositions arising from various quadruples of homology generators, which are also shown in various colors on the~homological diagrams. }
    \label{fig:symmetries_5_2}
\end{figure}
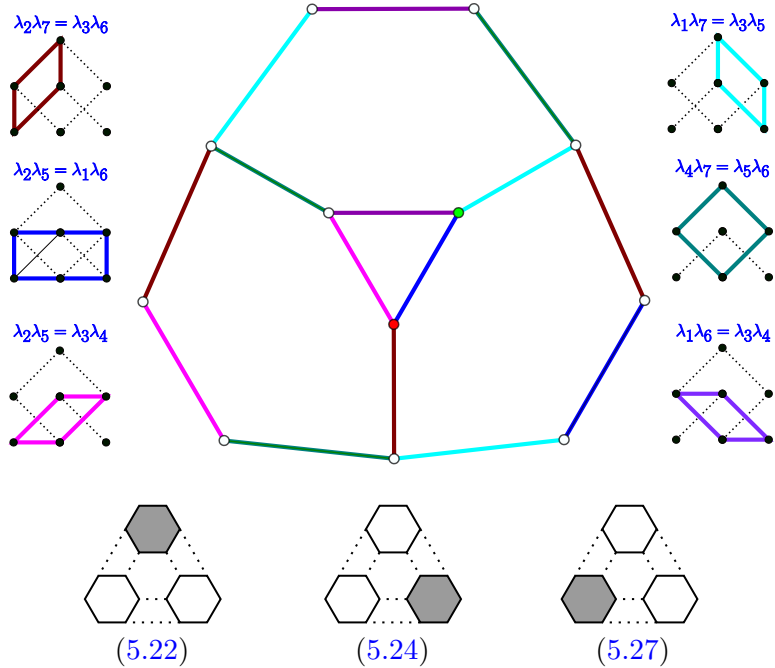

Following the~above analysis we find that the~permutohedra graph for $5_2$ has the~structure shown in figure~\ref{fig:symmetries_5_2}. The~permutohedron arising from 6 permutations associated to $(0,1)$-splitting of the~prequiver~\eqref{eq:5_2 prequiver v1} lies on the~top of the~graph. The~bottom-right $\Pi_3$ comes from all possible $(0,2)$-splittings of the~prequiver~\eqref{eq:5_2 prequiver v2}. Finally, $(1,3)$-splittings of~\eqref{eq:5_2 prequiver v3} lead to the~bottom-left hexagon. The~quiver~(\ref{eq:5_2 quiver}) (or its reordered form~(\ref{eq:5_2 quiver reordered})) is denoted by the~green dot. The~red dot represents the~quiver~(\ref{eq:5_2 from KRSS}) (or its reordered form~(\ref{eq:5_2 from KRSS reordered}))
found in~\cite{KRSS1707long}. The~symmetry connecting these two quivers is denoted by the~blue edge. Moreover, we find that each pair of permutohedra $\Pi_3$ identified above has 2 common quivers, which are connected by a~transposition that is also common to such two permutohedra. Altogether, the~permutohedra graph takes form of 3 permutohedra $\Pi_3$ glued along their edges, as shown in figure~\ref{fig:symmetries_5_2}. The~triangle in the~middle of the~graph represents two transpositions whose composition is also a~transposition (not a~3-cycle), so it does not contradict the~argument in section~\ref{sec:possible equivalences}. In the~figure we also show how various symmetries (transpositions of matrix elements that relate various equivalent quivers, which correspond to edges of the~permutohedra graph) arise from quadruples of homology generators, and denote them in various colors. According to conjecture~\ref{coj:necessary conditions}, we expect that figure~\ref{fig:symmetries_5_2} presents the~whole equivalence class of quivers.


\subsection{\texorpdfstring{$7_1$}{71} knot}\label{71 knot}

Another interesting example is $7_1$ knot. Applying  theorem~\ref{thm:main} systematically, we find 13~equivalent quivers, which we list explicitly in the~appendix~\ref{appendix_matrices}. More detailed analysis reveals that they form two permuthohedra $\Pi_3$ that share one common vertex (corresponding to a~common quiver), and each of these $\Pi_3$ in addition shares a~common vertex with one of the~two permutohedra $\Pi_2$.  

The generating function of colored superpolynomials takes the~form~\cite{FGSS1209,AFGS1203}
\begin{align}\label{eq:7_1 P(x)}
    P_{7_1}(x,a,q,t)=\sum_{r=0}^{\infty} \frac{x^r a^{6r}q^{-6r}}{(q^2;q^2)_r}\sum_{0 \leq k_3 \leq k_2 \leq k_1 \leq r} \left[\begin{array}{c}
    r\\
    k_1
    \end{array}\right]
    \left[\begin{array}{c}
    k_1\\
    k_2
    \end{array}\right]
    \left[\begin{array}{c}
    k_2\\
    k_3
    \end{array}\right]
    (-a^2q^{-2}t;q^2)_{k_1}\\
    \times \, q^{2[(2r+1)(k_1+k_2+k_3)-r k_1-k_1 k_2-k_2 k_3]}t^{2(k_1+k_2+k_3)}.    \nonumber
\end{align}
For $r = 1$ we get the~uncolored superpolynomial $P_1(a,q,t) = a^6 q^{-6} + a^6 q^{-2} t^2  + a^8 q^{-4} t^3 + a^6 q^{2} t^4 + a^8 t^5 + a^6 q^6 t^6 + a^8 q^4 t^7$. The~corresponding homological diagram consists of one zig-zag made of 7~nodes, see figure~\ref{fig:71homology}.
\begin{figure}[h!]
    \centering
    \tikzset{every picture/.style={line width=0.75pt}} 

\begin{tikzpicture}[x=0.75pt,y=0.75pt,yscale=-1,xscale=1]

\draw  [fill={rgb, 255:red, 0; green, 0; blue, 0 }  ,fill opacity=1 ][line width=1.5]  (1774.83,893.27) .. controls (1775.82,892.67) and (1777.12,892.99) .. (1777.71,893.99) .. controls (1778.31,894.99) and (1777.99,896.28) .. (1776.99,896.88) .. controls (1776,897.48) and (1774.7,897.15) .. (1774.1,896.16) .. controls (1773.51,895.16) and (1773.83,893.87) .. (1774.83,893.27) -- cycle ;
\draw  [fill={rgb, 255:red, 0; green, 0; blue, 0 }  ,fill opacity=1 ][line width=1.5]  (1855.59,893.27) .. controls (1856.59,892.67) and (1857.88,892.99) .. (1858.48,893.99) .. controls (1859.08,894.99) and (1858.75,896.28) .. (1857.76,896.88) .. controls (1856.76,897.48) and (1855.47,897.15) .. (1854.87,896.16) .. controls (1854.27,895.16) and (1854.59,893.87) .. (1855.59,893.27) -- cycle ;
\draw  [fill={rgb, 255:red, 0; green, 0; blue, 0 }  ,fill opacity=1 ][line width=1.5]  (1815.21,852.89) .. controls (1816.21,852.29) and (1817.5,852.61) .. (1818.1,853.61) .. controls (1818.69,854.6) and (1818.37,855.9) .. (1817.37,856.49) .. controls (1816.38,857.09) and (1815.09,856.77) .. (1814.49,855.77) .. controls (1813.89,854.78) and (1814.21,853.48) .. (1815.21,852.89) -- cycle ;
\draw  [fill={rgb, 255:red, 0; green, 0; blue, 0 }  ,fill opacity=1 ][line width=1.5]  (1895.97,852.89) .. controls (1896.97,852.29) and (1898.26,852.61) .. (1898.86,853.61) .. controls (1899.46,854.6) and (1899.14,855.9) .. (1898.14,856.49) .. controls (1897.14,857.09) and (1895.85,856.77) .. (1895.25,855.77) .. controls (1894.65,854.78) and (1894.98,853.48) .. (1895.97,852.89) -- cycle ;
\draw  [fill={rgb, 255:red, 0; green, 0; blue, 0 }  ,fill opacity=1 ][line width=1.5]  (1936.36,893.27) .. controls (1937.35,892.67) and (1938.65,892.99) .. (1939.24,893.99) .. controls (1939.84,894.99) and (1939.52,896.28) .. (1938.52,896.88) .. controls (1937.53,897.48) and (1936.23,897.15) .. (1935.63,896.16) .. controls (1935.04,895.16) and (1935.36,893.87) .. (1936.36,893.27) -- cycle ;
\draw  [dash pattern={on 0.84pt off 2.51pt}]  (1897.06,854.69) -- (1936.36,893.27) ;
\draw  [fill={rgb, 255:red, 0; green, 0; blue, 0 }  ,fill opacity=1 ][line width=1.5]  (1976.74,852.89) .. controls (1977.74,852.29) and (1979.03,852.61) .. (1979.63,853.61) .. controls (1980.22,854.6) and (1979.9,855.9) .. (1978.9,856.49) .. controls (1977.91,857.09) and (1976.62,856.77) .. (1976.02,855.77) .. controls (1975.42,854.78) and (1975.74,853.48) .. (1976.74,852.89) -- cycle ;
\draw  [fill={rgb, 255:red, 0; green, 0; blue, 0 }  ,fill opacity=1 ][line width=1.5]  (2017.12,893.27) .. controls (2018.12,892.67) and (2019.41,892.99) .. (2020.01,893.99) .. controls (2020.61,894.99) and (2020.28,896.28) .. (2019.29,896.88) .. controls (2018.29,897.48) and (2017,897.15) .. (2016.4,896.16) .. controls (2015.8,895.16) and (2016.12,893.87) .. (2017.12,893.27) -- cycle ;
\draw  [dash pattern={on 0.84pt off 2.51pt}]  (1816.29,854.69) -- (1855.59,893.27) ;
\draw  [dash pattern={on 0.84pt off 2.51pt}]  (1977.82,854.69) -- (2017.12,893.27) ;
\draw  [dash pattern={on 0.84pt off 2.51pt}]  (1937.44,895.07) -- (1977.82,854.69) ;
\draw  [dash pattern={on 0.84pt off 2.51pt}]  (1856.67,895.07) -- (1897.06,854.69) ;
\draw  [dash pattern={on 0.84pt off 2.51pt}]  (1775.91,895.07) -- (1816.29,854.69) ;

\draw (1775.91,885.07) node [anchor=south] [inner sep=0.75pt]  [color={rgb, 255:red, 0; green, 0; blue, 255 }  ,opacity=1 ]  {${\lambda _{1}}$};
\draw (1816.29,844.69) node [anchor=south] [inner sep=0.75pt]  [color={rgb, 255:red, 0; green, 0; blue, 255 }  ,opacity=1 ]  {${\lambda _{3}}$};
\draw (1856.67,885.07) node [anchor=south] [inner sep=0.75pt]  [color={rgb, 255:red, 0; green, 0; blue, 255 }  ,opacity=1 ]  {${\lambda _{2}}$};
\draw (1937.44,885.07) node [anchor=south] [inner sep=0.75pt]  [color={rgb, 255:red, 0; green, 0; blue, 255 }  ,opacity=1 ]  {${\lambda _{4}}$};
\draw (1897.06,844.69) node [anchor=south] [inner sep=0.75pt]  [color={rgb, 255:red, 0; green, 0; blue, 255 }  ,opacity=1 ]  {${\lambda _{5}}$};
\draw (1977.82,844.69) node [anchor=south] [inner sep=0.75pt]  [color={rgb, 255:red, 0; green, 0; blue, 255 }  ,opacity=1 ]  {${\lambda _{7}}$};
\draw (2018.2,885.07) node [anchor=south] [inner sep=0.75pt]  [color={rgb, 255:red, 0; green, 0; blue, 255 }  ,opacity=1 ]  {${\lambda _{6}}$};
\draw (1776.91,906.07) node [anchor=north] [inner sep=0.75pt]  [color={rgb, 255:red, 0; green, 0; blue, 0 }  ,opacity=1 ]  {$0$};
\draw (1817.29,865.69) node [anchor=north] [inner sep=0.75pt]  [color={rgb, 255:red, 0; green, 0; blue, 0 }  ,opacity=1 ]  {$3$};
\draw (1857.67,906.07) node [anchor=north] [inner sep=0.75pt]  [color={rgb, 255:red, 0; green, 0; blue, 0 }  ,opacity=1 ]  {$2$};
\draw (1938.44,906.07) node [anchor=north] [inner sep=0.75pt]  [color={rgb, 255:red, 0; green, 0; blue, 0 }  ,opacity=1 ]  {$4$};
\draw (1898.06,865.69) node [anchor=north] [inner sep=0.75pt]  [color={rgb, 255:red, 0; green, 0; blue, 0 }  ,opacity=1 ]  {$5$};
\draw (1978.82,865.69) node [anchor=north] [inner sep=0.75pt]  [color={rgb, 255:red, 0; green, 0; blue, 0 }  ,opacity=1 ]  {$7$};
\draw (2019.2,906.07) node [anchor=north] [inner sep=0.75pt]  [color={rgb, 255:red, 0; green, 0; blue, 0 }  ,opacity=1 ]  {$6$};

\end{tikzpicture}
    \caption{Homology diagram for $7_1$ knot; labels $\lambda_i$ are consistent with~(\ref{eq:7_1 quiver}).}
    \label{fig:71homology}
\end{figure}

First, we rewrite~(\ref{eq:7_1 P(x)}) as follows:
\begin{equation}\label{eq:7_1 prequiver v1}
\begin{split}
    P_{7_{1}}(x,a,q,t)&=\sum_{\boldsymbol{\check{d}}}(-q)^{\boldsymbol{\check{d}}\cdot\check{C}\cdot\boldsymbol{\check{d}}}\frac{\boldsymbol{\check{x}}^{\boldsymbol{\check{d}}}}{(q^{2};q^{2})_{\boldsymbol{\check{d}}}}(-a^{2}q^{-2}t;q^{2})_{\check{d}_{2}+\check{d}_{3}+\check{d}_{4}}\Big|_{\boldsymbol{\check{x}}=x\boldsymbol{\check{\lambda}}}\\
    \check{C}&=\left[\begin{array}{c:c:c:c}
    0 & 1 & 3 & 5\\
    \hdashline
    1 & 2 & 3 & 5\\
    \hdashline
    3 & 3 & 4 & 5\\
    \hdashline
    5 & 5 & 5 & 6
    \end{array}\right], \qquad  \qquad \boldsymbol{\check{\lambda}} = \left[\begin{array}{c}
    a^6q^{-6}\\
    a^6q^{-4}(-t)^2\\
    a^6q^{-2} (-t)^4\\
    a^6(-t)^6
    \end{array}\right].
\end{split}
\end{equation}
The (0,1)-splitting of the~last three nodes with trivial permutation, $h_1=0$, and $\kappa=-a^2q^{-3}t$ leads to
\begin{equation}
\label{eq:7_1 quiver}
    C=\left[\begin{array}{c:cc:cc:cc}
    0 & 1 & 1 & 3 & 3 & 5 & 5\\
    \hdashline
    1 & 2 & 2 & 3 & 3 & 5 & 5\\
    1 & 2 & 3 & 4 & 4 & 6 & 6\\
    \hdashline
    3 & 3 & 4 & 4 & 4 & 5 & 5\\
    3 & 3 & 4 & 4 & 5 & 6 & 6\\
    \hdashline
    5 & 5 & 6 & 5 & 6 & 6 & 6\\
    5 & 5 & 6 & 5 & 6 & 6 & 7\\
    \end{array}\right], \qquad   \boldsymbol{\lambda} = \left[\begin{array}{c}
    a^6q^{-6}\\
    a^6q^{-4}(-t)^2\\
    a^8q^{-7}(-t)^3\\
    a^6q^{-2} (-t)^4\\
    a^8q^{-5}(-t)^5\\
    a^6(-t)^6\\
    a^8q^{-3}(-t)^7
    \end{array}\right],
\end{equation}
which reproduces the~quiver from~\cite{KRSS1707long}. More generally, splitting these three nodes with all possible permutations yields one permutohedron $\Pi_3$.

Furthermore, we can also rewrite~(\ref{eq:7_1 P(x)}) as
\begin{equation}\label{eq:7_1 prequiver v2}
\begin{split}
    P_{7_{1}}(x,a,q,t)&=\sum_{\boldsymbol{\check{d}}}(-q)^{\boldsymbol{\check{d}}\cdot\check{C}\cdot\boldsymbol{\check{d}}}\frac{\boldsymbol{\check{x}}^{\boldsymbol{\check{d}}}}{(q^{2};q^{2})_{\boldsymbol{\check{d}}}}(-a^{2}q^{2r}t^{3};q^{2})_{\check{d}_{2}+\check{d}_{3}+\check{d}_{4}}\Big|_{\boldsymbol{\check{x}}=x\boldsymbol{\check{\lambda}}}\\
    \check{C}&=\left[\begin{array}{c:c:c:c}
6 & 5 & 5 & 5\\
\hdashline
5 & 0 & 1 & 3\\
\hdashline
5 & 1 & 2 & 3\\
\hdashline
5 & 3 & 3 & 4
    \end{array}\right], \qquad  \qquad \boldsymbol{\check{\lambda}} = \left[\begin{array}{c}
    a^6(-t)^6\\
    a^6q^{-6}\\
    a^6q^{-4}(-t)^2\\
    a^6q^{-2} (-t)^4
    \end{array}\right].    
\end{split}
\end{equation}
In this case $(1,3)$-splitting of the~last three nodes with permutation $\sigma=(2\ 4)$, $h_1=1$, and $\kappa=-a^2q^{-1}t^3$ gives a~rearrangment of the~quiver~\eqref{eq:7_1 quiver}:
\begin{equation}
\label{eq:7_1 quiver v2}
    C=\left[\begin{array}{c:cc:cc:cc}
    6 & 5 & 6 & 5 & 6 & 5 & 6\\
    \hdashline
    5 & 0 & 1 & 1 & 3 & 3 & 5\\
    6 & 1 & 3 & 2 & 4 & 4 & 6\\
    \hdashline
    5 & 1 & 2 & 2 & 3 & 3 & 5\\
    6 & 3 & 4 & 3 & 5 & 4 & 6\\
    \hdashline
    5 & 3 & 4 & 3 & 4 & 4 & 5\\
    6 & 5 & 6 & 5 & 6 & 5 & 7\\
    \end{array}\right], \qquad   \boldsymbol{\lambda} = \left[\begin{array}{c}
    a^6(-t)^6\\
    a^6q^{-6}\\
    a^8q^{-7}(-t)^3\\
    a^6q^{-4}(-t)^2\\
    a^8q^{-5}(-t)^5\\
    a^6q^{-2} (-t)^4\\
    a^8q^{-3}(-t)^7
    \end{array}\right],
\end{equation}
and analogous splittings with all other permutations give rise to another permutohedron~$\Pi_3$. Therefore we have identified two permutohedra that share a~common vertex, which represents the~quiver matrix ~(\ref{eq:7_1 quiver}) (or its reordered form~(\ref{eq:7_1 quiver v2})). Let us now focus on $\Pi_3$ arising from the~prequiver~\eqref{eq:7_1 prequiver v1}. One can check that almost all quivers represented by its other vertices cannot be obtained from other prequivers. The~only exception is
\begin{equation}
\label{eq:7_1 quiver v3}
    C=\left[\begin{array}{c:cc:cc:cc}
    0 & 1 & 1 & 3 & 3 & 5 & 5\\
    \hdashline
    1 & 2 & 2 & 3 & 4 & 5 & 6\\
    1 & 2 & 3 & 3 & 4 & 5 & 6\\
    \hdashline
    3 & 3 & 3 & 4 & 4 & 5 & 6\\
    3 & 4 & 4 & 4 & 5 & 5 & 6\\
    \hdashline
    5 & 5 & 5 & 5 & 5 & 6 & 6\\
    5 & 6 & 6 & 6 & 6 & 6 & 7\\
    \end{array}\right], \qquad   \boldsymbol{\lambda} = \left[\begin{array}{c}
    a^6q^{-6}\\
    a^6q^{-4}(-t)^2\\
    a^8q^{-7}(-t)^3\\
    a^6q^{-2} (-t)^4\\
    a^8q^{-5}(-t)^5\\
    a^6(-t)^6\\
    a^8q^{-3}(-t)^7
    \end{array}\right],
\end{equation}
that arises from (0,1)-splitting of~\eqref{eq:7_1 prequiver v1} with permutation $\sigma=(2\ 4)$. Indeed, $(0,1)$-splitting  of the~last two nodes of the~prequiver
\begin{equation}\label{eq:7_1 prequiver v3}
    \check{C}=\left[\begin{array}{ccc:c:c}
    0 & 1 & 5 & 1 & 3\\
    1 & 2 & 6 & 2 & 4\\
    5 & 6 & 7 & 6 & 6\\
    \hdashline
    1 & 2 & 6 & 3 & 4\\
    \hdashline
    3 & 4 & 6 & 4 & 5\\
    \end{array}\right], \qquad  \qquad \boldsymbol{\check{\lambda}} = \left[\begin{array}{c}
    a^6q^{-6}\\
    a^6q^{-4}(-t)^2\\
    a^8q^{-3}(-t)^7\\
    a^8q^{-7}(-t)^3\\
    a^8q^{-5}(-t)^5
    \end{array}\right],
\end{equation}
with permutation $\sigma=(4\ 5)$, $h_1=2,\,h_2=1,\,h_3=0$, and $\kappa=-a^{-2}q^5t$ leads to
\begin{equation}\label{eq:7_1 quiver v4}
    C=\left[\begin{array}{ccc:cc:cc}
    0 & 1 & 5 & 1 & 3 & 3 & 5\\
    1 & 2 & 6 & 2 & 3 & 4 & 5\\
    5 & 6 & 7 & 6 & 6 & 6 & 6\\
    \hdashline
    1 & 2 & 6 & 3 & 3 & 4 & 5\\
    3 & 3 & 6 & 3 & 4 & 4 & 5\\
    \hdashline
    3 & 4 & 6 & 4 & 4 & 5 & 5\\
    5 & 5 & 6 & 5 & 5 & 5 & 6\\
    \end{array}\right], \qquad  \qquad \boldsymbol{\lambda} = \left[\begin{array}{c}
    a^6q^{-6}\\
    a^6q^{-4}(-t)^2\\
    a^8q^{-3}(-t)^7\\
    a^8q^{-7}(-t)^3\\
    a^6q^{-2} (-t)^4\\
    a^8q^{-5}(-t)^5\\
    a^6q^{-6}
    \end{array}\right],
\end{equation}
which is a~rearrangement of~\eqref{eq:7_1 quiver v3}. This means that the~quiver~(\ref{eq:7_1 quiver v3}) (or its reordered form~(\ref{eq:7_1 quiver v4})) is a~gluing point of permutohedra $\Pi_3$ and $\Pi_2$.

\begin{figure}[h!]
    \centering
    \input{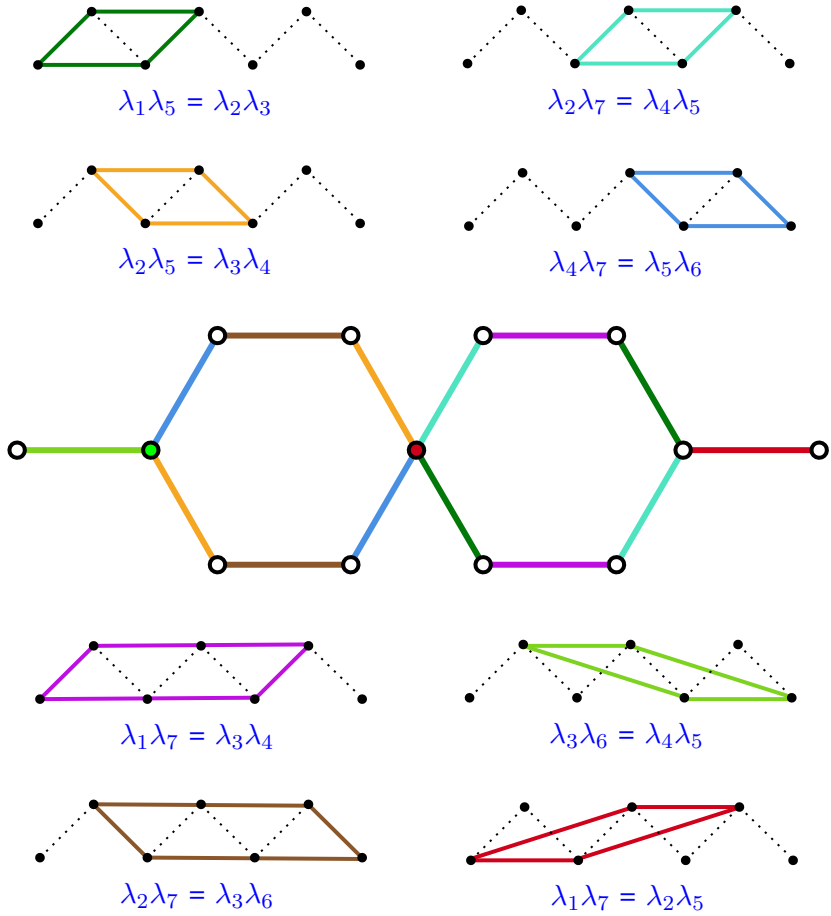}
    \caption{The permutohedra graph for $7_1$ knot consists of two $\Pi_3$ and two $\Pi_2$ appropriately glued. Altogether it has 13 vertices representing equivalent quivers, and 8 symmetries corresponding to various quadruples of homology generators (and represented by different colors of the~edges in the~graph).}
    \label{fig:graph71}
\end{figure}

An analogous phenomenon occurs for the~second $\Pi_3$, which is also connected to another permutohedron $\Pi_2$. Altogether, the~permutohedra graph consists of two $\Pi_3$ and two $\Pi_2$, as shown in figure~\ref{fig:graph71}. The~quiver~(\ref{eq:7_1 quiver}) (or equivalently~(\ref{eq:7_1 quiver v2})), also found in~\cite{KRSS1707long}, is common to the~two $\Pi_3$ and it is represented by the~red dot. The~$\Pi_3$ on the~left arises from the~prequiver~\eqref{eq:7_1 prequiver v1}, whereas the~one on the~right corresponds to the~prequiver~\eqref{eq:7_1 prequiver v2}. The~quiver~(\ref{eq:7_1 quiver v3}) (or its reordered form~(\ref{eq:7_1 quiver v4})) is represented by the~green node, and it glues the~left $\Pi_3$ with $\Pi_2$ coming from the~prequiver~\eqref{eq:7_1 prequiver v3}. Analogous gluing point is present on the~right-hand side of the~graph. In total we found 8 non-trivial symmetries shown in figure~\ref{fig:graph71} in various colors, and 13 equivalent quivers  that we list explicitly in the~appendix~\ref{appendix_matrices}. Using the~procedure described in section~\ref{sec:possible equivalences} we checked that there are no other equivalent quivers. According to conjecture~\ref{coj:necessary conditions}, we expect that figure~\ref{fig:graph71} presents the~whole equivalence class of quivers.


\subsection{\texorpdfstring{$6_1$}{61} knot}

\begin{figure}[h!]
     \centering
     \begin{subfigure}[b]{0.45\textwidth}
         \centering
         \includegraphics[width=\textwidth]{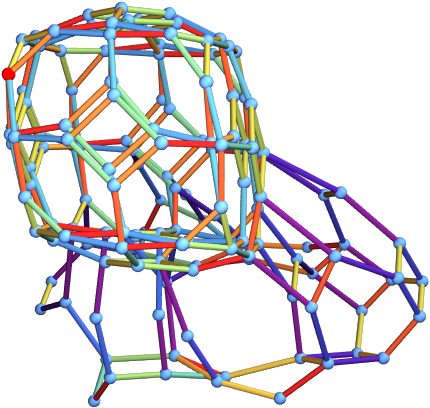}
         \label{fig:3Dgraph61}
     \end{subfigure}
     \hfill
     \begin{subfigure}[b]{0.45\textwidth}
         \centering
         \includegraphics[width=\textwidth]{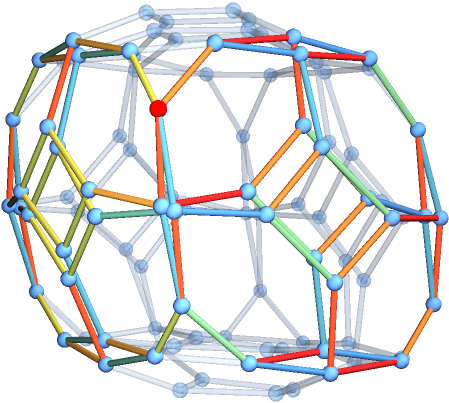}
         \label{fig:3DGraph61Cube}
     \end{subfigure}
        \caption{The permutohedra graph for $6_1$ knot has 141 vertices that represent equivalent quivers (left). Excluding symmetries that involve $\lambda_1$ reduces the~whole graph to a~cube-like shape (right). Each face of this cube is one $\Pi_4$ (a bit squashed), and neighboring $\Pi_4$'s are glued along a~square, which is a~common face to both $\Pi_4$'s. The~red vertex represents the~quiver~(\ref{6_1 quiver KRSS}) (or its reordered form~(\ref{6_1 quiver KRSS reordered})).}
        \label{fig:3DGraphs61}
\end{figure}

Another example that we consider is $6_1$ knot. We have found 141 equivalent quivers, which form quite complicated permutohedra graph, shown in figure~\ref{fig:3DGraphs61}. These quivers are related to each other by 16 symmetries (transpositions of various pairs of quiver matrices). 

The~generating function of colored superpolynomials for $6_1$ knot reads~\cite{FGSS1209}:
\begin{align}
    \label{eq:6_1 P(x)}
    P_{6_1}(x,a,q,t)=\sum_{r=0}^{\infty} \frac{x^r}{(q^2;q^2)_r}  \sum_{0\leq k_2\leq k_1 \leq r} & \left[\begin{array}{c}
    r\\
    k_1
    \end{array}\right]
    \left[\begin{array}{c}
    k_1\\
    k_2
    \end{array}\right]
    (-a^{-2}q^{2}t^{-1};q^{-2})_{k_1}(-a^{-2}q^{-2r}t^{-3};q^{-2})_{k_1} \nonumber\\
    &\times a^{2(k_1+k_2)} t^{2(k_1+k_2)} q^{2(k_1^2+k_2^2-k_1-k_2)}.
\end{align}
Linear order of this equation~gives the~uncolored superpolynomial $P_1(a,q,t)= 1 + a^{-2}t^{-2} + q^{2} t + q^{-2}t^{-1} + a^{2} t^{2} + 1 + a^{2} q^{2} t^{3} + a^{2} q^{-2}t + a^{4} t^{4}$. The~corresponding homological diagram, shown in figure~\ref{fig:61_homology}, consists of 2 diamonds and a~degenerate zig-zag made of one node that coincides with one vertex of the~upper diamond, so that $\lambda_1= \lambda_6$
.

\begin{figure}[h!]
    \centering
   \tikzset{every picture/.style={line width=0.75pt}} 

\begin{tikzpicture}[x=0.75pt,y=0.75pt,yscale=-1,xscale=1]

\draw  [dash pattern={on 0.84pt off 2.51pt}]  (147.04,507.18) -- (187.04,467.18) ;
\draw  [dash pattern={on 0.84pt off 2.51pt}]  (187.04,507.18) -- (227.04,547.18) ;
\draw  [dash pattern={on 0.84pt off 2.51pt}]  (147.04,547.18) -- (187.04,507.18) ;
\draw  [dash pattern={on 0.84pt off 2.51pt}]  (187.04,547.18) -- (227.04,507.18) ;
\draw  [dash pattern={on 0.84pt off 2.51pt}]  (147.04,507.18) -- (187.04,547.18) ;
\draw  [dash pattern={on 0.84pt off 2.51pt}]  (187.04,467.18) -- (227.04,507.18) ;
\draw  [fill={rgb, 255:red, 0; green, 0; blue, 0 }  ,fill opacity=1 ] (184.87,466.97) .. controls (184.99,465.77) and (186.06,464.89) .. (187.26,465.01) .. controls (188.46,465.13) and (189.33,466.2) .. (189.21,467.4) .. controls (189.09,468.6) and (188.02,469.47) .. (186.82,469.35) .. controls (185.62,469.24) and (184.75,468.17) .. (184.87,466.97) -- cycle ;
\draw  [fill={rgb, 255:red, 0; green, 0; blue, 0 }  ,fill opacity=1 ] (224.87,506.97) .. controls (224.99,505.77) and (226.06,504.89) .. (227.26,505.01) .. controls (228.46,505.13) and (229.33,506.2) .. (229.21,507.4) .. controls (229.09,508.6) and (228.02,509.47) .. (226.82,509.35) .. controls (225.62,509.24) and (224.75,508.17) .. (224.87,506.97) -- cycle ;
\draw  [fill={rgb, 255:red, 0; green, 0; blue, 0 }  ,fill opacity=1 ] (184.87,506.97) .. controls (184.99,505.77) and (186.06,504.89) .. (187.26,505.01) .. controls (188.46,505.13) and (189.33,506.2) .. (189.21,507.4) .. controls (189.09,508.6) and (188.02,509.47) .. (186.82,509.35) .. controls (185.62,509.24) and (184.75,508.17) .. (184.87,506.97) -- cycle ;
\draw  [fill={rgb, 255:red, 0; green, 0; blue, 0 }  ,fill opacity=1 ] (144.87,506.97) .. controls (144.99,505.77) and (146.06,504.89) .. (147.26,505.01) .. controls (148.46,505.13) and (149.33,506.2) .. (149.21,507.4) .. controls (149.09,508.6) and (148.02,509.47) .. (146.82,509.35) .. controls (145.62,509.24) and (144.75,508.17) .. (144.87,506.97) -- cycle ;
\draw  [fill={rgb, 255:red, 0; green, 0; blue, 0 }  ,fill opacity=1 ] (144.87,546.97) .. controls (144.99,545.77) and (146.06,544.89) .. (147.26,545.01) .. controls (148.46,545.13) and (149.33,546.2) .. (149.21,547.4) .. controls (149.09,548.6) and (148.02,549.47) .. (146.82,549.35) .. controls (145.62,549.24) and (144.75,548.17) .. (144.87,546.97) -- cycle ;
\draw  [fill={rgb, 255:red, 0; green, 0; blue, 0 }  ,fill opacity=1 ] (224.87,546.97) .. controls (224.99,545.77) and (226.06,544.89) .. (227.26,545.01) .. controls (228.46,545.13) and (229.33,546.2) .. (229.21,547.4) .. controls (229.09,548.6) and (228.02,549.47) .. (226.82,549.35) .. controls (225.62,549.24) and (224.75,548.17) .. (224.87,546.97) -- cycle ;
\draw  [fill={rgb, 255:red, 0; green, 0; blue, 0 }  ,fill opacity=1 ] (180.87,546.97) .. controls (180.99,545.77) and (182.06,544.89) .. (183.26,545.01) .. controls (184.46,545.13) and (185.33,546.2) .. (185.21,547.4) .. controls (185.09,548.6) and (184.02,549.47) .. (182.82,549.35) .. controls (181.62,549.24) and (180.75,548.17) .. (180.87,546.97) -- cycle ;
\draw  [dash pattern={on 0.84pt off 2.51pt}]  (187.04,587.18) -- (227.04,547.18) ;
\draw  [dash pattern={on 0.84pt off 2.51pt}]  (147.04,547.18) -- (187.04,587.18) ;
\draw  [fill={rgb, 255:red, 0; green, 0; blue, 0 }  ,fill opacity=1 ] (184.87,586.97) .. controls (184.99,585.77) and (186.06,584.89) .. (187.26,585.01) .. controls (188.46,585.13) and (189.33,586.2) .. (189.21,587.4) .. controls (189.09,588.6) and (188.02,589.47) .. (186.82,589.35) .. controls (185.62,589.24) and (184.75,588.17) .. (184.87,586.97) -- cycle ;
\draw  [fill={rgb, 255:red, 0; green, 0; blue, 0 }  ,fill opacity=1 ] (188.87,546.97) .. controls (188.99,545.77) and (190.06,544.89) .. (191.26,545.01) .. controls (192.46,545.13) and (193.33,546.2) .. (193.21,547.4) .. controls (193.09,548.6) and (192.02,549.47) .. (190.82,549.35) .. controls (189.62,549.24) and (188.75,548.17) .. (188.87,546.97) -- cycle ;

\draw (147.04,543.18) node [anchor=south] [inner sep=0.75pt]  [color={rgb, 255:red, 0; green, 0; blue, 255 }  ,opacity=1 ]  {$\lambda _{4}{}$};
\draw (227.04,543.18) node [anchor=south] [inner sep=0.75pt]  [color={rgb, 255:red, 0; green, 0; blue, 255 }  ,opacity=1 ]  {$\lambda _{3}$};
\draw (187.04,503.18) node [anchor=south] [inner sep=0.75pt]  [color={rgb, 255:red, 0; green, 0; blue, 255 }  ,opacity=1 ]  {$\lambda _{5}$};
\draw (227.04,503.18) node [anchor=south] [inner sep=0.75pt]  [color={rgb, 255:red, 0; green, 0; blue, 255 }  ,opacity=1 ]  {$\lambda _{7}$};
\draw (147.04,503.18) node [anchor=south] [inner sep=0.75pt]  [color={rgb, 255:red, 0; green, 0; blue, 255 }  ,opacity=1 ]  {$\lambda _{8}$};
\draw (186.82,465.35) node [anchor=south] [inner sep=0.75pt]  [color={rgb, 255:red, 0; green, 0; blue, 255 }  ,opacity=1 ]  {$\lambda _{9}$};
\draw (187.04,543.18) node [anchor=south] [inner sep=0.75pt]  [color={rgb, 255:red, 0; green, 0; blue, 255 }  ,opacity=1 ]  {$\lambda _{1} ,\lambda _{6}$};
\draw (147.04,553.18) node [anchor=north] [inner sep=0.75pt]  [color={rgb, 255:red, 0; green, 0; blue, 0 }  ,opacity=1 ]  {$-1$};
\draw (187.04,513.18) node [anchor=north] [inner sep=0.75pt]  [color={rgb, 255:red, 0; green, 0; blue, 0 }  ,opacity=1 ]  {$2$};
\draw (227.04,553.18) node [anchor=north] [inner sep=0.75pt]  [color={rgb, 255:red, 0; green, 0; blue, 0 }  ,opacity=1 ]  {$1$};
\draw (187.04,550.18) node [anchor=north] [inner sep=0.75pt]  [color={rgb, 255:red, 0; green, 0; blue, 0 }  ,opacity=1 ]  {$0,0$};
\draw (187.04,473.18) node [anchor=north] [inner sep=0.75pt]  [color={rgb, 255:red, 0; green, 0; blue, 0 }  ,opacity=1 ]  {$4$};
\draw (227.04,513.18) node [anchor=north] [inner sep=0.75pt]  [color={rgb, 255:red, 0; green, 0; blue, 0 }  ,opacity=1 ]  {$3$};
\draw (147.04,513.18) node [anchor=north] [inner sep=0.75pt]  [color={rgb, 255:red, 0; green, 0; blue, 0 }  ,opacity=1 ]  {$1$};
\draw (187.04,593.18) node [anchor=north] [inner sep=0.75pt]  [color={rgb, 255:red, 0; green, 0; blue, 0 }  ,opacity=1 ]  {$-2$};
\draw (187.04,582.18) node [anchor=south] [inner sep=0.75pt]  [color={rgb, 255:red, 0; green, 0; blue, 255 }  ,opacity=1 ]  {$\lambda _{2}$};

\end{tikzpicture}
    \caption{Homology diagram for $6_1$ knot; labels $\lambda_i$ are consistent with~\eqref{6_1 quiver KRSS}. 
    }
    \label{fig:61_homology}
\end{figure}

First, we rewrite~(\ref{eq:6_1 P(x)}) as 
\begin{equation}\label{eq:6_1 prequiver v1}
\begin{split}
    P_{6_{1}}(x,a,q,t)&=\sum_{\boldsymbol{\check{d}}}(-q)^{\boldsymbol{\check{d}}\cdot\check{C}\cdot\boldsymbol{\check{d}}}\frac{\boldsymbol{\check{x}}^{\boldsymbol{\check{d}}}}{(q^{2};q^{2})_{\boldsymbol{\check{d}}}}(-a^{2}q^{-2}t;q^{2})_{\check{d}_{2}+\check{d}_{3}+\check{d}_{4}+\check{d}_{5}}\Big|_{\boldsymbol{\check{x}}=x\boldsymbol{\check{\lambda}}}\\
    \check{C}&=\left[\begin{array}{c:c:c:c:c}
 0 & -1 & -1 & -1 & -1\\
 \hdashline
-1 & -2 & -2 & -2 & -1\\
\hdashline
-1 & -2 & -1 & -2 & -1\\
\hdashline
-1 & -2 & -2 & 0 & 0\\
\hdashline
-1 & -1 & -1 & 0 & 1
    \end{array}\right], \phantom{\qquad}  \phantom{\qquad}  \boldsymbol{\check{\lambda}}  = \left[\begin{array}{c}
    1\\
    a^{-2} q^2 (-t)^{-2}\\
    q^{-1} (-t)^{-1}\\
    1 \\
    a^2 q^{-3} (-t)\\
    \end{array}\right].
\end{split}
\end{equation}
Then $(1,3)$-splitting of the~last four nodes with permuation $\sigma=(2\ 4\ 5\ 3)$,
$h_1=1$, and $\kappa=-a^{2} q^{-1} t^{3}$ leads to the~quiver found  in~\cite{KRSS1707long}:
\begin{equation}\label{6_1 quiver KRSS}
C=\left[\begin{array}{c:cc:cc:cc:cc}
  0& -1&  0& -1&  0& -1&  0& -1&  0\\
  \hdashline
 -1& -2& -1& -2& -1& -2&  0& -1&  0\\
  0& -1&  1&  0&  1& -1&  1&  1&  2\\
  \hdashline
 -1& -2&  0& -1&  0& -2&  0& -1&  1\\
  0& -1&  1&  0&  2& -1&  1&  0&  2\\
  \hdashline
 -1& -2& -1& -2& -1&  0&  1&  0&  1\\
  0&  0&  1&  0&  1&  1&  3&  2&  3\\
  \hdashline
 -1& -1&  1& -1&  0&  0&  2&  1&  2\\
  0&  0&  2&  1&  2&  1&  3&  2&  4
\end{array}\right], \phantom{\qquad}  \phantom{\qquad}   
\boldsymbol{\lambda}  = \left[\begin{array}{c}
    1 \\
    a^{-2} q^2 (-t)^{-2}\\
    q (-t)\\
    q^{-1} (-t)^{-1}\\
    a^{2} q^{-2} (-t)^{2}\\
    1 \\    
    a^{2} q^{-1} (-t)^{3}\\
    a^{2} q^{-3} (-t)\\
    a^{4} q^{-4} (-t)^{4}
\end{array}\right].
\end{equation}

\begin{figure}[h!]
    \centering
    \includegraphics[scale=0.26]{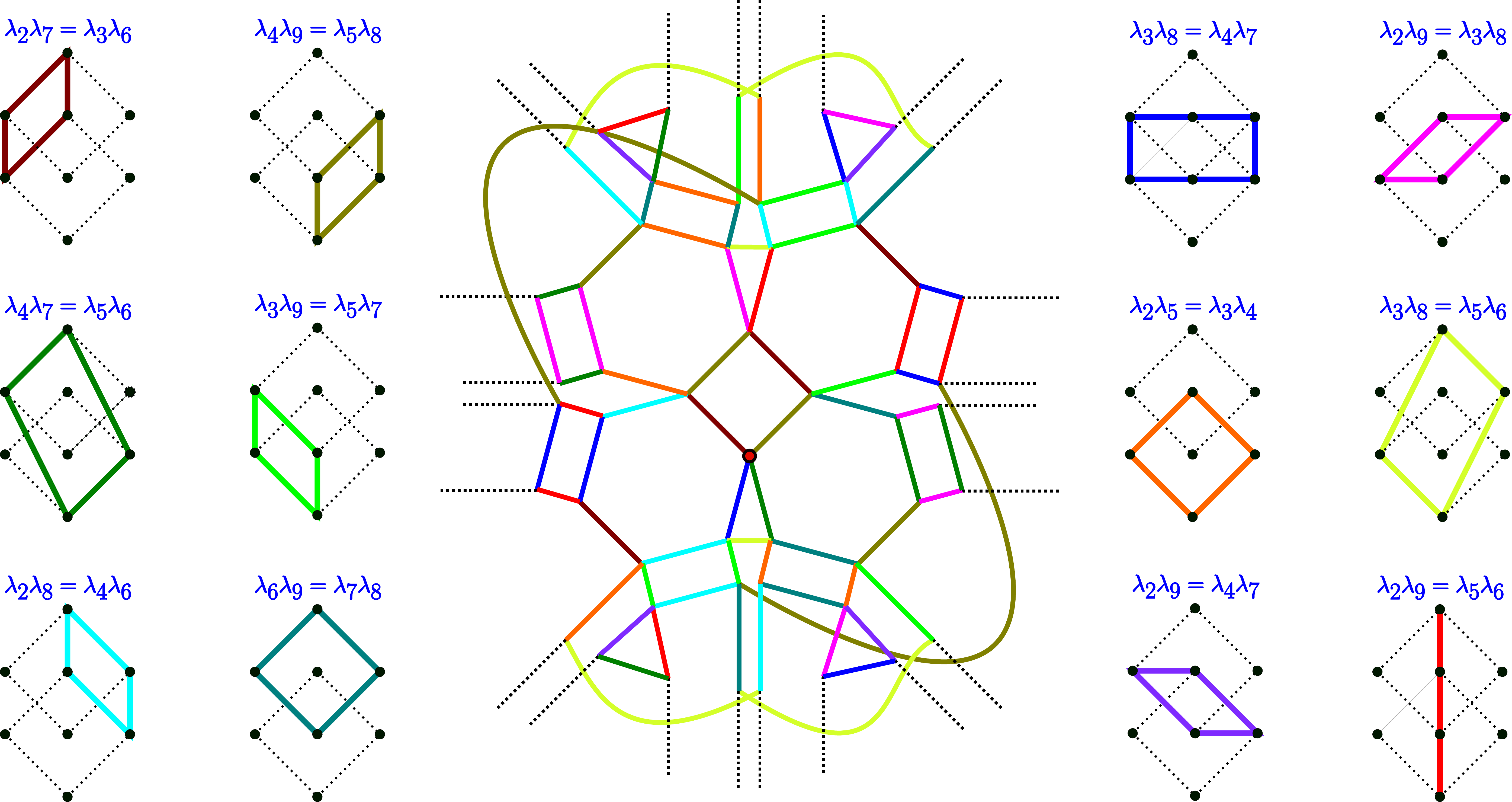}
    \vspace{0.5cm}
    \tikzset{every picture/.style={line width=0.75pt}} 

\begin{tikzpicture}[x=0.75pt,y=0.75pt,yscale=-1,xscale=1]

\draw  [fill={rgb, 255:red, 255; green, 255; blue, 255 }  ,fill opacity=1 ] (213.72,463.63) -- (286.57,536.48) -- (258.28,564.77) -- (185.43,491.92) -- cycle ;
\draw  [fill={rgb, 255:red, 155; green, 155; blue, 155 }  ,fill opacity=1 ] (258.28,463.63) -- (185.43,536.48) -- (213.72,564.77) -- (286.57,491.92) -- cycle ;
\draw  [fill={rgb, 255:red, 255; green, 255; blue, 255 }  ,fill opacity=1 ] (398.28,464.63) -- (325.43,537.48) -- (353.72,565.77) -- (426.57,492.92) -- cycle ;
\draw  [fill={rgb, 255:red, 155; green, 155; blue, 155 }  ,fill opacity=1 ] (353.72,464.63) -- (426.57,537.48) -- (398.28,565.77) -- (325.43,492.92) -- cycle ;
\draw  [fill={rgb, 255:red, 208; green, 2; blue, 27 }  ,fill opacity=1 ] (372.84,530.5) .. controls (372.84,528.75) and (374.25,527.34) .. (376,527.34) .. controls (377.75,527.34) and (379.16,528.75) .. (379.16,530.5) .. controls (379.16,532.25) and (377.75,533.66) .. (376,533.66) .. controls (374.25,533.66) and (372.84,532.25) .. (372.84,530.5) -- cycle ;
\draw  [fill={rgb, 255:red, 208; green, 2; blue, 27 }  ,fill opacity=1 ] (232.84,529.5) .. controls (232.84,527.75) and (234.25,526.34) .. (236,526.34) .. controls (237.75,526.34) and (239.16,527.75) .. (239.16,529.5) .. controls (239.16,531.25) and (237.75,532.66) .. (236,532.66) .. controls (234.25,532.66) and (232.84,531.25) .. (232.84,529.5) -- cycle ;

\draw (258.28,473.63) node [anchor=north] [inner sep=0.75pt]  [color={rgb, 255:red, 255; green, 255; blue, 255 }  ,opacity=1 ]  {$\Pi _{4}$};
\draw (353.72,474.63) node [anchor=north] [inner sep=0.75pt]  [color={rgb, 255:red, 255; green, 255; blue, 255 }  ,opacity=1 ]  {$\Pi _{4}$};
\draw (236,579.5) node [anchor=north] [inner sep=0.75pt]    {(\ref{eq:6_1 prequiver v1})};
\draw (376,579.5) node [anchor=north] [inner sep=0.75pt]    {(\ref{eq:6_1 prequiver v2})};

\end{tikzpicture}
    \caption{Planar projection of a~part of the~permutohedra graph for $6_1$ knot. In homological diagrams (on left and right) it is indicated how some of its symmetries, corresponding to edges of the~graph, arise from quadruples of homology generators. The~positions of two two  permutohedra $\Pi_4$ mentioned in the~text are indicated schematically in the~bottom.}
    \label{fig:graph61}
\end{figure}
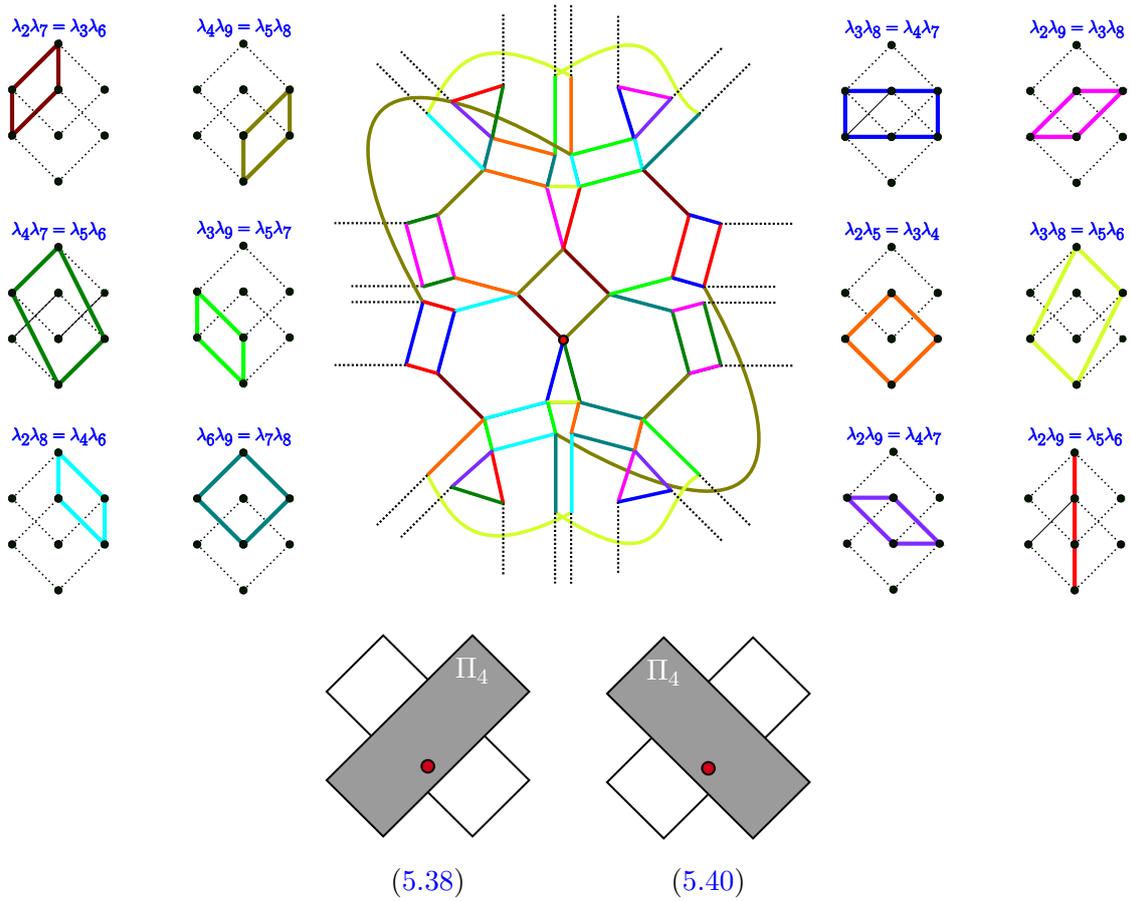

On the~other hand, we can rewrite~(\ref{eq:6_1 P(x)}) in the~form
\begin{equation}\label{eq:6_1 prequiver v2}
\begin{split}
    P_{6_{1}}(x,a,q,t)&=\sum_{\boldsymbol{\check{d}}}(-q)^{\boldsymbol{\check{d}}\cdot\check{C}\cdot\boldsymbol{\check{d}}}\frac{\boldsymbol{\check{x}}^{\boldsymbol{\check{d}}}}{(q^{2};q^{2})_{\boldsymbol{\check{d}}}}\Pi_{\check{d}_2,\check{d}_3,\check{d}_4,\check{d}_5}\Big|_{\boldsymbol{\check{x}}=x\boldsymbol{\check{\lambda}}}\\
    \Pi_{\check{d}_2,\check{d}_3,\check{d}_4,\check{d}_5}&=\sum\limits _{\alpha_{2}+\beta_{2}=\check{d}_{2}}\sum\limits _{\alpha_{3}+\beta_{3}=\check{d}_{3}}\sum\limits _{\alpha_{4}+\beta_{4}=\check{d}_{4}}\sum\limits _{\alpha_{5}+\beta_{5}=\check{d}_{5}} \prod_{i=2}^5 \frac{(a^{2}q^{-2}t^{2})^{\beta_{i}}(q^2;q^2)_{\check{d}_i}}{(q^2;q^2)_{\alpha_i}(q^2;q^2)_{\beta_i}} \\
    & \times (-q)^{2(\beta_{2}+\beta_{3}+\beta_{4}+\beta_{5})^{2}+2(\alpha_2\beta_3+\alpha_2\beta_4+\alpha_2\beta_5+\alpha_3 \beta_4+\alpha_3\beta_5+\alpha_4\beta_5)}\\
    \check{C}&=\left[\begin{array}{c:c:c:c:c}
 0 & -1 & -1 & 0 & 0\\
 \hdashline
-1 & -2 & -2 & -1 & -1\\
\hdashline
-1 & -2 & -1 & 0 & 0\\
\hdashline
0 & -1 & 0 & 1 & 1\\
\hdashline
0 & -1 & 0 & 1 & 2
    \end{array}\right], \phantom{\qquad}  \phantom{\qquad}  \boldsymbol{\check{\lambda}}  = \left[\begin{array}{c}
    1\\
    a^{-2} q^2 (-t)^{-2}\\
    q^{-1} (-t)^{-1}\\
    q (-t)\\
    a^{2} q^{-2} (-t)^{2}\\
    \end{array}\right],
\end{split}
\end{equation}
Then, $(0,2)$-splitting of the~last four nodes with permutation  $\sigma=(2\ 5)(3\ 4)$, $h_1=0$, and $\kappa=a^{2} q^{-2} t^{2}$ leads to
\begin{equation}\label{6_1 quiver KRSS reordered}
C=\left[\begin{array}{c:cc:cc:cc:cc}
  0& -1& -1& -1& -1&  0&  0&  0&  0\\
  \hdashline
 -1& -2& -2& -2& -1& -1&  0& -1&  0\\
 -1& -2&  0& -2&  0& -1&  1& -1&  1\\
  \hdashline
 -1& -2& -2& -1& -1&  0&  1&  0&  1\\
 -1& -1&  0& -1&  1&  0&  2&  0&  2\\
  \hdashline
  0& -1& -1&  0&  0&  1&  1&  1&  2\\
  0&  0&  1&  1&  2&  1&  3&  1&  3\\
  \hdashline
  0& -1& -1&  0&  0&  1&  1&  2&  2\\
  0&  0&  1&  1&  2&  2&  3&  2&  4
\end{array}\right], \phantom{\qquad}  \phantom{\qquad}   
\boldsymbol{\lambda}  = \left[\begin{array}{c}
    1 \\
    a^{-2} q^2 (-t)^{-2}\\
    1 \\  
    q^{-1} (-t)^{-1}\\
    a^{2} q^{-3} (-t)\\
    q (-t)\\  
    a^{2} q^{-1} (-t)^{3}\\
    a^{2} q^{-2} (-t)^{2}\\
    a^{4} q^{-4} (-t)^{4}
\end{array}\right].
\end{equation}
which is a~rearrangement of~\eqref{6_1 quiver KRSS}. This means that the~above quiver is common to two permutohedra $\Pi_4$, and it is represented by the~red dot in figure~\ref{fig:3DGraphs61} and~\ref{fig:graph61}. In figure~\ref{fig:graph61}, which shows a~planar projection of a~part of the~permutohedra graph, $\Pi_4$ coming from the~prequiver~\eqref{eq:6_1 prequiver v1} is oriented along axis $\nearrow$, whereas $\Pi_4$ oriented along $\nwarrow$ corresponds to the~prequiver~\eqref{eq:6_1 prequiver v2}. All other~quiver matrices that we found are listed in the~Mathematica file attached to the~arXiv submission. According to conjecture~\ref{coj:necessary conditions}, we expect that there are no more equivalent quivers and figure~\ref{fig:graph61} presents the~whole equivalence class.

\subsection{\texorpdfstring{$(2,2p+1)$}{(2,2p+1)} torus knots}

The last example we consider is a~series of $(2,2p+1)$ torus knots. For this class the~number of equivalent quivers grows rapidly; for $p=1,...,7$ we have found respectively 1, 3, 13, 68, 405, 2684 and 19557 equivalent quivers, which permutohedra graphs have an~interesting structure. For $p=1$ there is just one corresponding quiver, see section~\ref{ssec-31}; for $p>1$ the~permutohedra graph consists of two series of larger and larger permutohedra $\Pi_2,\ldots,\Pi_p$ (and several additional permutohedra of small size that do not belong to these series). In each of these two series each permutohedron $\Pi_i$ is connected to $\Pi_{i-1}$ and $\Pi_{i+1}$ (for $i=3,\ldots,p-1$), and the~two largest permutohedra $\Pi_p$ from both series are also connected. Such a~structure is present for $5_1$, $7_1$, $9_1$ and $11_1$ knots in figures~\ref{fig:5_1matrixplot},~\ref{fig:graph71},~\ref{fig:3DGraph91}, and~\ref{fig:3DGraph111}, respectively. In this section we explain how two largest permutohedra $\Pi_p$ for $(2,2p+1)$ torus knot arise. 

To start with, note that the~generating function of superpolynomials for $(2,2p+1)$-torus knot can be written, among others, in the~following two equivalent ways, which correspond to different grading conventions for the~$S^r$-colored HOMFLY-PT homologies~\cite{AFGS1203,FGSS1209}:
\begin{align}
P_{T_{2,2p+1}}(x,a,q,t) &=\sum_{r\geq 0} x^r a^{2pr}q^{-2pr}\sum_{0\leq k_{p}\leq\ldots\leq k_{2}\leq k_{1}\leq r} 
    \left[\begin{array}{c}
    r\\
    k_1
    \end{array}\right]
    \left[\begin{array}{c}
    k_1\\
    k_2
    \end{array}\right] \cdots 
    \left[\begin{array}{c}
    k_{p-1}\\
    k_{p}
    \end{array}\right]
    \label{eq:Torus knots a} \\
&\quad \times q^{2\sum_{i=1}^{p}((2r+1)k_{i}-k_{i-1}k_{i})}t^{2(k_{1}+k_{2}+\ldots+k_{p})}
(-a^2q^{-2}t;q^2)_{k_1} =
\nonumber 
\\
&= \sum_{r\geq 0} x^r a^{2pr}q^{-2pr}\sum_{0\leq k_{p}\leq\ldots\leq k_{2}\leq k_{1}\leq r} 
\left[\begin{array}{c}
    r\\
    k_1
    \end{array}\right]
    \left[\begin{array}{c}
    k_1\\
    k_2
    \end{array}\right] \cdots 
    \left[\begin{array}{c}
    k_{p-1}\\
    k_{p}
    \end{array}\right]
    \label{eq:Torus knots b}\\
&\quad \times q^{2\sum_{i=1}^{p}((2r+1)k_{i}-k_{i-1}k_{i})}t^{2(k_{1}+k_{2}+\ldots+k_{p})}
(-a^2q^{2r}t^3;q^2)_{r-k_p}.
\nonumber 
\end{align}
For $p=1$, i.e. $3_1$ knot, the~above expressions reduce to
\begin{equation}
\begin{split}
\sum_{0\leq k_{1}\leq r} \left[\begin{array}{c}
    r\\
    k_1
    \end{array}\right]
    \left[\begin{array}{c}
    k_1\\
    k_2
    \end{array}\right] q^{2k_1(r+1)}t^{2k_1} & (-a^2q^{-2}t;q^2)_{k_1}\\
&= \sum_{0\leq k_{1}\leq r} \left[\begin{array}{c}
    r\\
    k_1
    \end{array}\right]
    \left[\begin{array}{c}
    k_1\\
    k_2
    \end{array}\right] q^{2k_1(r+1)}t^{2k_1}(-a^2q^{2r}t^3;q^2)_{r-k_1},    
\end{split}
\end{equation}
and the~two permutohedra consist of one vertex. They are in fact identified, so that the~full permutohedra graph consists just of one $\Pi_1$. In general, both~(\ref{eq:Torus knots a}) and~(\ref{eq:Torus knots b}) can be rewritten in form~\eqref{PK-Pi} using the~formula
\begin{equation}
\left[\begin{array}{c}
    r\\
    k_1
    \end{array}\right]
    \left[\begin{array}{c}
    k_1\\
    k_2
    \end{array}\right] \cdots 
    \left[\begin{array}{c}
    k_{p-1}\\
    k_{p}
    \end{array}\right]=\frac{(q^2;q^2)_r}{(q^2;q^2)_{r-k_1}(q^2;q^2)_{k_1-k_2}\cdots(q^2;q^2)_{k_{p-1}-k_p}(q^2;q^2)_{k_p}}.
\end{equation}
In case of~\eqref{eq:Torus knots a} we set
\begin{align*}
\check{d}_1 &= r-k_1, &
\check{d}_2 &= k_1 - k_2, &
\check{d}_3 &= k_2 - k_3, &
\check{d}_4 &= k_3 - k_4,\\
 &\ldots &
\check{d}_{i+1} &= k_{i} - k_{i+1}, &
& \ldots &
\check{d}_{p+1} &= k_p,
\end{align*}
which leads to
\begin{align}\label{eq:2,2p+1 prequiver v1}
& P_{T_{2,2p+1}}(x,a,q,t) = \sum_{\boldsymbol{\check{d}}} (-q)^{\boldsymbol{\check{d}}\cdot \check{C}\cdot\boldsymbol{\check{d}}}\frac{\boldsymbol{\check{x}}^{\boldsymbol{\check{d}}}}{(q^2;q^2)_{\boldsymbol{\check{d}}}}(-a^2q^{-2}t;q^2)_{\check{d}_2+\dots+ \check{d}_{p+1}}\Big|_{\boldsymbol{\check{x}}=x\boldsymbol{\check{\lambda}}} \\
& \check{C}=\left[\begin{array}{c:c:c:c:c:c:c}
0 & 1 & 3 & 5 & \dots  & 2p-3 & 2p-1 \\
\hdashline
1 & 2 & 3 & 5 & \dots  & 2p-3 & 2p-1 \\
\hdashline
3 & 3 & 4 & 5 & \dots  & 2p-3 & 2p-1 \\
\hdashline
5 & 5 & 5 & 6 & \dots & 2p-3 & 2p-1 \\
\hdashline
\vdots & \vdots & \vdots & \vdots & \ddots & \vdots & \vdots \\
\hdashline
2p-3 & 2p-3  & 2p-3 & 2p-3 & \dots & 2p-2 & 2p-1  \\
\hdashline
2p-1 & 2p-1 & 2p-1 & 2p-1 & \dots & 2p-1 & 2p \\
\end{array}\right]
\left.
\begin{array}{c}
\check{d}_1 \\
\check{d}_2 \\
\check{d}_3 \\
\check{d}_4 \\
\vdots \\
\check{d}_{p} \\
\check{d}_{p+1}
\end{array}
\right.
, \quad \boldsymbol{\check{\lambda}} = \left[\begin{array}{c}
a^{2p}q^{-2p}\\
a^{2p}q^{-2(p-1)}(-t)^2\\
a^{2p}q^{-2(p-2)} (-t)^4\\
a^{2p}q^{-2(p-3)} (-t)^6\\
\vdots \\
a^{2p}q^{-2}(-t)^{2p+2} \\
a^{2p}(-t)^{2p}
\end{array}\right].  \nonumber  
\end{align}
The $(0,1)$-splitting of the~nodes corresponding to $\check{d}_2,\dots,\check{d}_{p+1}$ with trivial permutation, $h_1=0$, and $\kappa=\xi q^{-1}=-a^2q^{-3}t$ produces the~quiver found in~\cite{KRSS1707long}:
\begin{equation}\label{eq:quiver_torus_general}
C=\left[\begin{array}{c:cc:cc:c:cc}
0 & 1 & 1 & 3 & 3 & \dots & 2p-1 & 2p-1 \\
\hdashline
1 & 2 & 2 & 3 & 3 & \dots & 2p-1 & 2p-1  \\
1 & 2 & 3 & 4 & 4 & \dots & 2p & 2p \\
\hdashline
3 & 3 & 4 & 4 & 4 & \dots & 2p-1 & 2p-1 \\
3 & 3 & 4 & 4 & 5 & \dots & 2p & 2p \\
\hdashline
\vdots & \vdots & \vdots & \vdots & \vdots & \ddots & \vdots & \vdots \\
\hdashline
2p-1 & 2p-1 & 2p & 2p-1 & 2p & \dots & 2p & 2p  \\
2p-1 & 2p-1 & 2p & 2p-1 & 2p & \dots & 2p & 2p+1 \\
\end{array}\right],
\quad \boldsymbol{\lambda} = \left[\begin{array}{c}
a^{2p}q^{-2p}\\
a^{2p}q^{-2(p-1)}(-t)^2\\
a^{2(p+1)}q^{-2(p-1)-3}(-t)^3 \\
a^{2p}q^{-2(p-2)}(-t)^4\\
a^{2(p+1)}q^{-2(p-2)-3}(-t)^5\\
\vdots \\
a^{2p}(-t)^{2p} \\
a^{2(p+1)}q^{-3}(-t)^{2p+1}
\end{array}\right].    
\end{equation}
On the~other hand, for the~expression~(\ref{eq:Torus knots b}) we introduce
\begin{align*}
	& & \check{d}_1 &=  r-(r-k_p) = k_p, & & &\\
	\check{d}_{2} &= r-k_1, &
	\check{d}_{3} &= k_{1} - k_{2}, &
	&\ldots &
	\check{d}_{p+1} &=  k_{p-1}-k_p,
\end{align*}
and then find
\begin{align}\label{eq:2,2p+1 prequiver v2}
& P_{T_{2,2p+1}}(x,a,q,t) =\sum_{\boldsymbol{\check{d}}} (-q)^{\boldsymbol{\check{d}}\cdot \check{C}\cdot\boldsymbol{\check{d}}}\frac{\boldsymbol{\check{x}}^{\boldsymbol{\check{d}}}}{(q^2;q^2)_{\boldsymbol{\check{d}}}}(-a^2q^{2r}t^3;q^2)_{\check{d}_2+\dots+ \check{d}_{p+1}}\Big|_{\boldsymbol{\check{x}}=x\boldsymbol{\check{\lambda}}} \\
& 
\check{C}=\left[\begin{array}{c:c:c:c:c:c:c}
2p & 2p-1 & 2p-1 & 2p-1 & \dots & 2p-1 & 2p-1 \\
\hdashline
2p-1 & 0 & 1 & 3 & \dots & 2p-5 & 2p-3  \\
\hdashline
2p-1 & 1 & 2 & 3 & \dots & 2p-5 & 2p-3 \\
\hdashline
2p-1 & 3 & 3 & 4 & \dots & 2p-5 & 2p-3 \\
\hdashline
\vdots & \vdots & \vdots & \vdots  & \ddots & \vdots & \vdots \\
\hdashline
2p-1 & 2p-5 & 2p-5 & 2p-5 & \dots & 2p-4 & 2p-3  \\
\hdashline
2p-1 & 2p-3 & 2p-3 & 2p-3 & \dots & 2p-3 & 2p-2 \\
\end{array}\right]
\left.
\begin{array}{c}
\check{d}_1 \\
\check{d}_2 \\
\check{d}_3 \\
\check{d}_4 \\
\vdots \\
\check{d}_{p} \\
\check{d}_{p+1}
\end{array}
\right.
, \ \boldsymbol{\check{\lambda}} = \left[\begin{array}{c}
a^{2p}(-t)^{2p} \\
a^{2p}q^{-2p}\\
a^{2p}q^{-2(p-1)}(-t)^2\\
a^{2p}q^{-2(p-2)} (-t)^4 \\
\vdots \\
a^{2p}q^{-4}(-t)^{2p-4} \\
a^{2p}q^{-2}(-t)^{2p-2}
\end{array}\right]. \nonumber   
\end{align}
One can check that the~$(1,3)$-splitting of the~nodes corresponding to $\check{d}_2,\dots,\check{d}_{p+1}$ with permutation $\sigma=(2\ \ (p+1))$, $h_1=1$, and $\kappa=-a^2q^{-1}t^3$ yields the~same quiver as in~(\ref{eq:quiver_torus_general}). 

Note that both prequivers given above are the~same up to reordering of nodes, however the~two splittings are different. This is why we obtain two different permutohedra $\Pi_p$, respectively left (for~(\ref{eq:Torus knots a})) and right ((\ref{eq:Torus knots b})) in figures~\ref{fig:3DGraph91},~\ref{fig:3DGraph111},~\ref{fig:5_1matrixplot},  and ~\ref{fig:graph71}. These two permutohedra share the~quiver matrix~(\ref{eq:quiver_torus_general}), which can be obtained from appropriate splittings of corresponding prequivers, as explained above. An~interested reader may conduct careful analysis of other permutohedra in these graphs.

\section{Examples -- local structure}\label{sec:local symmetries}

In the~previous section we presented permutohedra graphs for simple knots and discussed in detail the~structure of glued permutohedra embedded in these graphs. In this section we take the~opposite perspective and study the~local structure: we choose some particular quiver and identify all equivalent quivers related to it by a~single transposition of matrix elements (a single symmetry, to which we refer as local). We also provide interpretation of such equivalences in terms of homological diagrams. We conduct such an~analysis for infinite families of $(2,2p+1)$ torus knots (also denoted $T_{2,2p+1}$), $TK_{2|p|+2}$ and $TK_{2p+1}$ twist knots, and in addition $6_2,6_3$ and $7_3$ knots. The~quivers that we analyze are those found in~\cite{KRSS1707long} (apart from the~quiver for $7_3$ knot that was found in~\cite{SW1711}), and they are indicated by red vertices in permutohedra graphs in figures ~\ref{fig:symmetries_4_1},~\ref{fig:5_1matrixplot},~\ref{fig:symmetries_5_2}, and~\ref{fig:graph61}. The~symmetries that we analyze in this section are represented by edges adjacent to these red vertices. 

Recall that:
\begin{itemize}
    \item Quiver matrices for $(2,2p+1)$ torus knots that we consider are given in~(\ref{eq:quiver_torus_general}). A~homological diagram for $(2,2p+1)$ torus knot consists of one zig-zag made of $2p+1$ generators.  
    \item Quiver matrices for twist knots $TK_{2|p|+2}$ (i.e. $4_1,6_1,8_1,\dots$ knots) are given in the~appendix~\ref{app-twist}.  
    A~homological diagram for $TK_{2|p|+2}$ knot consists of $p$ diamonds and a~zig-zag made of one generator, so altogether it has $4p+1$ generators.
    \item Quiver matrices for twist knots $TK_{2p+1}$ (i.e. $3_1,5_2,7_2,\dots$ knots) are also given in the~appendix~\ref{app-twist}. A~homological diagram for $TK_{2p+1}$ knot 
    consists of $p-1$ diamonds and a~zig-zag of length 3, so altogether it has $4p-1$ generators.
\end{itemize}

\begin{figure}[h!]
    \centering
    \input{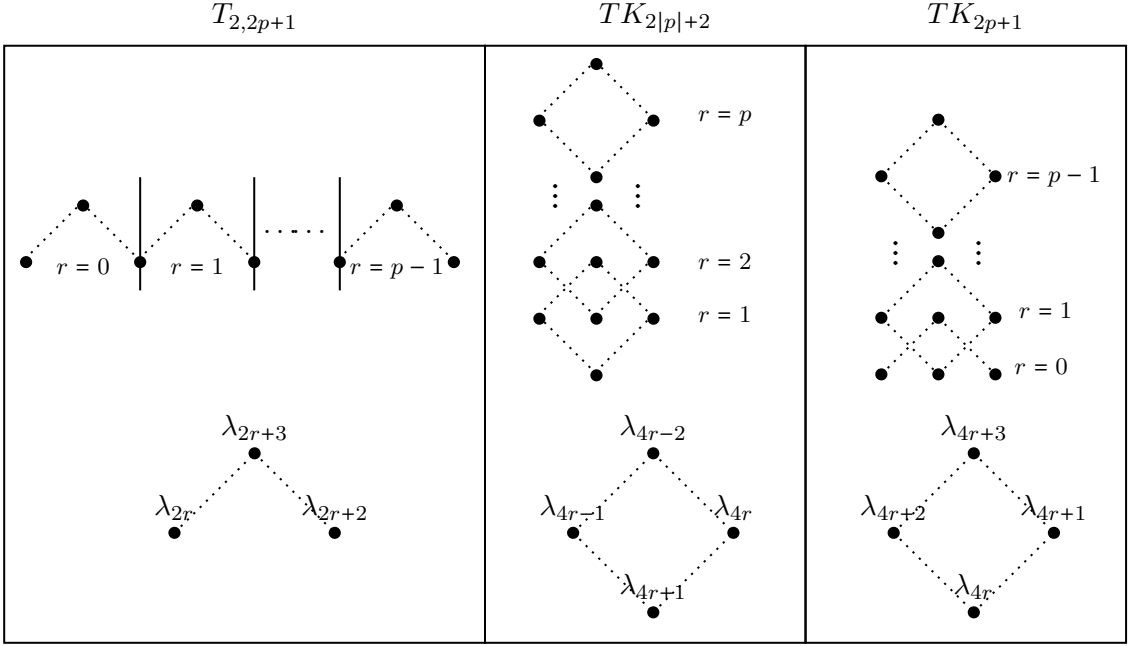}
    \caption{Enumeration of wedges and diamonds in the~homology diagrams, from left to right: $T_{2,2p+1}$, $TK_{2|p|+2}$, $TK_{2p+1}$.}
    \label{fig:diamonds_orientation}
\end{figure}

In this section we fix the~ordering of homological generators (and correspondingly quiver nodes) as shown in figure~\ref{fig:diamonds_orientation}. In what follows we call a~part of a~zig-zag consisting of three consecutive nodes that form a~shape $\wedge$ a~wedge. We enumerate diamonds and wedges by $r,r',r'',r''',\dots$, such that $r\leq r'\leq r''\leq r'''\dots$; for a~wedge or a~zig-zag labeled by $r$, we enumerate the~generators it consists of as in the~bottom of figure~\ref{fig:diamonds_orientation}. We write pairings $\lambda_a\lambda_b=\lambda_c\lambda_d$ as column vectors with entries $a,b,c,d$. Recall that we call such a~paring a~symmetry if quiver matrices with elements $C_{ab}$ and $C_{cd}$ exchanged are equivalent. We also call the~requirements $C_{ai}+C_{bi}=C_{ci}+C_{di}$ (for $i\neq a,b,c,d$) \emph{spectator constraints}.

\begin{thm}\label{thm:knots} For infinite families of knots $T_{2,2p+1},TK_{2|p|+2},TK_{2p+1},p=1,2,3,\dots$, quiver matrices 
given respectively in~(\ref{eq:quiver_torus_general}) and in appendix~\ref{app-twist}, 
have the~following local symmetries:
\begin{equation}
\begin{aligned}
T_{2,2p+1}: &\
\begin{bmatrix}
2r \\
2r'+3 \\
2r+3 \\
2r' 
\end{bmatrix},
\
\begin{bmatrix}
2r+3 \\
2r'+2 \\
2r+2 \\
2r'+3 
\end{bmatrix}
\\
TK_{2|p|+2}: &\
\begin{bmatrix}
4r-1 \\
4r' \\
4r \\
4r'-1 
\end{bmatrix},
\
\begin{bmatrix}
4r-1 \\
4r'-2 \\
4r-2 \\
4r'-1 
\end{bmatrix},
\
\begin{bmatrix}
4r+1 \\
4r' \\
4r \\
4r'+1 
\end{bmatrix},
\
\begin{bmatrix}
4r+1 \\
4r''-2 \\
4r'+1 \\
4r'-2 
\end{bmatrix},
\
\begin{bmatrix}
4 \\
4p-1 \\
5 \\
4p-2 
\end{bmatrix}
\\
TK_{2p+1}: &\ 
\begin{bmatrix}
2 \\
4r'+3 \\
3 \\
4r'+2 
\end{bmatrix},
\ 
\begin{bmatrix}
2 \\
4r'+1 \\
1 \\
4r'+2 
\end{bmatrix},
\
\begin{bmatrix}
2 \\
4p+1 \\
3 \\
4p 
\end{bmatrix}
\bigcup \mathcal{T}\left(TK_{2|p|+2} \setminus 
\begin{bmatrix}
4 \\
4p-1 \\
5 \\
4p-2 
\end{bmatrix}
\right)
\end{aligned}
\end{equation}
where $r'=r+1$, $r''=r+2$, and
\begin{equation}
  \mathcal{T}\left(TK_{2|p|+2} \setminus 
\begin{bmatrix}
4 \\
4p-1 \\
5 \\
4p-2 
\end{bmatrix}
\right)
:=
\begin{bmatrix}
4r+2 \\
4r'+1 \\
4r+1 \\
4r'+2 
\end{bmatrix},
\
\begin{bmatrix}
4r \\
4r'+1 \\
4r+1 \\
4r' 
\end{bmatrix},
\
\begin{bmatrix}
4r+2 \\
4r'+3 \\
4r+3 \\
4r'+2 
\end{bmatrix},
\
\begin{bmatrix}
4r \\
4r''+3 \\
4r' \\
4r'+3 
\end{bmatrix}.
\end{equation}
\end{thm}


\begin{figure}[h!]
    \centering
    \fbox{
    \tikzset{every picture/.style={line width=0.75pt}} 

\begin{tikzpicture}[x=0.75pt,y=0.75pt,yscale=-1,xscale=1]

\draw  [dash pattern={on 0.84pt off 2.51pt}]  (2023,2993.58) -- (2063,2953.58) ;
\draw  [dash pattern={on 0.84pt off 2.51pt}]  (2301,2953.58) -- (2341,2993.58) ;
\draw  [dash pattern={on 0.84pt off 2.51pt}]  (2221,2953.58) -- (2261,2993.58) ;
\draw  [dash pattern={on 0.84pt off 2.51pt}]  (1943,2993.58) -- (1983,2953.58) ;
\draw [color={rgb, 255:red, 0; green, 0; blue, 0 }  ,draw opacity=1 ][line width=1.5]    (1943,2993.58) -- (2023,2993.58) ;
\draw [color={rgb, 255:red, 0; green, 0; blue, 0 }  ,draw opacity=1 ][line width=1.5]    (1983,2953.58) -- (2063,2953.58) ;
\draw [color={rgb, 255:red, 0; green, 0; blue, 0 }  ,draw opacity=1 ][line width=1.5]    (2261,2993.58) -- (2341,2993.58) ;
\draw [color={rgb, 255:red, 0; green, 0; blue, 0 }  ,draw opacity=1 ][line width=1.5]    (2221,2953.58) -- (2301,2953.58) ;
\draw [color={rgb, 255:red, 0; green, 0; blue, 0 }  ,draw opacity=1 ][line width=1.5]    (1943,2993.58) -- (1983,2953.58) ;
\draw [color={rgb, 255:red, 0; green, 0; blue, 0 }  ,draw opacity=1 ][line width=1.5]    (2023,2993.58) -- (2063,2953.58) ;
\draw [color={rgb, 255:red, 0; green, 0; blue, 0 }  ,draw opacity=1 ][line width=1.5]    (2221,2953.58) -- (2261,2993.58) ;
\draw [color={rgb, 255:red, 0; green, 0; blue, 0 }  ,draw opacity=1 ][line width=1.5]    (2301,2953.58) -- (2341,2993.58) ;
\draw  [dash pattern={on 0.84pt off 2.51pt}]  (2063,2953.58) -- (2103,2993.58) ;
\draw  [dash pattern={on 0.84pt off 2.51pt}]  (1983,2953.58) -- (2023,2993.58) ;
\draw  [dash pattern={on 0.84pt off 2.51pt}]  (2261,2993.58) -- (2301,2953.58) ;
\draw  [dash pattern={on 0.84pt off 2.51pt}]  (2181,2993.58) -- (2221,2953.58) ;
\draw  [fill={rgb, 255:red, 0; green, 0; blue, 0}  ,fill opacity=0 ] (1944.63,2996.3) .. controls (1943.13,2997.2) and (1941.19,2996.71) .. (1940.29,2995.21) .. controls (1939.39,2993.71) and (1939.87,2991.77) .. (1941.37,2990.87) .. controls (1942.87,2989.97) and (1944.81,2990.46) .. (1945.71,2991.96) .. controls (1946.61,2993.45) and (1946.13,2995.4) .. (1944.63,2996.3) -- cycle ;
\draw  [fill={rgb, 255:red, 0; green, 0; blue, 0}  ,fill opacity=0 ] (1984.63,2956.3) .. controls (1983.13,2957.2) and (1981.19,2956.71) .. (1980.29,2955.21) .. controls (1979.39,2953.71) and (1979.87,2951.77) .. (1981.37,2950.87) .. controls (1982.87,2949.97) and (1984.81,2950.46) .. (1985.71,2951.96) .. controls (1986.61,2953.45) and (1986.13,2955.4) .. (1984.63,2956.3) -- cycle ;
\draw  [fill={rgb, 255:red, 0; green, 0; blue, 0}  ,fill opacity=0 ] (2064.63,2956.3) .. controls (2063.13,2957.2) and (2061.19,2956.71) .. (2060.29,2955.21) .. controls (2059.39,2953.71) and (2059.87,2951.77) .. (2061.37,2950.87) .. controls (2062.87,2949.97) and (2064.81,2950.46) .. (2065.71,2951.96) .. controls (2066.61,2953.45) and (2066.13,2955.4) .. (2064.63,2956.3) -- cycle ;
\draw  [fill={rgb, 255:red, 0; green, 0; blue, 0}  ,fill opacity=0 ] (2104.57,2996.33) .. controls (2103.06,2997.2) and (2101.13,2996.67) .. (2100.26,2995.16) .. controls (2099.39,2993.64) and (2099.91,2991.71) .. (2101.43,2990.84) .. controls (2102.94,2989.97) and (2104.87,2990.49) .. (2105.74,2992.01) .. controls (2106.61,2993.52) and (2106.09,2995.46) .. (2104.57,2996.33) -- cycle ;
\draw  [fill={rgb, 255:red, 0; green, 0; blue, 0}  ,fill opacity=0 ] (2024.57,2996.33) .. controls (2023.06,2997.2) and (2021.13,2996.67) .. (2020.26,2995.16) .. controls (2019.39,2993.64) and (2019.91,2991.71) .. (2021.43,2990.84) .. controls (2022.94,2989.97) and (2024.87,2990.49) .. (2025.74,2992.01) .. controls (2026.61,2993.52) and (2026.09,2995.46) .. (2024.57,2996.33) -- cycle ;
\draw  [fill={rgb, 255:red, 0; green, 0; blue, 0}  ,fill opacity=0 ] (2182.57,2996.33) .. controls (2181.06,2997.2) and (2179.13,2996.67) .. (2178.26,2995.16) .. controls (2177.39,2993.64) and (2177.91,2991.71) .. (2179.43,2990.84) .. controls (2180.94,2989.97) and (2182.87,2990.49) .. (2183.74,2992.01) .. controls (2184.61,2993.52) and (2184.09,2995.46) .. (2182.57,2996.33) -- cycle ;
\draw  [fill={rgb, 255:red, 0; green, 0; blue, 0}  ,fill opacity=0 ] (2262.57,2996.33) .. controls (2261.06,2997.2) and (2259.13,2996.67) .. (2258.26,2995.16) .. controls (2257.39,2993.64) and (2257.91,2991.71) .. (2259.43,2990.84) .. controls (2260.94,2989.97) and (2262.87,2990.49) .. (2263.74,2992.01) .. controls (2264.61,2993.52) and (2264.09,2995.46) .. (2262.57,2996.33) -- cycle ;
\draw  [fill={rgb, 255:red, 0; green, 0; blue, 0}  ,fill opacity=0 ] (2342.57,2996.33) .. controls (2341.06,2997.2) and (2339.13,2996.67) .. (2338.26,2995.16) .. controls (2337.39,2993.64) and (2337.91,2991.71) .. (2339.43,2990.84) .. controls (2340.94,2989.97) and (2342.87,2990.49) .. (2343.74,2992.01) .. controls (2344.61,2993.52) and (2344.09,2995.46) .. (2342.57,2996.33) -- cycle ;
\draw  [fill={rgb, 255:red, 0; green, 0; blue, 0}  ,fill opacity=0 ] (2222.57,2956.33) .. controls (2221.06,2957.2) and (2219.13,2956.67) .. (2218.26,2955.16) .. controls (2217.39,2953.64) and (2217.91,2951.71) .. (2219.43,2950.84) .. controls (2220.94,2949.97) and (2222.87,2950.49) .. (2223.74,2952.01) .. controls (2224.61,2953.52) and (2224.09,2955.46) .. (2222.57,2956.33) -- cycle ;
\draw  [fill={rgb, 255:red, 0; green, 0; blue, 0}  ,fill opacity=0 ] (2302.57,2956.33) .. controls (2301.06,2957.2) and (2299.13,2956.67) .. (2298.26,2955.16) .. controls (2297.39,2953.64) and (2297.91,2951.71) .. (2299.43,2950.84) .. controls (2300.94,2949.97) and (2302.87,2950.49) .. (2303.74,2952.01) .. controls (2304.61,2953.52) and (2304.09,2955.46) .. (2302.57,2956.33) -- cycle ;

\draw (1943-10,2988.58) node [anchor=south] [inner sep=0.75pt]  [font=\small,color={rgb, 255:red, 0; green, 0; blue, 255 }  ,opacity=1 ]  {$\lambda _{2r}$};
\draw (1986.26,2946.63) node [anchor=south] [inner sep=0.75pt]  [font=\small,color={rgb, 255:red, 0; green, 0; blue, 255 }  ,opacity=1 ]  {$\lambda _{2r+3}$};
\draw (2023,2988.58) node [anchor=south] [inner sep=0.75pt]  [font=\small,color={rgb, 255:red, 0; green, 0; blue, 255 }  ,opacity=1 ]  {$\lambda _{2r+2} =\lambda _{2r'}$};
\draw (2063,2948.58) node [anchor=south] [inner sep=0.75pt]  [font=\small,color={rgb, 255:red, 0; green, 0; blue, 255 }  ,opacity=1 ]  {$\lambda _{2r'+3}$};
\draw (2103,2988.58) node [anchor=south] [inner sep=0.75pt]  [font=\small,color={rgb, 255:red, 0; green, 0; blue, 255 }  ,opacity=1 ]  {$\lambda _{2r'+2}$};
\draw (2301,2948.58) node [anchor=south] [inner sep=0.75pt]  [font=\small,color={rgb, 255:red, 0; green, 0; blue, 255 }  ,opacity=1 ]  {$\lambda _{2r'+3}$};
\draw (2221,2948.58) node [anchor=south] [inner sep=0.75pt]  [font=\small,color={rgb, 255:red, 0; green, 0; blue, 255 }  ,opacity=1 ]  {$\lambda _{2r+3}$};
\draw (2341+10,2988.58) node [anchor=south] [inner sep=0.75pt]  [font=\small,color={rgb, 255:red, 0; green, 0; blue, 255 }  ,opacity=1 ]  {$\lambda _{2r'+2}$};
\draw (2181,2988.58) node [anchor=south] [inner sep=0.75pt]  [font=\small,color={rgb, 255:red, 0; green, 0; blue, 255 }  ,opacity=1 ]  {$\lambda _{2r}$};
\draw (2262.57,2991.33) node [anchor=south] [inner sep=0.75pt]  [font=\small,color={rgb, 255:red, 0; green, 0; blue, 255 }  ,opacity=1 ]  {$\lambda _{2r+2} =\lambda _{2r'}$};

\end{tikzpicture}
    }
    \caption{The local symmetries for $T_{2,2p+1}$ torus knots, $r=0,\dots,p-1$
    (the~symmetry exists only for $r'=r+1$)}
    \label{fig:symmetries_torus}
\end{figure}
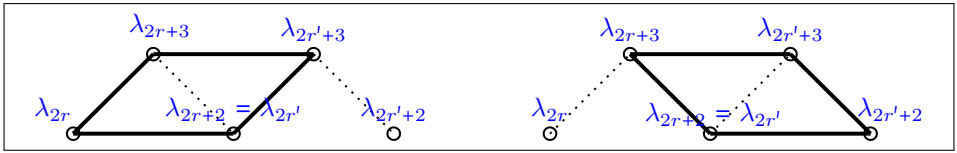

\begin{figure}[h!]
    \centering
\fbox{
    \input{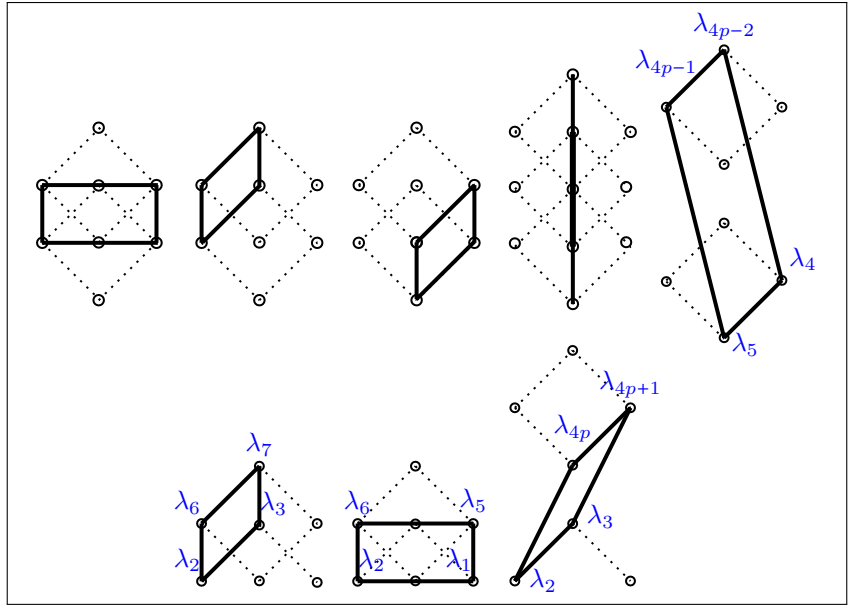}
    }
    \caption{The local symmetries for twist knots. The~symmetries which are shared between $TK_{2|p|+2}$ and $TK_{2p+1}$ twist knots do not have blue labels (any choice of $\lambda$'s ordering from figure~\ref{fig:diamonds_orientation} is valid for them). The~top-right symmetry is signature for the~$TK_{2|p|+2}$ twist knots, whereas the~three bottom ones -- for $TK_{2p+1}$ twist knots.}
    \label{fig:symmetries_twist}
\end{figure}

Recall again that entries of the~vectors given above are labels of appropriate quadruples of quiver nodes or homology generators. For $(2,2p+1)$ torus knots, the~condition $r'=r+1$ means that these generators belong to two consecutive wedges, see figure~\ref{fig:symmetries_torus}. For twist knots, generators that encode a~symmetry belong to various diamonds or the~wedge, see figure~\ref{fig:symmetries_twist}. Below we give a~proof of theorem~\ref{thm:knots} divided into three parts, each corresponding to one of the~infinite families of knots. It is followed by the~analysis of $6_2,6_3,7_3$ knots.

\begin{figure}[H]
    \centering
    \includegraphics[scale=0.24]{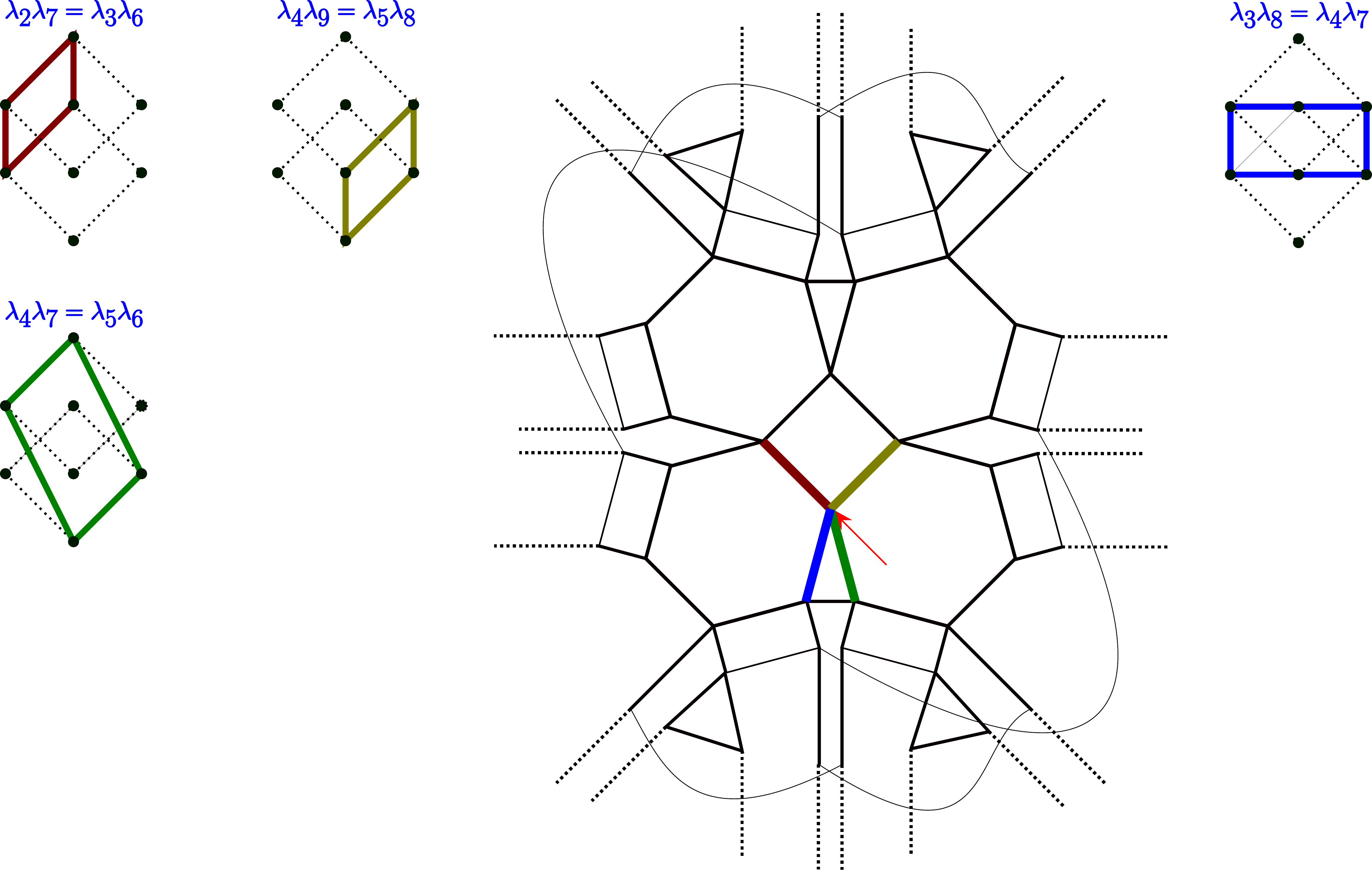}
    \caption{The four local symmetries of quiver~\eqref{eq:6_1 P(x)} corresponding to $6_1$ knot, shown as the~colorful thick edges}
    \label{fig:6_1_local_symmetries}
\end{figure}

\subsection{\texorpdfstring{$(2,2p+1)$}{(2,2p+1)} torus knots}

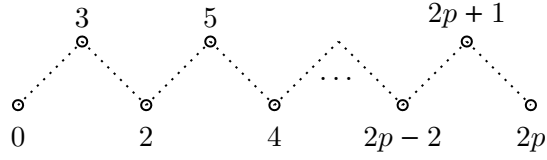
\begin{figure}[h!]
    \centering
    \tikzset{every picture/.style={line width=0.75pt}} 

\begin{tikzpicture}[x=0.75pt,y=0.75pt,yscale=-0.8,xscale=0.8]

\draw  [dash pattern={on 0.84pt off 2.51pt}]  (1013.52,2926.76) -- (1053.52,2886.76) ;
\draw  [dash pattern={on 0.84pt off 2.51pt}]  (1093.52,2926.76) -- (1133.52,2886.76) ;
\draw  [dash pattern={on 0.84pt off 2.51pt}]  (1173.52,2926.76) -- (1213.52,2886.76) ;
\draw  [dash pattern={on 0.84pt off 2.51pt}]  (1053.52,2886.76) -- (1093.52,2926.76) ;
\draw  [dash pattern={on 0.84pt off 2.51pt}]  (1133.52,2886.76) -- (1173.52,2926.76) ;
\draw  [dash pattern={on 0.84pt off 2.51pt}]  (1253.52,2926.76) -- (1293.52,2886.76) ;
\draw  [dash pattern={on 0.84pt off 2.51pt}]  (1293.52,2886.76) -- (1333.52,2926.76) ;
\draw  [fill={rgb, 255:red,0; green,0; blue, 0 }  ,fill opacity=0 ] (1091.89,2924.05) .. controls (1093.39,2923.15) and (1095.33,2923.63) .. (1096.23,2925.13) .. controls (1097.13,2926.63) and (1096.64,2928.57) .. (1095.14,2929.47) .. controls (1093.65,2930.37) and (1091.7,2929.89) .. (1090.8,2928.39) .. controls (1089.9,2926.89) and (1090.39,2924.95) .. (1091.89,2924.05) -- cycle ;
\draw  [fill={rgb, 255:red,0; green,0; blue, 0 }  ,fill opacity=0 ] (1011.89,2924.05) .. controls (1013.39,2923.15) and (1015.33,2923.63) .. (1016.23,2925.13) .. controls (1017.13,2926.63) and (1016.64,2928.57) .. (1015.14,2929.47) .. controls (1013.65,2930.37) and (1011.7,2929.89) .. (1010.8,2928.39) .. controls (1009.9,2926.89) and (1010.39,2924.95) .. (1011.89,2924.05) -- cycle ;
\draw  [fill={rgb, 255:red,0; green,0; blue, 0 }  ,fill opacity=0 ] (1051.89,2884.05) .. controls (1053.39,2883.15) and (1055.33,2883.63) .. (1056.23,2885.13) .. controls (1057.13,2886.63) and (1056.64,2888.57) .. (1055.14,2889.47) .. controls (1053.65,2890.37) and (1051.7,2889.89) .. (1050.8,2888.39) .. controls (1049.9,2886.89) and (1050.39,2884.95) .. (1051.89,2884.05) -- cycle ;
\draw  [fill={rgb, 255:red,0; green,0; blue, 0 }  ,fill opacity=0 ] (1171.89,2924.05) .. controls (1173.39,2923.15) and (1175.33,2923.63) .. (1176.23,2925.13) .. controls (1177.13,2926.63) and (1176.64,2928.57) .. (1175.14,2929.47) .. controls (1173.65,2930.37) and (1171.7,2929.89) .. (1170.8,2928.39) .. controls (1169.9,2926.89) and (1170.39,2924.95) .. (1171.89,2924.05) -- cycle ;
\draw  [fill={rgb, 255:red,0; green,0; blue, 0 }  ,fill opacity=0 ] (1131.89,2884.05) .. controls (1133.39,2883.15) and (1135.33,2883.63) .. (1136.23,2885.13) .. controls (1137.13,2886.63) and (1136.64,2888.57) .. (1135.14,2889.47) .. controls (1133.65,2890.37) and (1131.7,2889.89) .. (1130.8,2888.39) .. controls (1129.9,2886.89) and (1130.39,2884.95) .. (1131.89,2884.05) -- cycle ;
\draw  [fill={rgb, 255:red,0; green,0; blue, 0 }  ,fill opacity=0 ] (1251.89,2924.05) .. controls (1253.39,2923.15) and (1255.33,2923.63) .. (1256.23,2925.13) .. controls (1257.13,2926.63) and (1256.64,2928.57) .. (1255.14,2929.47) .. controls (1253.65,2930.37) and (1251.7,2929.89) .. (1250.8,2928.39) .. controls (1249.9,2926.89) and (1250.39,2924.95) .. (1251.89,2924.05) -- cycle ;
\draw  [fill={rgb, 255:red,0; green,0; blue, 0 }  ,fill opacity=0 ] (1331.89,2924.05) .. controls (1333.39,2923.15) and (1335.33,2923.63) .. (1336.23,2925.13) .. controls (1337.13,2926.63) and (1336.64,2928.57) .. (1335.14,2929.47) .. controls (1333.65,2930.37) and (1331.7,2929.89) .. (1330.8,2928.39) .. controls (1329.9,2926.89) and (1330.39,2924.95) .. (1331.89,2924.05) -- cycle ;
\draw  [fill={rgb, 255:red,0; green,0; blue, 0 }  ,fill opacity=0 ] (1291.89,2884.05) .. controls (1293.39,2883.15) and (1295.33,2883.63) .. (1296.23,2885.13) .. controls (1297.13,2886.63) and (1296.64,2888.57) .. (1295.14,2889.47) .. controls (1293.65,2890.37) and (1291.7,2889.89) .. (1290.8,2888.39) .. controls (1289.9,2886.89) and (1290.39,2884.95) .. (1291.89,2884.05) -- cycle ;
\draw  [dash pattern={on 0.84pt off 2.51pt}]  (1213.52,2886.76) -- (1253.52,2926.76) ;

\draw (1013.52,2938.76) node [anchor=north] [inner sep=0.75pt]  [font=\normalsize]  {$0$};
\draw (1093.52,2938.76) node [anchor=north] [inner sep=0.75pt]  [font=\normalsize]  {$2$};
\draw (1053.52,2879.76) node [anchor=south] [inner sep=0.75pt]  [font=\normalsize]  {$3$};
\draw (1133.52,2879.76) node [anchor=south] [inner sep=0.75pt]  [font=\normalsize]  {$5$};
\draw (1173.52,2938.76) node [anchor=north] [inner sep=0.75pt]  [font=\normalsize]  {$4$};
\draw (1253.52,2938.76) node [anchor=north] [inner sep=0.75pt]  [font=\normalsize]  {$2p-2$};
\draw (1333.52,2938.76) node [anchor=north] [inner sep=0.75pt]  [font=\normalsize]  {$2p$};
\draw (1293.52,2879.76) node [anchor=south] [inner sep=0.75pt]  [font=\normalsize]  {$2p+1$};
\draw (1213.52,2909.76) node    {$\dotsc $};

\end{tikzpicture}
    \caption{Homology diagram for $(2,2p+1)$ torus knot and labeling of its generators (the labeling of wedges is shown in figure~\ref{fig:diamonds_orientation}).}
    \label{fig:torus_homologial_saw}
\end{figure}

For this family of knots, the~homology diagram is a~chain of $p$ wedges joined together. The~wedges are labeled by $r=0,1,2,\dots,p-1$ as in figure~\ref{fig:diamonds_orientation}, and the~labeling of all generators is shown explicitly in figure~\ref{fig:torus_homologial_saw}. Note that what we label as the~zeroth node corresponds to the~quiver series parameter $x_1$, while the~$i$-th node for $i>1$ corresponds to $x_i$. This notation is convenient, since in the~formulas we can let $r=0$ referring to the~first wedge, so we do not have to treat it separately.
If $r$ and $r'$ label two wedges and $r'=r+1$, they share the~common node labeled by $2r+2=2r'$.

Note that the~quiver matrix~(\ref{eq:quiver_torus_general}) (its special cases are given in~(\ref{eq:3_1 quiver},~\ref{eq:5_1 quiver v1},~\ref{eq:7_1 quiver})) has elements $C_{ij}$ such that
\begin{equation}
\begin{aligned}   \label{Cij-torus}
\text{$i,j$ both odd or even:} \qquad &\ C_{ij} = j-1, & i = j: \qquad &\ C_{jj} = j, \\
\text{$i$ odd, $j$ even:} \qquad &\ C_{ij} = j, & \text{$j$ even:} \qquad &\ C_{1j} = j-1, \\
\text{$i$ even, $j$ odd:} \qquad &\ C_{ij} = j - 2 + \delta_{i+1,j},\qquad & \text{$j$ odd:} \qquad &\ C_{1j} = j-2. \\
\end{aligned}
\end{equation}

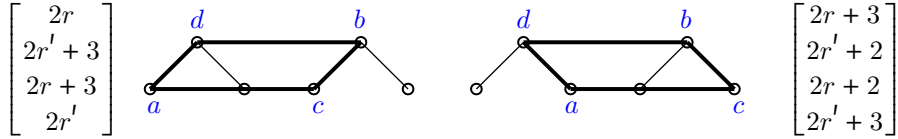
\begin{figure}[h!]
    \centering
    \tikzset{every picture/.style={line width=0.75pt}} 

\begin{tikzpicture}[x=0.75pt,y=0.75pt,yscale=-0.7,xscale=0.7]

\draw [fill={rgb, 255:red,0; green, 0; blue, 0}  ,draw opacity=1 ][line width=1.5]    (1440.18,3231.53) -- (1556.93,3231.53) ;
\draw [fill={rgb, 255:red,0; green, 0; blue, 0}  ,draw opacity=1 ][line width=1.5]    (1590.53,3265.13) -- (1556.93,3231.53) ;
\draw [fill={rgb, 255:red,0; green, 0; blue, 0}  ,draw opacity=1 ][line width=1.5]    (1473.78,3265.13) -- (1440.18,3231.53) ;
\draw [fill={rgb, 255:red,0; green, 0; blue, 0}  ,draw opacity=1 ][line width=1.5]    (1173.9,3265.13) -- (1207.5,3231.53) ;
\draw [fill={rgb, 255:red,0; green, 0; blue, 0}  ,draw opacity=1 ][line width=1.5]    (1290.66,3265.13) -- (1324.26,3231.53) ;
\draw [fill={rgb, 255:red,0; green, 0; blue, 0}  ,draw opacity=1 ][line width=1.5]    (1290.66,3265.13) -- (1173.9,3265.13) ;
\draw [fill={rgb, 255:red,0; green, 0; blue, 0}  ,draw opacity=1 ][line width=1.5]    (1473.78,3265.13) -- (1590.53,3265.13) ;
\draw[line width=0.5]     (1207.5,3231.53) -- (1241.1,3265.13) ;
\draw[line width=0.5]     (1324.26,3231.53) -- (1357.86,3265.13) ;
\draw[line width=0.5]    (1556.93,3231.53) -- (1523.33,3265.13) ;
\draw[line width=0.5]     (1440.18,3231.53) -- (1406.58,3265.13) ;
\draw  [fill={rgb, 255:red,0; green, 0; blue, 0}  ,fill opacity=0 ] (1171.94,3261.87) .. controls (1173.74,3260.79) and (1176.08,3261.38) .. (1177.16,3263.18) .. controls (1178.24,3264.98) and (1177.66,3267.31) .. (1175.86,3268.39) .. controls (1174.06,3269.47) and (1171.72,3268.89) .. (1170.64,3267.09) .. controls (1169.56,3265.29) and (1170.14,3262.95) .. (1171.94,3261.87) -- cycle ;
\draw  [fill={rgb, 255:red,0; green, 0; blue, 0}  ,fill opacity=0 ] (1239.14,3261.87) .. controls (1240.94,3260.79) and (1243.28,3261.38) .. (1244.36,3263.18) .. controls (1245.44,3264.98) and (1244.85,3267.31) .. (1243.05,3268.39) .. controls (1241.25,3269.47) and (1238.92,3268.89) .. (1237.84,3267.09) .. controls (1236.76,3265.29) and (1237.34,3262.95) .. (1239.14,3261.87) -- cycle ;
\draw  [fill={rgb, 255:red,0; green, 0; blue, 0}  ,fill opacity=0 ] (1205.54,3228.27) .. controls (1207.34,3227.19) and (1209.68,3227.78) .. (1210.76,3229.58) .. controls (1211.84,3231.38) and (1211.25,3233.71) .. (1209.46,3234.79) .. controls (1207.66,3235.87) and (1205.32,3235.29) .. (1204.24,3233.49) .. controls (1203.16,3231.69) and (1203.74,3229.35) .. (1205.54,3228.27) -- cycle ;
\draw  [fill={rgb, 255:red,0; green, 0; blue, 0}  ,fill opacity=0 ] (1322.3,3228.27) .. controls (1324.1,3227.19) and (1326.44,3227.78) .. (1327.52,3229.58) .. controls (1328.6,3231.38) and (1328.01,3233.71) .. (1326.21,3234.79) .. controls (1324.41,3235.87) and (1322.08,3235.29) .. (1321,3233.49) .. controls (1319.92,3231.69) and (1320.5,3229.35) .. (1322.3,3228.27) -- cycle ;
\draw  [fill={rgb, 255:red,0; green, 0; blue, 0}  ,fill opacity=0 ] (1288.7,3261.87) .. controls (1290.5,3260.79) and (1292.84,3261.38) .. (1293.92,3263.18) .. controls (1295,3264.98) and (1294.41,3267.31) .. (1292.61,3268.39) .. controls (1290.81,3269.47) and (1288.48,3268.89) .. (1287.4,3267.09) .. controls (1286.32,3265.29) and (1286.9,3262.95) .. (1288.7,3261.87) -- cycle ;
\draw  [fill={rgb, 255:red,0; green, 0; blue, 0}  ,fill opacity=0 ] (1355.9,3261.87) .. controls (1357.7,3260.79) and (1360.04,3261.38) .. (1361.12,3263.18) .. controls (1362.2,3264.98) and (1361.61,3267.31) .. (1359.81,3268.39) .. controls (1358.01,3269.47) and (1355.68,3268.89) .. (1354.6,3267.09) .. controls (1353.52,3265.29) and (1354.1,3262.95) .. (1355.9,3261.87) -- cycle ;
\draw [fill={rgb, 255:red,0; green, 0; blue, 0}  ,draw opacity=1 ][line width=1.5]    (1324.26,3231.53) -- (1207.5,3231.53) ;
\draw  [fill={rgb, 255:red,0; green, 0; blue, 0}  ,fill opacity=0 ] (1592.49,3261.87) .. controls (1590.69,3260.79) and (1588.36,3261.38) .. (1587.28,3263.18) .. controls (1586.2,3264.98) and (1586.78,3267.31) .. (1588.58,3268.39) .. controls (1590.38,3269.47) and (1592.71,3268.89) .. (1593.79,3267.09) .. controls (1594.87,3265.29) and (1594.29,3262.95) .. (1592.49,3261.87) -- cycle ;
\draw  [fill={rgb, 255:red,0; green, 0; blue, 0}  ,fill opacity=0 ] (1525.29,3261.87) .. controls (1523.49,3260.79) and (1521.16,3261.38) .. (1520.08,3263.18) .. controls (1519,3264.98) and (1519.58,3267.31) .. (1521.38,3268.39) .. controls (1523.18,3269.47) and (1525.51,3268.89) .. (1526.59,3267.09) .. controls (1527.67,3265.29) and (1527.09,3262.95) .. (1525.29,3261.87) -- cycle ;
\draw  [fill={rgb, 255:red,0; green, 0; blue, 0}  ,fill opacity=0 ] (1558.89,3228.27) .. controls (1557.09,3227.19) and (1554.76,3227.78) .. (1553.68,3229.58) .. controls (1552.6,3231.38) and (1553.18,3233.71) .. (1554.98,3234.79) .. controls (1556.78,3235.87) and (1559.11,3235.29) .. (1560.19,3233.49) .. controls (1561.27,3231.69) and (1560.69,3229.35) .. (1558.89,3228.27) -- cycle ;
\draw  [fill={rgb, 255:red,0; green, 0; blue, 0}  ,fill opacity=0 ] (1442.13,3228.27) .. controls (1440.33,3227.19) and (1438,3227.78) .. (1436.92,3229.58) .. controls (1435.84,3231.38) and (1436.42,3233.71) .. (1438.22,3234.79) .. controls (1440.02,3235.87) and (1442.35,3235.29) .. (1443.43,3233.49) .. controls (1444.51,3231.69) and (1443.93,3229.35) .. (1442.13,3228.27) -- cycle ;
\draw  [fill={rgb, 255:red,0; green, 0; blue, 0}  ,fill opacity=0 ] (1475.73,3261.87) .. controls (1473.93,3260.79) and (1471.6,3261.38) .. (1470.52,3263.18) .. controls (1469.44,3264.98) and (1470.02,3267.31) .. (1471.82,3268.39) .. controls (1473.62,3269.47) and (1475.95,3268.89) .. (1477.03,3267.09) .. controls (1478.11,3265.29) and (1477.53,3262.95) .. (1475.73,3261.87) -- cycle ;
\draw  [fill={rgb, 255:red,0; green, 0; blue, 0}  ,fill opacity=0 ] (1408.53,3261.87) .. controls (1406.73,3260.79) and (1404.4,3261.38) .. (1403.32,3263.18) .. controls (1402.24,3264.98) and (1402.82,3267.31) .. (1404.62,3268.39) .. controls (1406.42,3269.47) and (1408.75,3268.89) .. (1409.83,3267.09) .. controls (1410.91,3265.29) and (1410.33,3262.95) .. (1408.53,3261.87) -- cycle ;

\draw (1169.62,3272.37) node [anchor=north west][inner sep=0.75pt]  [font=\small]  {$\textcolor{blue}{a}$};
\draw (1317.1,3206.25) node [anchor=north west][inner sep=0.75pt]  [font=\small]  {$\textcolor{blue}{b}$};
\draw (1287.38,3272.37) node [anchor=north west][inner sep=0.75pt]  [font=\small]  {$\textcolor{blue}{c}$};
\draw (1200.34,3206.25) node [anchor=north west][inner sep=0.75pt]  [font=\small]  {$\textcolor{blue}{d}$};
\draw (1467.5,3272.37) node [anchor=north west][inner sep=0.75pt]  [font=\small]  {$\textcolor{blue}{a}$};
\draw (1549.77,3206.25) node [anchor=north west][inner sep=0.75pt]  [font=\small]  {$\textcolor{blue}{b}$};
\draw (1587.25,3272.37) node [anchor=north west][inner sep=0.75pt]  [font=\small]  {$\textcolor{blue}{c}$};
\draw (1433.02,3206.25) node [anchor=north west][inner sep=0.75pt]  [font=\small]  {$\textcolor{blue}{d}$};
\draw (1095.9-25,3198.13) node [anchor=north west][inner sep=0.75pt]  [font=\small]  {$\begin{bmatrix}
2r\\
2r'+3\\
2r+3\\
2r'
\end{bmatrix}$};
\draw (1604.53+25,3198.13) node [anchor=north west][inner sep=0.75pt]  [font=\small]  {$\begin{bmatrix}
2r+3\\
2r'+2\\
2r+2\\
2r'+3
\end{bmatrix}$};

\end{tikzpicture}
    \caption{Pairings between the~two wedges: type 2A.1 (left) and 2A.2 (right)}.
    \label{fig:2wedges}
\end{figure}

\begin{figure}[h!]
    \centering
    \input{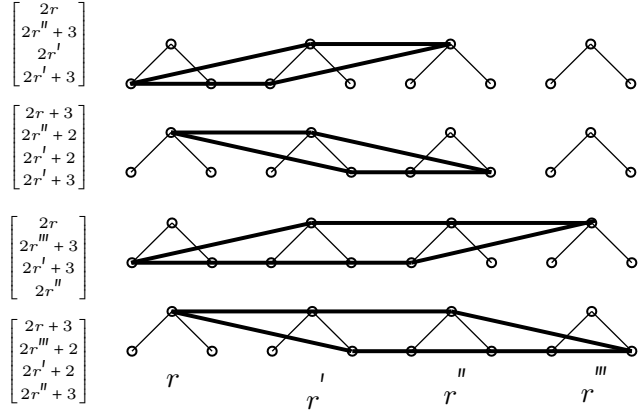}
    \caption{Pairings between 3 and 4 wedges, which are not symmetries for the~quiver matrix~(\ref{eq:quiver_torus_general}). From top to bottom: 3A.1, 3A.2, 4A.1, 4A.2.}
    \label{fig:wedges}
\end{figure}

We now use theorem~\ref{thm:main} to determine symmetries of this quiver. First, suppose that a~pairing is made of generators from only two wedges, which are located in a~generic position and not necessarily joined together, see figure~\ref{fig:2wedges}. A~direct check of conditions from theorem~\ref{thm:main} shows that the~two pairings in figure~\ref{fig:2wedges}
are the~symmetries if 
$r'=r+1$, see figure~\ref{fig:2wedge_good}.
In order to confirm that there are no other symmetries, we label the~four wedges by $r,r',r'',r'''$ such that $r<r'<r''<r'''$ (see figure~\ref{fig:wedges}).
In consequence equation~(\ref{Cij-torus}) leads to the~following pairings:
\begin{equation*}
\begin{aligned}
    \text{3A.1:} &\quad C_{ab} = 2r''+1,\; C_{cd} = 2r'+1 &
    \text{3A.2:} &\quad C_{ab} = 2r''+2,\; C_{cd} = 2r'+1 \\
    \text{4A.1:} &\quad C_{ab} = 2r'''+1,\; C_{cd} = \begin{cases} 2r'+1,\; r''=r'+1 \\ 2r'',\; r''>r'+1 \end{cases} &
    \text{4A.2:} &\quad C_{ab} = 2r'''+2,\; C_{cd} = 2r''+1
\end{aligned}
\end{equation*}
It follows that the~condition
$|C_{ab}-C_{cd}|=1$ from theorem~\ref{thm:main} cannot be met in all these cases, so the~only symmetries are indeed those in figure~\ref{fig:symmetries_torus}. 

\begin{figure}[h!]
    \centering
\fbox{
\begin{minipage}{.45\textwidth}
\footnotesize
\begin{center}
	\underline{Pairing $b=c+3,d=a+3$}
$$
	\begin{aligned}
	a = &\ a~\\
	b = &\ c+3 \\
	c = &\ c \\
	d = &\ a+3
	\end{aligned}
	\qquad
	\begin{pmatrix}
	a & \boxed{b-2} & c-1 & d-2 \\
	\boxed{b-2} & b & b-2 & b-1 \\
	c-1 & b-2 & c & c \\
	d-2 & b-1 & c & d \\
	\end{pmatrix}
$$
$$
	\begin{aligned}\label{extra_conds_exterior}
	s<a,\  \text{odd} \quad &\    a~+ (b-1)  =   c + (d-1) \\
	s<a,\  \text{even} \quad &\  (a-1) + (b-2)  = \\
	&\ (c-1) + (d-2) \\
	s>b,\  \text{odd} \quad &\   (s-2) + (s-1)  = \\
	&\ (s-2) + (s-1) \\
	s>b,\  \text{even} \quad &\  (s-1) + s  =  (s-1) + s \\
	d<s<c,\  \text{odd} \quad &\  (s-2) + (b-1)  \neq   c + (s-1) \\
	d<s<c,\  \text{even} \quad &\  (s-1) + (b-2)  \neq  (c-1) + d \\
	s=a+2 \quad &\  (i+1) + (j-2)  \neq \\
	&\ (k-1) + (l-1) \\
	s=c+1 \quad &\   (s-2) + (j-1)  \neq \\
	&\ (s-1) + (s-1) \\
	\end{aligned}
$$
\end{center}
\end{minipage}
\vline
\begin{minipage}{.45\textwidth}
\footnotesize
\begin{center}
\underline{Pairing $b=c+1,d=a+1$}
$$
\begin{aligned}
a = &\ a~\\
b = &\ c+1 \\
c = &\ c \\
d = &\ a+1
\end{aligned}
\qquad
\begin{pmatrix}
a & b-2 & c-1 & d-1 \\
b-2 & b & b-1 & b-1 \\
c-1 & b-1 & c & \boxed{c} \\
d-1 & b-1 & \boxed{c} & d \\
\end{pmatrix}     
$$
$$
\begin{aligned}
s<d,\ \text{odd} \quad &\  a~+ (b-1)  =  c + (d-1) \\
s<d,\ \text{even} \quad &\  (a-1) + (b-2)  = \\
&\ (c-1) + (d-2) \\
s>c,\ \text{odd} \quad &\  (s-2) + (s-1)  = \\
&\ (s-2) + (s-1) \\
s>c,\ \text{even} \quad &\  (s-1) + s  =  (s-1) + s \\
a<s<b,\ \text{odd} \quad &\  (s-2) + (b-1)  \neq  c + (s-1) \\
a<s<b,\ \text{even} \quad &\  (s-1) + (b-2)  \neq  (c-1) + d \\
\end{aligned}
$$
\end{center}
\end{minipage}
}
    \caption{The local symmetries of quivers~(\ref{eq:quiver_torus_general}) for $T_{2,2p+1}$ torus knots.}
    \label{fig:2wedge_good}
\end{figure}

\newpage

\subsection{Twist knots \texorpdfstring{$TK_{2|p|+2}$: $4_1,6_1,8_1,\dots$}{TK2p+2}}

We now conduct analogous analysis for a~family of twist knots $TK_{2|p|+2}$. Recall that a~homological diagram for such a~knot -- for a~given $p$ -- consists of $p$ diamonds and an~extra dot.
Consider a~quadruple of diamonds with labels $(r,r',r'',r''')$, such that $1\leq r\leq r'\leq r''\leq r''' \leq p$. We classify all pairings by the~number of diamonds and their relative position.
Tables in figure~\ref{fig:all_diamond_pairings} provide such classification, while all possible pairings between two diamonds are depicted in
figure~\ref{fig:2diamonds_all}.

\begin{figure}[h!]
\begin{footnotesize}
\centering

\underline{2 diamonds}

\begin{tabular}{|c|c|c|c|c|c|c|c|c|c|c|}
\hline   
    $a$ & \textcolor{Green}{$4r-1$}    & \textcolor{Green}{$4r-1$}      & \textcolor{Green}{$4r+1$}    & $4r+1$  & \textcolor{Green}{$4r+1$}   & $4r$    & $4r+1$   & $4r-1$   & $4r$     & $4r+1$  \\
    $b$ & \textcolor{Green}{$4r'$}     & \textcolor{Green}{$4r'-2$}     & \textcolor{Green}{$4r'$}     & $4r'-2$ & \textcolor{Green}{$4r'-2$}  & $4r'-2$ & $4r'-1$  & $4r'$    & $4r'-1$  & $4r'-2$    \\
    $c$ & \textcolor{Green}{$4r$}      & \textcolor{Green}{$4r-2$}      & \textcolor{Green}{$4r$}      & $4r-1$  & \textcolor{Green}{$4r$}     & $4r-2$  & $4r-1$   & $4r-2$   & $4r-2$   & $4r-2$    \\
    $d$ & \textcolor{Green}{$4r'-1$}   & \textcolor{Green}{$4r'-1$}     & \textcolor{Green}{$4r'+1$}   & $4r'$    & \textcolor{Green}{$4r'-1$}  & $4r'$   & $4r'+1$  & $4r'+1$  & $4r'+1$ & $4r'+1$       \\
    \hline
\end{tabular}
\\ \ \\
\vspace{0.3cm}
\underline{3 diamonds, equally distant}
\begin{tabular}{|c|c|c|c|c|c|c|c|c|c|c|}
\hline
    $4r-2$  & $4r-2$   & $4r-1$    & $4r-1$      & $4r$      & $4r-2$      & $4r+1$    & $4r$       & $4r-2$      \\
    $4r''$  & $4r''-1$ & $4r''+1$  & $4r''$      & $4r''+1$  & $4r''+1$    & $4r''-2$  & $4r''-1$   & $4r''+1$  \\
    $4r'-2$ & $4r'-2$  & $4r'-1$   & $4r'-1$     & $4r'$     & $4r'-1$     & $4r'-1$   & $4r'+1$    & $4r'+1$ \\
  $4r'$   & $4r'-1$  & $4r'+1$   & $4r'$       & $4r'+1$   & $4r'$       & $4r'$     & $4r'-2$    & $4r'-2$ \\
    \hline
   $4r-1$    & $4r$      & $4r$      & $4r-1$   &  \textcolor{Green}{$4r+1$}   & & & &\\
  $4r''-2$  & $4r''-1$  & $4r''-2$  & $4r''$   &  \textcolor{Green}{$4r''-2$} & & & &      \\
    $4r'-1$   & $4r'$     & $4r'$     & $4r'+1$  &  \textcolor{Green}{$4r'+1$}  & & & &    \\
   $4r'-2$   & $4r'-1$   & $4r'-2$   & $4r'-2$  &  \textcolor{Green}{$4r'-2$}  & & & &   \\
    \hline
\end{tabular}
\\ \ \\
\vspace{0.3cm}
\underline{3 diamonds, shifted up / down}
\begin{tabular}{|c|c|c|c|c|c|c|c|c|c|c|}
\hline
    $4r+1$    & $4r+1$      & $4r-1$      & $4r$     & $4r+1$   & $4r-2$    & $4r$      & $4r-2$     & $4r-2$      & $4r-1$ \\
    $4r''+1$  & $4r''$      & $4r''+1$    & $4r''+1$ & $4r''-1$ & $4r''-2$  & $4r''-2$  & $4r''-1$   & $4r''$      & $4r''-2$     \\
    $4r'-1$   & $4r'$       & $4r'-1$     & $4r'$    & $4r'-1$  & $4r'-1$   & $4r'+1$   & $4r'+1$    & $4r'+1$     & $4r'-1$ \\
   $4r'$     & $4r'-2$     & $4r'-2$     & $4r'-2$  & $4r'-2$  & $4r'$     & $4r'$     & $4r'-1$    & $4r'$       & $4r'+1$\\
\hline
\end{tabular}
\\ \ \\
\vspace{0.3cm}
\underline{4 diamonds, equally distant}

\begin{tabular}{|c|c|c|c|c|c|c|c|c|c|}
\hline
    $4r-1$     & $4r-1$         & $4r$       & $4r+1$         & $4r+1$      & $4r-1$    & $4r+1$   & $4r$ \\
     $4r'''$    & $4r'''+1$      & $4r'''+1$  & $4r'''-2$      & $4r'''-2$.  & $4r'''$   & $4r'''-2$ & $4r'''$ \\
    $4r'$      & $4r'+1$        & $4r'+1$    & $4r'$          & $4r'-1$     & $4r'+1$   & $4r'$ & $4r'$  \\
     $4r''-1$   & $4r''-1$       & $4r''$     & $4r''-1$       & $4r''$      & $4r''-2$  & $4r''-1$ & $4r''$ \\
\hline
    $4r-1$     & $4r$       & $4r$        & $4r+1$      & $4r+1$    & $4r$       & $4r-1$ &\\
    $4r'''-2$  & $4r'''-1$  & $4r'''-2$   & $4r'''-1$   & $4r'''$   & $4r'''-1$  & $4r'''$  &    \\
    $4r'-2$    & $4r'-1$    & $4r'-2$     & $4r'-1$     & $4r'$     & $4r'+1$    & $4r'-2$&\\
    $4r''-1$   & $4r''$     & $4r''$      & $4r''+1$    & $4r''+1$  & $4r''-2$   & $4r''+1$  &    \\
    \hline
     $4r-2$     & $4r-2$   & $4r-2$      & $4r-2$      & $4r+1$    & $4r+1$    & $4r-1$      &  \\
    $4r'''-1$  & $4r'''$  & $4r'''+1$   & $4r'''+1$   & $4r'''-2$ & $4r'''-2$ & $4r'''-1$    &       \\
    $4r'-1$    & $4r'$    & $4r'-1$     & $4r'$      & $4r'+1$   & $4r'-2$   & $4r'-1$     &  \\
     $4r''-2$   & $4r''-2$ & $4r''$      & $4r''-1$   & $4r''-2$  & $4r''+1$  & $4r''-1$     &  \\
    \hline
\end{tabular}
\\ \ \\
\vspace{0.3cm}
\underline{4 diamonds, shifted up / down} \\
\begin{tabular}{|c|c|c|c|c|c|c|c|c|c|}
\hline
$ \begin{matrix} 
4r-1 \\
4r''' \\
4r'+1 \\
4r''+1
\end{matrix}  $ &
$ \begin{matrix} 
4r-2 \\
4r''' \\
4r' \\
4r''+1
\end{matrix}  $ &
$ \begin{matrix} 
4r-2 \\
4r''' \\
4r'+1 \\
4r''
\end{matrix}  $ &
$ \begin{matrix} 
4r-2 \\
4r'''-2 \\
4r'-1 \\
4r''
\end{matrix}  $ &
$ \begin{matrix} 
4r-1 \\
4r'''-2 \\
4r'-1 \\
4r''+1
\end{matrix}  $ &
$ \begin{matrix} 
4r \\
4r'''-1 \\
4r'+1 \\
4r''+1
\end{matrix}  $ &
$ \begin{matrix} 
4r+1 \\
4r''' \\
4r'-2 \\
4r''
\end{matrix}  $ 
&
$ \begin{matrix} 
4r-2 \\
4r'''-2 \\
4r' \\
4r''-1
\end{matrix}  $ &
$ \begin{matrix} 
4r-2 \\
4r'''-1 \\
4r'+1 \\
4r''-1
\end{matrix}  $ &
$ \begin{matrix} 
4r-2 \\
4r'''-1 \\
4r'-1 \\
4r''+1
\end{matrix}  $ 
\\
\hline
$ \begin{matrix} 
4r \\
4r'''-1 \\
4r'-2 \\
4r''-2
\end{matrix}  $ &
$ \begin{matrix} 
4r \\
4r'''+1 \\
4r'-2 \\
4r''
\end{matrix}  $ &
$ \begin{matrix} 
4r \\
4r'''+1 \\
4r' \\
4r''-2
\end{matrix}  $ &
$ \begin{matrix} 
4r+1 \\
4r'''+1 \\
4r' \\
4r''-1
\end{matrix}  $ &
$ \begin{matrix} 
4r+1 \\
4r'''-1 \\
4r'-2 \\
4r''-1
\end{matrix}  $ &
$ \begin{matrix} 
4r \\
4r'''-2 \\
4r' \\
4r''+1
\end{matrix}  $ 
&
$ \begin{matrix} 
4r-1 \\
4r'''+1 \\
4r'-1 \\
4r''-2
\end{matrix}  $ &
$ \begin{matrix} 
4r-1 \\
4r'''+1 \\
4r'-2 \\
4r''-1
\end{matrix}  $ &
$ \begin{matrix} 
4r-1 \\
4r''' \\
4r'-2 \\
4r''-2
\end{matrix}  $ &
\\
\hline
\end{tabular}
\\ \ \\
\caption{The complete classification of pairings between diamonds in a~homological diagram. Only~\textcolor{Green}{green} pairings, which arise for some specific values of $r,r',r''$ for a~given $p$, produce local symmetries shown in figure~\ref{fig:symmetries_twist}.}
\label{fig:all_diamond_pairings}
\end{footnotesize}
\end{figure}

\newpage

\begin{figure}[hbt!]
\centering
\input{Figures/2diamonds_all.tikz}
\caption{All pairings between two homology diamonds $(r,r')$}
\label{fig:2diamonds_all}
\end{figure}

\noindent
\begin{figure}[h!]
\centering
\fbox{
\begin{minipage}{.45\textwidth}
\footnotesize
$$
\begin{aligned}
a = &\ 4r-1 \\
b = &\ 4r' \\
c = &\ 4r \\
d = &\ 4r'-1
\end{aligned}
\qquad
\begin{pmatrix}
2r-3 & 2r-2 & 2r-2 & 2r-3 \\
2r-2 & 2r'-1 & 2r-1 & 2r'-2 \\
2r-2 & 2r-1 & 2r-1 & \boxed{2r-1} \\
2r-3 & 2r'-2 & \boxed{2r-1} & 2r'-3
\end{pmatrix}    
$$
$$
\begin{aligned}
& \underline{r<r''<r':} \\
4r''-2\ &\ 2r-1+2r''-1 =  2r+2r''-2 \\
4r''-1\ &\ 2r-3+2r''-2 \neq  2r-1 + 2r''-3 \\
4r''\ &\   2r-2+2r''-1 =  2r-1+2r''-2  \\
4r''+1\ &\ 2r-4+2r''-2 =  2r-3+2r''-3  \\
& \underline{r<r'<r'':} \\
4r''-2\ &\ 2r-1 + 2r' =  2r + 2r'-1 &\  \\
4r''-1\ &\ 2r-3 + 2r'-1 =  2r-1 + 2r'-3 \\
4r''\ &\ 2r-2 + 2r'-1 =  2r-1 + 2r'-2  \\
4r''+1\ &\ 2r-4 + 2r'-3 =  2r-3 + 2r'-4 &\  \\
& \underline{r''<r<r':} \\
4r''-2\ &\ 2r''-4 + 2r''-1 =  2r''-1 + 2r''-4 &\  \\
4r''-1\ &\ 2r''-3 + 2r''-2 =  2r''-2 + 2r''-3 \\
4r''\ &\ 2r''-1 + 2r''-1 =  2r''-1 + 2r''-1  \\
4r''+1\ &\ 2r''-4 + 2r''-2 =  2r''-2 + 2r''-4 &\  \\
\end{aligned}
$$
\end{minipage}
\vline
\begin{minipage}{0.48\textwidth}
\footnotesize
$$
\begin{aligned}
a = &\ 4r+1 \\
b = &\ 4r'-2 \\
c = &\ 4r+5 \\
d = &\ 4r'-6
\end{aligned}  
\qquad 
\begin{pmatrix}
2r-4 & 2r-2 & 2r-4 & 2r-2 \\
2r-2 & 2r' & 2r & 2r'-2 \\
2r-4 &  2r & 2r-2 & \boxed{2r-1} \\
2r-2 & 2r'-2 & \boxed{2r-1} & 2r'-2
\end{pmatrix}     
$$
$$
\begin{aligned}
& \underline{\text{The left vertical axis:}} \\
\dots,4r-9,4r-5\ &\ 2r-6 + 2r-3 =  2r-6 + 2r-3 &\  \\
4r-1\ &\ 2r-4 + 2r-1 =  2r-4 + 2r-1 \\
4r+3\ &\ 2r-3 + 2r+1 = 2r-2 + 2r  \\
4r+7,4r+11,\dots\ &\ 2r-3 + 2r+2 = 2r-1 + 2r &\  \\
& \underline{\text{The right vertical axis:}} \\
\dots,4r-8,4r-4\ &\ 2r-5 + 2r-2 = 2r-5 + 2r-2 &\  \\
4r\ &\ 2r-3 + 2r = 2r-3 + 2r \\
4r+4\ &\ 2r-2 + 2r+2 =  2r-1 + 2r+1  \\
4r+8,4r+12,\dots\ &\ 2r-2 + 2r+3 =  2r + 2r+1 &\  \\
& \underline{\text{The middle vertical axis:}} \\
4r''-2, r''\leq p\ &\ 2r''-3 + 2r'' = 2r''-3 + 2r'' &\  \\
4r+9,4r+13,\dots\ &\ 2r-4 + 2r+1 =  2r-2 + 2r-1 &\  \\
\end{aligned}
$$
\end{minipage}
}
\caption{The two pairings which are symmetries only when $r'=r+1$}
\label{fig:symmetries_a}
\end{figure}

\newpage

\noindent
\begin{figure}[h!]
\centering
\fbox{
\begin{minipage}{.45\textwidth}
\footnotesize
\centering
$$
\begin{aligned}
a = &\ 4r-1 \\
b = &\ 4r'-2 \\
c = &\ 4r-2 \\
d = &\ 4r'-1
\end{aligned}  
\qquad 
\begin{pmatrix}
2r-3 & \boxed{2r-1} & 2r-2 & 2r-3 \\
\boxed{2r-1} & 2r' & 2r & 2r'-2 \\
2r-2 & 2r & 2r & 2r-2 \\
2r-3 & 2r'-2 & 2r-2 & 2r'-3
\end{pmatrix}     
$$
$$
\begin{aligned}
& \underline{r''<r<r':} \\
4r''-2\ &\ 2r''-4 + 2r''-2 = 2r''-2 + 2r''-4  \\
4r''-1\ &\ 2r''-3 + 2r''-1 = 2r''-1 + 2r''-3 \\
4r''\ &\ 2r''-1 + 2r'' =  2r'' + 2r''-1  \\
4r''+1\ &\ 2r''-4 + 2r''-2 = 2r''-2 + 2r''-4   \\
& \underline{r<r''<r':} \\
4r''-2\ &\ 2r-1 + 2r'' {\neq}  2r + 2r''-2   \\
4r''-1\ &\ 2r-3 + 2r''-1 {\neq} 2r-2 + 2r''-3 \\
4r''\ &\ 2r-2 + 2r'' =  2r-1 + 2r''-1  \\
4r''+1\ &\ 2r-4 + 2r''-2 {\neq}  2r-1 + 2r''-3  \\
& \underline{r<r'<r'':} \\
4r''-2\ &\ 2r-1 + 2r' =  2r + 2r'-1  \\
4r''-1\ &\ 2r-3 + 2r'-2 =  2r-2 + 2r'-3 \\
4r''\ &\ 2r-2 + 2r'-2 =  2r-2 + 2r'-2  \\
4r''+1\ &\ 2r-4 + 2r'-3 =  2r-3 + 2r'-4   \\
\end{aligned}
$$
\end{minipage}
\vline
\begin{minipage}{.45\textwidth}
\footnotesize
\centering
$$
\begin{aligned}
a = &\ 4r+1 \\
b = &\ 4r' \\
c = &\ 4r \\
d = &\ 4r'+1
\end{aligned}  
\qquad 
\begin{pmatrix}
2r-4 & \boxed{2r-2} & 2r-3 & 2r-4 \\
\boxed{2r-2} & 2r'-1 & 2r-1 & 2r'-3 \\
2r-3 & 2r-1 & 2r-1 & 2r-3 \\
2r-4 & 2r'-3 & 2r-3 & 2r'-4
\end{pmatrix}     
$$
$$
\begin{aligned}
& \underline{r''<r<r':} \\
\ 4r''-2\ &\ 2r''-3 + 2r''-1 =   2r''-1 + 2r''-3  \\
\ 4r''-1\ &\ 2r''-4 + 2r''-2 =  2r''-2 + 2r''-4 \\
\ 4r''\ &\ 2r''-3 + 2r''-1 =   2r''-1 + 2r''-3  \\
\ 4r''+1\ &\ 2r''-4 + 2r''-2 =   2r''-2 + 2r''-4 \\
& \underline{r<r''<r':} \\
\ 4r''-2\ &\ 2r-2 + 2r''-3 \neq   2r + 2r''-3  \\
\ 4r''-1\ &\ 2r-3 + 2r''-2 =  2r-1 + 2r''-4 \\
\ 4r''\ &\ 2r-2 + 2r''-1 \neq  2r-1 + 2r''-3  \\
\ 4r''+1\ &\ 2r-4 + 2r''-2 \neq  2r-3 + 2r''-4  \\
& \underline{r<r'<r'':} \\
\ 4r''-2\ &\ 2r-2 + 2r'' = 2r + 2r''-2  \\
\ 4r''-1\ &\ 2r-3 + 2r''-1 =  2r-1 + 2r''-3 \\
\ 4r''\ &\ 2r-2 + 2r''-1 =  2r-1 + 2r''-2  \\
\ 4r''+1\ &\ 2r-4 + 2r''-3 =  2r-3 + 2r''-4  \\
\end{aligned}
$$
\end{minipage}
}
\caption{Another two pairings which are symmetries only when $r'=r+1$}
\label{fig:symmetries_b}
\end{figure}
We  now show that the~green pairings in figure
\ref{fig:all_diamond_pairings} are indeed local symmetries.
The detailed analysis of four of them is given in figure~\ref{fig:symmetries_a} and~\ref{fig:symmetries_b}.
Notice that the~rightmost pairing in figure~\ref{fig:symmetries_b} is a~particular case of \begin{equation}\label{typeCstd}
    \begin{bmatrix}
    4r+1 \\
    4r''-2 \\
    4r'+1 \\
    4r'-2
    \end{bmatrix}
\end{equation}
Indeed, from the~sub-matrix
\begin{equation}
\begin{aligned}
a = &\ 4r+1 \\
b = &\ 4r'-2 \\
c = &\ 4r+5 \\
d = &\ 4r'-6
\end{aligned}  
\qquad 
\begin{pmatrix}
2r-4 & 2r-2 & 2r-4 & 2r-2-\delta_{r+1,r'} \\
2r-2 & 2r' & 2r & 2r'-2 \\
2r-4 &  2r & 2r-2 & 2r-\delta_{r+2,r'} \\
2r-2-\delta_{r+1,r'} & 2r'-2 & 2r-\delta_{r+2,r'} & 2r'-2
\end{pmatrix}     
\end{equation}
we see that $r'=r+2$ is the~only candidate for a~symmetry (otherwise the~condition $|C_{ab}-C_{cd}|=1$ fails).
To stress again, in the~examples above (figure~\ref{fig:symmetries_a} and~\ref{fig:symmetries_b}) the~crucial condition for the~symmetry is $r'=r+1$, i.e. pairing of the~two neighboring diamonds.

Among the~green candidates in table~\ref{fig:all_diamond_pairings} there is only one case left: 

\begin{footnotesize}
\begin{equation}\label{typeD}
\begin{aligned}
a = &\ 4r+1 \\
b = &\ 4r'-2 \\
c = &\ 4r \\
d = &\ 4r'-1
\end{aligned}  
\qquad 
\begin{pmatrix}
2r-4 & 2r-2 & 2r-3 & 2r-3 \\
2r-2 & 2r' & 2r & 2r'-2 \\
2r-3 & 2r & 2r-1 & \boxed{2r-1} \\
2r-3 & 2r'-2 & \boxed{2r-1} & 2r'-3
\end{pmatrix}     
\end{equation}
$$
\begin{aligned}
& s=4\ &\ (-1) + (2)\ \textcolor{red}{\neq}\ & (1) + (1) \\
& s=5\ &\ (-2) + (0)\ \textcolor{red}{\neq}\ & (0) + (-1) \\
& \underline{r<r''<r':} \\
& s=4r-1\ &\ (2r-4) + (2r-1) = &\ (2r-2) + (2r-3) \\
& s=4r-2\ &\ (2r-3) + (2r) = &\ (2r-1) + (2r-2) \\
& s=4r'\ &\ (2r-2) + (2r'-1) = &\ (2r-1) + (2r'-2) \\
& s=4r'+1\ &\ (2r-4) + (2r'-3) = &\ (2r-3) + (2r'-4) \\
& s=4r''\ &\ (2r-2) + (2r'') = &\ (2r-1) + (2r''-1) \\
& \underline{r<r'<r'':} \\
& s=4r''-1\ &\ (2r-3) + (2r'-2)\ \textcolor{red}{\neq}\ &\  (2r-1) + (2r'-3) \\
& s=4r''-2\ &\ (2r-2) + (2r')\ \textcolor{red}{\neq}\ &\  (2r) + (2r'-1) \\
\end{aligned}
$$
\end{footnotesize}
Due to the~failure of the~four spectators ($\textcolor{red}{\neq}$), the~case
(\ref{typeD}) gives a~symmetry if and only if $r=1$ and $r'=p$, which means that the~bottom diamond interacts with the~top diamond.
For example, if $r=p=1$, the~pairing~(\ref{typeD}) turns into the~only symmetry for $4_1$ knot, figure~\ref{fig:symmetries_4_1}.

We have thus shown that all five cases in the~first row of figure~\ref{fig:symmetries_twist} are indeed non-trivial symmetries.
It turns out that all other pairings listed in table~\ref{fig:all_diamond_pairings} fail to be a~(non-trivial) symmetry.
This happens because of the~two reasons: when $C_{ab}\neq C_{cd}$, either the~condition $|C_{ab}-C_{cd}|=1$ fails in general, or it is satisfied only when some diamonds collide, which brings us back to the~case of two diamonds. On the~other hand, any pairing between two diamonds which is not in our ``top five'' fails due to spectator constraints (which we verified in Mathematica). To sum up, only five cases give a~symmetry: four of them involve a~pair of diamonds, and one involves a~triple (the ``vertical'' pairing).

\subsection{Twist knots \texorpdfstring{$TK_{2p+1}$: $3_1,5_2,7_2,9_2,\dots$}{TK2p+1}}

For this family of twist knots, a~large portion of symmetries determined by the~pairings originating from diamonds is the~same as for the~previous family of twist knots $TK_{2|p|+2}$. The~reason is a~structural similarity between their HOMFLY-PT homologies. To be more specific, the~main building blocks (diamonds) are the~same for both families. The~difference is in the~form of a~zig-zag, which for $TK_{2|p|+2}$ knots is degenerated to a~dot, while for $TK_{2p+1}$ knot it takes form of a~single wedge (of length 3). Therefore, at this stage we only need to study how this wedge interacts with diamonds.
In total, there are five potential pairings:
\begin{equation}
\label{wedge_diamond_potential_pairings}
\begin{bmatrix}
2 \\
4r+1 \\
3 \\
4r 
\end{bmatrix},
\
\begin{bmatrix}
1 \\
4r+2 \\
3 \\
4r 
\end{bmatrix},
\ 
\begin{bmatrix}
2 \\
4r+3 \\
3 \\
4r+2 
\end{bmatrix},
\
\begin{bmatrix}
1 \\
4r+3 \\
3 \\
4r+1 
\end{bmatrix},
\
\begin{bmatrix}
1 \\
4r+2 \\
2 \\
4r+1 
\end{bmatrix},
\end{equation}
where $r=1,\ldots p-1$ enumerates diamonds.
One of these cases turns out to be trivial:
\begin{equation}
\begin{aligned}
a = &\ 1 \\
b = &\ 4r+2 \\
c = &\ 3\\
d = &\ 4r
\end{aligned}  
\qquad 
\begin{pmatrix}
2 & 1 & 2 & 1 \\
1 & 2r-2 & 2 & 2r-3 \\
2 & 2 & 3 & 1 \\
1 & 2r-3 & 1 & 2r-3
\end{pmatrix}     
\end{equation}
The other four cases are investigated below in detail, see tables in figure~\ref{fig:zigzag_vs_diamond_pairings}.
For the~top-left case the~only possibility for a~symmetry is $r=p-1$.
This proves the~bottom-right symmetry in figure
\ref{fig:symmetries_twist}.
Another non-trivial symmetry arises from the~top-right case in figure~\ref{fig:zigzag_vs_diamond_pairings}. The~spectator constraints are satisfied for $1<r<r'$, so we get the~symmetry between the~wedge and the~first diamond, which is depicted in figure~\ref{fig:symmetries_twist} (bottom-left). Likewise, the~rightmost pairing in~(\ref{wedge_diamond_potential_pairings}) is 
a symmetry as well, see figure~\ref{fig:symmetries_twist} (bottom-middle). However, $\lambda_1\lambda_{4r+3}=\lambda_{3}\lambda_{4r+1}$
does not lead to a~symmetry because of the~spectator constraint for $s=2$. That
is why we end up with only three local symmetries between the~wedge and a~diamond.

\begin{figure}[H]
\begin{footnotesize}
\fbox{
\begin{minipage}{.45\textwidth}
\begin{center}
\underline{Pairing $(2,4r+1,3,4r)$:}
$$
\begin{aligned}
a = &\ 2 \\
b = &\ 4r+1 \\
c = &\ 3 \\
d = &\ 4r
\end{aligned}
\qquad
\begin{pmatrix}
0 & \boxed{2} & 1 & 0 \\
\boxed{2} & 2r & 3 & 2r-2 \\
1 & 3 & 3 & 1 \\
0 & 2r-2 & 1 & 2r-3
\end{pmatrix}
$$
$$
\begin{aligned}\label{xxx2}
s = &\ 1 & 1+2= &2+1 \\
s = &\ 4r+2 & 0+2r+1=&2+2r-1 \\
s = &\ 4r+3 & 1+2r+2=&3+2r\\
& \underline{1<r'<r:} \\
s = &\ 4r' & 0+2r'-2 =& 1+2r'-3 \\
s = &\ 4r'+1 & 2 + 2r'=& 3+2r'-1 \\
s = &\ 4r'+2 & 0+2r' =& 2+2r'-2\\
s = &\ 4r'+3 & 1+2r'+1 =& 3+2r'-1\\
& \underline{r<r':} \\
s=  &\ 4r' & 0+2r-1{\color{red}\neq } & 1+2r-3\\
\end{aligned}
$$
\end{center}
\end{minipage}
\vline
\begin{minipage}{.45\textwidth}
\begin{center}
\underline{Pairing $(2,4r+3,3,4r+2)$:}
$$
\begin{aligned}
a = &\ 2  \\
b = &\ 4r+3 \\
c = &\ 3 \\
d = &\ 4r+2 
\end{aligned}
\qquad
\begin{pmatrix}
0 & 1 & 1 & 0 \\
1 & 2r+3 & 3 & 2r+1 \\
1 & 3 & 3 & \boxed{2} \\
0 & 2r+1 & \boxed{2} & 2r\\
\end{pmatrix}     
$$
$$
\begin{aligned}\label{yyy2}
s = &\ 1 & 1+2= & 2+1 \\
s = &\ 4r & 0+2r= & 1+2r-1 \\
s = &\ 4r+1 & 2+2r+2= & 3+2r+1\\
& \underline{1<r'<r:} \\
s = &\ 4r' & 0+2r' = & 1+2r'-1 \\
s = &\ 4r'+1 & 2 + 2r'+2= & 3+2r'+1 \\
s = &\ 4r'+2 & 0+2r'+1 {\color{red}\neq} & 2+2r'\\
s = &\ 4r'+3 & 1+2r'+3 {\color{red}\neq} & 3+2r'+2\\
& \underline{r<r':} \\
s = &\ 4r' & 0+2r-1 = & 1+2r-2 \\
s = &\ 4r'+1 & 2 + 2r+1= & 3+2r \\
s = &\ 4r'+2 & 0+2r = & 2+2r-2\\
s = &\ 4r'+3 & 1+2r+1 = & 3+2r-1\\
\end{aligned}
$$
\end{center}
\end{minipage}
}
\end{footnotesize}
\noindent
\begin{footnotesize}
\fbox{
\begin{minipage}{.45\textwidth}
\begin{center}
\underline{Pairing $(2,4r+1,1,4r+2)$:}
$$
\begin{aligned}
a = &\ 2 \\
b = &\ 4r+1 \\
c = &\ 1 \\
d = &\ 4r+2
\end{aligned}
\qquad
\begin{pmatrix}
0 & \boxed{2} & 1 & 0 \\
\boxed{2} & 2r+2 & 2 & 2r+1 \\
1 & 2 & 2 & 1 \\
0 & 2r+1 & 1 & 2r\\
\end{pmatrix}
$$
$$
\begin{aligned}\label{xxx4}
s = &\ 3 & 1+3= & 2+2 \\
s = &\ 4r & 0+2r= & 1+2r-1 \\
s = &\ 4r+3 & 1+2r+2= & 2+2r+1\\
& \underline{1<r'<r:} \\
s= &\ 4r' & 0+2r'-2= & 1+2r'-3\\
s= &\ 4r'+1 & 2+2r'{\color{red}\neq} & 2+2r'-2\\
s= &\ 4r'+2 & 0+2r'{\color{red}\neq} &  1+2r'-2\\
s= &\ 4r'+3 & 1+3r'+1= & 2+2r'~\\
& \underline{r<r':} \\
s= &\ 4r' & 0+2r-1= & 1+2r-2\\
s= &\ 4r'+1 & 2+2r= & 2+2r\\
s= &\ 4r'+2 & 0+2r-1= & 1+2r-2\\
s=&\ 4r'+3 & 1+2r= & 2+2r-1~
\end{aligned}
$$
\end{center}
\end{minipage}
\vline
\begin{minipage}{.45\textwidth}
\begin{center}
\underline{Pairing $(1,4r+3,3,4r+1)$:}
$$
\begin{aligned}
a = &\ 1  \\
b = &\ 4r+3 \\
c = &\ 3 \\
d = &\ 4r+1 
\end{aligned}
\qquad
\begin{pmatrix}
2 & 2 & 2 & 2 \\
2 & 2r+3 & 3 & 2r+2 \\
2 & 3 & 3 & \boxed{3} \\
2 & 2r+2 & \boxed{3} & 2r+2\\
\end{pmatrix}     
$$
$$
\begin{aligned}\label{yyy5}
s = &\ 2\qquad 1+1 \textcolor{red}{\neq} 1+2\\
\end{aligned}
$$
\end{center}
\end{minipage}
}
\end{footnotesize}
\caption{The non-trivial pairings between the~wedge and a~diamond}
\label{fig:zigzag_vs_diamond_pairings}
\end{figure}


\subsection{\texorpdfstring{$6_2,6_3,7_3$}{62, 63, 73} knots}

Finally, with the~support of the~attached Mathematica code, we determine local symmetries for three other knots $6_2, 6_3$ and $7_3$, for some particular quivers found in~\cite{KRSS1707long} and~\cite{SW1711}.

\paragraph{\texorpdfstring{$6_2$}{62} knot}

\begin{figure}[h!]
    \centering
    \input{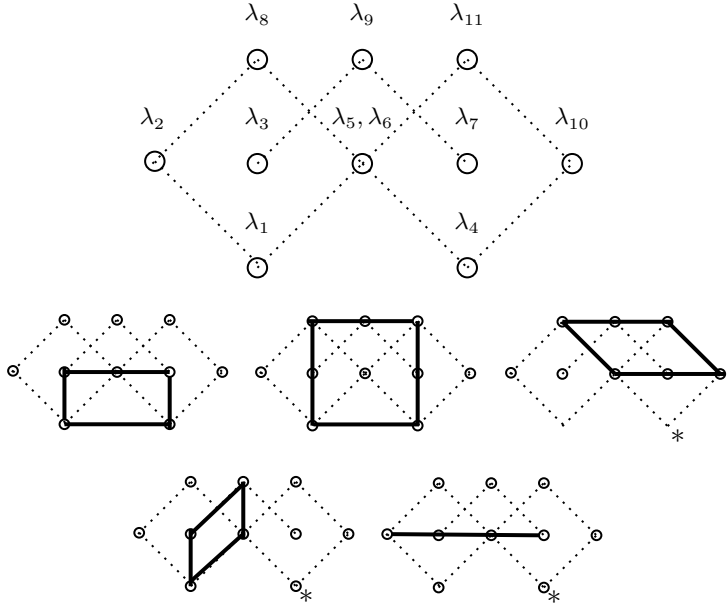}
    \caption{Homology diagram and local symmetries for $6_2$ knot, each picture marked with * corresponds to two symmetries, due to double-valued nodes $\lambda_5$ and $\lambda_6$.}
    \label{fig:homology62}
\end{figure}

Let us start from the~knot $6_2$. The~quiver from~\cite{KRSS1707long} reads
\begin{equation}
\label{eq:c62}
   C=\left[
\begin{array}{ccccccccccc}
 -2 & -2 & -1 & -1 & -1 & -1 & 0 & -1 & 1 & 1 & 1 \\
 -2 & -1 & -1 & 0 & 0 & 0 & 1 & 0 & 1 & 2 & 2 \\
 -1 & -1 & 0 & 1 & 0 & 0 & 1 & 0 & 1 & 2 & 2 \\
 -1 & 0 & 1 & 0 & 0 & 0 & 1 & 0 & 2 & 1 & 1 \\
 -1 & 0 & 0 & 0 & 1 & 1 & 1 & 1 & 2 & 2 & 2 \\
 -1 & 0 & 0 & 0 & 1 & 1 & 1 & 1 & 2 & 2 & 2 \\
 0 & 1 & 1 & 1 & 1 & 1 & 2 & 1 & 2 & 2 & 2 \\
 -1 & 0 & 0 & 0 & 1 & 1 & 1 & 2 & 2 & 3 & 3 \\
 1 & 1 & 1 & 2 & 2 & 2 & 2 & 2 & 3 & 3 & 3 \\
 1 & 2 & 2 & 1 & 2 & 2 & 2 & 3 & 3 & 3 & 3 \\
 1 & 2 & 2 & 1 & 2 & 2 & 2 & 3 & 3 & 3 & 4 \\
\end{array}
\right],\qquad \qquad \boldsymbol{\lambda} = \left[\begin{array}{c}
    q^{-2} (-t)^{-2}\\
    a^2 q^{-4} (-t)^{-1}\\
    a^2 q^{-2} \\
    q^2\\
    a^2 (-t)\\
    a^2 (-t)\\
    a^2 q^2 (-t)^2\\
    a^4 q^{-2} (-t)^2\\
    a^4 (-t)^3\\
    a^2 q^4 (-t)^3\\
    a^4 q^{2} (-t)^4\\
    \end{array}\right].
\end{equation}
%
%
There are eight local symmetries associated to~(\ref{eq:c62}) for the~following pairings:
\begin{align*}
 \lambda_1\lambda_7&=\lambda_3\lambda_4,  &
 \lambda_1\lambda_{11}&=\lambda_4\lambda_8,
 &
 \lambda_5\lambda_{11}&=\lambda_8\lambda_{10}, &
 \lambda_6\lambda_{11}&=\lambda_8\lambda_{10},\\
 \lambda_1\lambda_9&=\lambda_3\lambda_5, &
 \lambda_1\lambda_9&=\lambda_3\lambda_6,&
 \lambda_2\lambda_7&=\lambda_3\lambda_5, &
 \lambda_2\lambda_7&=\lambda_3\lambda_6.
\end{align*}
Their graphical representation, together with the~homology diagram, is given in figure~\ref{fig:homology62}.

\paragraph{\texorpdfstring{$6_3$}{63} knot}

\begin{figure}[h!]
    \centering
    \input{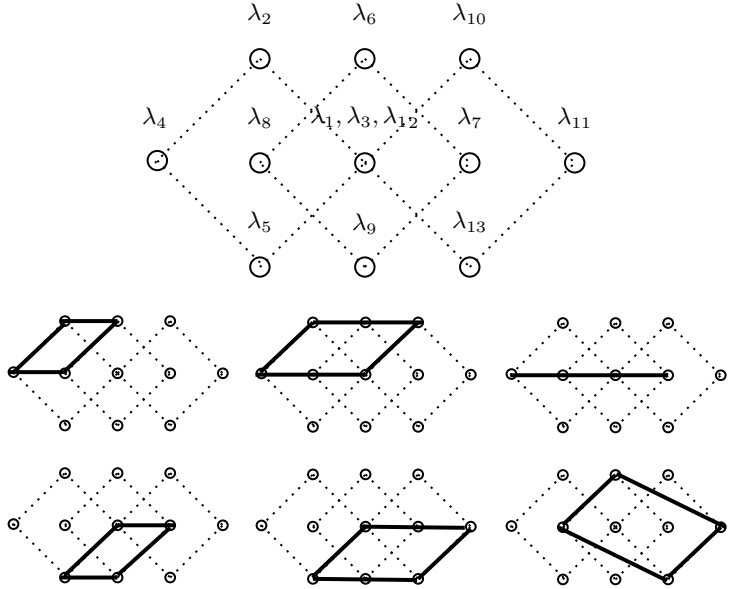}
    \caption{Homology diagram and local symmetries for $6_3$ knot.}
    \label{fig:homology63}
\end{figure}

For $6_3$ the~quiver matrix from~\cite{KRSS1707long} is given by 
\begin{equation}
\label{eq:c63}
    C=\left[
\begin{array}{ccccccccccccc}
 0 & 0 & 0 & -1 & -1 & 0 & 0 & -1 & -1 & 0 & 0 & -1 & -1 \\
 0 & 1 & 0 & -1 & -2 & 1 & 0 & -1 & -2 & 1 & 1 & 0 & -1 \\
 0 & 0 & 0 & -1 & -2 & 1 & 0 & 0 & -2 & 1 & 1 & 0 & 0 \\
 -1 & -1 & -1 & -2 & -3 & 0 & -1 & -2 & -3 & -1 & 0 & -2 & -2 \\
 -1 & -2 & -2 & -3 & -3 & -1 & -1 & -2 & -3 & -1 & -1 & -2 & -2 \\
 0 & 1 & 1 & 0 & -1 & 2 & 1 & 0 & -1 & 2 & 1 & 1 & -1 \\
 0 & 0 & 0 & -1 & -1 & 1 & 1 & 0 & -1 & 2 & 1 & 1 & 0 \\
 -1 & -1 & 0 & -2 & -2 & 0 & 0 & -1 & -2 & 0 & 0 & -1 & -2 \\
 -1 & -2 & -2 & -3 & -3 & -1 & -1 & -2 & -2 & 0 & -1 & -1 & -2 \\
 0 & 1 & 1 & -1 & -1 & 2 & 2 & 0 & 0 & 3 & 2 & 1 & 0 \\
 0 & 1 & 1 & 0 & -1 & 1 & 1 & 0 & -1 & 2 & 2 & 1 & 0 \\
 -1 & 0 & 0 & -2 & -2 & 1 & 1 & -1 & -1 & 1 & 1 & 0 & -1 \\
 -1 & -1 & 0 & -2 & -2 & -1 & 0 & -2 & -2 & 0 & 0 & -1 & -1 \\
\end{array}
\right],\quad  \boldsymbol{\lambda} = \left[\begin{array}{c}
    1\\
    a^2 q^{-2} (-t)\\
    1\\
    q^{-4} (-t)^{-2}\\
    a^{-2} q^{-2} (-t)^{-3}\\
    a^2 (-t)^2\\
    q^2 (-t)\\
    q^{-2} (-t)^{-1}\\
    a^{-2} (-t)^{-2}\\
    a^2 q^2 (-t)^3\\
    q^{4} (-t)^2\\
    1\\
    a^{-2} q^{2} (-t)^{-1}
    \end{array}\right].
\end{equation}
For~(\ref{eq:c63}) there are six local symmetries for the~following pairings:
\begin{align*}
 \lambda_2\lambda_8&=\lambda_4\lambda_6, &
 \lambda_2\lambda_{12}&=\lambda_4\lambda_{10}, &
 \lambda_3\lambda_8&=\lambda_4\lambda_7, \\
 \lambda_3\lambda_9&=\lambda_5\lambda_7, &
 \lambda_3\lambda_{13}&=\lambda_5\lambda_{11}, &
 \lambda_6\lambda_{13}&=\lambda_8\lambda_{11},
\end{align*}
which graphical representation, together with the~homology diagram, is given in figure~\ref{fig:homology63}.

\paragraph{\texorpdfstring{$7_3$}{73} knot}

\begin{figure}[h!]
    \centering
    \input{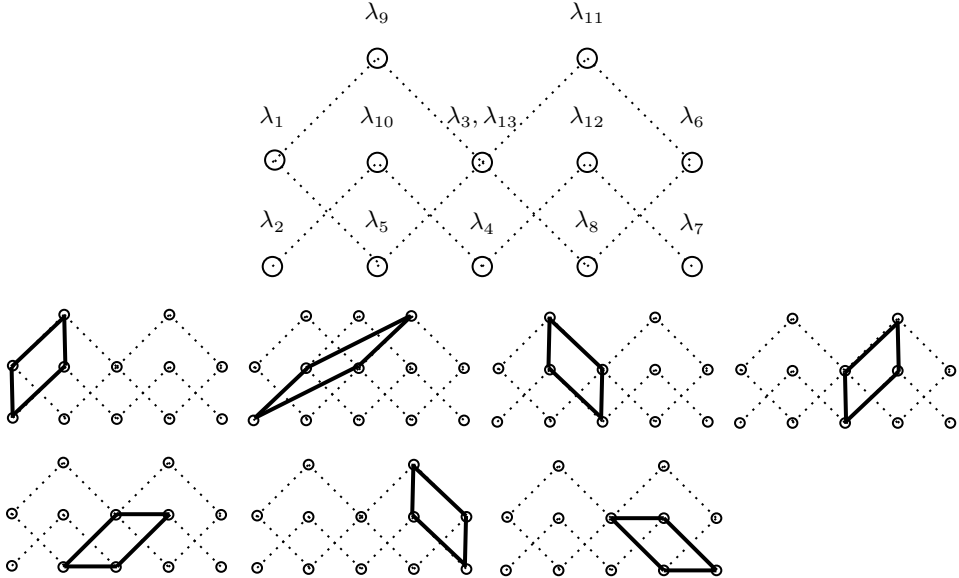}
    \caption{Homology diagram and local symmetries for $7_3$ knot.}
    \label{fig:homology73}
\end{figure}

As the~last isolated example we consider the~$7_3$ knot.
The~quiver from~\cite{SW1711} reads
\begin{equation}
\label{eq:c73}
  C = \left[
\begin{array}{ccccccccccccc}
 2 & 0 & 3 & 2 & 1 & 5 & 4 & 3 & 3 & 2 & 5 & 4 & 3 \\
 0 & 0 & 1 & 1 & 0 & 3 & 3 & 2 & 1 & 1 & 3 & 3 & 2 \\
 3 & 1 & 4 & 2 & 2 & 5 & 4 & 4 & 4 & 2 & 5 & 4 & 4 \\
 2 & 1 & 2 & 2 & 1 & 3 & 3 & 3 & 3 & 2 & 3 & 3 & 3 \\
 1 & 0 & 2 & 1 & 1 & 3 & 2 & 2 & 2 & 1 & 3 & 2 & 2 \\
 5 & 3 & 5 & 3 & 3 & 6 & 4 & 4 & 6 & 4 & 6 & 4 & 4 \\
 4 & 3 & 4 & 3 & 2 & 4 & 4 & 3 & 5 & 4 & 5 & 4 & 3 \\
 3 & 2 & 4 & 3 & 2 & 4 & 3 & 3 & 4 & 3 & 5 & 4 & 3 \\
 3 & 1 & 4 & 3 & 2 & 6 & 5 & 4 & 5 & 3 & 6 & 5 & 4 \\
 2 & 1 & 2 & 2 & 1 & 4 & 4 & 3 & 3 & 3 & 4 & 4 & 3 \\
 5 & 3 & 5 & 3 & 3 & 6 & 5 & 5 & 6 & 4 & 7 & 5 & 5 \\
 4 & 3 & 4 & 3 & 2 & 4 & 4 & 4 & 5 & 4 & 5 & 5 & 4 \\
 3 & 2 & 4 & 3 & 2 & 4 & 3 & 3 & 4 & 3 & 5 & 4 & 4 \\
\end{array}
\right],\qquad \qquad \boldsymbol{\lambda} = \left[\begin{array}{c}
    a^6 q^{-4} (-t)^2\\
    a^4 q^{-4} \\
    a^6 (-t)^4\\
    a^4 (-t)^2\\
    a^{4} q^{-2} (-t)\\
    a^6 q^4 (-t)^6\\
    a^4 q^4 (-t)^4\\
    a^4 q^2 (-t)^3\\
    a^{8} q^{-2} (-t)^5\\
    a^{6} q^{-2} (-t)^3\\
    a^{8} q^{2} (-t)^7\\
    a^{6} q^{2} (-t)^5\\
    a^{6} (-t)^4\\
    \end{array}\right].
\end{equation}
For~(\ref{eq:c73}) there are seven local symmetries for the~following pairings:
\begin{align*}
 \lambda_1\lambda_{10}&=\lambda_2\lambda_9, &
 \lambda_2\lambda_{11}&=\lambda_3\lambda_{10}, &
 \lambda_3\lambda_{10}&=\lambda_4\lambda_9, &
 \lambda_3\lambda_{12}&=\lambda_4\lambda_{11}, \\
 \lambda_4\lambda_{13}&=\lambda_5\lambda_{12}, &
 \lambda_6\lambda_{12}&=\lambda_7\lambda_{11}, &
 \lambda_7\lambda_{13}&=\lambda_8\lambda_{12}. &
\end{align*}
Their graphical representation, together with the~homology diagram, is given in figure~\ref{fig:homology73}.

\section{\texorpdfstring{$F_K(x,a,q)$}{FK} invariants and knot complement quivers} \label{sec-FK}

In the~last section we broaden our perspective and show that the~equivalence criteria from theorem~\ref{thm:main} can be used to relate quivers that we considered so far to another type of quivers, which in~\cite{Kuch2005} have been associated to $F_K(x,a,q)$ invariants of knot complements, constructed in~\cite{GM1904,Park1909,GGKPS20xx}. In this section we focus on $T_{2,2p+1}$ torus knots and show that for each $p$, a~quiver associated to $F_K(x,a,q)$ invariant is equivalent to a~subquiver of a~quiver for unreduced colored HOMFLY-PT polynomials construced in~\cite{KRSS1707long}. 

Before presenting this relation, let us recall how the~knots-quivers correspondence works in the~unreduced normalization (which we denote by adding a~bar to all quantities) defined for HOMFLY-PT generating functions by
\begin{equation}
    \bar{P}_{K}(x,a,q)=\sum_{r=0}^{\infty}x^r a^{-r}q^r \frac{(a^2;q^2)_r}{(q^2;q^2)_r}P_{r}(a,q).
\end{equation}
The presence of $(a^2;q^2)_r$ in the~numerator in the~summand (relative to the~reduced normalization) implies that the~unreduced quiver matrix $\bar{C}_{IJ}$ can be obtained from the~reduced one (given by $C_{ij}$) by the~following relation~\cite{KRSS1707long}\footnote{In our convention $\alpha_i \leftrightarrow \beta_i$ with respect to~\cite{KRSS1707long}.}:
\begin{equation}
    \sum_{I, J = 1}^{2 m} \bar{C}_{IJ} \bar{d}_I \bar{d}_J = \sum_{i, j = 1}^m \left[C_{ij} \alpha_i \alpha_j + (C_{ij}  + 1)  \beta_i \beta_j \right] + 2 \sum_{i \leq j} C_{ij} \alpha_i \beta_j + 2\sum_{i < j} (C_{ij} + 1) \alpha_i \beta_j,
\end{equation}
where $\alpha_i$ and $\beta_i$ are the~new summation indices for the~quiver motivic generating series. They are related to the~summation indices of the~reduced normalization by $d_i = \alpha_i + \beta_i$ and $\bar{d}_I$ can be thought of as the~entries of a~vector
\begin{equation}
    \boldsymbol{\bar{d}} = (\alpha_1, \alpha_2, \dots, \alpha_m, \beta_1, \beta_2, \dots, \beta_m). 
\end{equation}
Then the~unreduced quiver matrix takes the~form of a~$2 m \times 2 m$ block matrix 
\begin{equation}\label{UnredQuiverMat}
    \bar{C} = \left[\begin{array}{l|r}
        C \; & \; C \\
        \hline
        C & C
    \end{array}\right]
    +
    \left[\begin{array}{l|r}
        0 \; & \; 0 \\
        \hline
        0 & 1
    \end{array}\right]
    +
    \left[\begin{array}{c | c}
        \; 0 \; & \; \theta \\
        \hline
        \theta^T & \; 0 
    \end{array}\right]
    \begin{array}{l}
\} \alpha \\
\} \beta \\
\end{array}
\end{equation}
where 1 and 0 are the~matrices with only ones or zeros respectively, and the~matrix $\theta$ is defined as
\begin{equation}\label{ThetaMat}
    \theta_{ij} = \begin{cases}
    0, & j \geq i \\
    1, & j < i
    \end{cases}
    \quad {\rm with} \quad i, \, j = 1, 2, \dots, m.
\end{equation}
Note that going from $d_i$ to $\alpha_i$ and $\beta_i$ can be understood as an~example of splitting. It follows from the~fact that switching between the~reduced and unreduced normalization corresponds to multiplication by $a^{-r}q^r (a^2;q^2)_r$. Since $r=\sum_i d_i$, we split all nodes, and $a^{-r}q^r$ enters the~change of variables. The~only difference with splitting presented in section~\ref{sec:global structure} lies in the~ordering. There we put $\alpha_i$ next to $\beta_i$, here we start from all alphas and then write all betas to match the~convention in~\cite{KRSS1707long}.

\subsection{Trefoil knot complement}
Let us focus on the~simplest example of the~trefoil. The~``standard" and knot complement quivers are given by:
\begin{equation}\label{trefoil_matrix_homfly}
\bar{C}_{3_1} = 
\left[
\begin{array}{ccc|ccc}
0 & 1 & 1 & 0 & 2 & 2 \\
1 & 2 & 2 & 1 & 2 & 3 \\
1 & 2 & 3 & 1 & 2 & 3 \\
\hline
0 & 1 & 1 & 1 & 2 & 2 \\
2 & 2 & 2 & 2 & 3 & 3 \\
2 & 3 & 3 & 2 & 3 & 4
\end{array}
\right],
\qquad\qquad
C_{F_{3_1}} = 
\left[
\begin{array}{cccc}
3 & 2 & 3 & 2 \\
2 & 2 & 3 & 2 \\
3 & 3 & 4 & 3 \\
2 & 2 & 3 & 3
\end{array}
\right].
\end{equation}
Let us exchange $x_2 \leftrightarrow x_4$ in $\bar{C}_{3_1}$ and then remove the~first pair of nodes (interestingly, they look like the~redundant pair of nodes~\cite{EKL1910}, but they have a~different change of variables). After relabeling its vertices to $(x'_1,x'_2,x'_3,x'_4)$, we permute it into $(x'_3,x'_2,x'_4,x'_1)$. This gives:
\begin{equation}\label{trefoil_quiver_transform}
\begin{array}{c}
x_1 \\
x_2 \\
x_3 \\
x_4 \\
x_5 \\
x_6
\end{array}
\left[
\begin{array}{ccc|ccc}
0 & 1 & 1 & 0 & 2 & 2 \\
1 & 2 & 2 & 1 & 2 & 3 \\
1 & 2 & 3 & 1 & 2 & 3 \\
\hline
0 & 1 & 1 & 1 & 2 & 2 \\
2 & 2 & 2 & 2 & 3 & 3 \\
2 & 3 & 3 & 2 & 3 & 4
\end{array}
\right]
\rightsquigarrow
\begin{array}{c}
x_1 \\
x_4 \\
x_3 \\
x_2 \\
x_5 \\
x_6
\end{array}
\left[
\begin{array}{cccccc}
0 & 0 & 1 & 1 & 2 & 2 \\
0 & 1 & 1 & 1 & 2 & 2 \\
1 & 1 & 3 & 2 & 2 & 3 \\
1 & 1 & 2 & 2 & 2 & 3 \\
2 & 2 & 2 & 2 & 3 & 3 \\
2 & 2 & 3 & 3 & 3 & 4
\end{array}
\right]
\rightsquigarrow
\begin{array}{c}
x'_1 \\
x'_2 \\
x'_3 \\
x'_4
\end{array}
\left[
\begin{array}{cccccc}
3 & 2 & 2 & 3 \\
2 & 2 & 2 & 3 \\
2 & 2 & 3 & 3 \\
3 & 3 & 3 & 4
\end{array}
\right]
\rightsquigarrow
\begin{array}{c}
x'_3 \\
x'_2 \\
x'_4 \\
x'_1
\end{array}
\left[
\begin{array}{cccc}
3 & 2 & 3 & 2 \\
2 & 2 & 3 & 2 \\
3 & 3 & 4 & 3 \\
2 & 2 & 3 & 3
\end{array}
\right].
\end{equation}
After framing by $-3$, the~rightmost quiver in~(\ref{trefoil_quiver_transform}) agrees with the~quiver associated to the~trefoil complement in~\cite{Kuch2005}.
We can also illustrate this relation at the~level of formulas. The~$F_K$~invariant reads~\cite{GGKPS20xx}
\begin{equation}\label{eq:F31}
\begin{split}
F_{3_{1}}(x,a,q)&=\sum_{k=0}^{\infty}x^{k} q^{2k}\frac{(x;q^{-2})_{k}(a^2q^{-2};q^2)_{k}}{(q^2;q^2)_{k}}\\
&=\sum_{k=0}^{\infty}x^{2k} q^{3k}\frac{(x^{-1};q^{2})_{k}(a^2q^{-2};q^2)_{k}}{(q^2;q^2)_{k}}(-q)^{-k^2}.
\end{split}
\end{equation}
On the~other hand, the~unreduced HOMFLY-PT generating function is given by~\cite{FGS1205}
\begin{equation}\label{eq:P31unreduced}
\begin{split}
    \bar{P}_{3_1}(x,a,q)&=\sum_{r=0}^{\infty}x^ra^rq^{-r} \frac{(a^2;q^2)_r}{(q^2;q^2)_r}\sum_{k=0}^{r} \left[\begin{array}{c}
    r\\
    k
    \end{array}\right] q^{2k(r+1)} (a^2q^{-2};q^2)_k\\
    &=\sum_{k=0}^{\infty}x^{k}a^{k}q^{k} \frac{(a^2;q^2)_k (a^2q^{-2};q^2)_k} {(q^2;q^2)_k} q^{2k^2} \sum_{l=0}^{\infty}x^{l}a^{l}q^{-l}\frac{(a^2 q^{2k};q^2)_l } {(q^2;q^2)_l } q^{2kl}.
\end{split}
\end{equation}    
Comparing~\eqref{eq:P31unreduced} with~\eqref{eq:F31}, we can see that the~structure of $q$-Pochhammers indexed by $k$ is exactly the~same. The~net difference $-3k^2$ in $q$ powers corresponds to the~framing change, whereas all powers linear in $k$ enter the~change of variables and do not interfere with the~general structure. Finally, the~whole sum over $l=r-k$ contributes to the~removed pair of nodes.

\subsection{Cinquefoil knot complement}

For the~knot $5_1$, the~two quivers are given by
\begin{equation}
\bar{C}_{5_1} = 
\left[
\begin{array}{ccccc|ccccc}
0 & 1 & 1 & 3 & 3 & 0 & 2 & 2 & 4 & 4 \\
1 & 2 & 2 & 3 & 3 & 1 & 2 & 3 & 4 & 4 \\
1 & 2 & 3 & 4 & 4 & 1 & 2 & 3 & 5 & 5 \\
3 & 3 & 4 & 4 & 4 & 3 & 3 & 4 & 4 & 5 \\
3 & 3 & 4 & 4 & 5 & 3 & 3 & 4 & 4 & 5 \\
\hline
0 & 1 & 1 & 3 & 3 & 1 & 2 & 2 & 4 & 4 \\
2 & 2 & 2 & 3 & 3 & 2 & 3 & 3 & 4 & 4 \\
2 & 3 & 3 & 4 & 4 & 2 & 3 & 4 & 5 & 5 \\
4 & 4 & 5 & 4 & 4 & 4 & 4 & 5 & 5 & 5 \\
4 & 4 & 5 & 5 & 5 & 4 & 4 & 5 & 5 & 6
\end{array}
\right],
\qquad\qquad
C_{F_{5_1}} = \left[
\begin{array}{cccccccc}
5 & 4 & 5 & 4 & 4 & 4 & 5 & 4 \\
4 & 4 & 5 & 4 & 3 & 3 & 5 & 4 \\
5 & 5 & 6 & 5 & 4 & 4 & 5 & 5 \\
4 & 4 & 5 & 5 & 3 & 3 & 4 & 4 \\
4 & 3 & 4 & 3 & 3 & 2 & 3 & 2 \\
4 & 3 & 4 & 3 & 2 & 2 & 3 & 2 \\
5 & 5 & 5 & 4 & 3 & 3 & 4 & 3 \\
4 & 4 & 5 & 4 & 2 & 2 & 3 & 3
\end{array}
\right]
\end{equation}
We repeat similar steps as in the~trefoil case, exchanging $x_2 \leftrightarrow x_6$ in $\bar{C}_{5_1}$ and permuting 
\begin{equation}
(x'_1,x'_2,x'_3,x'_4,x'_5,x'_6,x'_7,x'_8) \mapsto (x'_7,x'_2,x'_8,x'_3,x'_5,x'_4,x'_6,x'_1)
\end{equation}
to obtain:
\begin{equation}\label{cinquefoil_matrix_gm}
\begin{array}{c}
x_1 \\
x_2 \\
x_3 \\
x_4 \\
x_5 \\
x_6 \\
x_7 \\
x_8 \\
x_9 \\
x_{10}
\end{array}
\left[
\begin{array}{ccccc|ccccc}
0 & 1 & 1 & 3 & 3 & 0 & 2 & 2 & 4 & 4 \\
1 & 2 & 2 & 3 & 3 & 1 & 2 & 3 & 4 & 4 \\
1 & 2 & 3 & 4 & 4 & 1 & 2 & 3 & 5 & 5 \\
3 & 3 & 4 & 4 & 4 & 3 & 3 & 4 & 4 & 5 \\
3 & 3 & 4 & 4 & 5 & 3 & 3 & 4 & 4 & 5 \\
\hline
0 & 1 & 1 & 3 & 3 & 1 & 2 & 2 & 4 & 4 \\
2 & 2 & 2 & 3 & 3 & 2 & 3 & 3 & 4 & 4 \\
2 & 3 & 3 & 4 & 4 & 2 & 3 & 4 & 5 & 5 \\
4 & 4 & 5 & 4 & 4 & 4 & 4 & 5 & 5 & 5 \\
4 & 4 & 5 & 5 & 5 & 4 & 4 & 5 & 5 & 6
\end{array}
\right]
\rightsquigarrow
\begin{array}{c}
x'_1 \\
x'_2 \\
x'_3 \\
x'_4 \\
x'_5 \\
x'_6 \\
x'_7 \\
x'_8
\end{array}
\left[
\begin{array}{cccccccc}
3 & 4 & 4 & 2 & 2 & 3 & 5 & 5 \\
4 & 4 & 4 & 3 & 3 & 4 & 4 & 5 \\
4 & 4 & 5 & 3 & 3 & 4 & 4 & 5 \\
2 & 3 & 3 & 2 & 2 & 3 & 4 & 4 \\
2 & 3 & 3 & 2 & 3 & 3 & 4 & 4 \\
3 & 4 & 4 & 3 & 3 & 4 & 5 & 5 \\
5 & 4 & 4 & 4 & 4 & 5 & 5 & 5 \\
5 & 5 & 5 & 4 & 4 & 5 & 5 & 6
\end{array}
\right]
\rightsquigarrow
\begin{array}{c}
x'_7 \\
x'_2 \\
x'_8 \\
x'_3 \\
x'_5 \\
x'_4 \\
x'_6 \\
x'_1
\end{array}
\left[
\begin{array}{cccccccc}
{5} & 4 & 5 & 4 & 4 & 4 & 5 & \myboxy{5} \\
4 & {4} & 5 & 4 & 3 & 3 & \myboxy{4} & 4
\\
5 & 5 & 6 & 5 & 4 & 4 & 5 & 5 \\
4 & 4 & 5 & 5 & 3 & 3 & 4 & 4 \\
4 & 3 & 4 & 3 & 3 & 2 & 3 & 2 \\
4 & 3 & 4 & 3 & 2 & 2 & 3 & 2 \\
5 & \myboxy{4} & 5 & 4 & 3 & 3 & {4} & 3 \\
\myboxy{5} & 4 & 5 & 4 & 2 & 2 & 3 & {3}
\end{array}
\right]
\end{equation}
If we now subtract the~result from $C_{F_{5_1}}$, we get
\begin{equation}
\left[
\begin{array}{cccc|cccc}
0 & 0 & 0 & 0 & 0 & 0 & 0 & 1 \\
0 & 0 & 0 & 0 & 0 & 0 & -1 & 0 \\
0 & 0 & 0 & 0 & 0 & 0 & 0 & 0 \\
0 & 0 & 0 & 0 & 0 & 0 & 0 & 0 \\
\hline
0 & 0 & 0 & 0 & 0 & 0 & 0 & 0 \\
0 & 0 & 0 & 0 & 0 & 0 & 0 & 0 \\
0 & -1 & 0 & 0 & 0 & 0 & 0 & 0 \\
1 & 0 & 0 & 0 & 0 & 0 & 0 & 0
\end{array}
\right]
\end{equation}
The two quivers would agree if we swap $C_{2,7}\leftrightarrow C_{1,8}$ in the~rightmost matrix of~(\ref{cinquefoil_matrix_gm}).
Fortunately, it turns out to be an~example of the~quiver equivalence from theorem~\ref{thm:main}, so the~relation between two kinds of quivers holds.

\subsection{General \texorpdfstring{$T_{2,2p+1}$}{T(2,2p+1)} knot complement}

We now compare the~two recursive formulas for $T_{2,2p+1}$ torus knots.
Starting with the~``standard" quiver defined using  unreduced colored HOMFLY-PT polynomials $\bar{C}_{T_{2,2p+1}}$,
we propose an~algorithm of transforming it into a~quiver $C_{F_{T_{2,2p+1}}}$ associated to the~respective knot complement:
\begin{enumerate}
    \item Label the~vertices of $\bar{C}_{T_{2,2p+1}}$ upside down as $x_1,\dots,x_{4p+2}$.
    \item Exchange $x_2 \leftrightarrow x_{2p+2}$.
    \item Remove the~first two nodes $(x_1,x_{2p+2})$ with the~smallest number of self-loops.
    \item The~resulting diagonal is of the~form:
       $(3,\dots,2p+1,\  2,3,\dots,2p+1,2p+2)$.
    Relabel these entries as $(x'_1,\dots,x'_{4p})$.
    \item Permute the~$x'_i$'s:
        $(x'_1,\dots,x'_{4p}) \mapsto (x'_{4p-1},\dots, x'_1)$
    in order to get the~diagonal:\\
        $(2p+1, 2p, 2p+2, 2p+1,\ \dots \ ,\ 3, 2, 4, 3)$.
    Such permutation is fixed uniquely for each~$p$. For example, $p = 1,2,3$ leads respectively to:
    \begin{equation}
    \begin{aligned}
        (x'_1,\dots, x'_4) \mapsto & \ (x'_4, x'_2, x'_3, x'_1),\\
        (x'_1,\dots, x'_8) \mapsto & \ (x'_7,x'_2,x'_8,x'_3,x'_5,x'_4,x'_6,x'_1), \\
        (x'_1, \dots, x'_{12}) \mapsto & \ (x'_{11}, x'_4, x'_{12}, x'_5, x'_9, x'_2, x'_{10}, x'_3, x'_7, x'_6, x'_8, x'_1). \\
    \end{aligned}
    \end{equation}
 \end{enumerate}

\begin{figure}[h!]
\centering
\input{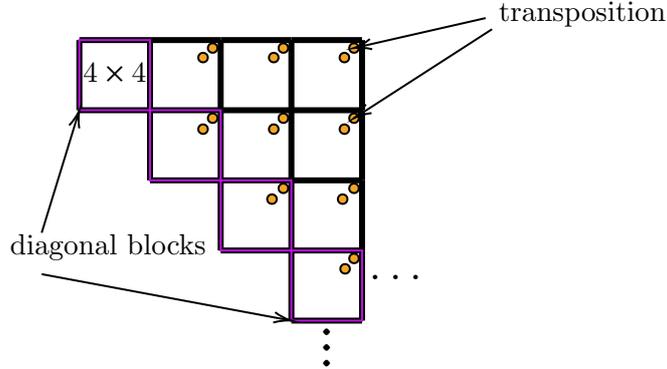}
\caption{The block structure and transpositions that relate the~``standard" sub-quiver based on unreduced HOMFLY-PT polynomials for $T_{2,2p+1}$ torus knots, to the~knot complement quiver  (only the~upper part is shown, since it is symmetric).}
\label{fig:upper_triangular}
\end{figure}

After these steps, we compare the~resulting quiver matrix to $C_{F_{T_{2,2p+1}}}$.
It turns out that the~results \emph{almost} agree, up to transpositions of certain non-diagonal entries, indicated in figure~\ref{fig:upper_triangular}. 
Each block in this figure has size $4\times 4$: the~diagonal blocks represent framed knot complement quivers for the~trefoil, while the~off-diagonal part differs from them by a~transposition of elements, each time appearing in the~top-right corner of each upper-diagonal block, and extending to lower diagonal blocks by symmetry.
This suggests that the~two formulas agree, up to the~quiver equivalence relation. Another argument comes from the~fact that transforming the~quiver from reduced to unreduced normalization corresponds to splitting all nodes, which (as discussed in section~\ref{sec:global structure}) can be done in many ways, all of which lead to equivalent quivers. 

We checked that transpositions depicted in figure~\ref{fig:upper_triangular} are indeed symmetries for $T_{2,2p+1}$ torus knots up to $p=3$. We conjecture that it is always the~case, which means that in the~equivalence class of quivers corresponding to the~$T_{2,2p+1}$ torus knot in the~unreduced normalization there exists a~representative such that the~knot complement quiver is its subquiver.

\section*{Acknowledgements}
 
We thank Tobias Ekholm, Angus Gruen, Sergei Gukov, Pietro Longhi, Sunghyuk Park, and Marko Sto\v{s}i\'c for insightful discussions and comments on the~manuscript. The~work of J.J. was supported by the~Polish National Science Centre (NCN) grant 2016/23/D/ST2/03125. P.K. is supported by the~Polish Ministry of Education and Science through its programme Mobility Plus (decision no.~1667/MOB/V/2017/0).
The~work of H.L., D.N. and P.S. is supported by the~TEAM programme of the~Foundation for Polish Science co-financed by the~European Union under the~European Regional Development Fund (POIR.04.04.00-00-5C55/17-00).

\newpage
\appendix

\section{Equivalent quivers for knots \texorpdfstring{$5_2$}{52} and \texorpdfstring{$7_1$}{71}}\label{appendix_matrices}

In this appendix we present equivalent quivers that we found for knots  $5_2$ and $7_1$. Quiver matrices given below correspond to appropriate vertices in the~permutohedra graphs, as indicated by their labels; the~same labeling is used in the~attached Mathematica file.

\subsection*{\texorpdfstring{$5_2$}{52} knot}

\begin{figure}[h!]
    \centering
    \includegraphics[scale=0.35]{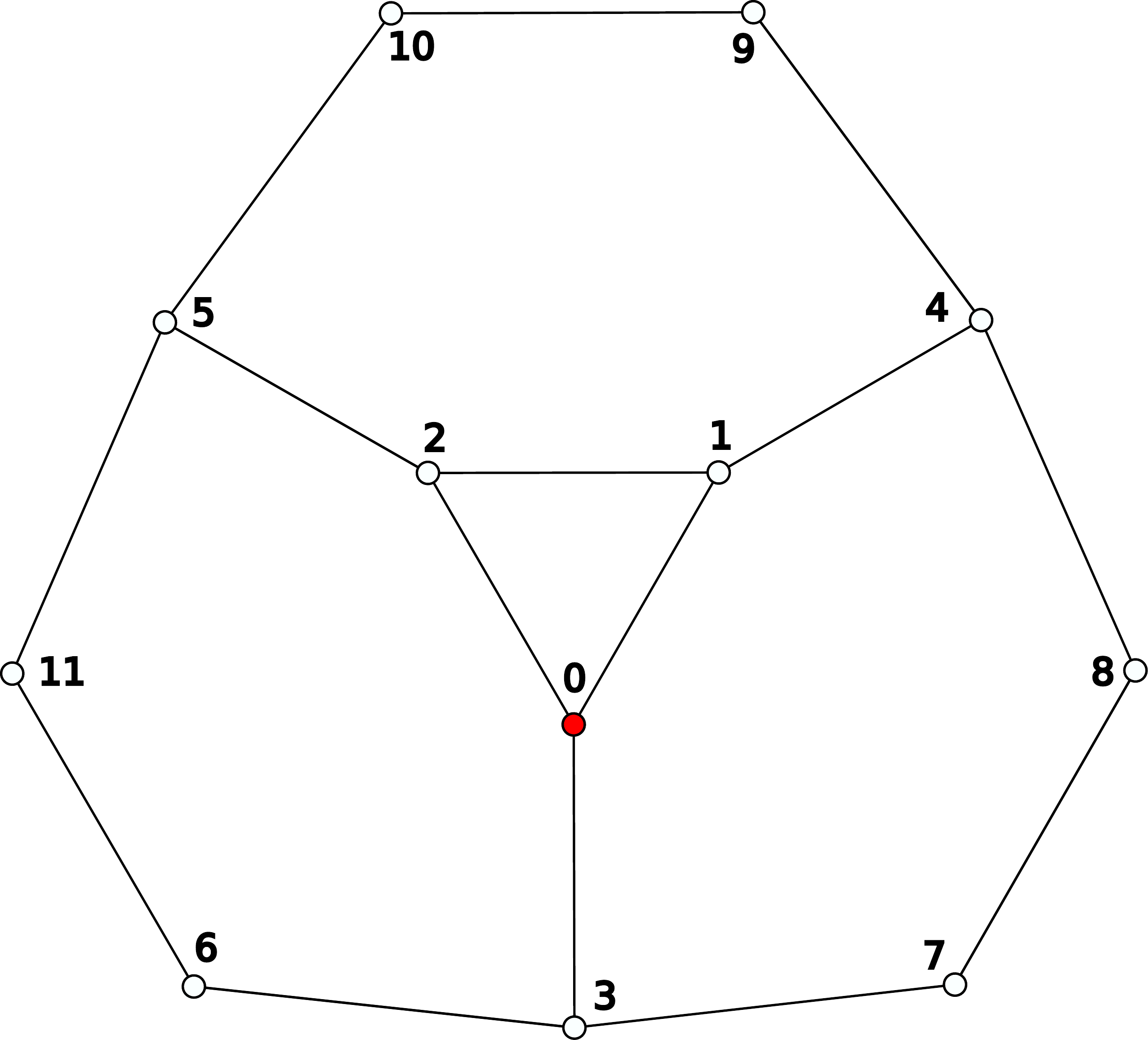}
\begin{scriptsize}
\begin{tabular}{c c c c c c}
\\ \\
    $C_0=\left[
\begin{array}{ccccccc}
 2 & 1 & 2 & 1 & 2 & 1 & 2 \\
 1 & 0 & 1 & 0 & 2 & 0 & 1 \\
 2 & 1 & 3 & 1 & 3 & 2 & 3 \\
 1 & 0 & 1 & 1 & 2 & 1 & 2 \\
 2 & 2 & 3 & 2 & 4 & 3 & 4 \\
 1 & 0 & 2 & 1 & 3 & 2 & 3 \\
 2 & 1 & 3 & 2 & 4 & 3 & 5 \\
\end{array}
\right]$
     &
         $C_1=\left[
\begin{array}{ccccccc}
 2 & 1 & 2 & 1 & 2 & 2 & 2 \\
 1 & 0 & 1 & 0 & 1 & 0 & 1 \\
 2 & 1 & 3 & 1 & 3 & 2 & 3 \\
 1 & 0 & 1 & 1 & 2 & 1 & 2 \\
 2 & 1 & 3 & 2 & 4 & 3 & 4 \\
 2 & 0 & 2 & 1 & 3 & 2 & 3 \\
 2 & 1 & 3 & 2 & 4 & 3 & 5 \\
\end{array}
\right]$
&
$C_2=\left[
\begin{array}{ccccccc}
 2 & 1 & 2 & 1 & 2 & 1 & 2 \\
 1 & 0 & 1 & 0 & 1 & 0 & 1 \\
 2 & 1 & 3 & 2 & 3 & 2 & 3 \\
 1 & 0 & 2 & 1 & 2 & 1 & 2 \\
 2 & 1 & 3 & 2 & 4 & 3 & 4 \\
 1 & 0 & 2 & 1 & 3 & 2 & 3 \\
 2 & 1 & 3 & 2 & 4 & 3 & 5 \\
\end{array}
\right]$
&
$C_3=\left[
\begin{array}{ccccccc}
 2 & 1 & 2 & 1 & 2 & 1 & 2 \\
 1 & 0 & 1 & 0 & 2 & 0 & 2 \\
 2 & 1 & 3 & 1 & 3 & 1 & 3 \\
 1 & 0 & 1 & 1 & 2 & 1 & 2 \\
 2 & 2 & 3 & 2 & 4 & 3 & 4 \\
 1 & 0 & 1 & 1 & 3 & 2 & 3 \\
 2 & 2 & 3 & 2 & 4 & 3 & 5 \\
\end{array}
\right]
$
     \\ \\
     $C_4=\left[
\begin{array}{ccccccc}
 2 & 1 & 2 & 1 & 2 & 2 & 3 \\
 1 & 0 & 1 & 0 & 1 & 0 & 1 \\
 2 & 1 & 3 & 1 & 2 & 2 & 3 \\
 1 & 0 & 1 & 1 & 2 & 1 & 2 \\
 2 & 1 & 2 & 2 & 4 & 3 & 4 \\
 2 & 0 & 2 & 1 & 3 & 2 & 3 \\
 3 & 1 & 3 & 2 & 4 & 3 & 5 \\
\end{array}
\right]
$
 & 
     $C_5=\left[
\begin{array}{ccccccc}
 2 & 1 & 2 & 1 & 2 & 1 & 2 \\
 1 & 0 & 1 & 0 & 1 & 0 & 1 \\
 2 & 1 & 3 & 2 & 3 & 2 & 3 \\
 1 & 0 & 2 & 1 & 2 & 1 & 3 \\
 2 & 1 & 3 & 2 & 4 & 2 & 4 \\
 1 & 0 & 2 & 1 & 2 & 2 & 3 \\
 2 & 1 & 3 & 3 & 4 & 3 & 5 \\
\end{array}
\right]
$
 & 
     $C_6=\left[
\begin{array}{ccccccc}
 2 & 1 & 2 & 1 & 2 & 1 & 2 \\
 1 & 0 & 1 & 0 & 2 & 0 & 2 \\
 2 & 1 & 3 & 1 & 3 & 1 & 3 \\
 1 & 0 & 1 & 1 & 2 & 1 & 3 \\
 2 & 2 & 3 & 2 & 4 & 2 & 4 \\
 1 & 0 & 1 & 1 & 2 & 2 & 3 \\
 2 & 2 & 3 & 3 & 4 & 3 & 5 \\
\end{array}
\right]
$
& 
     $C_7=\left[
\begin{array}{ccccccc}
 2 & 1 & 2 & 1 & 2 & 1 & 3 \\
 1 & 0 & 1 & 0 & 2 & 0 & 2 \\
 2 & 1 & 3 & 1 & 2 & 1 & 3 \\
 1 & 0 & 1 & 1 & 2 & 1 & 2 \\
 2 & 2 & 2 & 2 & 4 & 3 & 4 \\
 1 & 0 & 1 & 1 & 3 & 2 & 3 \\
 3 & 2 & 3 & 2 & 4 & 3 & 5 \\
\end{array}
\right]
$
\\
\\
     $C_8=\left[
\begin{array}{ccccccc}
 2 & 1 & 2 & 1 & 2 & 2 & 3 \\
 1 & 0 & 1 & 0 & 1 & 0 & 2 \\
 2 & 1 & 3 & 1 & 2 & 1 & 3 \\
 1 & 0 & 1 & 1 & 2 & 1 & 2 \\
 2 & 1 & 2 & 2 & 4 & 3 & 4 \\
 2 & 0 & 1 & 1 & 3 & 2 & 3 \\
 3 & 2 & 3 & 2 & 4 & 3 & 5 \\
\end{array}
\right]
$
&
     $C_9=\left[
\begin{array}{ccccccc}
 2 & 1 & 2 & 1 & 2 & 2 & 3 \\
 1 & 0 & 1 & 0 & 1 & 0 & 1 \\
 2 & 1 & 3 & 1 & 2 & 2 & 3 \\
 1 & 0 & 1 & 1 & 2 & 1 & 3 \\
 2 & 1 & 2 & 2 & 4 & 2 & 4 \\
 2 & 0 & 2 & 1 & 2 & 2 & 3 \\
 3 & 1 & 3 & 3 & 4 & 3 & 5 \\
\end{array}
\right]
$
&
     $C_{10}=\left[
\begin{array}{ccccccc}
 2 & 1 & 2 & 1 & 2 & 1 & 3 \\
 1 & 0 & 1 & 0 & 1 & 0 & 1 \\
 2 & 1 & 3 & 2 & 2 & 2 & 3 \\
 1 & 0 & 2 & 1 & 2 & 1 & 3 \\
 2 & 1 & 2 & 2 & 4 & 2 & 4 \\
 1 & 0 & 2 & 1 & 2 & 2 & 3 \\
 3 & 1 & 3 & 3 & 4 & 3 & 5 \\
\end{array}
\right]
$
&
    $ C_{11}=\left[
\begin{array}{ccccccc}
 2 & 1 & 2 & 1 & 2 & 1 & 3 \\
 1 & 0 & 1 & 0 & 1 & 0 & 1 \\
 2 & 1 & 3 & 2 & 2 & 2 & 3 \\
 1 & 0 & 2 & 1 & 2 & 1 & 3 \\
 2 & 1 & 2 & 2 & 4 & 2 & 4 \\
 1 & 0 & 2 & 1 & 2 & 2 & 3 \\
 3 & 1 & 3 & 3 & 4 & 3 & 5 \\
\end{array}
\right]
$
\end{tabular}
\end{scriptsize}
    \label{fig:52_graph_matrices}
\end{figure}

\newpage

\subsection*{\texorpdfstring{$7_1$}{71} knot}

\begin{figure}[h!]
    \centering
    \includegraphics[scale=0.35]{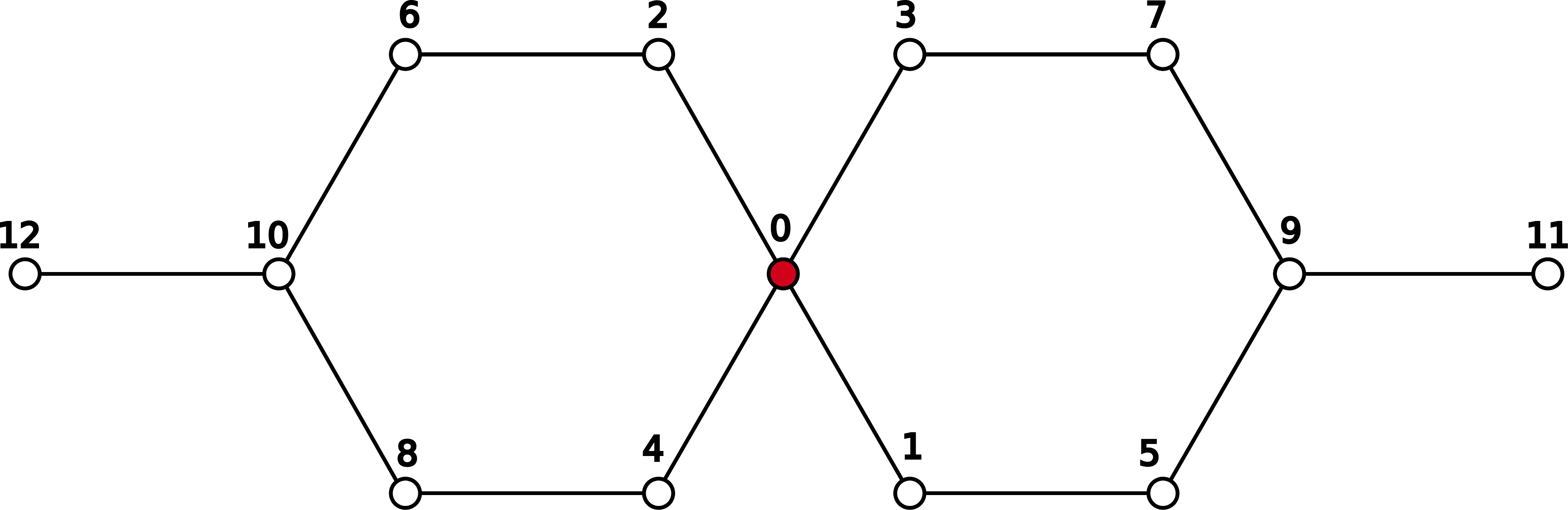}
    \begin{scriptsize}
\begin{tabular}{c c c c c c}
\\ \\
    $C_0=\left[
\begin{array}{ccccccc}
 0 & 1 & 1 & 3 & 3 & 5 & 5 \\
 1 & 2 & 2 & 3 & 3 & 5 & 5 \\
 1 & 2 & 3 & 4 & 4 & 6 & 6 \\
 3 & 3 & 4 & 4 & 4 & 5 & 5 \\
 3 & 3 & 4 & 4 & 5 & 6 & 6 \\
 5 & 5 & 6 & 5 & 6 & 6 & 6 \\
 5 & 5 & 6 & 5 & 6 & 6 & 7 \\
\end{array}
\right]$ 
&
$C_1=\left[
\begin{array}{ccccccc}
 0 & 1 & 1 & 3 & 2 & 5 & 5 \\
 1 & 2 & 3 & 3 & 3 & 5 & 5 \\
 1 & 3 & 3 & 4 & 4 & 6 & 6 \\
 3 & 3 & 4 & 4 & 4 & 5 & 5 \\
 2 & 3 & 4 & 4 & 5 & 6 & 6 \\
 5 & 5 & 6 & 5 & 6 & 6 & 6 \\
 5 & 5 & 6 & 5 & 6 & 6 & 7 \\
\end{array}
\right]$ 
&
$C_2=\left[
\begin{array}{ccccccc}
 0 & 1 & 1 & 3 & 3 & 5 & 5 \\
 1 & 2 & 2 & 3 & 4 & 5 & 5 \\
 1 & 2 & 3 & 3 & 4 & 6 & 6 \\
 3 & 3 & 3 & 4 & 4 & 5 & 5 \\
 3 & 4 & 4 & 4 & 5 & 6 & 6 \\
 5 & 5 & 6 & 5 & 6 & 6 & 6 \\
 5 & 5 & 6 & 5 & 6 & 6 & 7 \\
\end{array}
\right]
$
&
$C_3=\left[
\begin{array}{ccccccc}
 0 & 1 & 1 & 3 & 3 & 5 & 5 \\
 1 & 2 & 2 & 3 & 3 & 5 & 4 \\
 1 & 2 & 3 & 4 & 4 & 6 & 6 \\
 3 & 3 & 4 & 4 & 5 & 5 & 5 \\
 3 & 3 & 4 & 5 & 5 & 6 & 6 \\
 5 & 5 & 6 & 5 & 6 & 6 & 6 \\
 5 & 4 & 6 & 5 & 6 & 6 & 7 \\
\end{array}
\right]
$
\\ \\
$C_4=\left[
\begin{array}{ccccccc}
 0 & 1 & 1 & 3 & 3 & 5 & 5 \\
 1 & 2 & 2 & 3 & 3 & 5 & 5 \\
 1 & 2 & 3 & 4 & 4 & 6 & 6 \\
 3 & 3 & 4 & 4 & 4 & 5 & 6 \\
 3 & 3 & 4 & 4 & 5 & 5 & 6 \\
 5 & 5 & 6 & 5 & 5 & 6 & 6 \\
 5 & 5 & 6 & 6 & 6 & 6 & 7 \\
\end{array}
\right]$ 
&
$C_5=\left[
\begin{array}{ccccccc}
 0 & 1 & 1 & 3 & 2 & 5 & 4 \\
 1 & 2 & 3 & 3 & 3 & 5 & 5 \\
 1 & 3 & 3 & 5 & 4 & 6 & 6 \\
 3 & 3 & 5 & 4 & 4 & 5 & 5 \\
 2 & 3 & 4 & 4 & 5 & 6 & 6 \\
 5 & 5 & 6 & 5 & 6 & 6 & 6 \\
 4 & 5 & 6 & 5 & 6 & 6 & 7 \\
\end{array}
\right]$
&
$C_6=\left[
\begin{array}{ccccccc}
 0 & 1 & 1 & 3 & 3 & 5 & 5 \\
 1 & 2 & 2 & 3 & 4 & 5 & 6 \\
 1 & 2 & 3 & 3 & 4 & 5 & 6 \\
 3 & 3 & 3 & 4 & 4 & 5 & 5 \\
 3 & 4 & 4 & 4 & 5 & 6 & 6 \\
 5 & 5 & 5 & 5 & 6 & 6 & 6 \\
 5 & 6 & 6 & 5 & 6 & 6 & 7 \\
\end{array}
\right]$
&
$C_7=\left[
\begin{array}{ccccccc}
 0 & 1 & 1 & 3 & 3 & 5 & 4 \\
 1 & 2 & 2 & 3 & 3 & 5 & 4 \\
 1 & 2 & 3 & 5 & 4 & 6 & 6 \\
 3 & 3 & 5 & 4 & 5 & 5 & 5 \\
 3 & 3 & 4 & 5 & 5 & 6 & 6 \\
 5 & 5 & 6 & 5 & 6 & 6 & 6 \\
 4 & 4 & 6 & 5 & 6 & 6 & 7 \\
\end{array}
\right]$
\\ \\
$C_8=\left[
\begin{array}{ccccccc}
 0 & 1 & 1 & 3 & 3 & 5 & 5 \\
 1 & 2 & 2 & 3 & 3 & 5 & 6 \\
 1 & 2 & 3 & 4 & 4 & 5 & 6 \\
 3 & 3 & 4 & 4 & 4 & 5 & 6 \\
 3 & 3 & 4 & 4 & 5 & 5 & 6 \\
 5 & 5 & 5 & 5 & 5 & 6 & 6 \\
 5 & 6 & 6 & 6 & 6 & 6 & 7 \\
\end{array}
\right]$
&
$C_9=\left[
\begin{array}{ccccccc}
 0 & 1 & 1 & 3 & 3 & 5 & 4 \\
 1 & 2 & 2 & 3 & 3 & 5 & 6 \\
 1 & 2 & 3 & 5 & 4 & 4 & 6 \\
 3 & 3 & 5 & 4 & 5 & 5 & 5 \\
 3 & 3 & 4 & 5 & 5 & 6 & 6 \\
 5 & 5 & 4 & 5 & 6 & 6 & 6 \\
 4 & 6 & 6 & 5 & 6 & 6 & 7 \\
\end{array}
\right]$
&
$C_{10}=\left[
\begin{array}{ccccccc}
 0 & 1 & 1 & 3 & 3 & 5 & 5 \\
 1 & 2 & 2 & 3 & 4 & 5 & 6 \\
 1 & 2 & 3 & 3 & 4 & 5 & 6 \\
 3 & 3 & 3 & 4 & 4 & 5 & 6 \\
 3 & 4 & 4 & 4 & 5 & 5 & 6 \\
 5 & 5 & 5 & 5 & 5 & 6 & 6 \\
 5 & 6 & 6 & 6 & 6 & 6 & 7 \\
\end{array}
\right]$
&
$C_{11}=\left[
\begin{array}{ccccccc}
 0 & 1 & 1 & 3 & 3 & 5 & 3 \\
 1 & 2 & 2 & 3 & 4 & 5 & 6 \\
 1 & 2 & 3 & 5 & 4 & 4 & 6 \\
 3 & 3 & 5 & 4 & 5 & 5 & 5 \\
 3 & 4 & 4 & 5 & 5 & 6 & 6 \\
 5 & 5 & 4 & 5 & 6 & 6 & 6 \\
 3 & 6 & 6 & 5 & 6 & 6 & 7 \\
\end{array}
\right]$
\\ \\
$C_{12}=\left[
\begin{array}{ccccccc}
 0 & 1 & 1 & 3 & 3 & 5 & 5 \\
 1 & 2 & 2 & 3 & 4 & 5 & 6 \\
 1 & 2 & 3 & 3 & 4 & 5 & 6 \\
 3 & 3 & 3 & 4 & 4 & 5 & 5 \\
 3 & 4 & 4 & 4 & 5 & 6 & 6 \\
 5 & 5 & 5 & 5 & 6 & 6 & 6 \\
 5 & 6 & 6 & 5 & 6 & 6 & 7 \\
\end{array}
\right]$
\end{tabular}
\end{scriptsize}
    \label{fig:71_graph_matrices}
\end{figure}

\newpage 

\section{Quiver matrices for twist knots}   \label{app-twist}

In this appendix we provide quiver matrices for twist knots, which were found in~\cite{KRSS1707long}. 
Interestingly, for each of the~two families of twist knots $TK_{2|p|+2}$ and $TK_{2p+1}$, such a~matrix can be presented in a~universal way.

The quiver matrix for $TK_{2|p|+2}$ twist knot found in~\cite{KRSS1707long} takes form
\begin{equation}
C^{TK_{2|p|+2}}=\left[\begin{array}{ccccccc}
F_{0} & F & F & F & \cdots & F & F\\
F^{T} & D_{1} & R_{1} & R_{1} & \cdots & R_{1} & R_{1}\\
F^{T} & R_{1}^{T} & D_{2} & R_{2} & \cdots & R_{2} & R_{2}\\
F^{T} & R_{1}^{T} & R_{2}^{T} & D_{3} & \cdots & R_{3} & R_{3}\\
\vdots & \vdots & \vdots & \vdots & \ddots & \vdots & \vdots\\
F^{T} & R_{1}^{T} & R_{2}^{T} & R_{3}^{T} & \cdots & D_{|p|-1} & R_{|p|-1}\\
F^{T} & R_{1}^{T} & R_{2}^{T} & R_{3}^{T} & \cdots & R_{|p|-1}^{T} & D_{|p|}
\end{array}\right],
\end{equation}
where
\begin{equation}
F_{0}=\left[0\right],\qquad\qquad F=\left[\begin{array}{cccc}
0 & -1 & 0 & -1\end{array}\right],
\end{equation}
and
\begin{equation}
D_{k}=\left[\begin{array}{cccc}
2k & 2k-2 & 2k-1 & 2k-3\\
2k-2 & 2k-3 & 2k-2 & 2k-4\\
2k-1 & 2k-2 & 2k-1 & 2k-3\\
2k-3 & 2k-4 & 2k-3 & 2k-4
\end{array}\right], \qquad
R_{k}=\left[\begin{array}{cccc}
2k & 2k-2 & 2k-1 & 2k-3\\
2k-1 & 2k-3 & 2k-2 & 2k-4\\
2k & 2k-1 & 2k-1 & 2k-3\\
2k-2 & 2k-3 & 2k-2 & 2k-4
\end{array}\right].
\end{equation}
The element $F_0$ represents a~zig-zag of length 1, i.e. a~single homology generator, while the~diagonal blocks  $D_k$ represent  diamonds (up to a~permutation of homology generators and an~overall shift). The~identification with $\lambda_i$'s in figure~\ref{fig:diamonds_orientation} is as follows:
    \begin{equation}
        \begin{array}{c|cccc}
 & \textcolor[rgb]{0.82,0.01,0.11}{\lambda_{4r-2}} & \textcolor[rgb]{0.82,0.01,0.11}{\lambda_{4r-1}} & \textcolor[rgb]{0.82,0.01,0.11}{\lambda_{4r}} & \textcolor[rgb]{0.82,0.01,0.11}{\lambda_{4r+1}}\\
 \hline
\textcolor[rgb]{0.82,0.01,0.11}{\lambda_{4r-2}} & 2r & 2r-2 & 2r-1 & 2r-3\\
\textcolor[rgb]{0.82,0.01,0.11}{\lambda_{4r-1}} & 2r-2 & 2r-3 & 2r-2 & 2r-4\\
\textcolor[rgb]{0.82,0.01,0.11}{\lambda_{4r}} & 2r-1 & 2r-2 & 2r-1 & 2r-3\\
\textcolor[rgb]{0.82,0.01,0.11}{\lambda_{4r+1}} & 2r-3 & 2r-4 & 2r-3 & 2r-4
\end{array} \qquad
\begin{array}{c|cccc}
 & \textcolor[rgb]{0.82,0.01,0.11}{\lambda_{4r-2}} & \textcolor[rgb]{0.82,0.01,0.11}{\lambda_{4r-1}} & \textcolor[rgb]{0.82,0.01,0.11}{\lambda_{4r}} & \textcolor[rgb]{0.82,0.01,0.11}{\lambda_{4r+1}}\\
 \hline
\textcolor[rgb]{0.82,0.01,0.11}{\lambda_{4r'-2}} & 2r & 2r-1 & 2r & 2r-2\\
\textcolor[rgb]{0.82,0.01,0.11}{\lambda_{4r'-1}} & 2r-2 & 2r-3 & 2r-1 & 2r-3\\
\textcolor[rgb]{0.82,0.01,0.11}{\lambda_{4r'}} & 2r-1 & 2r-2 & 2r-1 & 2r-2\\
\textcolor[rgb]{0.82,0.01,0.11}{\lambda_{4r'+1}} & 2r-3 & 2r-4 & 2r-3 & 2r-4
\end{array}
\end{equation}
This means that $D_k$ encodes interactions of nodes nodes within one diamond, while
$R_k$ encodes interactions of nodes from two diamonds labelled by $r,r'$.

Quiver matrices for $TK_{2p+1}$ twist knots found in~\cite{KRSS1707long} read
\begin{equation}
C^{TK_{2p+1}}=\left[\begin{array}{ccccccc}
D_{1} & R_{1} & R_{1} & R_{1} & \cdots & R_{1} & R_{1}\\
R_{1}^{T} & D_{2} & R_{2} & R_{2} & \cdots & R_{2} & R_{2}\\
R_{1}^{T} & R_{2}^{T} & D_{3} & R_{3} & \cdots & R_{3} & R_{3}\\
R_{1}^{T} & R_{2}^{T} & R_{3}^{T} & D_{4} & \cdots & R_{4} & R_{4}\\
\vdots & \vdots & \vdots & \vdots & \ddots & \vdots & \vdots\\
R_{1}^{T} & R_{2}^{T} & R_{3}^{T} & R_{4}^{T} & \cdots & D_{p-1} & R_{p-1}\\
R_{1}^{T} & R_{2}^{T} & R_{3}^{T} & R_{4}^{T} & \cdots & R_{p-1}^{T} & D_{p}
\end{array}\right],
\label{C-TK2p1}
\end{equation}
where the~block elements in the~first row and column are
\begin{equation}
D_{1}=\left[\begin{array}{ccc}
2 & 1 & 2\\
1 & 0 & 1\\
2 & 1 & 3
\end{array}\right],\qquad \qquad 
R_{1}=\left[\begin{array}{cccc}
1 & 2 & 1 & 2\\
0 & 2 & 0 & 1\\
1 & 3 & 2 & 3
\end{array}\right],    \label{D1R1-TK2p1}
\end{equation}
and all other elements, for $k>1$, take the~form
\begin{equation}
D_{k}=\left[\begin{array}{cccc}
2k-3 & 2k-2 & 2k-3 & 2k-2\\
2k-2 & 2k & 2k-1 & 2k\\
2k-3 & 2k-1 & 2k-2 & 2k-1\\
2k-2 & 2k & 2k-1 & 2k+1
\end{array}\right], \qquad \quad
R_{k}=\left[\begin{array}{cccc}
2k-3 & 2k-2 & 2k-3 & 2k-2\\
2k-1 & 2k & 2k-1 & 2k  \\
2k-2 & 2k & 2k-2 & 2k-1\\
2k-1 & 2k+1 & 2k & 2k+1
\end{array}\right].
\end{equation}
In this case $D_1$ represents a~zig-zag of the~same form as for the~trefoil knot, and $D_k$ (for $k>1$) represent diamonds (up to a~permutation of homology generators and an~overall constant shift).

\newpage

\bibliography{refs}

\def\cprime{$'$}
\providecommand{\href}[2]{#2}\begingroup\raggedright\begin{thebibliography}{10}

\bibitem{KRSS1707short}
P.~Kucharski, M.~Reineke, M.~Stosic, and P.~Sulkowski, {\it {BPS states, knots
  and quivers}},  {\em Phys. Rev.} {\bf D96} (2017), no.~12 121902,
  [\href{http://arxiv.org/abs/1707.02991}{{\tt arXiv:1707.02991}}].

\bibitem{KRSS1707long}
P.~Kucharski, M.~Reineke, M.~Stosic, and P.~Sulkowski, {\it Knots-quivers
  correspondence},  {\em Adv. Theor. Math. Phys.} {\bf 23} (2019), no.~7
  1849--1902, [\href{http://arxiv.org/abs/1707.04017}{{\tt arXiv:1707.04017}}].

\bibitem{EKL1811}
T.~Ekholm, P.~Kucharski, and P.~Longhi, {\it Physics and geometry of
  knots-quivers correspondence},  {\em Commun. Math. Phys.} {\bf 379} (2020),
  no.~2 361--415, [\href{http://arxiv.org/abs/1811.03110}{{\tt
  arXiv:1811.03110}}].

\bibitem{EKL1910}
T.~Ekholm, P.~Kucharski, and P.~Longhi, {\it Multi-cover skeins, quivers, and
  3d $\mathcal{N}=2$ dualities},  {\em JHEP} {\bf 02} (2020) 018,
  [\href{http://arxiv.org/abs/1910.06193}{{\tt arXiv:1910.06193}}].

\bibitem{PS1811}
M.~Panfil and P.~Sulkowski, {\it {Topological strings, strips and quivers}},
  {\em JHEP} {\bf 01} (2019) 124, [\href{http://arxiv.org/abs/1811.03556}{{\tt
  arXiv:1811.03556}}].

\bibitem{Kimura:2020qns}
T.~Kimura, M.~Panfil, Y.~Sugimoto, and P.~Sulkowski, {\it {Branes, quivers and
  wave-functions}},  {\em SciPost Phys.} {\bf 10} (2021) 51,
  [\href{http://arxiv.org/abs/2011.06783}{{\tt arXiv:2011.06783}}].

\bibitem{Bousseau:2020fus}
P.~Bousseau, A.~Brini, and M.~Van~Garrel, {\it {Stable maps to Looijenga
  pairs}},  \href{http://arxiv.org/abs/2011.08830}{{\tt arXiv:2011.08830}}.

\bibitem{PSS1802}
M.~Panfil, P.~Sulkowski, and M.~Stosic, {\it {Donaldson-Thomas} invariants,
  torus knots, and lattice paths},  {\em Phys. Rev.} {\bf D98} (2018), no.~2
  026022, [\href{http://arxiv.org/abs/1802.04573}{{\tt arXiv:1802.04573}}].

\bibitem{SW1711}
M.~Stosic and P.~Wedrich, {\it {Rational links and DT invariants of quivers}},
  {\em Int. Math. Res. Not.} {\bf rny289} (2019)
  [\href{http://arxiv.org/abs/1711.03333}{{\tt arXiv:1711.03333}}].

\bibitem{SW2004}
M.~Stosic and P.~Wedrich, {\it Tangle addition and the knots-quivers
  correspondence},  \href{http://arxiv.org/abs/2004.10837}{{\tt
  arXiv:2004.10837}}.

\bibitem{Kuch2005}
P.~Kucharski, {\it Quivers for 3-manifolds: the correspondence, {BPS} states,
  and 3d $ \mathcal{N} $ = 2 theories},  {\em JHEP} {\bf 09} (2020) 075,
  [\href{http://arxiv.org/abs/2005.13394}{{\tt arXiv:2005.13394}}].

\bibitem{AFGS1203}
H.~Awata, S.~Gukov, H.~Fuji, and P.~Sulkowski, {\it Volume conjecture: Refined
  and categorified},  {\em Adv. Theor. Math. Phys.} {\bf 16} (2012) 1669--1777,
  [\href{http://arxiv.org/abs/1203.2182}{{\tt arXiv:1203.2182}}].

\bibitem{FGSS1209}
H.~Fuji, S.~Gukov, P.~Sulkowski, and M.~Stosic, {\it {3d analogs of
  Argyres-Douglas theories and knot homologies}},  {\em JHEP} {\bf 01} (2013)
  175, [\href{http://arxiv.org/abs/1209.1416}{{\tt arXiv:1209.1416}}].

\bibitem{Chung:2014qpa}
H.-J. Chung, T.~Dimofte, S.~Gukov, and P.~Sulkowski, {\it {3d-3d Correspondence
  Revisited}},  {\em JHEP} {\bf 04} (2016) 140,
  [\href{http://arxiv.org/abs/1405.3663}{{\tt arXiv:1405.3663}}].

\bibitem{GM1904}
S.~Gukov and C.~Manolescu, {\it A two-variable series for knot complements},
  \href{http://arxiv.org/abs/1904.06057}{{\tt arXiv:1904.06057}}.

\bibitem{Park1909}
S.~Park, {\it Higher rank $\hat{Z}$ and ${F}_{K}$},  {\em SIGMA} {\bf 16}
  (2020) 044, [\href{http://arxiv.org/abs/1909.13002}{{\tt arXiv:1909.13002}}].

\bibitem{GGKPS20xx}
T.~Ekholm, A.~Gruen, S.~Gukov, P.~Kucharski, S.~Park, and P.~Sulkowski, {\it
  {$\hat{Z}$} at large {$N$}: from curve counts to quantum modularity},
  \href{http://arxiv.org/abs/2005.13349}{{\tt arXiv:2005.13349}}.

\bibitem{DGR0505}
N.~M. Dunfield, S.~Gukov, and J.~Rasmussen, {\it The superpolynomial for knot
  homologies},  {\em Experiment. Math.} {\bf 15} (2006), no.~2 129--159,
  [\href{http://arxiv.org/abs/math/0505662}{{\tt math/0505662}}].

\bibitem{GS}
S.~Gukov and M.~Stosic, {\it Homological algebra of knots and {BPS} states},
  {\em Proc. Symp. Pure Math.} {\bf 85} (2012) 125--172,
  [\href{http://arxiv.org/abs/1112.0030}{{\tt arXiv:1112.0030}}].

\bibitem{KS1006}
M.~Kontsevich and Y.~Soibelman, {\it Cohomological {Hall} algebra, exponential
  {Hodge} structures and motivic {Donaldson-Thomas} invariants},  {\em Commun.
  Num. Theor. Phys.} {\bf 5} (2011) 231--352,
  [\href{http://arxiv.org/abs/1006.2706}{{\tt arXiv:1006.2706}}].

\bibitem{Efi12}
A.~I. Efimov, {\it Cohomological {Hall} algebra of a symmetric quiver},  {\em
  Compos. Math.} {\bf 148} (2012), no.~4 1133--1146,
  [\href{http://arxiv.org/abs/1103.2736}{{\tt arXiv:1103.2736}}].

\bibitem{OV9912}
H.~Ooguri and C.~Vafa, {\it {Knot invariants and topological strings}},  {\em
  Nucl. Phys.} {\bf B577} (2000) 419--438,
  [\href{http://arxiv.org/abs/hep-th/9912123}{{\tt hep-th/9912123}}].

\bibitem{Wit92}
E.~Witten, {\it {Chern-Simons gauge theory as a string theory}},  {\em Prog.
  Math.} {\bf 133} (1995) 637--678,
  [\href{http://arxiv.org/abs/hep-th/9207094}{{\tt hep-th/9207094}}].

\bibitem{ES1901}
T.~Ekholm and V.~Shende, {\it Skeins on branes},
  \href{http://arxiv.org/abs/1901.08027}{{\tt arXiv:1901.08027}}.

\bibitem{KS0811}
M.~Kontsevich and Y.~Soibelman, {\it Stability structures, motivic
  {Donaldson-Thomas} invariants and cluster transformations},
  \href{http://arxiv.org/abs/0811.2435}{{\tt arXiv:0811.2435}}.

\bibitem{ziegler_lectures_1995}
G.~M. Ziegler, {\em Lectures on Polytopes}, vol.~152 of {\em Graduate Texts in
  Mathematics}.
\newblock Springer New York, 1995.

\bibitem{aguiar_hopf_2017}
M.~Aguiar and F.~Ardila, {\it Hopf monoids and generalized permutahedra},
  \href{http://arxiv.org/abs/1709.07504}{{\tt arXiv:1709.07504}}.

\bibitem{FGS1205}
H.~Fuji, S.~Gukov, and P.~Sulkowski, {\it {Super-A-polynomial for knots and BPS
  states}},  {\em Nucl. Phys. B} {\bf 867} (2013) 506,
  [\href{http://arxiv.org/abs/1205.1515}{{\tt arXiv:1205.1515}}].

\end{thebibliography}\endgroup
\bibliographystyle{JHEP}


\end{document}